\documentclass[a4paper,onecolumn,10pt,openright,twoside]{book}
\usepackage[dvips]{graphicx}

\usepackage{doc}
\usepackage[utf8]{inputenc}
\usepackage[english]{babel}
\usepackage{color}
\usepackage{fancyhdr}
\usepackage[hang,small,sc,bf,tt,oneline]{caption}
\usepackage{ae,aecompl,aeguill} 
\usepackage{amsmath,amssymb,amsfonts}
\usepackage{wasysym}
\usepackage{hyperref}
\usepackage{hypcap}
\usepackage{pstricks}
\usepackage{makeidx,showidx}
\usepackage[toc,page]{appendix}
\usepackage{multirow,rotating}

\newcommand{\erf}{\mathrm{Erf}}
\newcommand{\tu}[1]{\texttt{#1}}             % typeset for units!
\newcommand{\tn}[1]{\textnormal{#1}}         % abbr for \textnormal
\newcommand{\mr}[1]{\mathrm{#1}}             % abbr for \mathrm
\newcommand{\mb}[1]{\mathbf{#1}}             % abbr for \mathbf
\newcommand{\ics}{\langle\eta\sigma\rangle}  % inclusive cross section symbol
\newcommand{\ie}{{\it i.e. }}
\newcommand{\roots}{{\sqrt{s}}}

\newcommand{\lD}{\lambda_{\tn{D}}}
\newcommand{\lDM}{\lD^{\max}}
\newcommand{\fluxe}{\Phi_{e^+}}

\newcommand{\tpow}[1]{\times 10^{#1}} % 10^{x}
\newcommand{\mass}[1]{ m_{#1} }
\newcommand{\ener}[1]{ \varepsilon_{#1} } 
\newcommand{\kin}[1]{ \tn{T}_{#1} }
\newcommand{\mom}[1]{ p_{#1} }

\newcommand{\dnde}[1]{\frac{dn}{d\ener{#1}}}
\newcommand{\curr}[1]{ \mathcal{J}_{#1} }
\newcommand{\CS}{\tn{CS}} % cross section
\newcommand{\lab}{\tn{Lab}}
\newcommand{\cms}{\tn{CMS}}
\newcommand{\ism}{\tn{ISM}}
\newcommand{\CR}{\tn{CR}}
\newcommand{\bsm}{\tn{BSM}}
\newcommand{\SM}{\tn{SM}} % standard model
\newcommand{\THDM}{\tn{TDHM}} % two higgs doublet model
\newcommand{\dm}{\tn{DM}}
\newcommand{\cdm}{\tn{CDM}}
\newcommand{\ecm}{\tn{ECM}}
\newcommand{\MD}{\tn{MD}} % multiplicity distrib
\newcommand{\nue}{\nu_{e}}
\newcommand{\nuebar}{\bar{\nu}_{e}}
\newcommand{\numu}{\nu_{\mu}}
\newcommand{\numubar}{\bar{\nu}_{\mu}}
\newcommand{\nutau}{\nu_{\tau}}
\newcommand{\nutaubar}{\bar{\nu}_{\tau}}

\newcommand{\fref}[1]{{\small\texttt{\ref{#1}}}}
\newcommand{\citeeq}[1]{{\small\texttt{Equation}}~\fref{#1}}
\newcommand{\citeeqq}[2]{{\small\texttt{Equations}}~\ref{#1} and \fref{#2}}
\newcommand{\citefig}[1]{{\small\texttt{Figure}}~\fref{#1}}
\newcommand{\citefigg}[2]{{\small\texttt{Figure}}~\fref{#1} and \fref{#2}}
\newcommand{\citetab}[1]{{\small\texttt{Table}}~\fref{#1}}
\newcommand{\citecha}[1]{{\small\texttt{Chapter}}~\fref{#1}}
\newcommand{\citeapp}[1]{{\small\texttt{Appendix}}~\fref{#1}}

\newenvironment{fig}{\begin{figure}[tpb]\vspace{1ex}\begin{minipage}{\textwidth}\centering}{\end{minipage}\end{figure}}
\newenvironment{tab}{\begin{table}[bpt]\vspace{1ex}\begin{minipage}{\textwidth}\centering}{\end{minipage} \end{table}}
\newenvironment{prechap}{\begin{center}\begin{minipage}{0.85\textwidth}}{\end{minipage}\end{center}}
\newenvironment{eq}{\begin{eqnarray}}{\end{eqnarray}}

\let\oldsqrt\sqrt
\def\sqrt{\mathpalette\DHLhksqrt}
\def\DHLhksqrt#1#2{%
\setbox0=\hbox{$#1\oldsqrt{#2\,}$}\dimen0=\ht0
\advance\dimen0-0.2\ht0
\setbox2=\hbox{\vrule height\ht0 depth -\dimen0}%
{\box0\lower0.4pt\box2}}

\graphicspath{{./figs/}{./figs/f1/}{./figs/f2/}{./figs/f3/}{./figs/f4/}{./figs/f5/}{./figs/apps/}}

\makeindex

\begin{document}

\sloppy
\frenchspacing

\begin{titlepage}

\begin{center}
	\begin{pspicture}(0,0)(14, 3)
		%\psgrid
   	\rput[mt](0,3.5){\includegraphics[width=2.8cm]{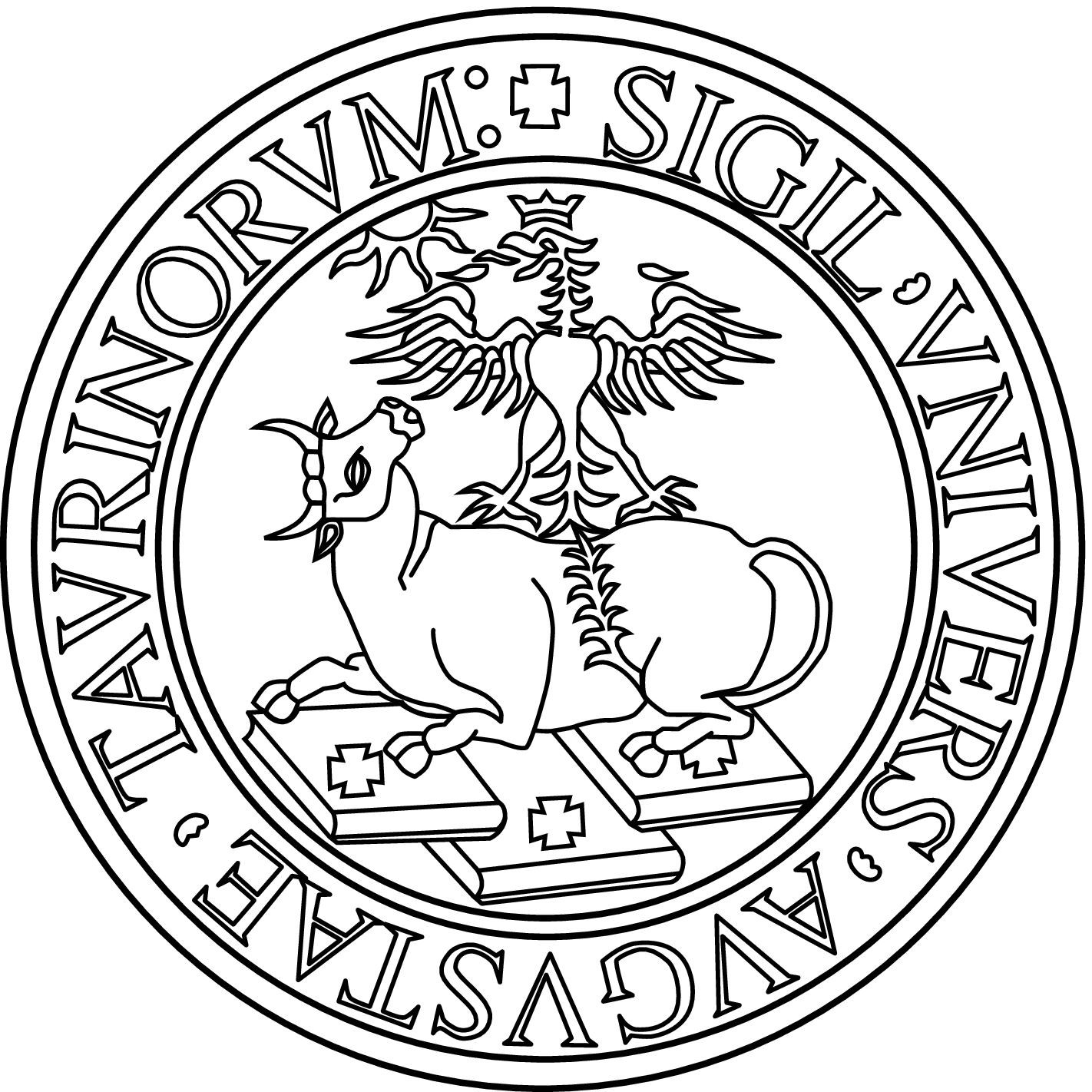}}
	\rput[mt](13.5,3.5){\includegraphics[width=2.8cm]{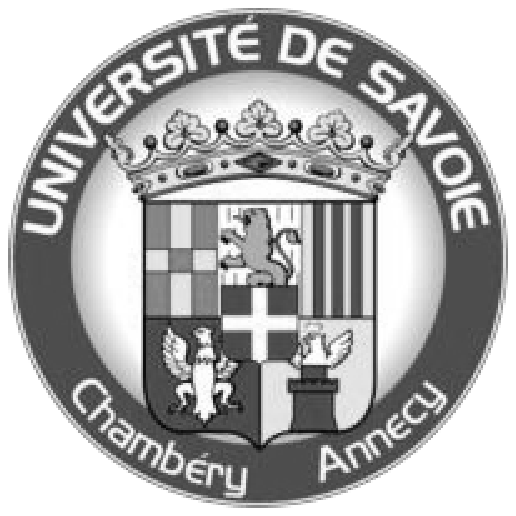}}
		\rput[lt](1.0,3.5){
			\begin{minipage}{0.8\textwidth}
				\centering
				\Large \textsc{Universit\`{a} Degli Studi Di Torino}  \\
				\normalsize \textsc{Facolt\`{a} di Scienze Matematiche Fisiche e Naturali} \\
				\textsc{Dipartimento di Fisica Teorica}\\[1ex]
				\Large \textsc{Universit\'{e} De Savoie}\\ 
				\normalsize \textsc{\`{E}cole Doctorale -- DCT Physique Th\`{e}orique}
			\end{minipage}}
	\end{pspicture}
\end{center}

\vspace{0.5cm}

\begin{center}
	\large \textbf{Dottorato di Ricerca in Fisica ed Astrofisica} \\
	\large \textbf{Ciclo XXI} \\
\end{center}

\vspace{1cm}

\begin{center}
\begin{minipage}[t]{0.9\textwidth}
\begin{center}
\Large{\textbf{\textsc{STUDY OF POSITRONS FROM \mbox{COSMIC RAYS} INTERACTIONS AND COLD DARK MATTER ANNIHILATIONS IN THE GALACTIC ENVIRONMENT}}}
%\Large{\textbf{\textsc{STUDY OF RELATIVISTIC POSITRON COSMIC RAYS PRODUCED IN GALACTIC COLD DARK MATTER ANNIHILATION PROCESSES}}}
\end{center}
\end{minipage}
\end{center}
\vspace{1.5cm}

\begin{center}
TESI PRESENTATA DA:\\
Roberto Alfredo Lineros Rodriguez
\end{center}

\vspace{1cm}

\begin{flushright}
  \begin{tabular}{rcl}
     \textsc{RELATORI} & : & Prof. Nicolao Fornengo, U. Torino.  \\
     & & Prof. Pierre Salati, U. Savoie.\\ \\
     \textsc{CONTRORELATORI} & : & Prof. Joseph Silk, U. Oxford.  \\
     & & Prof. G\"unther Sigl, U. Hamburg.\\ \\
		\textsc{COORDINATORE DEL DOTTORATO} &:& Prof. Stefano Sciuto, U. Torino\\
  \end{tabular}
\end{flushright}

\vspace{1cm}
\begin{center}
  \textbf{Settore Scientifico-Disciplinare di Afferenza: FIS/02, FIS/05}\\
	\textbf{Anni Accademici: 2005-2008}
\end{center}
\setcounter{page}{0}
\end{titlepage}

\begin{titlepage}
	\begin{flushleft}
		\begin{minipage}{50ex}
			\centering
			\textsl{Asking questions, search for clues}\\
			\textsl{The answer's been right in front of you.}\\[3ex]
			``The Answer Lies Within'' -- John Petrucci
		\end{minipage}
	\end{flushleft}

 \vspace{0.55\textheight}
 \resizebox{0.8\textwidth}{!}{
 \begin{tabular}{rl}
  \textbf{Publication type}:& Doctoral thesis.\\[1.5ex]
  \textbf{Author}:& Roberto Alfredo Lineros Rodriguez.\\[1.5ex]
  \multirow{2}{*}{\textbf{Title}:}&Study of positrons from cosmic rays interactions and \\
   & cold dark matter annihilations in the galactic environment.\\[1.5ex]
  \textbf{Department}:& University of Turin and University of Savoie, Theoretical Physics Department.\\[1.5ex]
  \multirow{2}{*}{\textbf{Advisors}:}& Professor Nicolao Fornengo (U. Torino) and \\ 
    & Professor Pierre Salati (U. Savoie).\\[1.5ex]
  \multirow{2}{*}{\textbf{Opponents}:}& Professor Joseph Silk (U. Oxford.) and  \\
    & Professor G\"unther Sigl (U. Hamburg).\\[1.5ex]
  \textbf{Defense}:& December 12, 2008.\\[1.5ex]
  \textbf{Keywords}:& cosmic--rays, dark--matter, secondary positrons.\\[1.5ex]
  \textbf{ArXiv}:&\href{http://arxiv.org/abs/0812.4272}{0812.4272}\\
  \textbf{Last modification}:&18/12/2008\\
 \end{tabular}}
 \begin{center}
  {\small Copyright \textcopyright 2008 -- Roberto Alfredo Lineros Rodriguez}
 \end{center}
 
	\setcounter{page}{0}
\end{titlepage}

\frontmatter
\cleardoublepage
\addcontentsline{toc}{chapter}{Abstract}
\chapter*{Abstract}

% \noindent\tred{Intro:}\\
Positron and Electron Cosmic Rays represent just a fraction of the Cosmic Ray species that arrive to the Earth. 
These particles have the potentiality to only reveal nearby sources because they are affected more by energy loss processes than other types of cosmic--rays.\\

% Context
% \noindent\tred{Context:}\\
Galactic positrons are mainly produced by the interaction of nuclei cosmic rays with the interstellar gas.
Cosmic rays observations in the positron--electron signal reveal the possible presence of an unexpected feature for energies above 10~\tu{GeV}: the positron fraction appears larger in the high--energy range with respect to current theoretical predictions.
% in the positron fraction respect to current theoretical predictions. 
%
During the years, many explanations have been proposed to elucidate this feature. Those are based on new physics, such as dark matter particles annihilations, or aditional standard astrophysics processes.\\

% Aims
% \noindent\tred{Aims:}\\
Cosmic--ray experiments as HEAT, AMS, PAMELA among others, have provided high--statistic positron--electron data for range of energies from $\sim 500\;\tu{MeV}$ to 100~\tu{GeV}. That gives important restrictions in the theoretical constraints for positron--electron cosmic--ray physics.
The contribution of dark--matter to the signal, which would appear as deviations with respect to the standard prediction, and some of the dark--matter properties would be also constrained.
A detailed study of the positron background is important to estimate the capabilities to discriminate a possible new signal present in the experimental data.\\

% Methods
% \noindent\tred{Methods:}\\
The propagation of cosmic--ray in the galactic environment can be treated in many ways. In the case of positrons and electrons, we model the propagation according to the Two--Zone Propagation Model in which sources related to dark--matter annihilation and secondary production have been considered. The positron--electron transport equation is solved by analytical methods taking into account the uncertainties related to the propagation.
In addition, the energy spectra of positrons and electrons are described by nuclear and particle physics, where we study different production models and their uncertainties.\\

% Results
% \noindent\tred{Results:}\\
The feature in the positron fraction, originally seen by the HEAT experiment, can be reproduced in the dark--matter annihilation scenario. As well, we showed the effect of propagation uncertainties on the dark--matter signal and we studied the potentiality to discover a new signal for AMS02 and PAMELA experiment.
The secondary production of positrons was also studied, taking care on the theoretical uncertainties related to nuclear cross section and propagation. The estimations for the positron flux reproduce current available experimental data. 
The positron fraction is calculated on the basis of our results on positron flux and fits performed on electron flux data. We obtain that, depending on the electron flux used, the fraction may sizeably change in the high--energy range stressing more or less the necessity of a ``positron excess'' feature.\\

%the ``positron excess'' feature.\\ 

% Conclusions
% \noindent\tred{Conclusions:}\\
Finally, we give promising results to disentangle a dark--matter component signal from the positron background by studying the uncertainties in the positron--electron propagation.
As well, from the study of secondary positrons, we reproduced the observations and stressed the importance of the electron signal.
Furthermore, PAMELA observations and the forthcoming AMS02 mission will soon allow much better constraints on the cosmic--ray transport parameters, and are likely to drastically reduce theoretical uncertainties.\\

%%%%%%%%%%%%%%%%%%%%%%%%%%%%%%%%%%%%%%%%%%%%%%%%
\pagestyle{fancy}
\fancyhf{}
\fancyhead{}
\fancyfoot[RO,LE]{\thepage}
\fancyhead[RO,LE]{\bfseries ABSTRACT}

%
% 
% 
% 
%  ITALIANO
% 
% 
% 
\cleardoublepage
\addcontentsline{toc}{chapter}{Sintesi}
\chapter*{Sintesi}

%Intro
I raggi cosmici di elettroni e di positroni costituiscono solo una frazione dei raggi cosmici che raggiungono la Terra. Queste particelle sono potenzialmente in grado di rivelare le sorgenti vicine perch\'e sono pi\`u affette da processi di perdita di energia rispetto ad altri tipi di raggi cosmici.\\

%Contesto\\
I positroni galattici sono principalmente prodotti da interazioni dei raggi cosmici nucleare con il gas interstellare. 
Le osservazioni di positroni ed elettroni provenienti da raggi cosmici evidenziano una possibile peculiarit\`a per energie superiori ai 10~\tu{GeV}: la frazione di positroni (\emph{positron fraction}) \`e maggiore rispetto all'attuale predizione teorica in questo intervallo di energie.
Nel corso degli anni, molte spiegazioni sono state proposte per chiarire questa particolarit\`a. Questo \`e alla base di una nuova fisica, come l'annichilazione della materia oscura, o ulteriori processi astrofisici. \\

%Obiettivi\\
Gli esperimenti sui raggi cosmici, tra cui HEAT, AMS e PAMELA, hanno fornito grandi quantit\`a di dati su elettroni e positroni nell'intervallo di energie che va da $\sim500$~\tu{MeV} a 100~\tu{GeV}.
Questo impone importanti restrizioni sulla teoria dei raggi cosmici. Il contributo da materia oscura a questo segnale, che appare come una deviazione rispetto alla previsione standard, ed alcune propriet\`a della materia oscura potrebbero anche essere limitati.
Un dettagliato studio del fondo di positroni \`e importante per stimare la capacit\`a di discriminare un possibile nuovo segnale nei dati sperimentali.\\

%Metodi\\
La propagazione dei raggi cosmici galattici nell'ambiente pu\`o essere trattata in molti modi. 
Nel caso di elettroni e positroni, abbiamo modellizzato la propagazione secondo il Modello di Propagazione a Due--Zone (\emph{Two--Zone Propagation Model}), all'interno del quale sono state considerate le sorgenti relative all'annichilazione ed alla produzione secondaria.
L'equazione per il trasporto di elettroni e positroni \`e stata risolta con metodi analitici, tenendo conto delle incertezze relative alla propagazione.
Inoltre, gli spettri in energia di elettroni e positroni sono stati descritti dalla fisica nucleare e  da quella delle particelle, dove sono stati studiati diversi modelli di produzione e le incertezze di questi.\\

%Risultati\\
La peculiarit\`a nella frazione di positroni, originariamente osservata nell'esperimento HEAT, pu\`o essere riprodotta nel contesto di annichilazione della materia oscura.
Inoltre, abbiamo mostrato gli effetti delle incertezze nella propagazione del segnale a causa di materia oscura ed abbiamo studiato la potenzialit\`a di rivelare un nuovo segnale negli esperimenti PAMELA ed AMS02.
La produzione secondaria di positroni \`e stata anche studiata, in considerazione delle incertezze relative alla sezioni d'urto teoriche e di propagazione.
Le stime per il flusso di positroni riproduce i dati sperimentali disponibili. La frazione di positroni \`e stata calcolata sulla base sia dei nostri risultati nel calcolo del flusso di queste particelle che dei fit dal flusso di dati di elettroni.
In conclusione, abbiamo mostrato che, a seconda del flusso di elettroni utilizzati, la frazione pu\`o variare in modo significativo nell'intervallo di alta energia accentuando una maggiore o minore necessit\`a di un ``eccesso di positroni".\\

%Conclusioni\\
Infine, abbiamo ottenuto risultati promettenti per distinguere un componente del segnale dovuta a materia oscura, in considerazione delle incertezze del fondo nella propagazione.
A sua volta, dallo studio di positroni abbiamo riprodotto le osservazioni e abbiamo sottolineato l'importanza del segnale degli elettroni.
Le osservazioni di PAMELA e la futura missione AMS02 permetter\`a delle migliori restrizioni nei parametri di trasporto dei raggi cosmici, che permetteranno di ridurre le incertezze teoriche.\\

%%%%%%%%%%%%%%%%%%%%%%%%%%%%%%%%%%%%%%%%%%%%%%%%
\pagestyle{fancy}
\fancyhf{}
\fancyhead{}
\fancyfoot[RO,LE]{\thepage}
\fancyhead[RO,LE]{\bfseries SINTESI}

% FRANCES
% 
% 
% 
% 
% 
% 
% 
\cleardoublepage
\addcontentsline{toc}{chapter}{R\'esum\'e}
\chapter*{R\'esum\'e}
%Contexte\\
Les rayons cosmiques de positrons et d'\'electrons ne constituent q'une fraction des rayons cosmiques qui arrivent \`a la Terre. 
Ces particules ont le potentiel de seulement r\'ev\'eler des sources proches parce qu'elles sont affect\'ees par des processus de perte d'\'energie plus que d'autres classes de rayons cosmiques.\\

%Objectifs\\
Les positrons galactiques sont principalement produits par l'interaction de rayons cosmiques nucl\'eaires avec le gaz interstellaire. 
Les observations des rayons cosmiques dans le signal positron--\'electron montrent la possible pr\'esence d'une particularit\'e inattendue aux \'energies au dessus de 10~\tu{GeV}.
La fraction des positrons (\emph{positron fraction}) semble plus haute \`a hautes \'energies que les actuelles pr\'edictions th\'eoriques indiquent.
Pendant des ann\'ees, beaucoup d'explications ont \'et\'e propos\'ees pour \'elucider cette particularit\'e. 
Celles-ci sont bas\'ees sur de la physique nouvelle, comme l'annihilation de mati\`ere fonc\'ee, ou sur des  processus astrophysiques additionnels. \\

%M\'ethodes\\
Des exp\'eriences de rayons cosmiques comme entre autres HEAT, AMS, PAMELA, ont offert une grande quantit\'e de donn\'ees sur les positrons--\'electrons dans l'intervalle entre $\sim500$\;\tu{MeV} et 100~\tu{GeV}.
Ceci impose de grandes restrictions aux limites th\'eoriques pour la physique de rayons cosmiques. La contribution de la mati\`ere fonc\'ee \`a ce signal-ci, qui apparaîtrait comme une d\'eviation de la pr\'evision normale, et quelques propri\'et\'es de la mati\`ere fonc\'ee, seraient aussi restreintes.
Une \'etude d\'etaill\'ee du fond de positrons est importante pour estimer les capacit\'es pour discriminer un possible nouveau signal existant dans les donn\'ees exp\'erimentales.\\

La propagation de rayons cosmiques dans l'atmosph\`ere galactique peut être trait\'e de beaucoup de mani\`eres. 
Dans le cas des positrons et des \'electrons, nous avons mod\'el\'e la propagation en accord avec le Mod\`ele de Propagation \`a Deux Zones (\emph{Two--Zone Propagation Model}) dans lequel les sources concernant l'annihilation et la production secondaire ont \'et\'e consid\'er\'ees.
L'\'equation de transport des positrons--\'electrons a \'et\'e r\'esolue par des m\'ethodes analytiques en prenant en consid\'eration les incertitudes del la propagation.
Aditionnellement, les spectres d'\'energie des positrons et des \'electrons ont \'et\'e d\'ecrits par la physique nucl\'eaire et la physique des particules, dans les quelles on a \'etudi\'e de diff\'erents mod\`eles de production et des incertitudes de ceux-ci. \\

%R\'esultats\\
La particularit\'e dans la fraction de positrons, initiellement vue dans l'exp\'erience HEAT, peut être reproduite dans le sc\'enario de l'annihilation de mati\`ere fonc\'ee.
En outre, nous avons montr\'e les effets des incertitudes de propagation du signal, \'etant donn\'e la mati\`ere fonc\'ee et nous avons \'etudi\'e la possibilit\`e de d\'ecouvrir un nouveau signal dans les exp\'eriences AMS02 et PAMELA.
Le potentiel secondaire de positrons a aussi \'et\'e \'etudi\'ee, consid\'erant des incertitudes th\'eoriques relat\'ees aux sections efficaces et \`a la propagation.
Les estimations pour le flux de positrons reproduisent les donn\'ees exp\'erimentales disponibles. La fraction de positrons est calcul\'ee sur base de nos r\'esultats dans le calcul des flux de ceux-ci et des ajustements faits sur les donn\'ees de flux d'\'electrons.
Nous avons obtenu que, suivant le flux d'\'electrons utilis\'e, la fraction peut consid\'erablement changer dans l'intervalle des hautes \'energies accentuant plus ou moins la particularit\'e d'un ``exc\`es de positrons''. \\ 

%Conclusions\\
Finalement, nous donnons des r\'esultats prometteurs pour d\'ecouvrir un composant du signal, \'etant donn\'e mati\`ere fonc\'ee, du fond en \'etudiant des incertitudes de la propagation.
À son tour, \`a partir de l'\'etude de positrons secondaires, nous avons reproduit les observations et nous avons insist\'e sur l'importance du signal d'\'electrons.
En outre, les observations de PAMELA et de la future mission AMS02 permettront d'obtenir de meilleures restrictions aux param\`etres de transport des rayons cosmiques, et seront prometteuses pour r\'eduire des incertitudes th\'eoriques. \\

%%%%%%%%%%%%%%%%%%%%%%%%%%%%%%%%%%%%%%%%%%%%%%%%
\pagestyle{fancy}
\fancyhf{}
\fancyhead{}
\fancyfoot[RO,LE]{\thepage}
\fancyhead[RO,LE]{\bfseries R\'ESUM\'E}

% CASTELLANO
% 
% 
% 
% 
% 
% 
\cleardoublepage
\addcontentsline{toc}{chapter}{Resumen}
\chapter*{Resumen}

%Contexto\\
Los rayos c\'osmicos de positrones y electrones constituyen solo una fracci\'on de los rayos c\'osmicos que llegan a la Tierra.
Estas part\'{\i}culas tienen la potencialidad de revelar solo fuentes cercanas porque son afectadas en mayor medida por procesos de p\'erdida de energ\'{\i}a que otras clases de rayos c\'osmicos .\\

%Objetivos\\
Positrones gal\'acticos son principalmente producidos por la interacci\'on de rayos c\'osmicos nucleares con el gas interestelar. 
Las observaciones de rayos c\'osmicos en la se\~nal positron--electr\'on muestran la posible presencia de una inesperada peculiaridad en el rango por sobre de los 10~\tu{GeV}:
la fracci\'on de positrones (\emph{positron fraction}) se muestra m\'as grande en el rango de altas energ\'{\i}as que las actuales prediciones te\'oricas. 
Durante a\~nos, muchas explicaciones han sido propuestas para dilucidar esta peculiaridad.
\'Estos est\'an basados en nueva f\'{\i}sica, como la aniquilaci\'on de mat\'eria oscura, o en procesos astrof\'{\i}sicos adicionales.\\

%M\'etodos\\
Experimentos de rayos c\'osmicos como HEAT, AMS, PAMELA entre otros, han brindado gran volumen de datos de positrones y electrones en el rango que va desde $\sim500$~\tu{MeV} hasta los 100~\tu{GeV}. 
\'Esto impone grandes restricciones a los l\'{\i}mites teoricos para la f\'{\i}sica de rayos c\'osmicos. La contribuci\'on de la materia oscura a esta se\~nal, que aparecer\'{\i}a como desviaciones respecto a la predicci\'on est\'andar, y algunas propiedades de la materia oscura ser\'{\i}an tambi\'en restringidas. 
Un estudio detallado del fondo de positrones resulta importante para estimar las capacidades de discriminar una posible nueva se\~nal existente en los datos experimentales.\\

La propagaci\'on de rayos c\'osmicos en el ambiente gal\'actico puede ser tratado de muchas formas. 
En el caso de positrones y electrones, hemos modelado la propagaci\'on de acuerdo al Modelo de Propagaci\'on a Dos Zonas (\emph{Two--Zone Propagation Model}) dentro del cual las fuentes relacionadas a la aniquilaci\'on y a la producci\'on secundaria han sido consideradas. 
La ecuaci\'on de transporte de positrones y electrones ha sido resuelta por m\'etodos anal\'{\i}ticos tomando en cuenta las incertezas relacionadas a la propagaci\'on. 
En a\~nadidura, los espectros de energ\'{\i}a de positrones y electrones han sido descritos por f\'{\i}sica nuclear y de part\'{\i}culas, donde han sido estudiados distintos modelos de producci\'on y las incertezas de \'estos.\\ 

%Resultados\\
La peculiaridad en la fracci\'on de positrones, originalmente observados en el experimento HEAT, puede ser reproducida en el escenario de la aniquilaci\'on de materia oscura. 
Adem\'as, hemos mostrado los efectos de las incertezas en la propagaci\'on de la se\~nal debida a materia oscura y hemos estudiado la potencialidad de descubrir una nueva se\~nal en los experimentos AMS02 y PAMELA. 
La producci\'on secundaria de positrones tambi\'en ha sido estudiada, considerando las incertezas te\'oricas relacionadas a las secciones eficaces y a la propagaci\'on. 
Las estimaciones para el flujo de positrones reproducen los datos experimentales disponibles. La fraci\'on de positrones es calculada en base a nuestros resultados en el c\'alculo de los flujos de \'estos y de ajustes hechos sobre datos de flujos de electrones. 
Obtuvimos que, dependiendo del flujo de electrones usado, la fracci\'on puede cambiar considerablemente en el rango de altas energ\'{\i}as acentuando en mayor o menor medida la peculiaridad de un ``exceso de positrones''.\\

%Conclusiones\\
Finalmente, damos prometedores resultados para desentra\~nar una componente en la se\~nal debida a materia oscura del fondo, estudiando las incertezas en la propagaci\'on. 
A su vez, del estudio de positrones secundarios, hemos reproducido las observaciones y hecho hincapi\'e en la importancia de la se\~nal de electrones. 
Adem\'as, las observaciones de PAMELA y de la futura misi\'on AMS02 permitir\'an mejores restriciones a los parametros de transporte de rayos c\'osmicos, y ser\'an prometedoras para reducir las incertezas te\'oricas.\\

%%%%%%%%%%%%%%%%%%%%%%%%%%%%%%%%%%%%%%%%%%%%%%%%
\pagestyle{fancy}
\fancyhf{}
\fancyhead{}
\fancyfoot[RO,LE]{\thepage}
\fancyhead[RO,LE]{\bfseries RESUMEN}

\cleardoublepage
\pagestyle{fancy}
\fancyhf{}
\fancyhead{}
\fancyfoot[RO,LE]{\thepage}
\fancyhead[LE]{\bfseries\leftmark}
\fancyhead[RO]{\bfseries\rightmark}

\tableofcontents
\mainmatter
\cleardoublepage 
\pagestyle{fancy}
\fancyhf{}
\fancyhead{}
\fancyfoot[RO,LE]{\thepage}
\fancyhead[LE]{\bfseries\leftmark}
\fancyhead[RO]{\bfseries\rightmark}

\cleardoublepage 
\chapter{Introduction}
\label{cha1}
\begin{prechap}
In the last decades, the astrophysical and cosmological evidence of Dark Matter and Dark Energy have created a revolution in the field of fundamental physics. Some models, which are candidates to replace the current Standard Model of particles physics, predict dark--matter particle candidates.
In similar way, the cosmic--ray physics have been stimulated with observations that show a much active universe that was thought. As well, cosmic--ray are promising proves in the understanding of the local environment. Those are genuine samples of the matter composition of the galaxy. 
In addition, the antimatter cosmic--rays component, which are less abundant than matter cosmic--rays component, gives crucial clues regarding to the non--standard contribution to the cosmic--ray signal.\\
%\tred{
% \dropping{2}{T}he quest for the identification of Dark--Matter, together with the comprehension of the nature of dark energy, is one of the most challenging problems in the understanding of the physical world.
%
% Many efforts in DM detection have been done in the last decade, and major breakthroughs are expected in the following years from the underground facilities and antimatter searches in space.
% Cosmic Ray part.
% On the other hand, Cosmic Rays observations are promising proves in the understanding of the local environment. Also, those are genuine samples of the matter composition of the galaxy. In addition, the antimatter CR component gives crucial clues regarding to the non--standard contribution to the CR signal.\\}
\end{prechap}

\section{Dark Matter}
The definition of Dark Matter (DM) comes out from the fact that this kind of matter does not emit or absorb electromagnetic radiation at any wavelength, its gravitational interactions dominate on scales from tiny dwarf galaxies, to large spirals such as the Milky Way, to clusters of galaxies, to the largest scales until now observed.\\

Spiral galaxies support the hypothesis of DM in which star dynamics suggest the presence of additional mass which is not detectable using electromagnetic radiation detection. Moving to larger scales, such as galaxy clusters, the evidence of DM comes from different experimental methods, such as gravitational lensing, X--ray gas temperatures and the motion of cluster member galaxies. In general, depending on the scale to which we look at, different methods of measuring directly or indirectly the presence of DM are employed (\citefig{f:dm_evids}).\\

As well, the Dark Matter component in the Standard Cosmological Model is fundamental to explain many of the current observations. The combined WMAP, SNIa and galaxy clusters measurements suggest the presence of DM on cosmological scales. Its contribution to the Universe energy content is~\cite{Spergel:2007,PDBook}
\begin{eq}
 \Omega_{\tn{DM}} \simeq 0.23 \;,
\end{eq}
which corresponds to the 23\% of the total energy density. A satisfactory description of most cosmological observations is obtained by the so called ``$\Lambda$CDM model'', which comes out from the best fit of the combined data analysis.\\

\begin{fig}
 \resizebox{\textwidth}{!}{\includegraphics[height=30ex]{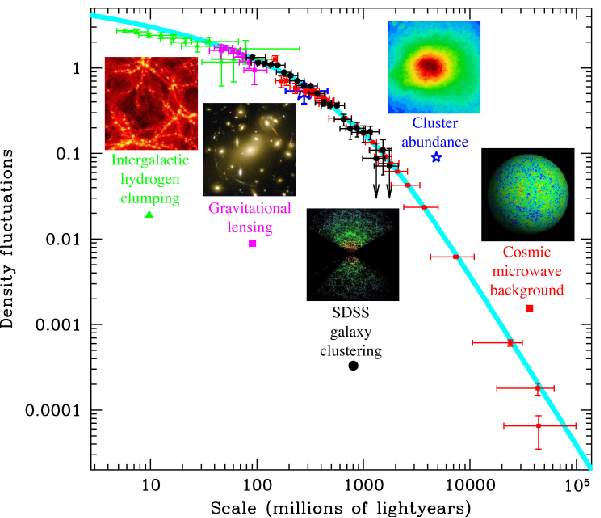} \hspace{3ex} \includegraphics[height=30ex]{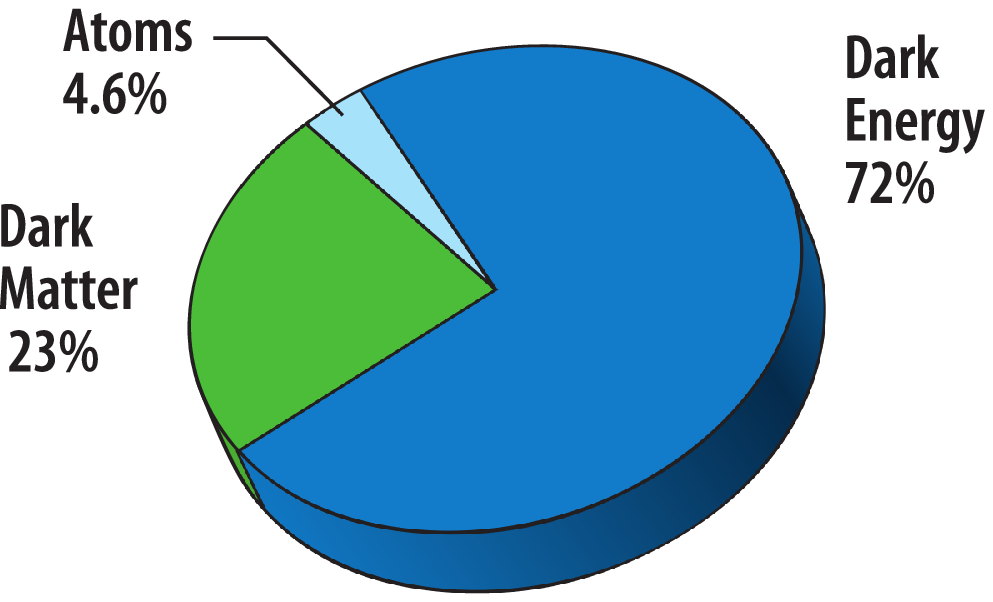}}
\caption{\label{f:dm_evids} Left panel: Indirect evidences of Dark Matter present at all scale. Right panel: Pie chart of composition of the observable Universe, Dark Matter is the 23\% at current time~\cite{Hinshaw:2008kr}.}
\end{fig}

\subsection{Dark Matter particle candidates}
Potentially the only indication compatible with cosmological measurements is that dark matter is composed of non-baryonic, neutral and weakly interacting particles. In the literature several candidates were proposed, the most relevant are:

\paragraph{Standard Model Neutrinos:}
The existence of a relic sea of neutrinos in number only slightly below that of relic photons that constitute the CMB, is a generic prediction of the standard hot Big Bang model. Their contribution to the matter density of the Universe is:
\begin{eq}
 \Omega_{\nu} = \frac{\sum_i \mass{i}}{93.14 h^2\;\tu{eV}}
\end{eq}
The requirement that $\Omega_{\nu} \lesssim \Omega_{m}\simeq 0.3$ imposes stringent limits on their masses. Indeed dark matter particles with a large velocity dispersion such as that of the neutrinos affect the evolution of the cosmological perturbations. This leads to a top--down scenarios which is not supported by the present observations since the galaxies seems older than cluster.\\

\paragraph{Heavy Neutrinos:} The allowed mass range is bounded from below from the Lee-Weinberg limit~\cite{Lee:1977}, which was $\mass{\nu} > 2\;\tu{GeV}$ at the time of their work. 
The current limits have been updated to $\mass{\nu}\ge1.3-4.2\;\tu{GeV}$ for Dirac neutrinos and $\mass{\nu}\ge4.9-13\;\tu{GeV}$ for Majorana ones \cite{Kolb:1986}.
A more stringent bound comes from colliders: neutrinos lighter than 45~\tu{GeV} are excluded by the total decay width of the $Z$ boson. 
For very heavy neutrinos the Yukawa coupling to the Higgs boson would be so strong that perturbative calculation become non-reliable~\cite{Dolgov:2002} and partial waves unitarity has to be imposed~\cite{Griest:1990}, leading to a stringent mass upper bound, $\mass{\nu} \lesssim 3\;\tu{TeV}$.
In the allowed mass range the cosmological properties can be interesting, however their interactions are quite strong, therefore they are mainly excluded by direct detection bounds and lead to a low relic abundance.\\

\paragraph{Sterile Neutrinos:} Those are similar to Standard Model neutrinos, but without Standard Model interaction apart from mixing~\cite{Dodelson:1994}. Stringent cosmological and astrophysical constraints on the sterile neutrinos come from the analysis of their cosmological abundance and the study of their decay products~\cite{Abazajian:2001}.\\

\paragraph{Axions:} Originally those have been introduced to explain the so called strong CP  violation problem~\cite{Peccei:1977,Peccei:1977b}. From different searches, it is expected that axions are very light and extremely weakly interacting with ordinary particles, which implies that they were not in thermal equilibrium in the early Universe. 
The calculation of the relic density is uncertain, nevertheless it is possible to find some range where axions satisfy the present constraints and represent a possible DM candidate~\cite{Rosenberg:2000}.\\

\paragraph{Supersymmetric Particles:} In models conserving $R$-parity the lightest supersymmetric particle (LSP), such as the neutralino, sneutrino, gravitino or axino could provide the right amount of dark matter density in the Universe. Which particle is the LSP, it depends on the supersymmetric models and on how supersymmetry is broken.\\

\paragraph{Kaluza-Klein states:} Those are excitations of the Standard Model fields which appears in models with extra dimensions~\cite{Cheng:2002,Agashe:2004,Cheung:2001mq,Servant:2003}.\\

\paragraph{Super--heavy dark matter:} It is composed by heavy stable particles with a mass in the range of $10^{12}$ to $10^{16}$~\tu{GeV}. These particles leads to scenarios for production of nonthermal dark matter~\cite{Kolb:1999}.

\paragraph{Light scalar dark matter:} A class of fermionic dark matter candidate, Lee and Weinberg concludes that relic density arguments preclude such a WIMP with a masses less than a few \tu{GeV}~\cite{Boehm:2003ha}. The dark matter is form by other stable dark matter species, as in the case of SUSY $N=2$ theories, where both types coexist at the same time.\\
%If the dark matter is made up by other types of particles, however, this limit could be evaded, leading to light dark matter particles.\\

\paragraph{Dark matter from Little Higgs models:} These models have been proposed in order to stabilize the weak scale and to solve the hierarchy problem as an extension of the Standard Model. This class of models possess discrete symmetries which results in the existence of stable weakly interacting particles~\cite{Birkedal:2004,Cheng:2003ju}.

\section{Cosmic Rays}

% intro to cosmic--rays
The cosmic rays (CR) are generically described as charged particles that travel across the Universe. Their origin is usually associated to the nuclear activity present in stars, galaxies, etc. 
Depending on the processes in which they are involved, they are present at different energies scales usually going from the \tu{eV}-- up to \tu{PeV}--scale. \\

A first classification of CR regards their production location:
%A very roughly classification regards their origin:
\begin{itemize}
\item \textbf{Solar CR.}\\ Also known as solar energetic particles (SEP), these are cosmic rays that originate from the Sun. The average composition is similar to that of the Sun itself~\cite{Veselovsky:2006}.\\
\item \textbf{Galactic CR.}\\ This type consists of those cosmic rays that enter the solar system from the outside. They are high-energy charged particles composed of protons, electrons, and fully ionized nuclei of light elements.\\
\item \textbf{Extragalactic CR.}\\ Unlike solar or galactic cosmic rays, little is known about the origins of extragalactic cosmic rays. This is largely due to a lack of statistics: only about 1 extragalactic cosmic ray particle per square meter per year reaches the Earth's surface.\\
\end{itemize}

Many other ways to classify them are possible. The nature of the cosmic--rays also establish a good rule to classify them. Generally, those are separated into: Nuclei CR and Electron CR. 
Each category involves different features. Nuclei CR are less affected by energy losses, unlike electron CR that can cover shorter distances. This characteristic allows to nuclei CR to travel longer distances and increase their chances to interact with the medium.\\

On the other hand, we have the antimatter CR, which are less abundant than matter CR. The origin of this type of particles are not well understood, even though a fraction of those are related to spallation process between CR and the interstellar gas.\\ 

\subsection{Electron and positron cosmic--rays.}

Let us focus on the electron and positron CR species. As we said before, those particles may travel shorter distance in opposition to nuclei ones.
This would give detailed information about the Earth's local environment and/or the presence of exotic component, because a shorter distance reduces the chances to interact more with the medium. \\

The production of this species is generally related to:
\begin{itemize}
\item \textbf{Supernovae and gas:}\\ Cosmic--rays are injected into the medium when supernova explosions occur. The effect of shockwaves allows the particles in the medium to gain energy and travel across space.\\

\item \textbf{Secondary production:}\\Primary cosmic--rays interact with the gas present in the interstellar and intergalactic space. The spallation processes allow the production of new CR, which start to propagate and contribute to the cosmic--ray signal~\cite{Maurin:2002ua}.\\

\item \textbf{Pulsars:}\\ A number of studies have been done involving pulsars as sources of electron CR~\cite{Shen:1970,Busching:2008} and recently of positrons CR as well~\cite{Yuksel:2008rf}. \\

\item \textbf{Exotic sources:}\\ Dark--matter particle annihilation~\cite{Cheng:2002,Hooper:2002nq,Delahaye:2007fr} and evaporation of Primordial Black Holes~\cite{Barrau:2002,Barrau:2003} make possible contributions to the CR signal that come from the galactic halo.\\
\end{itemize}

\subsection{Positron and electron observation.}

Most of the observations of galactic electron and positron cosmic--rays are performed by balloon and space-borne experiments. 
Some of the balloon experiments are HEAT~\cite{Barwick:1997ig}, CAPRICE~\cite{Boezio:2000}, MASS~\cite{Grimani:2002yz}, PPB-BETS~\cite{Torii:2001ApJ,Torii:2008xu} and ATIC-2~\cite{Chang:2008zz}.
The most well known space--borne experiments are AMS~\cite{Alcaraz:2000PhLB,Aguilar:2007}, which was attached to the space shuttle in one of its missions, and PAMELA~\cite{Boezio:2004jx} that is currently on flight since 2006.\\

% FIGURE
% 
% 
\begin{fig}
 \resizebox{\textwidth}{!}{\includegraphics[angle=270]{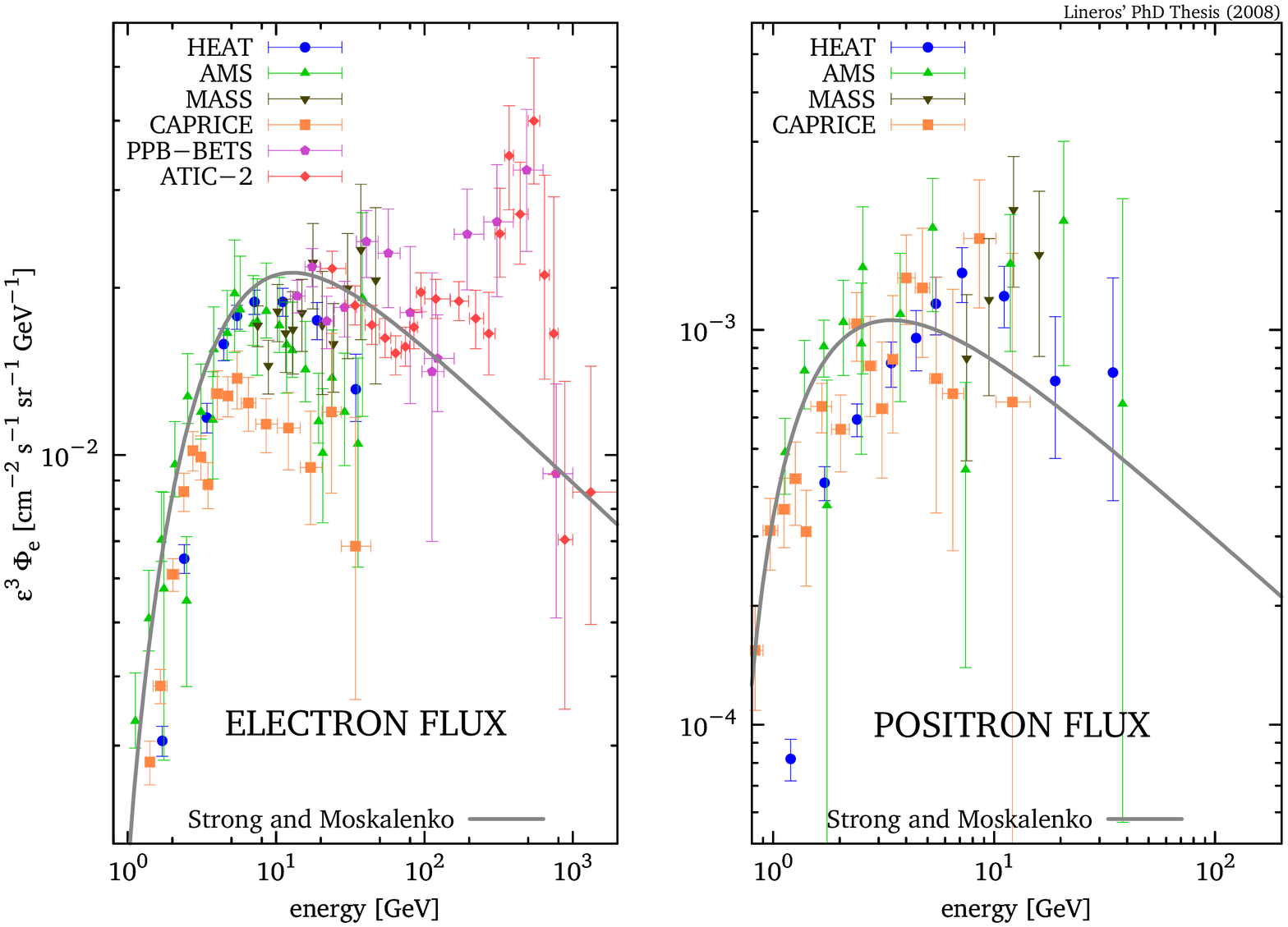}}
 \caption{\label{f:exp-flux} Electron (left) and positron (right) fluxes versus energy. Some of the lastest experimental results and the Strong and Moskalenko prediction~\cite{Moskalenko:1997gh,Baltz:1998xv} are shown. The data correspond to HEAT~\cite{Barwick:1997ig}, AMS~\cite{Alcaraz:2000PhLB,Aguilar:2007}, CAPRICE~\cite{Boezio:2000}, MASS~\cite{Grimani:2002yz}, PPB-BETS~\cite{Torii:2001ApJ,Torii:2008xu} and ATIC-2~\cite{Chang:2008zz} experiments.}
\end{fig}

The current status of electron and positron flux observations is shown in \citefig{f:exp-flux}. We observe that electrons arrive more abundant than positrons. As well, the effect of Solar activity on the low--energy range affect the fluxes reducing and deforming them with respect to the shape at the Solar System boundary.
The theoretical predictions done by Strong and Moskalenko~\cite{Moskalenko:1997gh} show agreement with respect to the observation in both signal.
A different situation occurs when the positron fraction, the ratio between the positrons and the total amount of positron and electrons, is plotted. In \citefig{f:exp-pf}, the positron fraction for different experiment is reported. 
We can observe that the Strong and Moskalenko prediction deviates in the high--energy tail.
This has produced a big revolution in the field, because there are not certain answers to this problem.\\

In this thesis, we study the positron and electron cosmic--ray signal. The work is oriented to solved and understand the ``positron excess'' feature.
In \citecha{cha2}, we study the different production mechanisms, that occur in the galactic environment, related to nuclear and particle physics.
We study the production of positron and electron in proton--proton interactions and from dark--matter annihilation--like processes. 
In \citecha{cha3}, we review the propagation model for ultrarelativistic cosmic--ray and improve the current solutions regarding the positron and electron case.\\

\citecha{cha4} is devoted to study the hypothesis that DM annihilation is the responsable of the ``positron excess'' feature. We analyze the DM annihilation signal from the context of generic--DM candidates and propagation uncertainties.
As generic--DM candidate, we consider some DM annihilation channels which correspond to typical signatures of candidates proposed by Beyond the Standard Model theories, as SUSY or Kaluza--Klein DM particles.
The positron background is studied in \citecha{cha5}. We study the secondary positrons produced by the interaction of proton and alpha particle CR with the interstellar gas present in the galactic plane. 
As well, we reanalyze the ``positron excess'' problem using the available experimental data. In this case, we found that the positron fraction is sensible to small variation in the positron flux.\\

% FIGURE
% 
% 
\begin{fig}
 \resizebox{0.8\textwidth}{!}{\includegraphics[angle=270]{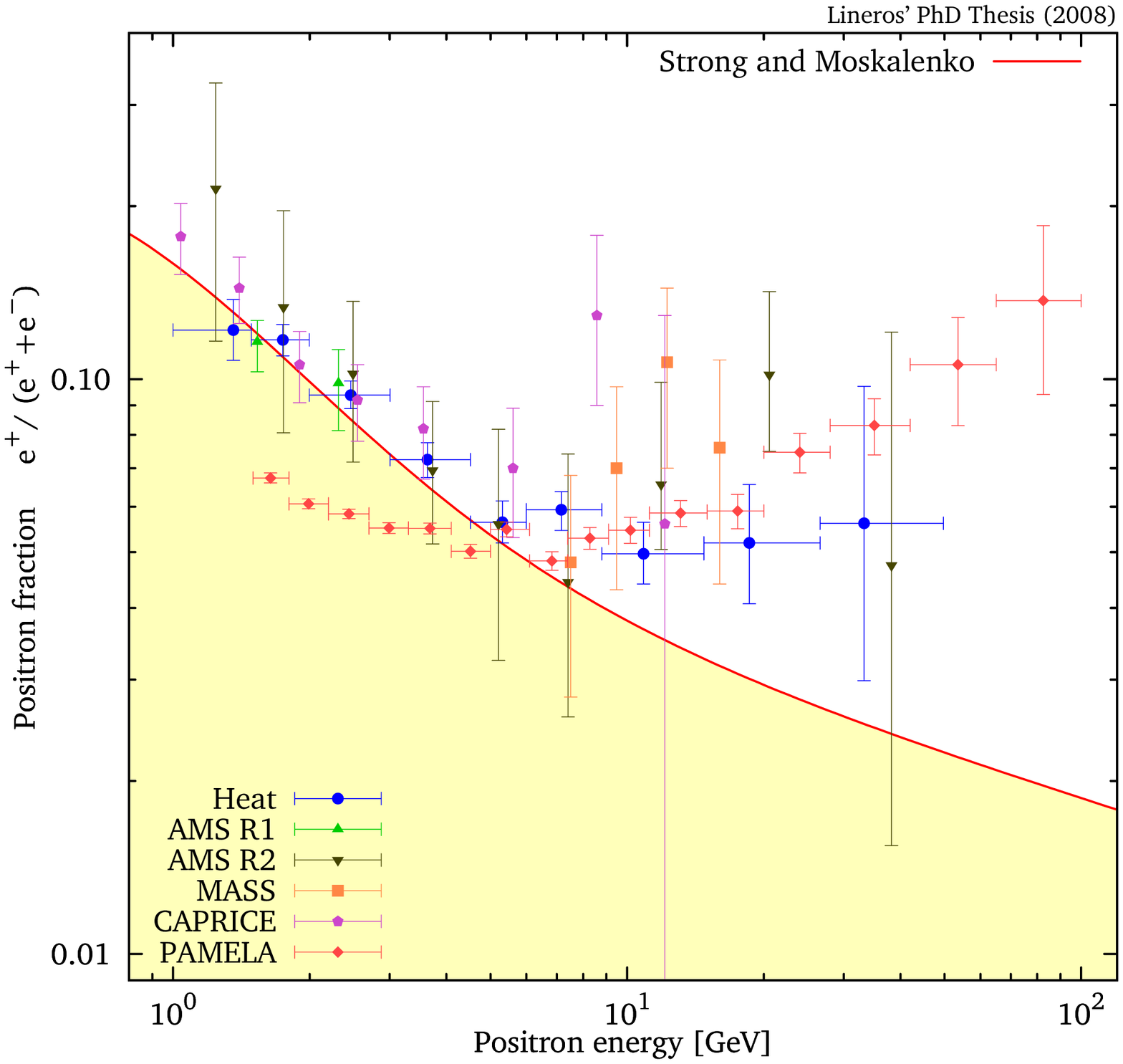}}
 \caption{\label{f:exp-pf} Positron fraction versus energy. The current observational status of the positron fraction and the prediction done by Strong and Moskalenko~\cite{Moskalenko:1997gh,Baltz:1998xv} are shown. The prediction shows the presence of a ``positron excess'' in the high--energy tail. The data correspond to HEAT~\cite{Barwick:1997ig}, AMS~\cite{Alcaraz:2000PhLB,Aguilar:2007}, CAPRICE~\cite{Boezio:2000}, MASS~\cite{Grimani:2002yz} and PAMELA~\cite{Adriani:2008zr} experiments.}
\end{fig}

\cleardoublepage
\chapter{Positron and Electron production}
\label{cha2}
\begin{prechap}
The production of electrons and positrons in galactic environment takes place in different ways. These processes are mainly described by standard particle physics. In this chapter, production processes related to proton--proton interactions and Dark Matter annihilation are reviewed and described.\\
\end{prechap}

\section{Overview}
The production of positrons and electrons is possible through different processes and in a wide range of energies. In the observable Universe, electrons are present in every place where atoms are. Electrons can be removed from atoms and be expelled into the interstellar space as result of supernova explosions, gas ionization processes and pulsars interactions with the medium. As well, electrons are produced by \CR\ interaction with the Interstellar Medium (\ism) \cite{Maurin:2002ua}.\\

The case of positrons is quite different. The principal contribution comes basically from \CR\ interaction with \ism\ \cite{Moskalenko:1997gh}. The matter/antimatter abundance asymmetry makes that mechanisms similar to the electron's ones become highly suppressed.\\

In the range of energies below 1 \tu{GeV}, \CR\ measurements, specially electron and positron \CR\ fluxes, can roughly be explained with our actual knowledge in nuclear and particle physics and nearby sources. Nevertheless, measurements above this limit may suggest the presence of undiscovered sources or new physical processes related to production \cite{Barwick:1997ig, Boezio:2000, Torii:2008xu, Torii:2001ApJ, Alcaraz:2000PhLB, Grimani:2002yz}.\\

An undiscovered category of sources may lay on Beyond the Standard Model (\bsm) theories context \cite{Bertone:2004pz}. These theories suggest that exotic \CR\ signals may be related to \cdm\ particles, where positrons and electrons are a consequence of their annihilations. In this category, sources are described and modeled in terms of initial and final states, formed by \bsm\ particles and \SM\ particles respectively. \\ 

In this chapter, electron and positron production related to proton--proton and \dm-like interaction are reviewed and explained. Let's warn that topics described in this chapter are crutial to understand the following ones.\\

\section{Production in Proton--Proton collision}

Physics behind cosmic rays sources are closely related to standard particle physics. In fact, most of processes where \CR\ production takes place are explained by the last one. Collisions between nuclei \CR\ and \ism\ are sources of secondary \CR. Secondary positrons and electrons are also produced in that way.\\

The \ism\ is composed mainly by Hydrogen and Helium with densities of $n_{\tn{H}} = 0.9 \ [\tu{part}/\tu{cm}^3]$ and $n_{\tn{He}} = 0.1 \ [\tu{part}/\tu{cm}^3]$ respectively \cite{Lyman:1998,Thorndike:1930, Field:1969, Ferriere:2001rg}.\\

Nuclei \CR\ are mainly constituted by protons and alpha particles. Many experiments are devoted to measure \CR\ fluxes. Generically, interstellar (IS) \CR\ fluxes are parameterized following a power--law form \cite{Shikaze:2006je}:
\begin{eq}
 \Phi^{\tn{IS}}(\mathcal{R}) = A \ \beta^{p_1} \ \left(\frac{\mathcal{R}}{1 \tu{GV}}\right)^{-p_2} \qquad \big(\tu{m}^{-2} \ \tu{s}^{-1} \ \tu{sr}^{-1} \ \left(\frac{\tu{GeV}}{\tu{n}}\right)^{-1} \big),
\end{eq}
which depends on the \CR\ rigidity ($\displaystyle \mathcal{R} = \frac{pc}{Ze}$) and Lorentz boost factor $\beta$.\\

The values of parameters depend on the \CR\ type. In the case of protons, those are:
\begin{eq}
 A = (1.94 \pm 0.13)\times 10^4, \  p_1 = 0.70 \pm 0.52 \ \tn{and} \ p_2 = 2.76 \pm 0.03 . 
\end{eq}
The values for alpha particles are:
\begin{eq}
 A = (7.10 \pm 0.56)\times 10^3, \ p_1 = 0.50 \pm 0.31 \ \tn{and} \ p_2 = 2.78 \pm 0.03 . 
\end{eq} \newline

Positrons (electrons) produced in \CR\ and \ism\ interactions will depend directly on the energy dependence of proton and alpha fluxes. As well, there is an intrinsic positron (electron) energy  distribution related to the collision itself. The intrinsic distribution is out of the astrophysics context, and it can be explained by using just particle and nuclear physics.\\

\subsection{Production through mesons decay}

The number of particles produced in hadronic collision, like in the proton-proton case, grows faster than in leptonic ones. In general, lighter mesons are produced in big quantities, specially pions and kaons. Those have decay modes which produce many types of particles, for example, neutral pions are well known as an efficient source of \mbox{gamma--rays}. As well, charged pions and kaons are efficient in production of positrons and electrons.\\

Positive and negative pions - or kaons - are produced in equal number due to conservation laws. However, positron and electron energy distributions are not equal. The difference is related to polarization effects in the muon production (\citeapp{app2}), the left-right asymmetry present in weak interactions is responsable of this. The result is that electrons are more energetically produced than positrons.\\

In general terms, positron and electron inclusive productions are not so different in form; the main difference lays on the muon's polarization effect. Apart of that, those are calculated in same way. The first ingredient is the inclusive cross sections (\CS) of charged mesons, which are used to guide the positrons (electrons) production (through the decay into positrons and electrons).\\

In the literature, there are many models to explain mesons production. However, we decided to use parameterizations of inclusive \CS, in order to obtain more realistic results. In the following, we explain the most used ones.\\

%%%
%%%
%%%
%%% Badhwar, TanNg and Stecker subsections
%%% sub file with ics parameterization

%-----------------------------------------------------------------------------------
\subsubsection{Badhwar--Stephens--Golden parameterization}

This was proposed by Badhwar, Stephens and Golden in the late 70's \cite{Badhwar:1977zf,Stephens:1981Ap}. Its aim is to explain and to get good agreement with accelerators and \CR\ data. Basically, it is a parameterization of the invariant \CS\ in p--p collisions.\\

The parameterization for production of charged pions is: 
\begin{eq}\label{e:bad-pion}
 \ener{} \frac{d^3 \ics}{dp^3} = A \kappa^{2 r} (1-\tilde{x})^{q(\mom{t})} \exp\left(-B \ \mom{t} \ \kappa^2\right) \qquad (\tu{mb}  \ \tu{GeV}^{-2} \ c^3) \; , 
\end{eq}
where $\mom{t}$ is the transverse momentum,  
\begin{eq}
\label{e:xt} q(\mom{t}) = \left(C_1 + C_2 \ \mom{t} + C_3 \ \mom{t}^2\right) \kappa \quad \tn{and} \quad \tilde{x} = \sqrt{x_{\parallel}^{2} + \frac{4 (\mom{t}^{2} + \mass{\pi}^{2})}{s}} \; .
\end{eq}

Values for the parameters are given in \citetab{c2:t1}. However, it is necessary to remark that this is a parameterization of differential inclusive \CS\ in the laboratory frame, that means:
\begin{eq}
 \gamma_{\lab} = \frac{\roots}{2 m_p} \ \tn{and} \quad \kappa^{-1} =  \sqrt{1 + \gamma_{\lab}^{-2}}  \ .
\end{eq} \newline 

\begin{tab}
	\begin{tabular}{|c|cc|cc|}
		\hline Particle & $\pi^{+}$ & $\pi^{-}$ & $K^{+}$ & $K^{-}$ \\
		\hline \hline A  &  153 & 127 & 8.85 & 9.3\\
		B &  5.55 & 5.3 & 4.05 & 3.8\\
		$r$ & 1 & 3 & \ldots & \ldots \\
		C  & \ldots & \ldots & 2.5 & 8.3 \\
		\hline $\tn{C}_1$ & 5.3667 & 7.0334 & \ldots & \ldots \\
		$\tn{C}_2$ & -3.5 & -4.5 & \ldots & \ldots\\
		$\tn{C}_3$ & 0.8334 & 1.667 & \ldots & \ldots\\
		\hline
	\end{tabular}
	\caption{\label{c2:t1} Parameters for charged Pion and Kaons inclusive \CS\ according to Badhwar--Stephens--Golden parameterization \cite{Badhwar:1977zf,Stephens:1981Ap}. The parameter units are such that the inclusive \CS\ are in units of $\tu{mb}  \ \tu{GeV}^{-2} \ \tu{c}^3$ (\citeeqq{e:bad-pion}{e:bad-kaon}).}
\end{tab}

\begin{tab}
	\begin{tabular}{|c|c|c|}
 	\hline Particle & Exclusive reaction & $\mass{X} \ (\tu{GeV} / \tn{c}^2)$\\
	\hline \hline $\pi^{+}$ & $pn\pi^{+}$ & 1.878 \\
	$\pi^{-}$ & $pp\pi^{+}\pi^{-}$ & 2.016  \\
	\hline $K^{+}$ & $p\Lambda^{0}K^{+}$ & 2.054  \\
	$K^{-}$ & $ppK^{+}K^{-}$ & 2.370 \\ \hline
	\end{tabular}
 	\caption{\label{c2:t0} Exclusive reactions in p-p collisions and the minimum configuration mass $\mass{X}$ (\citeeq{e:pcms_max}) for charged pions and kaons \cite{Tan:1984ha}. }
\end{tab}

The variable $x_{\parallel}$, which appears in \citeeq{e:xt}, is formally defined as the ratio among momenta in the center of mass system (\cms):
\begin{eq}
	x_{\parallel} = \frac{\displaystyle \mom{\parallel}^{\cms}}{\displaystyle \mom{\max}^{\cms}} \ ,
\end{eq}
where 
\begin{eq}
 \mom{\parallel}^{\cms} = \gamma_{\lab} (\mom{\parallel}^{\lab} - \beta_{\lab} \ener{}^{\lab}) \; ,
\end{eq}
\begin{eq}  
\label{e:pcms_max} \mom{\max}^{\cms} \rightarrow \ener{\max}^{\cms} = \frac{s + \mass{}^{2} - \mass{X}^{2}}{2 \roots} \; .
\end{eq} \newline

We need to specify that $\mass{}$ is the produced meson mass and $\mass{X}$ is the mass of minimal production configuration, which also gives infomation about thresholds for the energy of center of mass (\ecm) and meson energy (\citetab{c2:t0}).\\

A second parameterization is proposed, this time is for charged kaons and it is slightly different respect to the pions case:
\begin{eq} \label{e:bad-kaon}
 \ener{} \frac{d^3 \ics}{dp^3} = A (1-\tilde{x})^C \exp\left(-B \ \mom{t}\right) \qquad (\tu{mb}  \ \tu{GeV}^{-2} \ \tu{c}^3) \; , 
\end{eq}
where $A$, $B$ and $C$ are constants with values given in \citetab{c2:t1}. Let's emphasize that the parameterization fits good observational data. \\

%
%
%-----------------------------------------------------------------------------------------
\subsubsection{Tan--Ng parameterization}

This is a well known and used parameterization for the study of inclusive \CS's in p-p collisions. It reproduces and predicts with good precision the low and high energy data \cite{Tan:1983ICRC,Tan:1984ha}.\\

The invariant \CS\ is generically written as:
\begin{eq}\label{e:tan-par}
 \ener{} \frac{d^3 \ics}{dp^3} = f(x_\tn{R}) \exp\left(-A(x_\tn{R})p_t - B(x_\tn{R})p_t^2\right) \qquad (\tu{mb}  \ \tu{GeV}^{-2} \ \tu{c}^3) \ , 
\end{eq}
where 
\begin{eq}
	x_{\tn{R}} = \frac{\displaystyle E^{\cms}}{\displaystyle E_{\max}^{\cms}} 
\end{eq}
is the radial scaling variable, which is expressed in terms of meson energy and maximum possible energy for the minimal configuration in the \cms\ (\citetab{c2:t0}). Another scaling variable is 
\begin{eq}
 x_{\tn{T}} \equiv \frac{x_{\tn{R}} - x_{\tn{m}}}{1 - x_{\tn{m}}}\; ,
\end{eq}
where 
\begin{eq}
x_{\tn{m}} = \frac{mc^2}{E_{\max}^{\cms}} \; ,
\end{eq}
it helps to extend the parameterization to very low energy scales, almost arriving to threshold energy, with pretty good precision.\\

The resting terms are:
\begin{eq}
	f(x_\tn{R}) &=& a_1 \exp\big( -a_2 x_{\tn{R}} \big) \Theta(a_3 - x_{\tn{R}}) + (\sigma_{00} -a_1)(1-x_{\tn{R}})^{a_4} \\
	A(x_\tn{R}) &=& a_5 \exp\big( -a_6 x_\tn{R} \big) + a_7 \exp\big(a_8 x_\tn{R} \big) \nonumber \\
	B(x_\tn{R}) &=& a_9 \exp\big( -a_{10}(x_\tn{R}+a_{11})\big) (x_\tn{R} + a_{11})^{a_{12}} \nonumber
\end{eq}
where $\Theta$ is just the Heaviside step function.\\

For our purposes, we are interested in the production of charged pions and kaons, for which, the parameter values are given in \citetab{c2:t2}.\\

\begin{tab}
	\begin{tabular}{|c|cc|cc|}
		\hline Particle & $\pi^{+}$ & $\pi^{-}$ & $K^{+}$ & $K^{-}$ \\
		\hline \hline $\sigma_{00}$ & 163 & 163 & 7.33 & 7.33 \\
		       $a_{1}$  & 143   & 150 &2.61&6.97\\  
		       $a_{2}$  & 9.54  & 13.8 & 5.36 & 11.8\\ 
		       $a_{3}$  & 0.7   & 0.5 & 0.7 & 0.5 \\ 
		       $a_{4}$  & 2.72  & 3.38 & 2.49 & 3.27\\ 
		\hline $a_{5}$  & 3.53  & 4.20 & 1.96 & 2.27\\ 
		       $a_{6}$  & 3.48  & 4.82 & 1.03 $\times 10^{-3}$ & 4.32\\ 
		\hline $a_{7}$  & 1.88  & 1.07 & 1.67 & 4.89 $\times 10^{-2}$\\ 
		       $a_{8}$  & 0.212 & 0.984 & 0.263 & 6.03 \\ 
		       $a_{9}$  & 1.64 $\times 10^{-4}$ & 0.873 & 8.24 $\times 10^{3}$ & 22.6\\
		\hline $a_{10}$ & 0.779 & 0.455 & 6.41 & 0.894 \\ 
		       $a_{11}$ & 4.70 & 1.71 & 4.72 & 2.81\\ 
		       $a_{12}$ & 7.82 & 2.01 & 6.09 & 0.375\\
		\hline
	\end{tabular}
	\caption{\label{c2:t2} Parameter for charged Pion and Kaons inclusive \CS. These sets are for Tan--Ng parameterization \cite{Tan:1983ICRC,Tan:1984ha}. The parameter units are such that the inclusive \CS\ are in units of $\tu{mb}  \ \tu{GeV}^{-2} \ \tu{c}^3$ (\citeeq{e:tan-par}).}
\end{tab}

%-----------------------------------------------------------------------------------
\subsubsection{Stecker model}

The Stecker model was created to explain the pions production at very low energies in p--p collisions. The principal mechanism is through the production and decay of $\Delta$. Also, a secondary pion production mechanism is related to fireballs production, \ie thermal pion production from residual collision energy \cite{Stecker:1967SA,Stecker:1970Ap}. \\

In the p--p collision, production of $\Delta$ baryons produce charged and neutral pions in almost the same proportion. Originally, Stecker's model was used to estimate gamma--rays fluxes. However, his ideas are used also for the production of charged pions, and then positron and electrons. \\

The discussed process is:
\begin{eq}
 p + p \longrightarrow p + (\Delta^{+} \rightarrow n + \pi^{+}) \longrightarrow p + n + \pi^{+}  \ ,
\end{eq}
where $\Delta$ baryons are considered as resonances, \ie its effective mass varies following a Breit-Wigner distribution,
\begin{eq}
 \mathcal{F}(\mass{\Delta}) =  \frac{1}{\pi} \frac{\Gamma}{(\mass{\Delta}^{*} - \mass{\Delta})^{2} + \Gamma^{2}} \qquad (\tu{GeV}^{-1}),
\end{eq}
where $\mass{\Delta}^{*} =  1.232 \ \tu{GeV}$ and $\Gamma =  0.118 \ \tu{GeV}$ are the most recent determinations \cite{PDBook}. Some physical considerations reduce the $\mass{\Delta}$ range to:
\begin{eq}
 \mass{\Delta}^{\min} = \mass{n} + \mass{\pi}  \quad , \quad \mass{\Delta}^{\max} = \roots - \mass{p} \; .
\end{eq}\newline

The first bound corresponds to the lowest value of mass able to produce a physical pion and the second bound is related to the minimal configuration, \ie the production of a proton with a $\Delta$ and nothing else. \\

The $\Delta$ decay into charged pions may happen just in one way, and so, the pion energy distribution will be:
\begin{eq}
	f_{\pi, \Delta}(\ener{\pi}, \ener{\Delta}) = \left\{ \begin{array}{ccl}
 \delta (\ener{\pi} - \ener{\pi}^{*})  & & \tn{when} \ \Delta \ \tn{is at rest.}  \\[2ex]
\displaystyle \frac{1}{2 \gamma_{\Delta} \beta_{\Delta} \mom{\pi}^{*}} & & \tn{moving } \Delta \tn{ and} \ \ener{\pi}^{\min} < \ener{\pi} < \ener{\pi}^{\max}
\end{array}\right. \qquad (\tu{GeV}^{-1}) \; , 
\end{eq}
where 
\begin{eq}
 \ener{\pi}^{*} = \frac{\mass{\Delta}^{2} - \mass{n}^2 + \mass{\pi}^2}{2 \mass{\Delta}}
\end{eq}
 is the pion energy in the $\Delta$ rest frame. In the case of a moving $\Delta$, pion kinematic limits are:
\begin{eq}
  \ener{\pi}^{\min} = \gamma_{\Delta} (\ener{\pi}^{*}  - \beta_{\Delta} \mom{\pi}^{*}) \quad \tn{and} \quad \ener{\pi}^{\max} = \gamma_{\Delta} (\ener{\pi}^{*}  + \beta_{\Delta} \mom{\pi}^{*}) \; ,
\end{eq}
where
\begin{eq}
 \beta_{\Delta} = \frac{\mom{\Delta}}{\ener{\Delta}} \; ,
\end{eq}
the description of moving $\Delta$ is based on the boosted spectra method (see \citeapp{app1}).\\

\begin{fig}
	\includegraphics[width=0.5\textwidth]{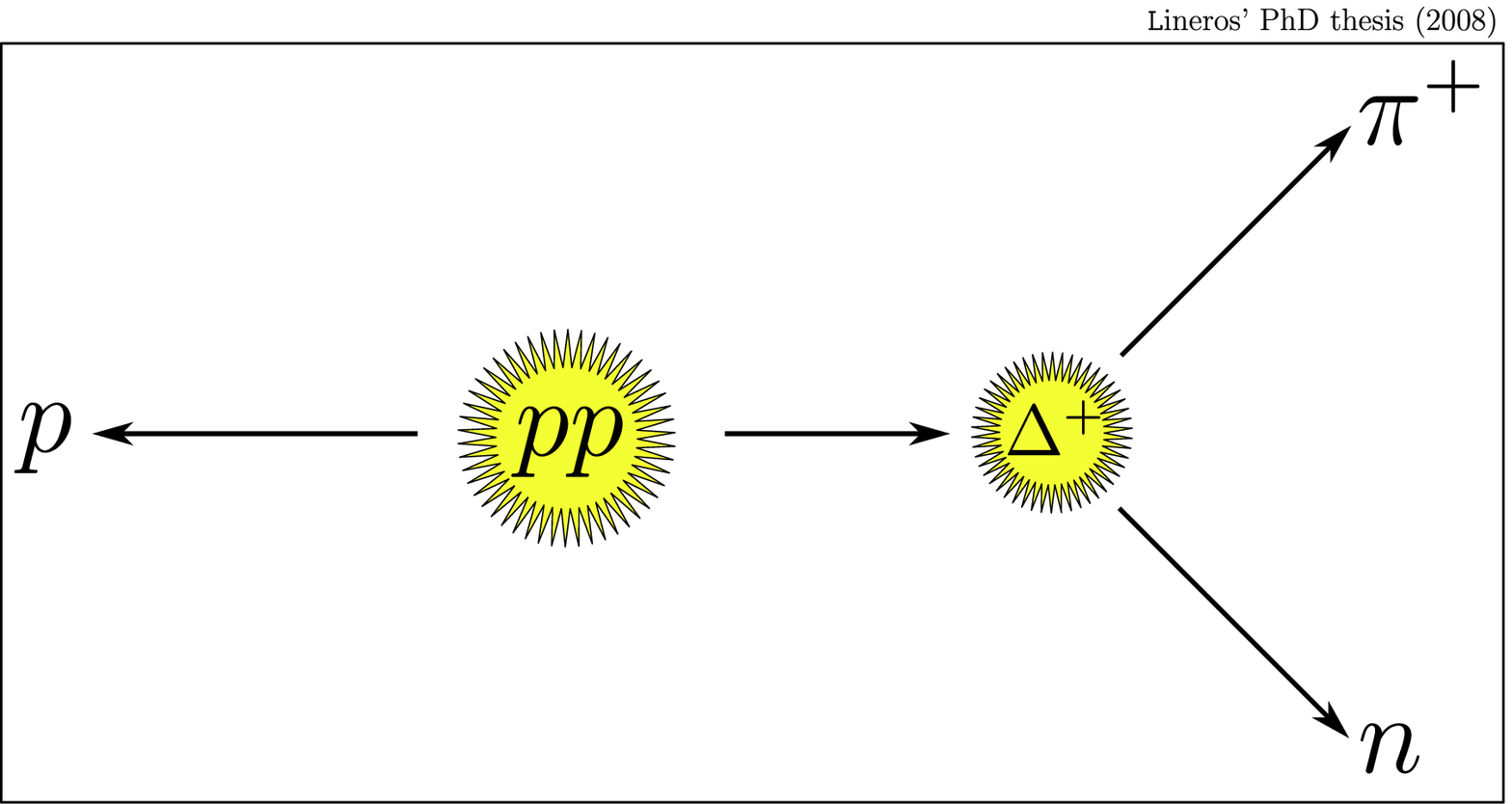}
	\caption{\label{c2:f1} Schematic represetantion of pion production resulting from p--p collision according to Stecker's model.}
\end{fig}

A third point to discuss about is the $\Delta$ production in p--p collisions. In the \cms, the energy distribution of $\Delta$ baryons is:
\begin{eq}
 f_{\Delta, pp}^{\tn{cms}} (\ener{\Delta}, \roots) = \delta( \ener{\Delta} - \ener{\Delta}^{*}) \qquad (\tu{GeV}^{-1})\;,
\end{eq}
where 
\begin{eq}
	\ener{\Delta}^{*} = \displaystyle \frac{s - \mass{p}^{2} + \mass{\Delta}^2}{2\roots}
\end{eq}
is solution in the \cms for 2--body production and an isotropic $\Delta$ production is assumed.\\

However, we are interested in the solution in the laboratory frame. Using the boosted spectra formalism (\citeapp{app1}), we found the equivalent energy distribution in the this frame:
\begin{eq}
 f_{\Delta, pp}^{\lab} (\ener{\Delta}, \ener{p}) = \frac{1}{2 \ \gamma_{\lab} \beta_{\lab} \ \mom{\Delta}^{*}} \qquad \forall \ \ener{\Delta} \in [\ener{\Delta}^{\min},\ener{\Delta}^{\max}] \qquad (\tu{GeV}^{-1}) \;, 
\end{eq}
also the kinematic limits change giving new bounds for the $\Delta$ energy:
\begin{eq}
 \ener{\Delta}^{\min} = \gamma_{\lab} (\ener{\Delta}^{*}  - \beta_{\lab} \mom{\Delta}^{*}) \quad \tn{and} \quad \ener{\Delta}^{\max} = \gamma_{\lab} (\ener{\Delta}^{*}  + \beta_{\lab} \mom{\Delta}^{*}) \ .
\end{eq}\newline

For a fixed $\Delta$ mass, the pion energy distribution is obtained as a composition of processes:
\begin{eq}
 f_{\pi,pp}^{\lab}(\ener{\pi},\ener{p},\mass{\Delta}) = \int_{\ener{\Delta}^{\min}}^{\ener{\Delta}^{\max}} d\ener{\Delta} \  f_{\Delta,pp}^{\lab}(\ener{\Delta},\ener{p}) \ \ f_{\pi,\Delta}(\ener{\pi},\ener{\Delta}) \; ,
\end{eq}
where after integration, we obtain:
\begin{eq}
 f_{\pi,pp}^{\lab}(\ener{\pi},\ener{p},\mass{\Delta}) = \frac{1}{2 \ \gamma_{\lab} \beta_{\lab} \mom{\Delta}^{*}} \times \frac{\mass{\Delta}}{2 \ \mom{\pi}^{*}} \log\left( \frac{\gamma_{\Delta}^{\max}(1 + \beta_{\Delta}^{\max})}{\bar{\gamma}_{\Delta}(\ener{\pi}) \ (1 + \bar{\beta}_{\Delta}(\ener{\pi}))} \right) \; ,
\end{eq}
where 
\begin{eq}
 \bar{\gamma}_{\Delta}(\ener{\pi}) = \max\left(\gamma_{\Delta}^{\min}, \ \frac{\ener{\pi}^{*}\ener{\pi} - \mom{\pi}^{*}\mom{\pi}}{\mass{\pi}^{2}} \right) \; ,
\end{eq}
which is a typical behavior related with the change of reference frame (\citeapp{app1}).\\

The next step is to include the $\Delta$--resonance. For that, an average respect to $\mass{\Delta}$ is needed:
\begin{eq}
 f_{\pi,pp}^{\lab}(\ener{\pi}, \ener{p}) =  N \int_{\mass{\Delta}^{\min}}^{\mass{\Delta}^{\max}} d\mass{\Delta} \mathcal{F}(\mass{\Delta}) \ f_{\pi,pp}^{\lab}(\ener{\pi},\ener{p},\mass{\Delta})
\end{eq}
where $N$ is a normalization constant for $\mathcal{F}(\mass{\Delta})$, in which, 
\begin{eq}
N^{-1} = \frac{1}{\pi} \Big(\arctan\left( \frac{\mass{\Delta}^{\max} - \mass{\Delta}^{*}}{\Gamma} \right) - \arctan\left( \frac{\mass{\Delta}^{\min} - \mass{\Delta}^{*}}{\Gamma} \right)  \Big) \ .
\end{eq}\newline

Unfortunately, this integral cannot be solved analytically and numerical integration should be performed \cite{Stecker:1967SA,Stecker:1970Ap}.\\

%%%
%%%
%%%
%%%

\subsubsection{Calculation of inclusive cross sections}

Inclusive \CS's for positron and electron are calculated as convolutions between mesons \CS's and their energy decay distributions into positrons and electrons. \\

The final positron (electron) inclusive \CS\ is obtained from contributions of pions and kaons:
\begin{eq}
 \frac{d\ics_{pp \rightarrow e}}{d\ener{e}}  = \frac{d\ics_{pp \rightarrow e}^{\pi}}{d\ener{e}} + \frac{d\ics_{pp \rightarrow e}^{K}}{d\ener{e}} \qquad (\tu{mb} \ \tu{GeV}^{-1}) \; ,
\end{eq}
where each contribution is generically calculated as:
\begin{eq}
 \frac{d\ics_{pp \rightarrow e}^{X}}{d\ener{e}}(\ener{e},\ener{p}) = \int d \ener{X} \frac{d \ics_{pp \rightarrow X}}{d\ener{X}}(\ener{X}, \ener{p}) \times f_{e, X}(\ener{e}, \ener{X}) \qquad (\tu{mb} \ \tu{GeV}^{-1}) \; ,
\end{eq}
where $X$ denotes pions or kaons as intermediate particles. Note that meson \CS\ are related to the parameterizations previously seen, through an integration over the transversal momentum:
\begin{eq}
 \frac{d\ics}{d \ener{}} (\ener{},\ener{p}) = \pi \int_{0}^{\ener{}^2 - m^2} \left(\ener{} \frac{d^3 \ics}{dp^3}\right)(\mom{t}, \mom{\parallel};\ener{p}) \ \frac{d (\mom{t}^2)}{\mom{\parallel}} \qquad (\tu{mb} \ \tu{GeV}^{-1}) \;.
\end{eq} \newline

The decay distributions $f_{e,X}$ have been analytically calculated including muon polarization effects. Pion decays are the simplest ones because they have just one dominant decay mode (details in \citeapp{app3}). On the contrary, Kaon decays are more complex to treat (details in \citeapp{app4}). \\

%%%
%%%
%%%
%%% Kamae et al. parameterization subsection
%% subfile this kamae parameterization explication
\subsection{Kamae et al. parameterization}

A parameterization for production of positrons, electrons and other particles are proposed by Kamae et al. \cite{Abe:2004gp,Kamae:2004xx,Kamae:2006bf}. The objective of this parameterization is to give an easy way to compute and estimate \CR\ fluxes, that comes from \ism\ interactions with nuclei \CR. \\

New processes are included, contributions from $\Delta(1238)$ and several hadron resonances around 1600 $[\tu{MeV}/\tu{c}^2]$ make it more accurate in the very low proton energy range. Another included process comes from diffractive dissociation which contributes in an intermediate energy range.\\

For the positron and electron cases, the inclusive differential \CS\ is described by:
\begin{eq}
 \frac{d \ics_{\tn{ND}}(\ener{e})}{d \log(\ener{e})} = r(y) \times F_{\tn{ND}}(x) \times F_{\tn{ND,kl}}(x)  \qquad (\tu{mb}) \;,
\end{eq}
where $y = \log_{10}(\kin{p}/1\tu{TeV})$ and $x = \log_{10}(\ener{e}/1\tu{GeV})$.\\

The function $r(y)$ is a rescaling of the Non-Diffractive (ND) contribution in order to reproduce experimental data. $F_{\tn{ND}}$ is the parameterization of ND \CS:
\begin{eq}
 F_{\tn{ND}}(x) &=& a_0 \exp\left\{-a_1\big(x - a_3 + a_2(x-a_3)^2\big)^2 \right\} +  \\
&& a_4 \exp\left\{-a_5 \big(x - a_8 + a_6(x - a_8)^2 + a_7 (x-a_8)^3\big)^2 \nonumber \right\} ,
\end{eq}
The parameter values and $r(y)$ functions are given in \citetab{t:kamae-pos} for positrons and in \citetab{t:kamae-ele} for electrons. $F_{\tn{ND,kl}}$ represents the conservation of energy-momentum: 
\begin{eq}
F_{\tn{ND,kl}}(x) = \frac{1}{\exp\big\{W_{\tn{ND,l}}(L_{\min} - x)\big\} + 1} \times \frac{1}{\exp\big\{W_{\tn{ND,h}}(x - L_{\max})\big\} + 1}. 
\end{eq}
where values of parameters are give in \citetab{t:kamae-limits}.\\

\begin{tab}
 \begin{tabular}{|c|cc|cc|}
	\hline
  Particle & $L_{\min}$ & $L_{\max}$ & $W_{\tn{ND,l}}$ & $W_{\tn{ND,h}}$ \\
	\hline \hline
	$e^{-}$ & -2.6 & $0.96 \times \log_{10}\left\{\kin{p}/1\tu{GeV}\right\}$ & 20&45 \\
	$e^{+}$ & -2.6 & $0.94 \times \log_{10}\left\{\kin{p}/1\tu{GeV}\right\}$ & 15&47 \\
	\hline
 \end{tabular}
\caption{\label{t:kamae-limits} Kinematics limits for Kamae et al. parameterization for electron and positron production.}
\end{tab}

Diffractive processes are parameterized as follows:
\begin{eq}
F_{\tn{Diff}}(x) = b_{0}\exp\left\{-b_{1}\Big(\frac{x - b_{2}}{1 + b_{3}(x - b_{2})}\Big)^{2}\right\} +  b_{4}\exp\left\{-b_{5}\Big(\frac{x - b_{6}}{1 + b_{7}(x - b_{6})}\Big)^{2}\right\},
\end{eq}
and the resonance contributions are:
\begin{eq}
F_{\tn{res}}(x) = c_{0}\exp\left\{-c_{1}\Big(\frac{x - c_{2}}{1 + c_{3}(x - c_{2}) + c_{4}(x - c_{2})^{2}}\Big)^{2}\right\}.
\end{eq}

As before, the energy-momentum conservation is a separate function and is described by:
\begin{eq}
F_{\tn{kl}}(x) = \frac{1}{\exp\big\{(W_{\tn{diff}}(x - L_{\max})\big\} + 1},
\end{eq}
where $W_{\tn{diff}} = 75$ and $L_{\max} = \log_{10}\left\{\kin{p}/1\tu{GeV}\right\}$. The inclusive \CS\ is composed as before, 
\begin{eq}
 \frac{d \ics_{\tn{ND}}(\ener{e})}{d \log(\ener{e})} = F_{\tn{Diff/res}}(x) \times F_{\tn{kl}}(x)  \qquad (\tu{mb}) ,
\end{eq}
although the rescaling $r(y)$ function is not included here.\\

Kamae et al. parameterizations work well for positron and electrons. However, there were changes in the parameter values respect to published ones \cite{kamae:comm}. Let's clarify that the new values are given in all the tables.\\ 

%%%%%%% kamae tables!!!!
\begin{tab}
\resizebox{\hsize}{!}{
\begin{tabular}{|c|c|l|}
\hline
\multicolumn{3}{|c|}{\large Parameterization for Electrons} \\
\hline
& Parameters & Formulae as functions of the proton kinetic energy ($y=\log_{10}(T_{p})$) in \tu{TeV}. \\
\hline \hline
\multirow{12}{*}{\begin{sideways}Non-Diffraction\end{sideways}} & $a_0$ & $\displaystyle -0.018639(y + 3.3) + 2.4315(y + 3.3)^{2} - 0.57719(y + 3.3)^{3} + 0.063435(y + 3.3)^{4}$ \\

& \multirow{2}{*}{$a_1$} & $\displaystyle 7.1827\tpow{-6} - 3.5067\tpow{-6}y + 1.3264\tpow{-6}y^{2} - 3.3481\tpow{-7}y^{3} + \ldots$ \\
& & $\displaystyle \ldots + 2.3551\tpow{-8}y^{4} + 3.4297\tpow{-8}y^{5}$ \\

& $a_2$ & \rule{0ex}{4ex}$\displaystyle 563.91 - 362.18\log_{10}\big\{2.7187(y + 3.4)\big\} - \frac{2.8924\tpow{4}}{(y + 7.9031)^{2}}$ \\
& $a_3$ & $0.52684 + 0.57717y + 0.0045336y^{2} - 0.0089066y^{3}$ \\

& \multirow{2}{*}{$a_4$} & $\displaystyle 0.36108(y + 3.32) + 1.6963(y + 3.32)^{2} - 0.074456(y + 3.32)^{3} - \ldots$ \\
& & $\displaystyle \ldots - 0.071455(y + 3.32)^{4} + 0.010473(y + 3.32)^{5}$\\

& \multirow{2}{*}{$a_5$} & $\displaystyle 9.7387\tpow{-5} + 7.8573\tpow{-5}\log_{10}\big\{0.0036055(y + 4.3)\big\} + \ldots $ \\
& & $\displaystyle \ldots + \frac{0.00024660}{y + 4.9390} - 3.8097\tpow{-7}y^{2}$\\

& $a_6$ & $\displaystyle -273.00 - 106.22\log_{10}\big\{0.34100(y + 3.4)\big\} + 89.037y - 12.546y^{2}$ \\

& $a_7$ & $\displaystyle 432.53 - 883.99\log_{10}\big\{0.19737(y + 3.9)\big\} - \frac{4.1938\tpow{4}}{(y + 8.5518)^2}$ \\
& $a_8$ & $\displaystyle -0.12756 + 0.43478y - 0.0027797y^{2} - 0.0083074y^{3}$ \\

\hline
& \multirow{2}{*}{$r(y)$} & \rule{0ex}{4ex} $\displaystyle 3.63\exp\left\{-106\Big(\frac{y+3.26}{1 + 9.21(y + 3.26)}\Big)^2\right\} - 0.182y - 0.175y^{2}$ for $T_{p}\leq 15.6$ $     $  \tu{GeV} \\
&       & $\displaystyle 1.01$ for $T_{p} > 15.6$ \tu{GeV} \\
\hline \hline 
\multirow{10}{*}{\begin{sideways}Diffraction\end{sideways}} & \multirow{2}{*}{$b_0$} & $\displaystyle 0.20463\tanh\big\{-6.2370(y + 2.2)\big\} - 0.16362(y + 1.6878)^{2} + \ldots$ \\ 
& &  $\ldots + 3.5183\tpow{-4}(y + 9.6400)^{4}$ \\

& $b_1$ & $\displaystyle 1.6537 + 3.8530\exp\left\{-3.2027\Big(\frac{y + 2.0154}{1.0 + 0.62779(y + 2.0154)}\Big)^{2}\right\}$ \\

& $b_2$ & $\displaystyle -10.722 - 0.082672\tanh\big\{-1.8879(y + 2.1)\big\} + 1.4895\tpow{-4}(y + 256.63)^{2}$ \\

& $b_3$ & $\displaystyle -0.023752 - 0.51734\exp\left\{-3.3087\Big(\frac{y + 1.9877}{1.0 + 0.40300(y + 1.9877)}\Big)^2\right\}$ \\
& $b_{0},\ldots,b_{3}$ &  0 for $T_{p} < 5.52$ \tu{GeV} \\

& $b_4$ & $\displaystyle 0.94921 + 0.12280(y + 2.9)^{2} - 7.1585\tpow{-4}(y + 2.9)^{4} + 0.52130\log_{10}\big\{y + 2.9\big\}$ \\

& $b_5$ & $\displaystyle -4.2295 - 1.0025\tanh\big\{9.0733(y + 1.9)\big\} - 0.11452(y - 62.382)$ \\
& $b_6$ & $\displaystyle 1.4862 + 0.99544y - 0.042763y^{2} - 0.0040065y^{3} + 0.0057987y^{4}$ \\
& $b_7$ & $\displaystyle 6.2629 + 6.9517\tanh\big\{-0.36480(y + 2.1)\big\} - 0.026033(y - 2.8542)$ \\

\hline \hline
\multirow{5}{*}{\begin{sideways}Res(1600)\end{sideways}} & $d_0$ & \rule{0ex}{4ex}$ \displaystyle 0.37790\exp\left\{-56.826\Big(\frac{y + 2.9537}{1.0 + 1.5221(y + 2.9537)}\Big)^{2}\right\} - 0.059458 + 0.0096583y^{2}$ \\
& $d_1$ & $\displaystyle -5.5135 - 3.3988y$ \\
& $d_2$ & $\displaystyle -7.1209 - 7.1850\tanh\big\{30.801(y + 2.1)\big\} + 0.35108y$ \\
& $d_3$ & $\displaystyle -6.7841 - 4.8385y - 0.91523y^{2}$ \\
& $d_4$ & $\displaystyle -134.03 - 139.63y - 48.316y^{2} - 5.5526y^{3}$ \\
\hline
\end{tabular}
}
\caption{\label{t:kamae-ele} Kamae et al. parameters describing electron spectra for arbitrary proton energy.}
\end{tab}

\begin{tab}
\resizebox{\hsize}{!}{
\begin{tabular}{|c|c|l|}
\hline
\multicolumn{3}{|c|}{\large Parameterization for Positrons} \\
\hline
& Parameters & Formulae as functions of the proton kinetic energy  ($y=\log_{10}(T_{p})$) in \tu{TeV}. \\
\hline\hline
 \multirow{13}{*}{\begin{sideways}Non-Diffraction\end{sideways}} & \multirow{2}{*}{$a_0$} & $\displaystyle -0.79606(y + 3.3) + 7.7496(y + 3.3)^{2} - 3.9326(y + 3.3)^{3} + \ldots$ \\
& & $\displaystyle \ldots + 0.80202(y + 3.3)^{4} - 0.054994(y + 3.3)^{5}$\\

& \multirow{2}{*}{$a_1$} & $\displaystyle 6.7943\tpow{-6} - 3.5345\tpow{-6}y + 6.0927\tpow{-7}y^{2} + \ldots$ \\
&      & $\displaystyle \ldots + 2.0219\tpow{-7}y^{3} + 5.1005\tpow{-8}y^{4} - 4.2622\tpow{-8}y^{5}$ \\

& $a_2$ & $\displaystyle 44.827 - 81.378\log_{10}\big\{0.027733(y + 3.5)\big\} - \frac{1.3886\tpow{4}}{(y + 8.4417)^2}$ \\

& $a_3$ & $\displaystyle 0.52010 + 0.59336y + 0.012032y^{2} - 0.0064242y^{3}$ \\

& \multirow{2}{*}{$a_4$} & $\displaystyle 2.1361(y + 3.32) + 1.8514(y + 3.32)^{2} - 0.47872(y + 3.32)^{3} + \ldots$ \\
&      & $\displaystyle \ldots + 0.0032043(y + 3.32)^{4} + 0.0082955(y + 3.32)^{5}$ \\

& \multirow{2}{*}{$a_5$} & $\displaystyle 1.0845\tpow{-6} + 1.4336\tpow{-6}\log_{10}\big\{0.0077255(y + 4.3)\big\} + \ldots$\\ 
& & $\displaystyle \ldots + \frac{1.3018\tpow{-4}}{(y + 4.8188)^{2}} + 9.3601\tpow{-8}y$ \\

& $a_6$ & $\displaystyle -267.74 + 14.175\log_{10}\big\{0.35391(y + 3.4)\big\} + 64.669y - 7.7036y^2$ \\

& $a_7$ & $\displaystyle 138.26 - 539.84\log_{10}\big\{0.12467(y + 3.9)\big\} - \frac{1.9869\tpow{4}}{(y + 7.6884)^2} + 1.0675y^{2}$ \\

& $a_8$ & $\displaystyle -0.14707 + 0.40135y + 0.0039899y^{2} - 0.0016602y^{3}$ \\

\hline
& \multirow{2}{*}{$r(y)$} & \rule{0ex}{4ex} $\displaystyle 2.22\exp\left\{-98.9\Big(\frac{y + 3.25}{1 + 10.4(y + 3.25)}\Big)^{2}\right\}$ for $T_{p}\leq 5.52$ \tu{GeV} \\
 &          & 1.0 for $T_{p} > 5.52$ \tu{GeV} \\

\hline \hline 
\multirow{10}{*}{\begin{sideways}Diffraction\end{sideways}} & $b_0$ & $\displaystyle 29.192\tanh\big\{-0.37879(y + 2.2)\big\} - 3.2196(y + 0.67500)^{2} + 0.0036687(y + 9.0824)^{4}$ \\

& $b_1$ & $\displaystyle -142.97 + 147.86\exp\left\{-0.37194\Big(\frac{y + 1.8781}{1.0 + 3.8389(y + 1.8781)}\Big)^{2}\right\}$ \\

& $b_2$ & $\displaystyle -14.487 - 4.2223\tanh\big\{-13.546(y + 2.2)\big\} + 1.6988\tpow{-4}(y + 234.65)^{2}$ \\

& $b_3$ & $\displaystyle -0.0036974 - 0.41976\exp\left\{-6.1527\Big(\frac{y + 1.8194}{1.0 + 0.99946(y + 1.8194)}\Big)^2\right\}$ \\
& $b_{0},\ldots,b_{3}$     & 0 for $T_{p} < 11.05$ \tu{GeV} \\

& \multirow{2}{*}{$b_4$} & $\displaystyle 1.8108 + 0.18545(y + 2.95)^{2} - 0.0020049(y + 2.9)^{4} + \ldots$ \\ 
& & $\ldots + 0.85084\exp\left\{-14.987\big(y + 2.29 - 0.18967(x + 2.29)\big)^{2}\right\}$\\

& $b_5$ & $\displaystyle 2.0404 - 0.51548\tanh\big\{2.2758(y + 1.9)\big\} - 0.035009(y - 6.6555)$ \\

& $b_6$ & $\displaystyle 1.5258 + 1.0132y - 0.064388y^{2} - 0.0040209y^{3} - 0.0082772y^{4}$ \\

& $b_7$ & $\displaystyle 3.0551 + 3.5240\tanh\big\{-0.36739(y + 2.1)\big\} - 0.13382(y - 2.7718)^2$ \\

\hline \hline
\multirow{5}{*}{\begin{sideways}$\Delta(1232)$\end{sideways}} & $c_0$ & \rule{0ex}{4ex}$\displaystyle 2.9841\exp\left\{-67.857\Big(\frac{y + 3.1272}{1.0 + 0.22831(y + 3.1272)}\Big)^{2}\right\} - 6.5855 - \frac{9.6984}{y} + 0.41256y^{2}$ \\

& $c_1$ & $\displaystyle 6.8276 + 5.2236y + 1.4630y^{2}$ \\

& $c_2$ & $\displaystyle -6.0291 - 6.4581\tanh\big\{5.0830(y + 2.1)\big\} + 0.46352y$ \\

& $c_3$ & $\displaystyle 0.59300 + 0.36093y$ \\

& $c_4$ & $\displaystyle 0.77368 + 0.44776y + 0.056409y^{2}$ \\

\hline \hline
\multirow{5}{*}{\begin{sideways}Res(1600)\end{sideways}} & $d_0$ & \rule{0ex}{4ex}$\displaystyle 1.9186\exp\left\{-56.544\Big(\frac{y + 2.9485}{1.0 + 1.2892(y + 2.9485)}\Big)^{2}\right\} - 0.23720 + 0.041315y^{2}$ \\

& $d_1$ & $\displaystyle -4.9866 - 3.1435y$ \\

& $d_2$ & $\displaystyle -7.0550 - 7.2165\tanh(31.033(y + 2.1)) + 0.38541y$ \\

& $d_3$ & $\displaystyle -2.8915 - 2.1495y - 0.45006y^{2}$ \\

& $d_4$ & $\displaystyle -1.2970 - 0.13947y - 0.41197y^{2} - 0.10641y^{3}$ \\
\hline
\end{tabular}}
\caption{\label{t:kamae-pos} Kamae et al. parameters describing positron spectra for arbitrary proton energy.}
\end{tab}
%%%
%%%
%%%
%%%

\subsection{Production uncertainties}

Since each inclusive \CS\ parameterization is based on physical assumptions and it reproduces experimental data, there are uncertainties related to parameter determination, which modify the asymptotical behavior at low and high energy ranges.\\

As it can be observed in \citefig{f:crs-comp}, at different proton energies, positron and electron \CS\ for the three parameterizations are closely similar in behavior. However, there are variations up to 80\% at proton energies of 20 \tu{GeV},  as in the case of Kamae et al. versus Tan and Ng parameterizations. Another feature is that Kamae's parameterization estimate a smaller electron \CS\ respect to the other two parameterizations.\\ 

% The low statistics for pions and kaons at very low energy produce that Badhwar's and Tan's parameterizations do not well behave for $\kin{p} < 6 \ \tu{GeV}$, although, total inclusive \CS's are in agreement with experimental data. 

Due to the low statistics at very low energy, Badhwar's and Tan's parametrizations tend to produce unphysical distributions for proton kinetic energies below $6 \ \tu{GeV}$. Nevertheless, the total inclusive \CS\ -- the integrated version of those -- are still in agreetment with the available experimental data. To fix this undesiderable feature, both parameterizations are patched by doing a smooth transition from 3 \tu{GeV} until 7 \tu{GeV} with the Stecker's model. Let's clarify that Kamae's parameterization also includes that feature, but considering more resonances.\\

Moreover, for proton energies above 100~\tu{GeV}, Badhwar's parameterization becomes unstable specially for the electron \CS\ case.\\

\begin{fig}
 \resizebox{\hsize}{!}{\includegraphics[angle=270, width=0.5\textwidth]{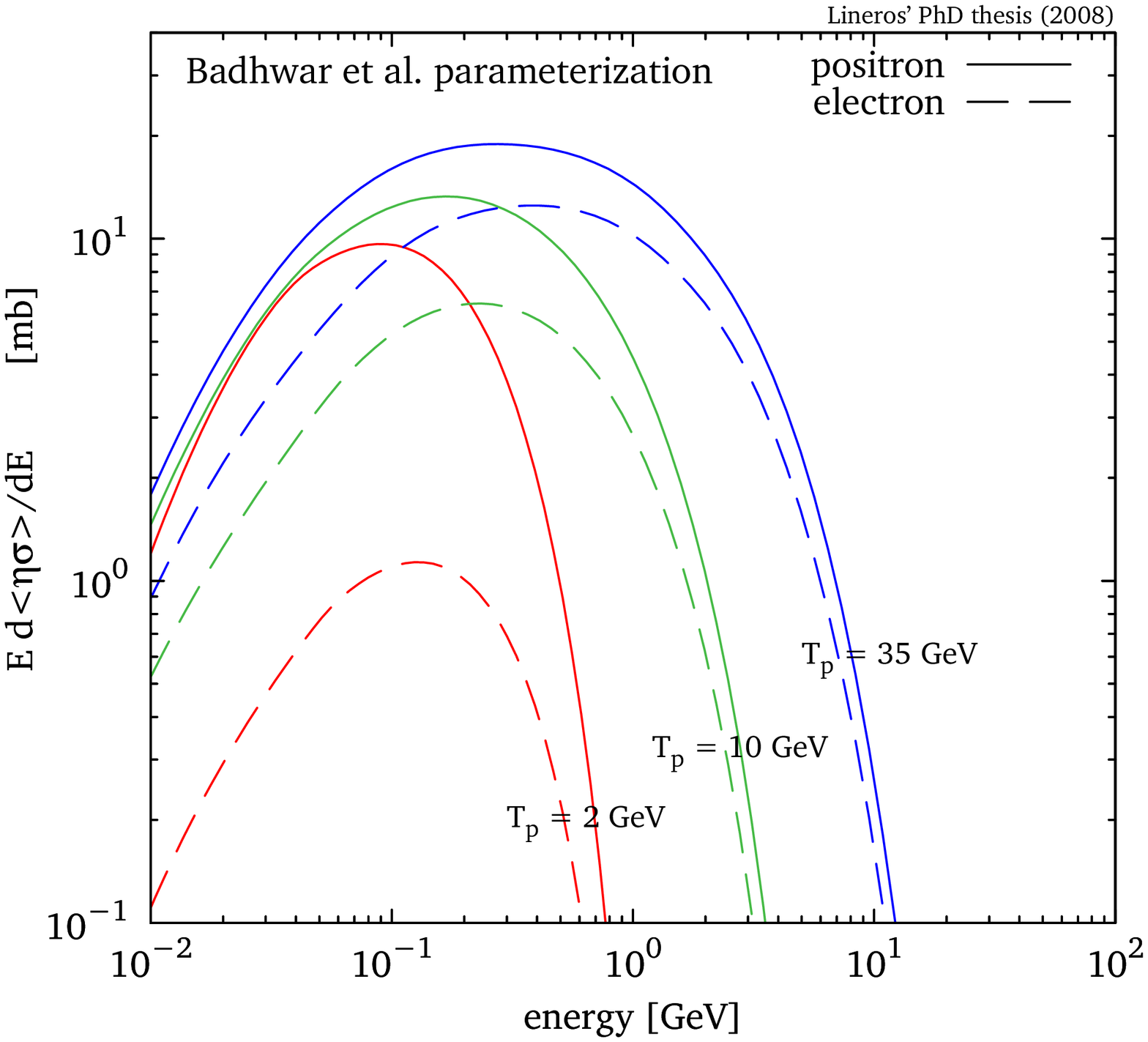} \includegraphics[angle=270, width=0.5\textwidth]{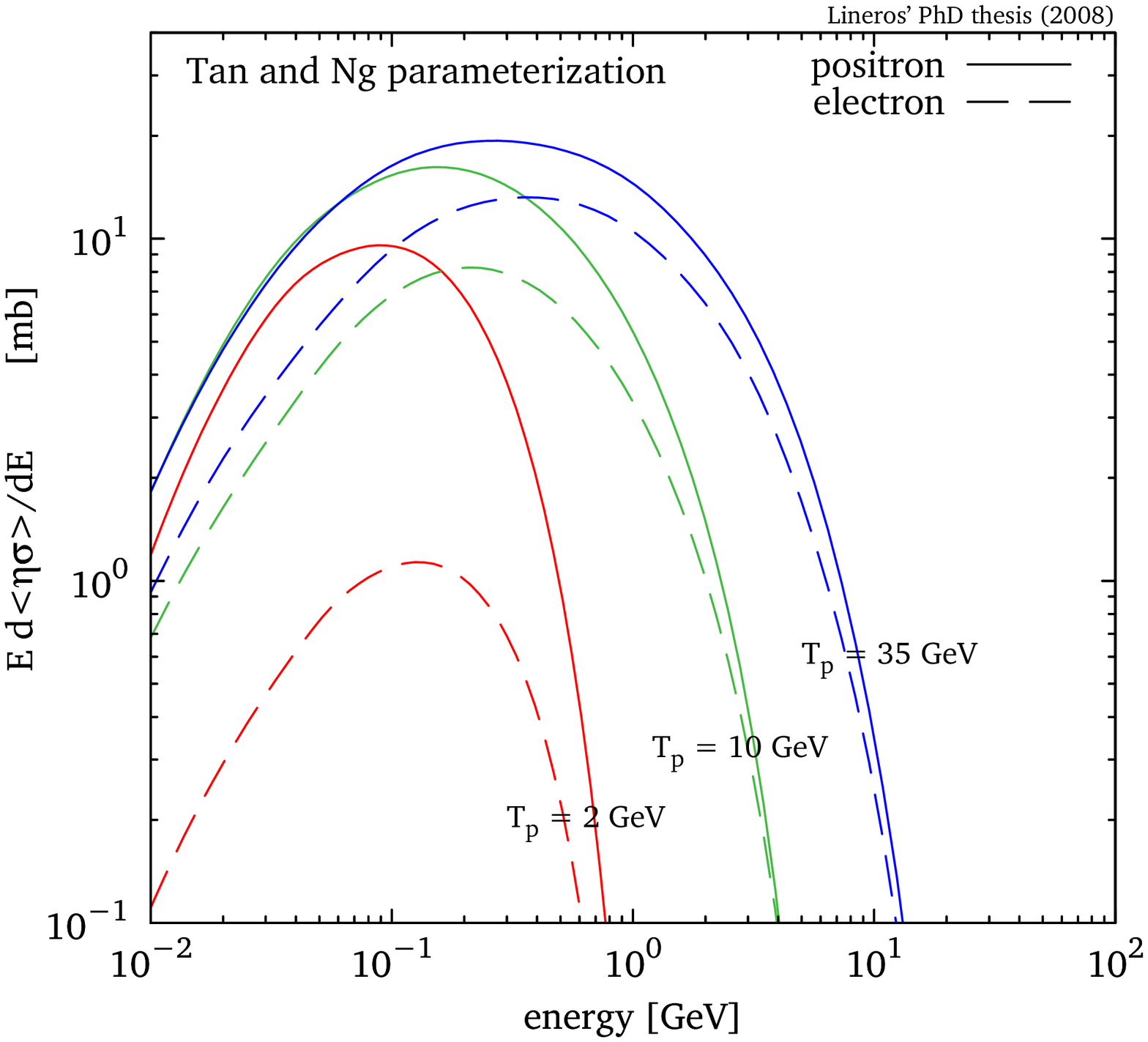}}\\
 \resizebox{\hsize}{!}{\includegraphics[angle=270, width=0.5\textwidth]{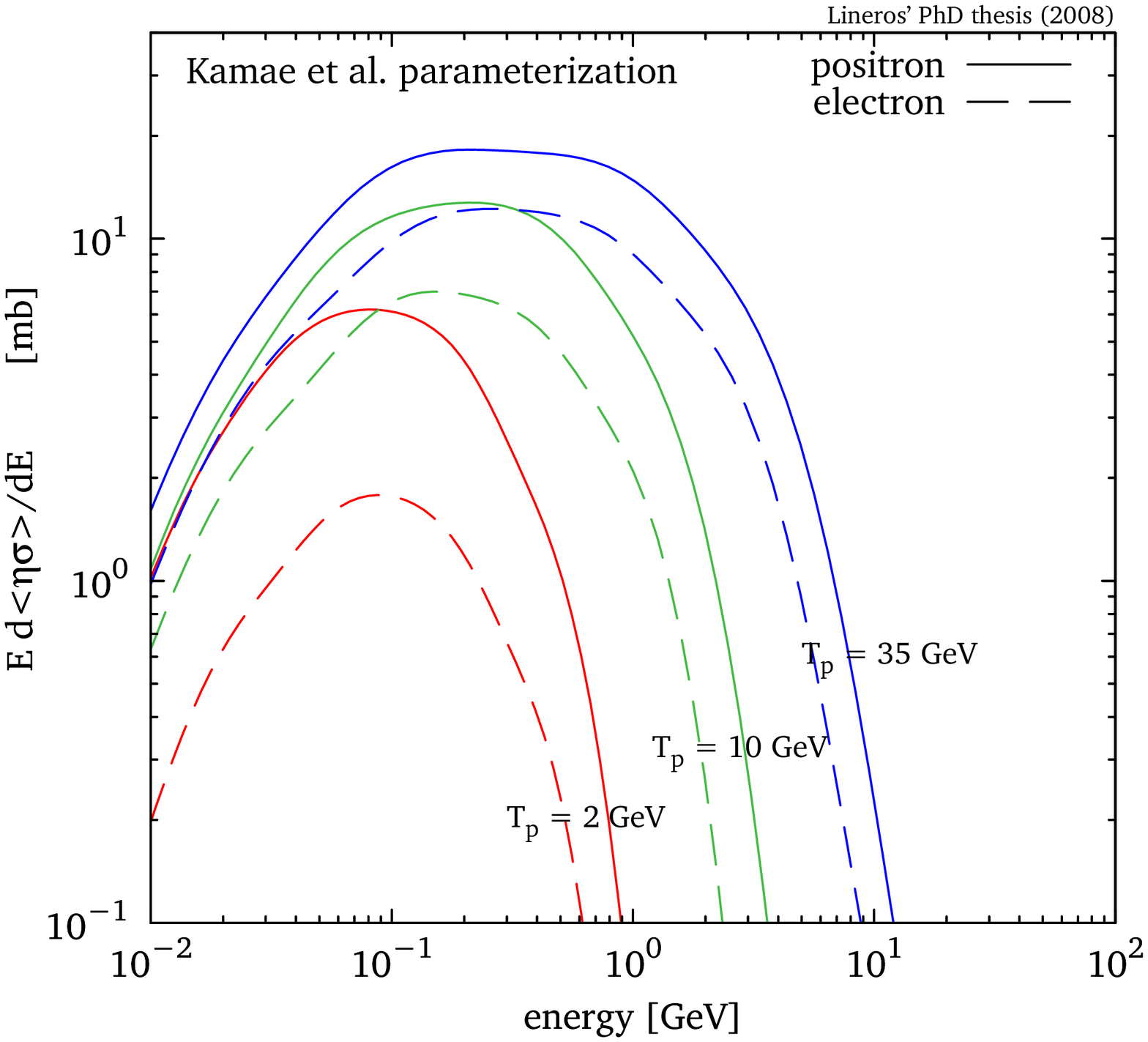} \includegraphics[angle=270, width=0.5\textwidth]{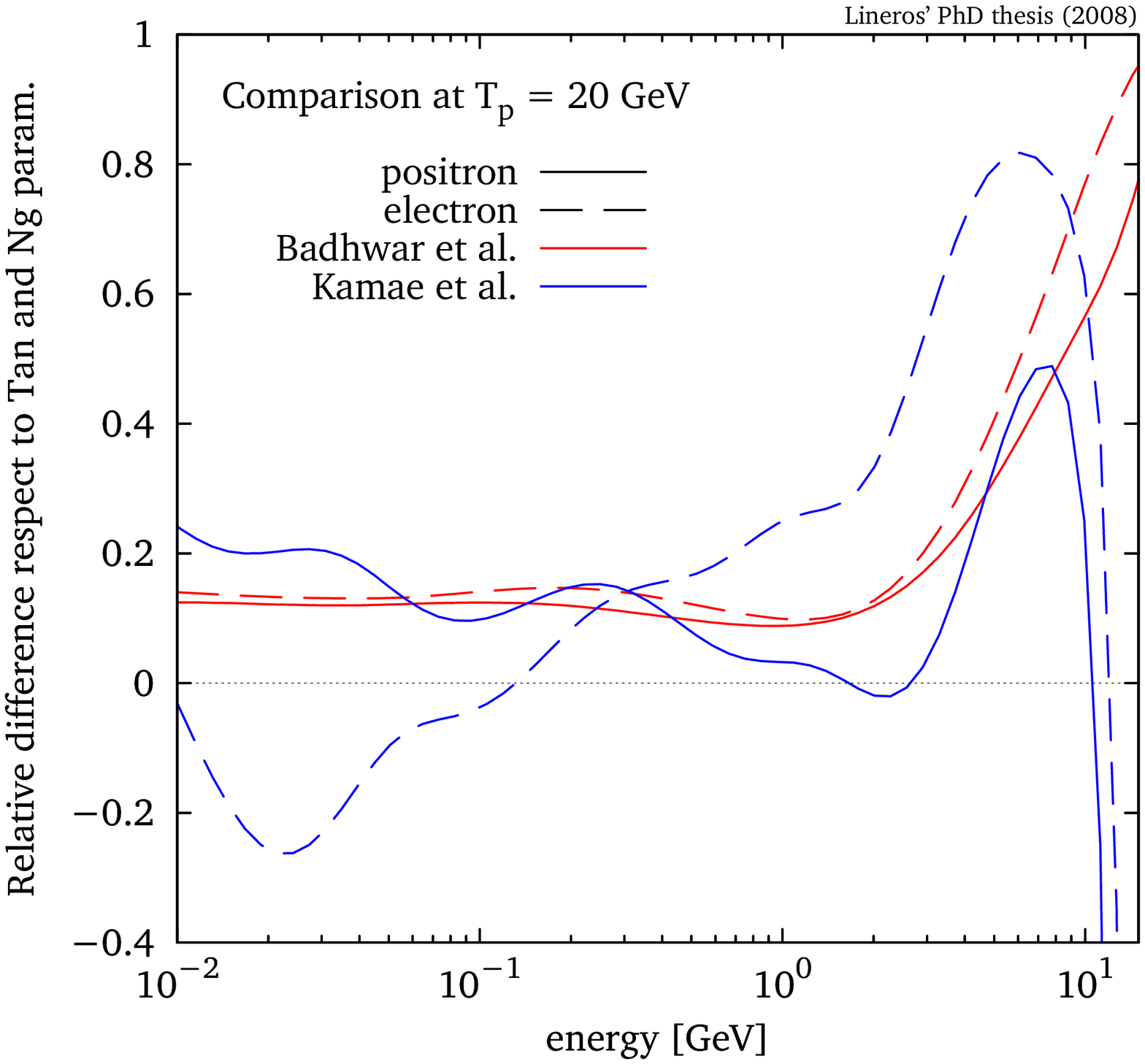}}
\caption{\label{f:crs-comp} Plots of positron and electron inclusive \CS\ versus energy for proton energies of 2, 10 and 35 \tu{GeV} are shown. Each plot is obtained by using Badhwar et al.~\cite{Badhwar:1977zf}, Tan and Ng~\cite{Tan:1984ha}, and Kamae et al.~\cite{Kamae:2004xx} parameterizations. Also the relative difference respect to Tan and Ng solution shows a mean difference around 15\% and 25\% for Badhwar et al. and Kamae et al. cases respectively.}
\end{fig}

\section{Production in annihilation processes}

Another phenomenon where positron and electron production takes place is in annihilation processes. Those particles can be produced directly and/or from the subsequent decays of other particles. For example, a muon eventualy produce an electron. As well, similar situation would happen when a quark hadronizes.\\
\begin{fig}
	\includegraphics[width=0.5\textwidth]{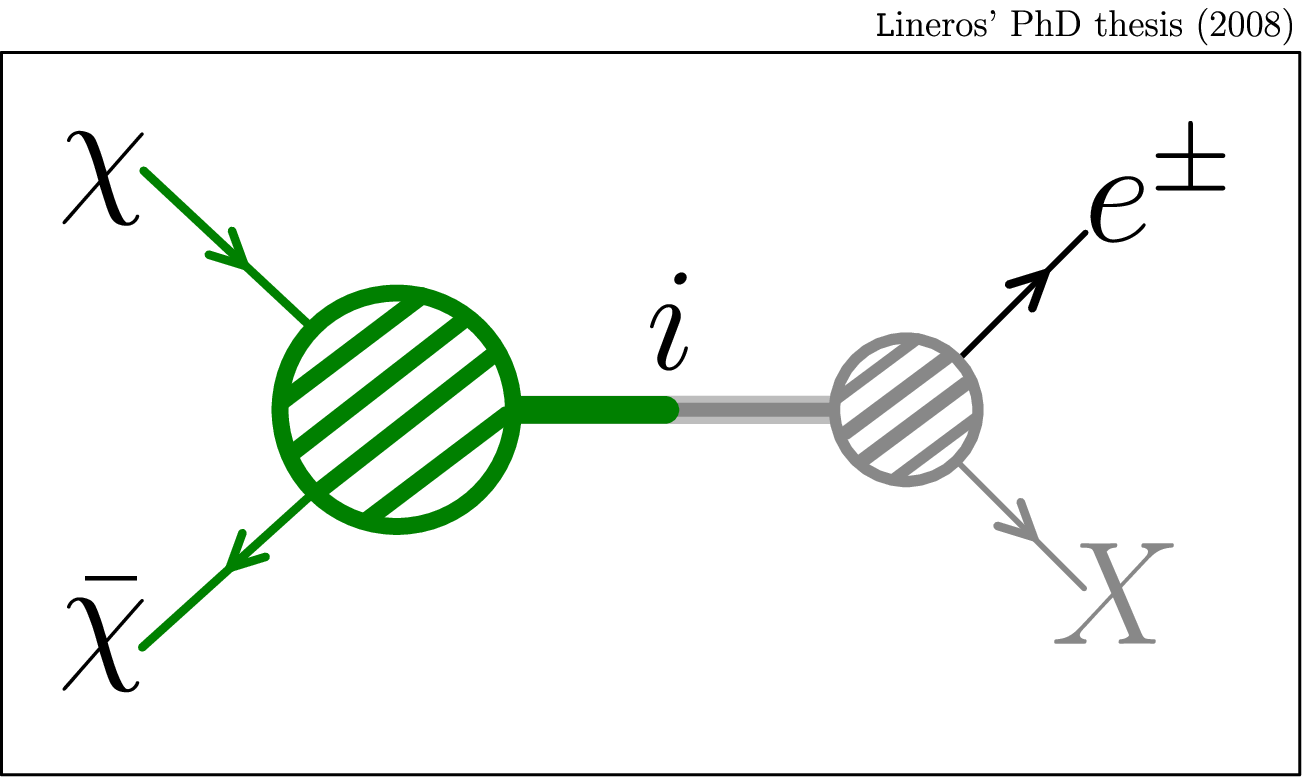}
\caption{\label{f:dm_ann1} Annihilation event which produces positrons and/or electrons is labeled by an intermediate state $i$.}
\end{fig}

An annihilation process is described by specific rules, which lay on some particle physics theory, for example, Supersymmetric models describe the Neutralino \dm\ physics \cite{Krauss:1985Ap, Gondolo:1990Nu, Griest:1990dm}. Other types of annihilations are not out of that situation, in fact, the annihilation can be labeled in terms of outgoing particles (\citefig{f:dm_ann1}). \\

In that sense, the positron (electron) multiplicity distribution (\MD), \ie the number of positrons (electrons) per unit of energy produced in a single annihilation event, is:
\begin{eq}
 \left(\frac{dn_e}{d\ener{}}\right)_{\chi \bar{\chi} \rightarrow e X} = \sum_{i} \tn{BR}\big(\chi \bar{\chi} \rightarrow i\big) \  \left(\frac{dn_e}{d\ener{}}\right)_{i \rightarrow e X} ,
\end{eq}
and it depends on branching ratios of intermediate states $i$, 
\begin{eq}
 \tn{BR}\big(\chi \bar{\chi} \rightarrow i\big) = \frac{\sigma\big(\chi \bar{\chi} \rightarrow i\big)}{\sigma_{\tn{total}}}.
\end{eq}
Note that those are described by the theory which describes the annihilation process. Also, it is needed the \MD\ that the $i$--state can generate, which is generally described by \SM\ processes.\\ 

In astrophysical scenarios, it is expected that \dm\ particles move at non--relativistic speeds, \ie \ecm\ should be closely equal to the double of \dm\ mass. Related to this, annihilation products should be composed by lighter particles which will correspond to \SM\ ones.\\

In the case of \dm\ annihilations, intermediate states are described by the production of two \SM\ particles which keep conserved quantities as in the annihilation; those are electrically neutral, colorless and unpolarized processes.\\ 

The \MD's can be calculated from analytical expressions in the simplest cases, but in complex ones, it is needed to use more sophisticated methods.\\

\subsection{Generation of Multiplicity Distributions}

Multiplicity Distributions for positrons and electrons are classified through their total electrical charge. As well, each one is function of \ecm\ and kinetic energy of the target particle, in our case positrons and electrons.\\

The method to generate them employed here is based mainly on the Lund's PYTHIA montecarlo generator \cite{Sjostrand:2000wi}, which is used for simulating decay chains and hadronization processes. With PYTHIA it is possible to generate a basic set of intermediate states (\citetab{t:md-pythia}), even though, extended set of states can be composed from the PYTHIA's ones. \\

\begin{tab}
	\begin{tabular}{|c|c|c|c|}
	\hline
	& \multicolumn{3}{|c|}{Intermediate state} \\
	\hline \hline
	Charge & Leptons & Quarks & Gauge Bosons \\
	\hline
	+1 & $(\nue e^{+}) \ (\numu \mu^{+}) \ (\nutau \tau^{+})$ & $(u\bar{d})\ (c\bar{s})\ (t\bar{b})$ & $(\gamma W^{+})\ (Z W^{+})$ \\
	0 & $(e^{-} e^{+})\ (\mu^{-} \mu^{+})\ (\tau^{-} \tau^{+})$ & $(u\bar{u}) \ (d\bar{d}) \ (c\bar{c}) \ (s\bar{s}) \ (b\bar{b})\ (t\bar{t}) $ & $(gg) \ (ZZ) \ (W^{-} W^{+}) $\\ 
	-1 & $(e^{-} \nuebar) \ (\mu^{-} \numubar) \ (\tau^{-} \nutaubar )$ & $(d\bar{u})\ (s\bar{c})\ (b\bar{t})$ & $(W^{-} \gamma)\ (W^{-} Z)$ \\
	\hline
	\end{tabular}
	\caption{\label{t:md-pythia} Intermediate states generated with PYTHIA for positron and electron \MD's.}
\end{tab}

\subsubsection{Generation from PYTHIA}

PYTHIA is used to simulate decay chains and hadronization processes from an initial (intermediate) state. Not all possible PYTHIA's initial states may produce positrons and electrons. \citetab{t:md-pythia} lists all processes; however, some states as $(e^{+} e^{-})$, $(\mu^{+} \mu^{-})$  and its combinations with neutrinos will produce \MD's which are obtained from analytical expressions (See \citeapp{app2}).\\

The method for \MD\ generation is based on histogram creation from simulated events. Histograms correspond to number of positrons (electrons) per kinetic energy bin, where each bin follows a Poisson distribution. Moreover, the number of total positrons (electrons) produced is $N_{\tn{part}} = 10^8$ to reduce as maximum as possible statistical uncertainties. As well, to reduce even more possible uncertainties each histogram is done using the Method of Average (See \citeapp{app5})\\

An extra feature, which is not included in the original PYTHIA program, is the addition of muon's polarization effect (See \citeapp{app2}). This is performed by a selection of muons -- in generated events -- which have been produced from spin--0 particles, as mesons and kaons (See \citeapp{app3} and \citeapp{app4}). This effect produces differences up to 10\% between positron and electron \MD's at the same kinetic energy.\\

Each state has been simulated for different \ecm. The energy range covered is from the threshold energy till 20 \tu{TeV}. In that way, most of possible interesting \dm\ mass range is covered. Although there are special values of \ecm, 
\begin{eq}
 \roots_{\tn{special}} = \mass{\tau},\ 2\mass{\tau},\ \mass{b},\ 2\mass{b},\ \mass{W},\ \mass{Z},\ 2\mass{W},\ \mass{t},\ 2\mass{Z},\ 2\mass{t},
\end{eq}
which are important keys in composed states production.\\

We generated 60 \MD's - in  the range previously described - for each intermediate state. Furthermore, interpolations are performed for the positron (electron) energy and \ecm. Interpolation variables are $x$ and $\tau$:
\begin{eq}
 x = \frac{\ener{e}}{\roots}, \quad \tau = \log\left(\frac{\roots}{1\tu{ GeV}}\right);
\end{eq}
variable $x$ usually goes from 0 to $\frac{1}{2}$. The interpolation is performed in two steps, where the first one is on $x$ and then on $\tau$. Let's specify that instead of interpolating the \MD\ values we interpolate the logarithm of this. In that way, the interpolation is stable and correctly reproduces \MD's for intermediate values of $x$ and $\tau$.\\

\begin{fig}
	\resizebox{\hsize}{!}{\includegraphics[angle=270, width=0.5\textwidth]{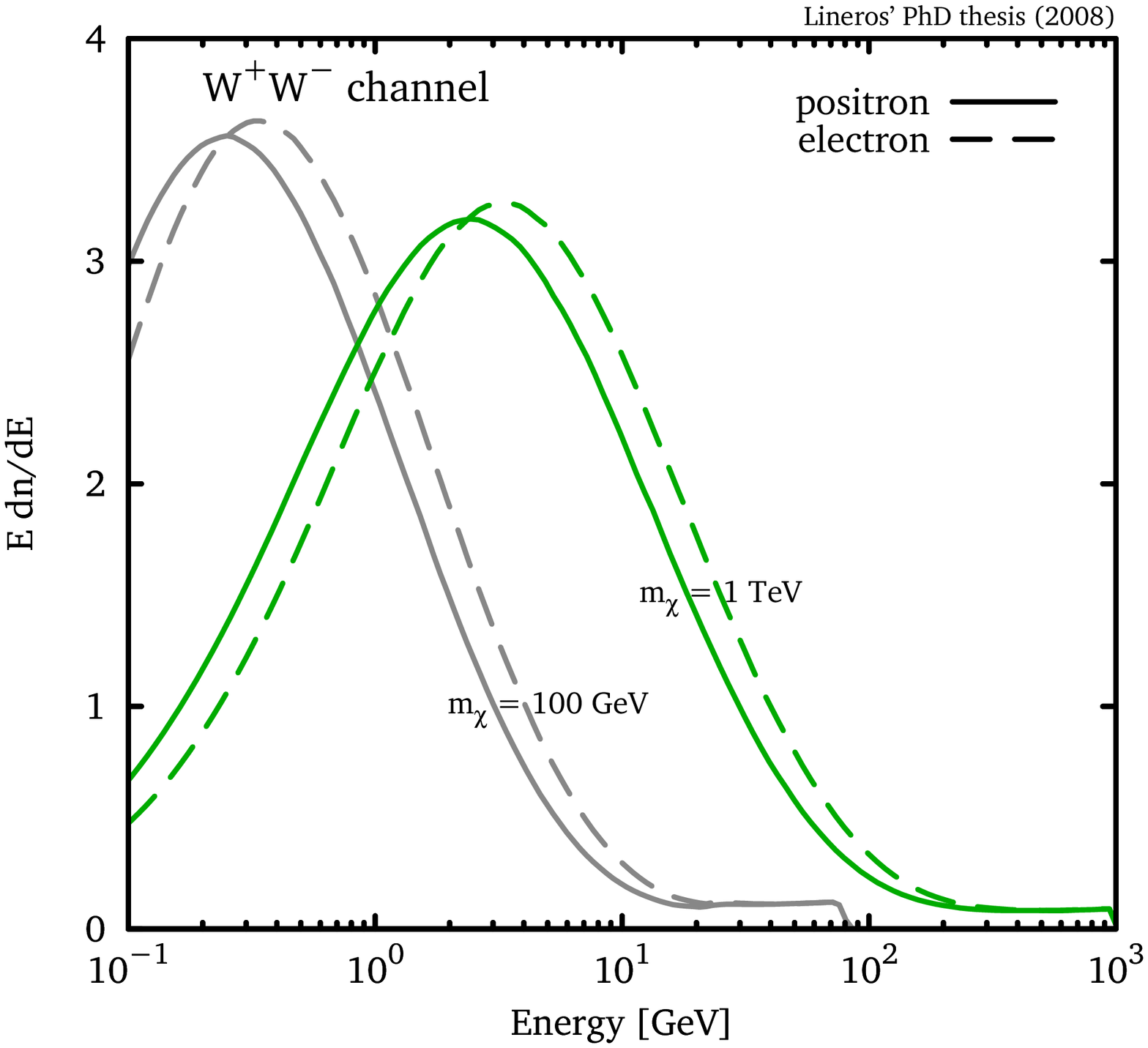}\includegraphics[angle=270, width=0.5\textwidth]{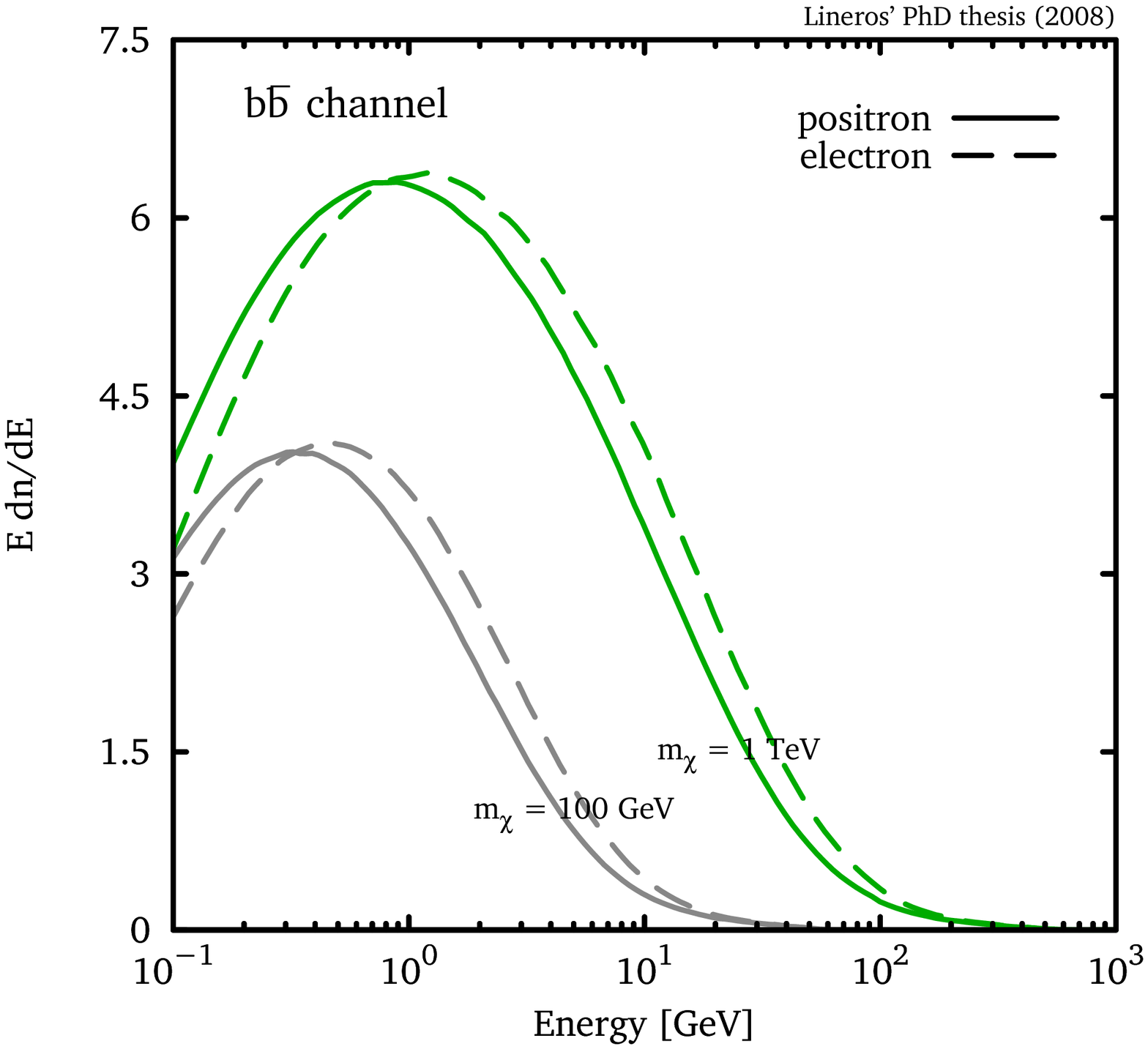}}
	\resizebox{\hsize}{!}{\includegraphics[angle=270, width=0.5\textwidth]{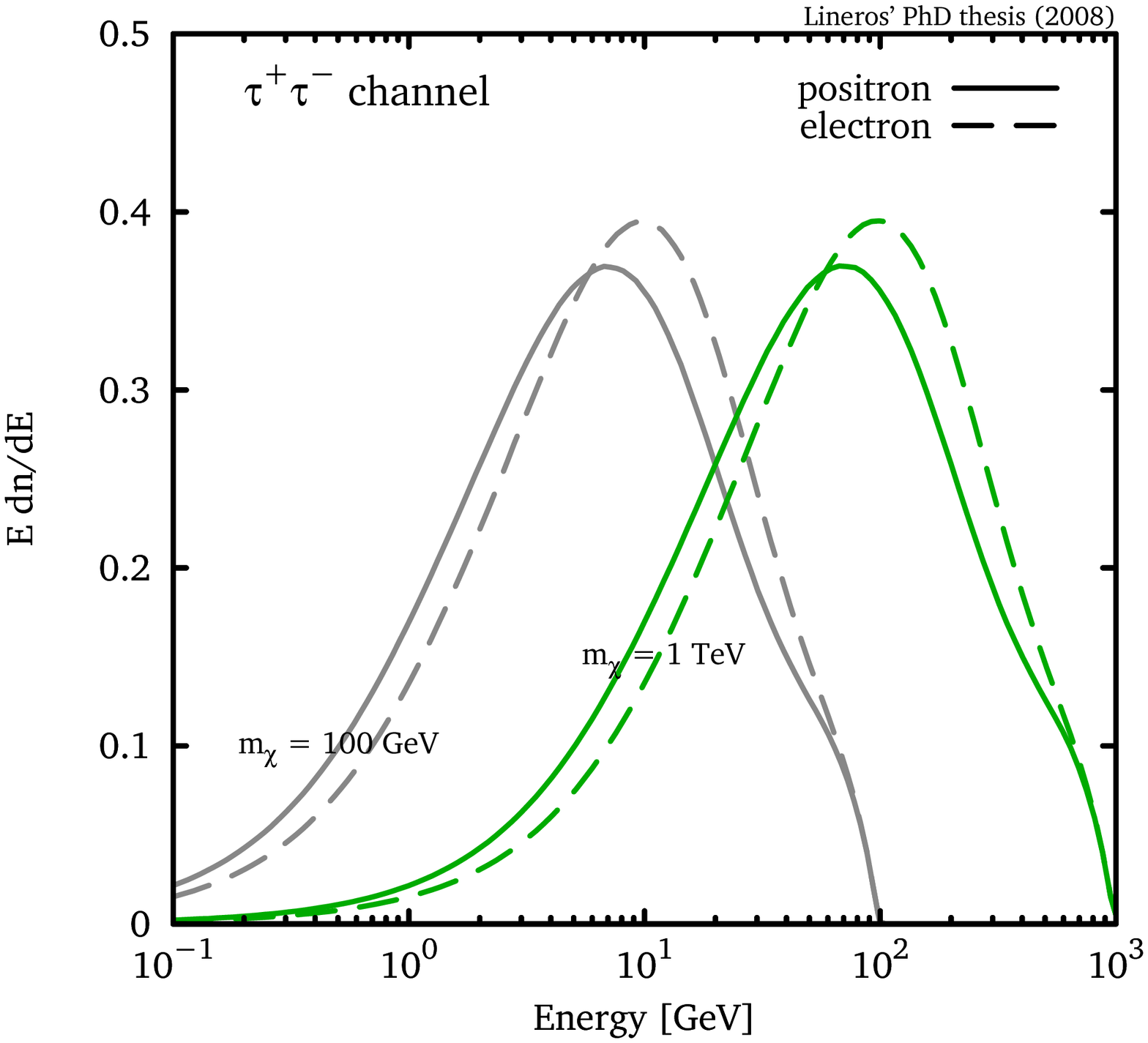}\includegraphics[angle=270, width=0.5\textwidth]{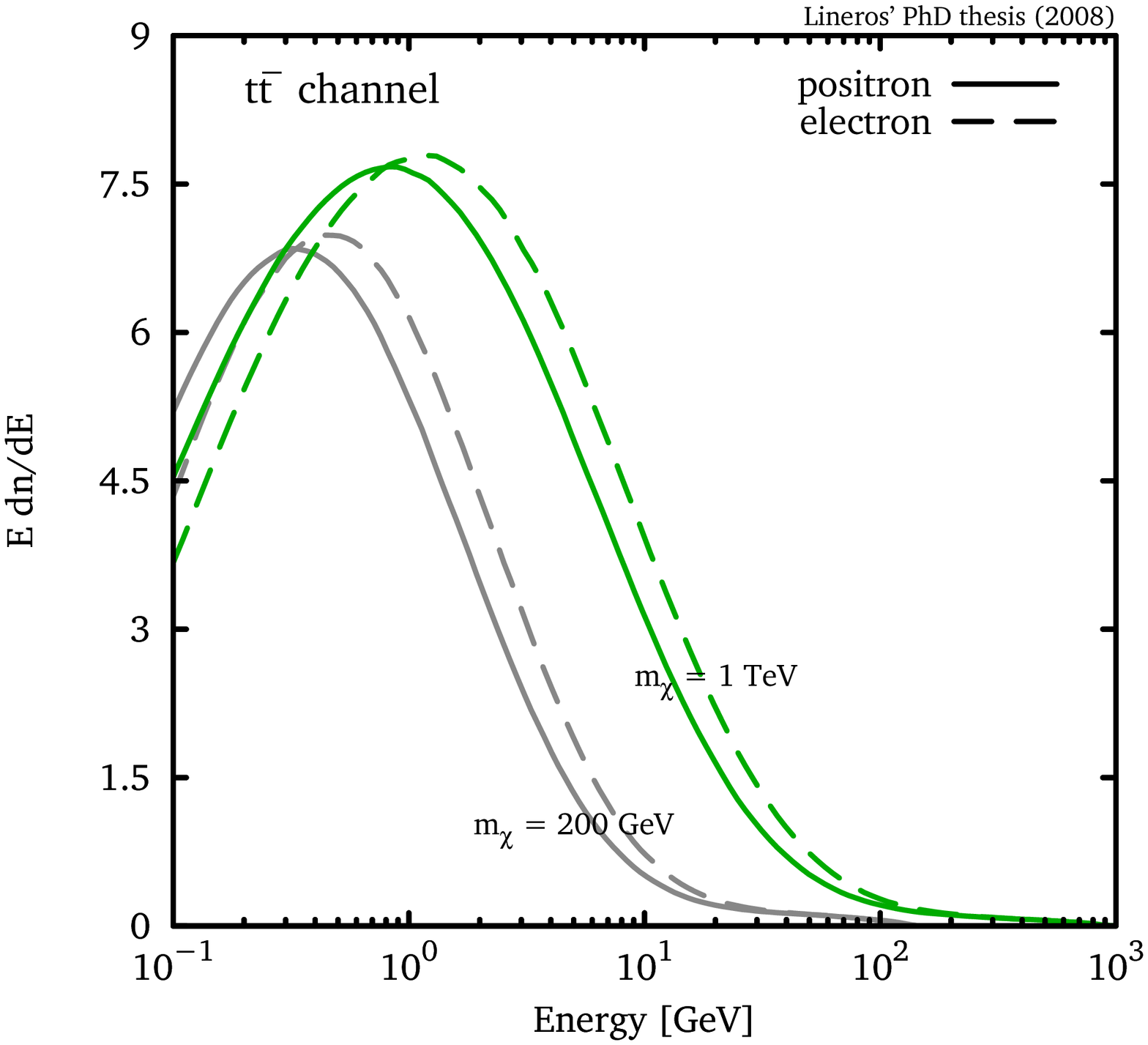}}
	\caption{\label{f:md-pythia} Multiplicity distribution for positrons and electrons obtained from PYTHIA. Plots made for \dm\ masses of 100 \tu{GeV} and 1 \tu{TeV} for $W^{+}W^{-}$, $\tau^{+}\tau^{-}$ and $b\bar{b}$ intermediate states (channels) and the $t\bar{t}$ channels with masses of 200 \tu{GeV} and 1 \tu{TeV}.}
\end{fig}

%
%
%% subfile for SM and THDM subsubsection
%%%%%%%%%%%%%%%%%%%%%%%%%%%%%%%%%%%%%%%%%%%%%%%%%%%%%%%%%%%%%%%%%%%%%%%%%%%%%%%%%%%%
\subsubsection{Standard Model composed states}

Montecarlo generator programs, as PYTHIA, are useful to construct \MD's, however, there are still states where those are not an efficient choice. For those cases, a solution is to compound them by using already known \MD's.\\

\begin{tab}
	\begin{tabular}{|c|c|c|}
		\hline
		& \multicolumn{2}{|c|}{\SM\ intermediate state} \\
		\hline \hline
		Charge & Higgs--Higgs & Higgs--Gauge \\
		\hline
		+1	&  -- & $(h W^{+})$ \\
		0 & $(h h)$ & $(h \gamma)\ (h Z)$ \\
		-1 &  -- & $(W^{-} h)$ \\
		\hline
	\end{tabular}
	\caption{\label{t:md-sm} Composed states for \SM.}
\end{tab}

In the \SM\ particles context, \MD\ related to higgs particles (\citetab{t:md-sm}) should be composed because their mass value is still unknown. The composition is based on higgs 2 body decay modes, which are dominant. \\

Decay widths, in which higgs goes to a couple of leptons or quarks, are computed from \SM\ Feynman rules:\\
\begin{eq}
 \Gamma(h\rightarrow f\bar{f}) = \frac{N_c g^2 \mass{f}^{2}}{32 \pi \mass{W}^{2}} \mass{h} \left(1 - \frac{4 \mass{f}^{2}}{\mass{h}^{2}} \right)^{3/2} , 
\end{eq}
where $N_c$ is the number of color and it takes values 1 for leptons and 3 for quarks.\\

Higgs can also decay into gauge bosons:
\begin{eq}
	\Gamma(h\rightarrow ZZ) &=&  \frac{g^2}{128 \pi} \frac{\mass{h}^3}{\mass{W}^2} \sqrt{1-x_Z} \left(1- x_Z + \frac{3}{4}x_Z^2\right) \\
	\Gamma(h\rightarrow W^{+}W^{-}) &=& \frac{g^2}{64 \pi} \frac{\mass{h}^3}{\mass{W}^2} \sqrt{1-x_W} \left( 1 - x_{W} + \frac{3}{4}x_{W}^2 \right)
\end{eq}
where $\displaystyle x_{Z/W} = 4 \mass{Z/W}^2 / \mass{h}^2$.\\

There are extra decays modes, such as $h \rightarrow \gamma \gamma$, $\gamma Z$ and $g g$. However those are one--loop processes, and can be safely neglected in most of cases \cite{Gunion:1990}.\\

The positron (electron) \MD\ for a single higgs at rest is composed by:\\
\begin{eq}
 \left(\dnde{e}\right)_{h}(\ener{e},\mass{h}) &=& \sum \tn{BR}(h\rightarrow i) \left(\dnde{e}\right)_{i \rightarrow e^{\pm}X}(\ener{e},\roots = \mass{h}) ,
\end{eq}
where $i$-states correspond to $l\bar{l}$, $q\bar{q}$, $ZZ$ or any allowed decay mode. To compose the higgs' \MD\ the \ecm\ should be equal to the higgs mass.\\

Nevertheless, it is almost improbable to produce a higgs at rest. To generalize the situation, the higgs \MD\ should be boosted to a reference frame where its energy matches the production energy. A proper procedure is to use the boosted spectra formalism (\citeapp{app1}), where for positrons and electrons is:
\begin{eq}
 \left(\dnde{e}\right)_{h}(\ener{e},\ener{h}) = \frac{1}{2\gamma\beta} \int_{\xi^{-}}^{\xi^{+}} \frac{d\xi_e}{\xi_e} \left(\dnde{e}\right)_{h}(\xi_{e},\mass{h}) ,
\end{eq}
where $\gamma = \ener{h}/\mass{h}$ and the integration limits are:
\begin{eq}
 \xi^{+} &=& \min\Big(\xi_{\max}, (\gamma + \beta)\ener{e})\Big) , \\
 \xi^{-} &=& (\gamma - \beta)\ener{e} . \nonumber
\end{eq}
Note that $\xi_{\max}$ is the maximum allowable energy in the higgs rest frame and it also helps to calculate the maximum energy in the boosted frame, 
\begin{eq}
 \ener{e,\max} = (\gamma + \beta)\xi_{\max}. 
\end{eq} \newline

Two--particle states \MD\ are created as direct sum of former single--particle \MD:
\begin{eq}
 \left(\dnde{e}\right)_{AB}(\ener{e},\roots) = \left(\dnde{e}\right)_{A}(\ener{e}, \ener{A}) + \left(\dnde{e}\right)_{B}(\ener{e},\ener{B}),
\end{eq}
each single-particle state is set to conserve energy and momentum: 
\begin{eq}
 \ener{A} = \frac{s +\mass{A}^2 - \mass{B}^2}{2\roots} & \tn{and} & \ener{B} = \frac{s +\mass{B}^2 - \mass{A}^2}{2\roots}.
\end{eq}\newline

For example, the state $hh$ is the simplest to be computed, as it is formed by two identical particles. Then, the positron (electron) \MD\ is:
\begin{eq}
 \left(\dnde{e}\right)_{hh}(\ener{e},\roots) = 2 \times  \left(\dnde{e}\right)_{h}(\ener{e}, \ener{h}=\frac{\roots}{2}), 
\end{eq}
that is just two times a single higgs \MD. \\

A different situation happens for states $hZ$ and $hW^{\pm}$. Those states are formed by 2 kind of particles. The $Z$ and $W^{\pm}$ single states are obtained from two--particle states $\gamma Z$ and $\gamma W^{\pm}$ by taking advance of that photons do not produce any positron or electron. The single--particle state and the two--particle one are related as follows:
\begin{eq}
	\left(\dnde{e}\right)_{Z/W}(\ener{e},\ener{Z/W}) =  \left(\dnde{e}\right)_{\gamma Z/\gamma W }(\ener{e},\roots = \ener{Z/W} + \mom{Z/W}) ,
\end{eq}
where $\gamma Z$ and $\gamma W^{\pm}$ are computed with PYTHIA.\\

The final step is to combine them by direct addition:
\begin{eq}
	\left(\dnde{e}\right)_{hZ/hW}(\ener{e},\roots) = \left(\dnde{e}\right)_{h}(\ener{e},\ener{h}) + \left(\dnde{e}\right)_{Z/W}(\ener{e},\ener{Z/W}) , 
\end{eq}
where 
\begin{eq}
 \ener{h} = \frac{s + \mass{h}^2 - \mass{Z/W}^2}{2\roots} &\tn{and}& \ener{Z/W} = \frac{s + \mass{Z/W}^2 - \mass{h}^2}{2\roots}.
\end{eq}
Let's emphasize that this method works well for most of the states which do not need quarks as single--particle \MD's.\\

For example, \citefig{f:md-sm} shows \MD's for composed states $hh$ and $hZ$, assuming a higgs mass value of 125 \tu{GeV}. The $hh$ state produces more positrons and electron than $hZ$ state. That is related to how they couple to fermions, especially quarks, which are major contributors to positron and electron production. Higgs bosons strength interaction is proportional to fermion mass, instead of $Z$ bosons that couple in the same way with all fermions. Then, it is expected that higgs bosons decay mainly in $b\bar{b}$ pairs, where hadronization processes have a strong effect in the production.\\ 

\begin{fig}
	\resizebox{\hsize}{!}{\includegraphics[angle=270, width=0.5\textwidth]{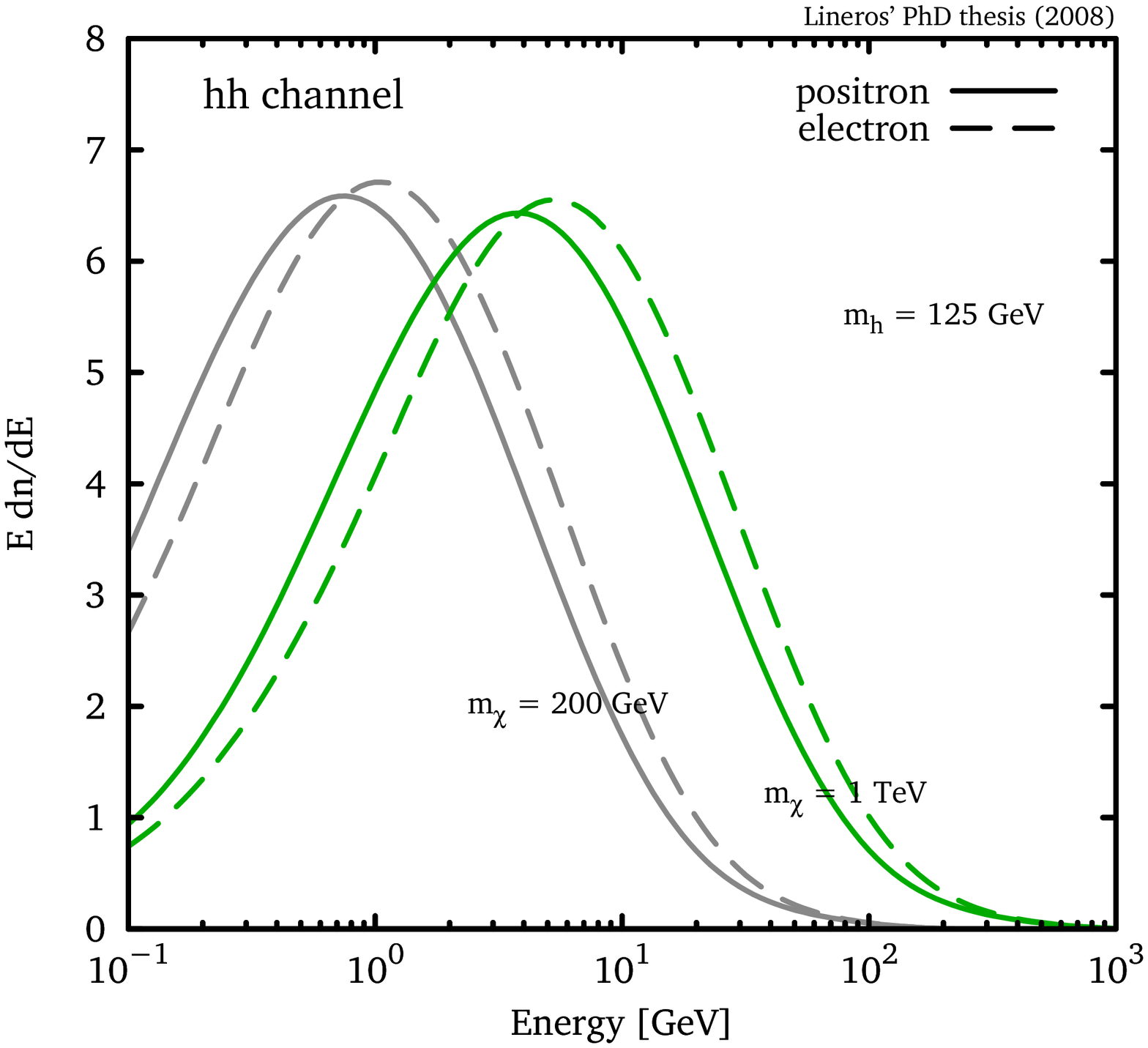}\includegraphics[angle=270, width=0.5\textwidth]{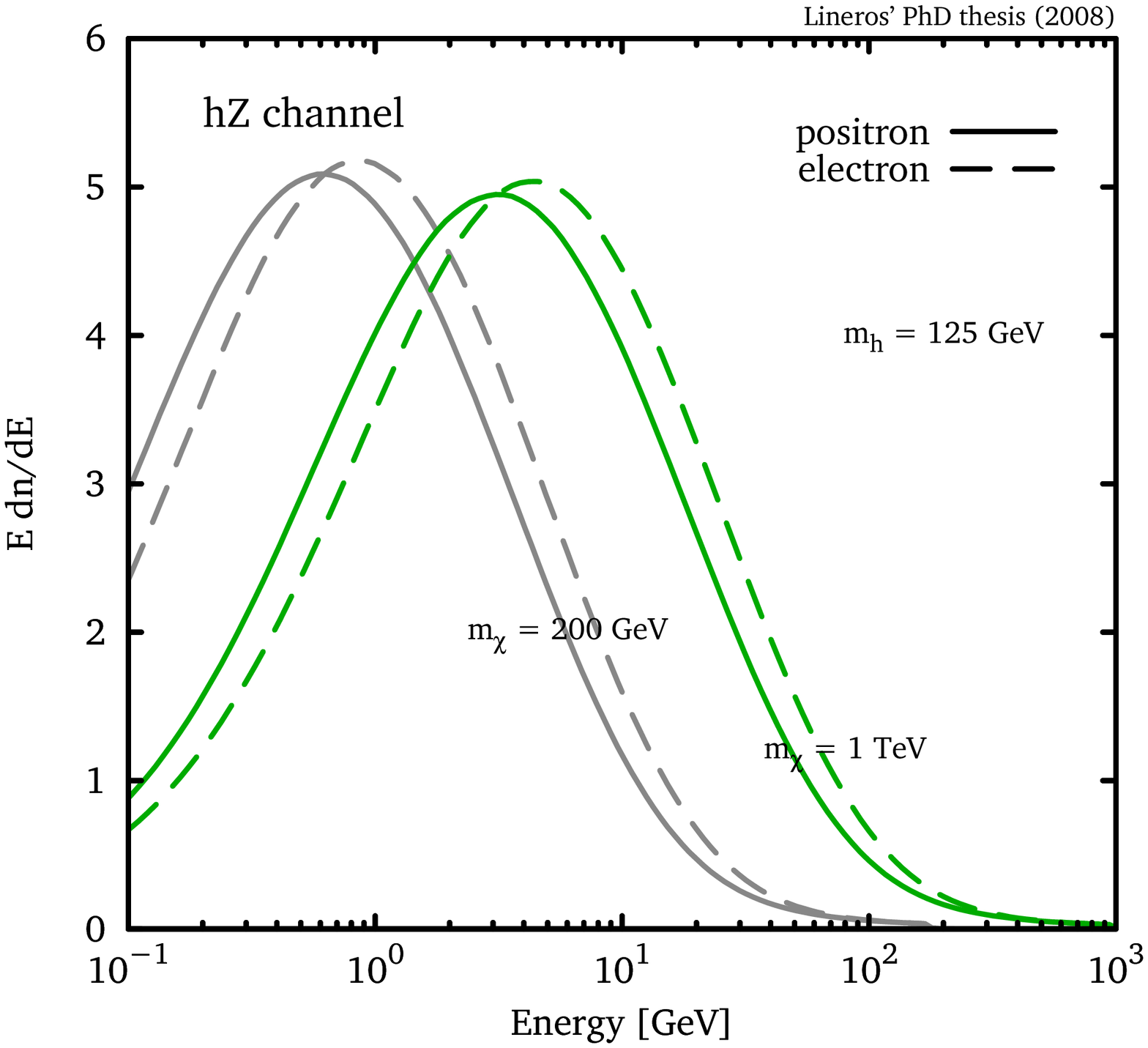}}
	\caption{\label{f:md-sm} Multiplicity distributions versus energy for \SM\ higgs-higgs and higgs-Z composed states. Examples with \dm\ masses of 200 \tu{GeV} and 1 \tu{TeV} are shown.}
\end{fig}

%%%%%%%%%%%%%%%%%%%%%%%%%%%%%%%%%%%%%%%%%%%%%%%%%%%%%%%%%%%%%%%%%%%%%%%%%%%%%%%%%%%%%
\subsubsection{Two Higgs Doublet Model composed states}

As an alternative for the \SM\ higgs sector is the Two Higgs Doublet Model (\THDM) which includes \SM\ fields and two higgs isospin doublet - instead of one. Those may have equal or opposite isospin charge. An advantage of this model is to describe a richer higgs sector as supersymmetric models do. \\

The observable particles are increased in number, where it appears three neutral and one charged higgs fields. This makes possible new \MD's (\citetab{t:md-thdm}). Following similar procedure as \SM\ cases, those are generated in terms of two body decay modes described by the model.\\

Two body decay widths are easily calculated from model's Feynman rules \cite{Gunion:1990}. Charged higgs decay are mainly dominated by decays into heavy fermions: 
\begin{eq}
	\Gamma(H^{+} \rightarrow t\bar{b}) &=& \frac{N_c g^{2} \lambda^{1/2}(\mass{H^{+}},\mass{t},\mass{b})}{32 \pi \mass{W}^2 \mass{H^{+}}^3} \\ &&\times \Big((\mass{H^{+}}^2 - \mass{b}^2 - \mass{t}^2)(\mass{b}^2 \tan^2\beta + \mass{t}^2 \cot^2\beta) - 4 \mass{b}^2 \mass{t}^2 \Big) \nonumber ,
\end{eq}
where $N_c$ is the number of colors, $\tan\beta$ is the ratio between vacuum expectation values of two doublets and $\lambda^{1/2}$ is a kinematical factor,
\begin{eq}
	\lambda^{1/2}(\mass{0}, \mass{1}, \mass{2}) = \sqrt{(\mass{1}^2 + \mass{2}^2 - \mass{3}^2) - 4\mass{1}^2\mass{2}^2} .
\end{eq}
Also there is another decay mode into a $W^{\pm}$ and a lightest neutral higgs:
\begin{eq}
 \Gamma(H^{+}\rightarrow W^{+} h_0) &=& \frac{g^{2} \cos^{2}(\beta-\alpha) \lambda^{1/2}(\mass{H^{+}},\mass{W},\mass{h_0})}{64\pi\mass{H^{+}}^3}  \\ 
&& \times \left( \mass{W}^2 - 2(\mass{H^{+}}^2 + \mass{h_0}^2) + \frac{(\mass{H^{+}}^2 - \mass{h_0}^2)^2}{\mass{W}^{2}}\right) , \nonumber
\end{eq}
where $\alpha$ is the mixing angle among neutral higgs fields. Furthermore, this mode is less important than previous ones because some regions in the parameter space produce physical configurations, for example charged higgs mass has to be bigger than the sum of lightest higgs and bottom quark masses.\\

Decay widths for neutral higgs are quite similar to \SM\ ones. These can be written in a compact way \cite{Gunion:1990}:
\begin{eq}
	\Gamma(h\rightarrow t\bar{t}) = \frac{N_c g^2 \mass{t}^2 d_h^2 \mass{h}}{32 \pi \mass{W}^2 \sin^2\beta} \left(1 - \frac{4 \mass{t}^2}{\mass{h}^2} \right)^{p},
\end{eq}
\begin{eq}
	\Gamma(h\rightarrow b\bar{b}) = \frac{N_c g^2 \mass{t}^2 e_h^2 \mass{h}}{32 \pi \mass{W}^2 \cos^2\beta} \left(1 - \frac{4 \mass{b}^2}{\mass{h}^2} \right)^{p},
\end{eq}
where $d_h$ and $e_h$ are defined as:
\begin{eq}
 d_h = \left\{\begin{array}{cc} -\sin\alpha & h=H_{0} \\ \cos\alpha & h=h_0 \\ \cos\beta & h=A_0 \end{array} \right. , \
 e_h = \left\{ \begin{array}{cc} \cos\alpha & h=H_0 \\ \sin\alpha & h=h_0 \\ -\sin\beta & h=A_0 \end{array}\right. ,
\end{eq}
and the power index $p$ is:
\begin{eq}
	p = \left\{ \begin{array}{cc} 3/2 & h = h_0 , H_0 \\ 1/2 & h = A_0 \end{array}\right. .
\end{eq}\newline

With most important decay widths already calculated, branching ratios are also known. Those help to weight each \MD\ for composing a single higgs \MD:
\begin{eq}
 \left(\frac{dn}{d\ener{e}}\right)_{h} = \sum_i \tn{BR}(h \rightarrow i)  \left(\frac{dn}{d\ener{e}}\right)_{i},
\end{eq}
where $h$ could be $h_0$, $H_0$, $A_0$ or $H^{\pm}$. Note that charged intermediate states are specially useful for computing charged higgs \MD, which increase accuracy of the final \MD.\\

Depending on the \dm\ candidate, different kind of higgs particles are involved. Comparing \SM\ with neutral \THDM\ higgs fields at equivalent production conditions (\citefig{f:md-thdm}), we see how \SM\ higgs produces more positrons and electrons, instead $h_0$ and $A_0$ higgses, for higher values of $\tan\beta$. This is related to neutral higgs field couplings, which depends directly of parameters $\beta$ and $\alpha$.

\begin{tab}
	\begin{tabular}{|c|c|c|}
		\hline
		& \multicolumn{2}{|c|}{\THDM\ intermediate state} \\
		\hline \hline
		Charge & Higgs--Higgs & Higgs--Gauge \\
		\hline
		\multirow{2}{*}{+1}	&  \multirow{2}{*}{$(h_0 H^{+})\ (H_0 H^{+})\ (A_0 H^{+})$}& $(h_0 W^{+})\ (H_0 W^{+})\ (A_0 W^{+}) $ \\ 
		& & $(\gamma H^{+})\ (Z H^{+})$ \\
		%\hline
		\multirow{2}{*}{0} &  $(h_0 h_0)\ (H_0 H_0)\ (A_0 A_0)$  & \rule{0ex}{4ex}$(W^{+} H^{-})\ (H^{+} W^{-})\ (h_0 \gamma)\ (H_0 \gamma)$ \\
		&  $(h_0 A_0)\ (H_0 A_0)\ (H^{+}H^{-})$ & $(A_0 \gamma)\ (h_0 Z)\ (H_0 Z)\ (A_0 Z)$\\
		%\hline
		\multirow{2}{*}{-1} & \multirow{2}{*}{$(h_0 H^{-})\ (H_0 H^{-})\ (A_0 H^{-})$} & \rule{0ex}{4ex}$(h_0 W^{-})\ (H_0 W^{-})\ (A_0 W^{-})$ \\
		&& $(\gamma H^{-})\ (Z H^{-})$\\
		\hline
	\end{tabular}
	\caption{\label{t:md-thdm} Composed states for \THDM.}
\end{tab}

\begin{fig}
	\resizebox{\hsize}{!}{\includegraphics[angle=270, width=0.5\textwidth]{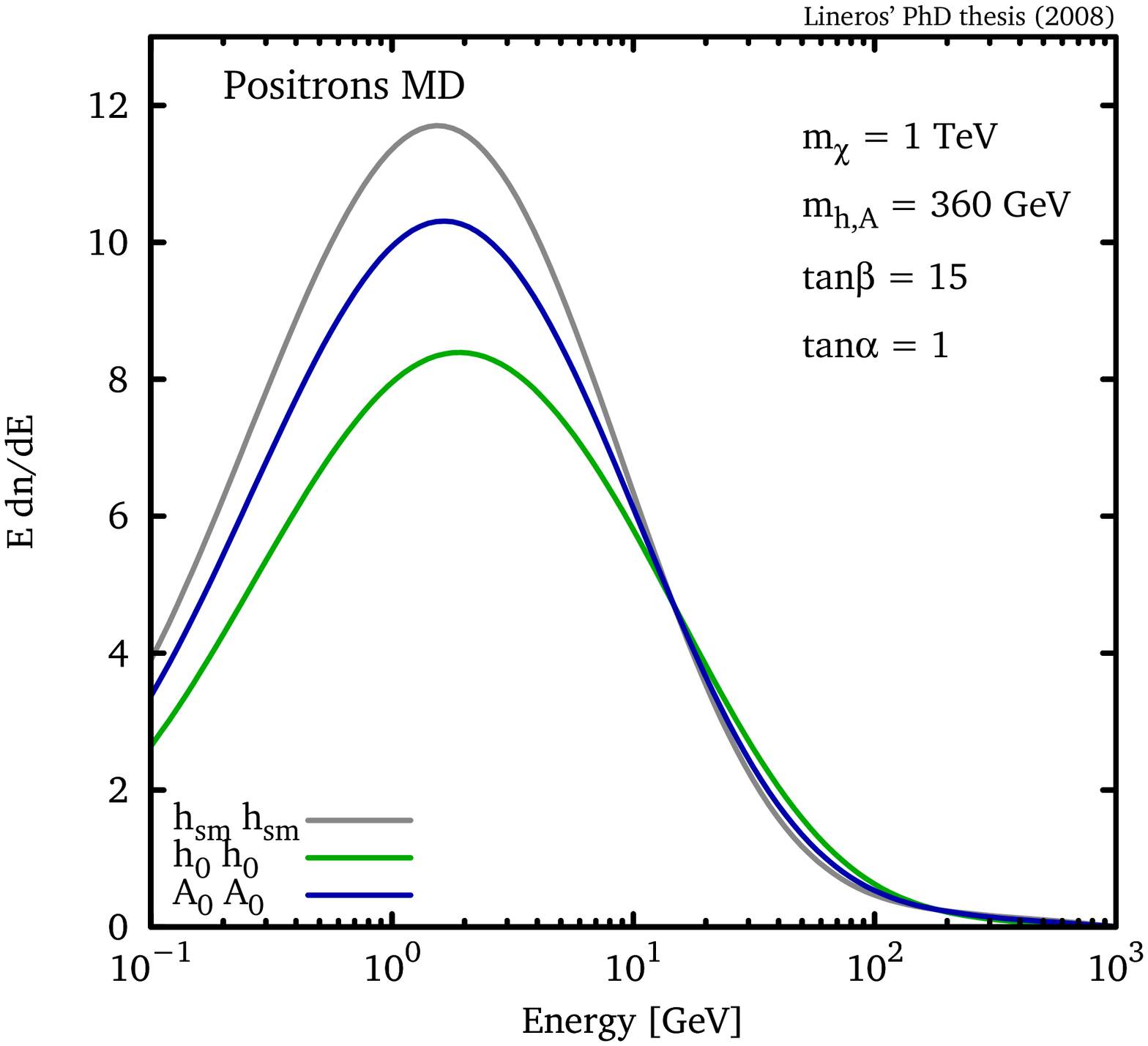}\includegraphics[angle=270, width=0.5\textwidth]{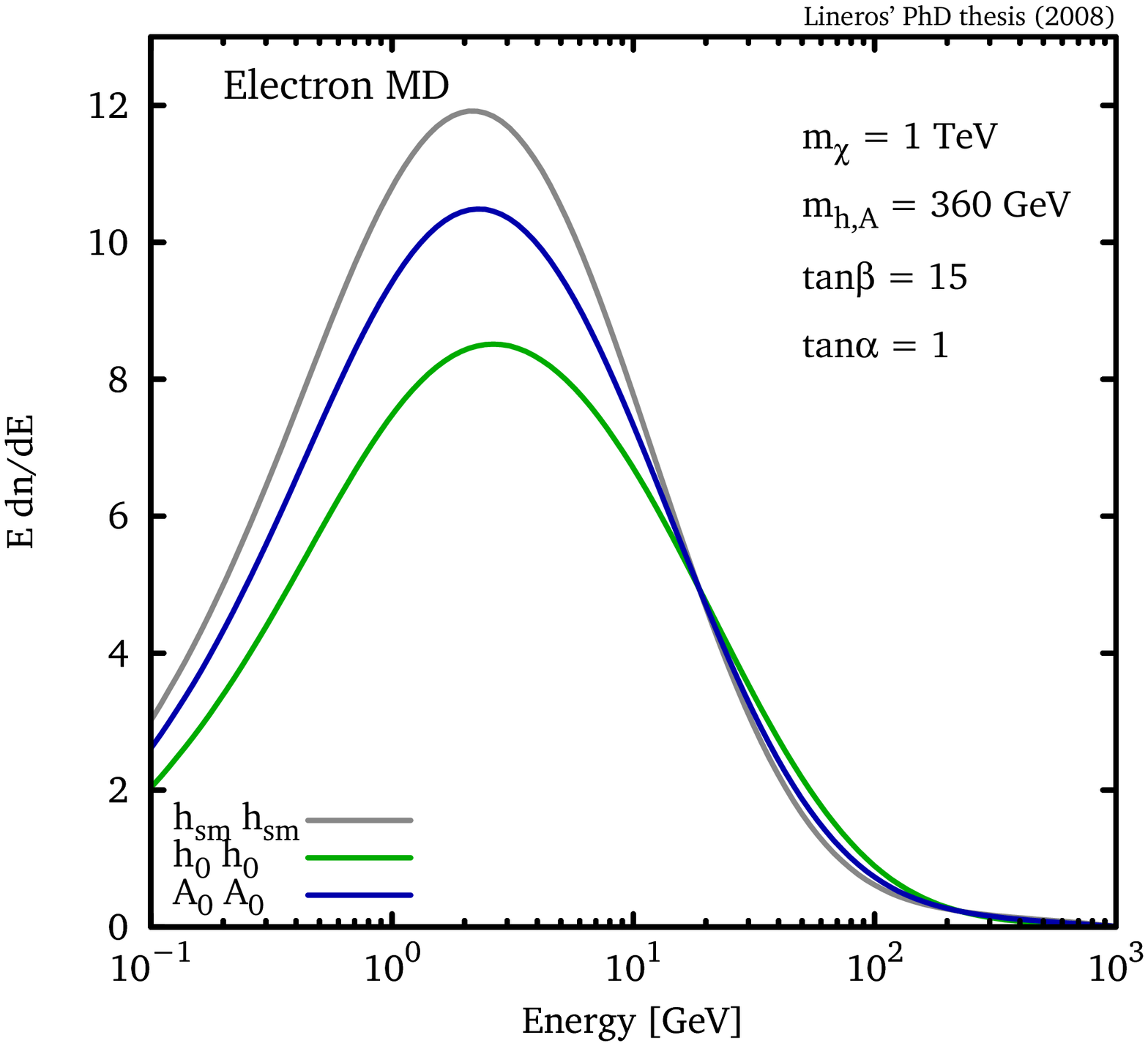}}
	\caption{\label{f:md-thdm} Multiplicity distribution for positrons and electrons versus energy. Comparison between \SM\ higgs and \THDM\ neutral higgs with equal mass value for a \dm\ mass of 1 \tu{TeV}.}
\end{fig}

\cleardoublepage
\chapter{Positron and Electron propagation in the galaxy}
\label{cha3}
\begin{prechap}
The propagation of positrons and electrons, and in general of any CR, in the Galaxy can be a hard problem to solve. There are many possible ways to deal with the propagation. The Two--Zone Propagation Model provides a good physical approach to describe - in a nice way - the travel between the CR source and the Solar System borders. Up to this point, the influence of the Sun becomes strong enough that CR need to be specially treated.\\
\end{prechap}

%
%
%%%%%%%%%%%%%%%%%%%%%%%%%%%%%%%%%%%%%%%%%%%%%%%%%%%%%%%%%%%%%%%%%%%%%%%%%%%%%%%%%%%%%%%%%%%%%%
\section{Overview}

In the travel across the Galaxy, cosmic rays are affected by many processes. In the low--energy range, one of the most important process is magnetic diffusion, which is produced by random magnetic regions that fill all the Galactic surrounding. These magnetic regions affect the cosmic ray propagation in such a way that makes impossible to trace back a cosmic ray up to its source. As well, cosmic rays interact with the gas and other particles present in the interstellar space. This could make cosmic rays loose (or gain) energy or to produce other types of cosmic rays.\\

Most of the sources of Galactic cosmic rays are in the Galactic Plane (GP). However, sources can be also located outside of it. In this way, a model to describe CR propagation should be able to represent all this possibilities.\\

Apart of the Galactic--scale effects, which rule the propagation, we cannot forget the effect associated to Solar activity. The Sun has 11--years cycles during which it increase and decrease periodically its activity. One of its manifestation is the increment of the Solar Wind flux, which acts on CR by pushing them away from the Solar System and reducing their energy.\\

In this chapter, the Two--Zone Propagation Model is described and solved for the case of positrons and electrons. General solutions of the transport equation are explained. As well, the model's space of parameters is discussed and the solar modulation problem and standard method to model it are also reviewed.\\

\section{Two--Zone Propagation Model}

This is a model for studying CR propagation in the Milky Way \cite{Ginzburg:1980, Maurin:2001sj}. In general terms, it is composed by a Propagation Zone (PZ) which demarcates the region where CR propagates and by the Transport Equation (TE), which models the physics of propagation.\\

The PZ is composed by two cylinders centered at the Galactic Center (\citefig{f:pz-cyl}). Both cylinder has a common radius equal to the galactic one (R = 20\;\tu{kpc}). However, their thickness are rather different.\\

The \emph{thick} cylinder has a height of $2 L$ and fills all the PZ. Its height $L$ is related to how much magnetic fields extend. CR measurement have constrainted its values to a range that goes from 1 to 20 \tu{kpc} \cite{Maurin:2001sj}. Moreover, in the thick cylinder, processes as magnetic diffusion and energy losses -- related to interaction with magnetic field and light -- take place.\\

The second cylinder is a \emph{thin disk} with height equal to $2 h_z$, where $h_z = 100\;\tu{pc}$.
The interstellar medium, cosmic-ray sources and interactions, as energy losses related to ionization or Bremsstrahlung, are contained inside the thin disk. 
Also, CR reacceleration processes related to supernova explosions and shockwaves take place there.\\

In the PZ, the Solar System is placed inside the thin disk at 8.5 \tu{kpc} from the GC and lays in the Galactic Plane.\\

% FIGURE
% 
% 
\begin{fig}
 \includegraphics[width=0.6\textwidth]{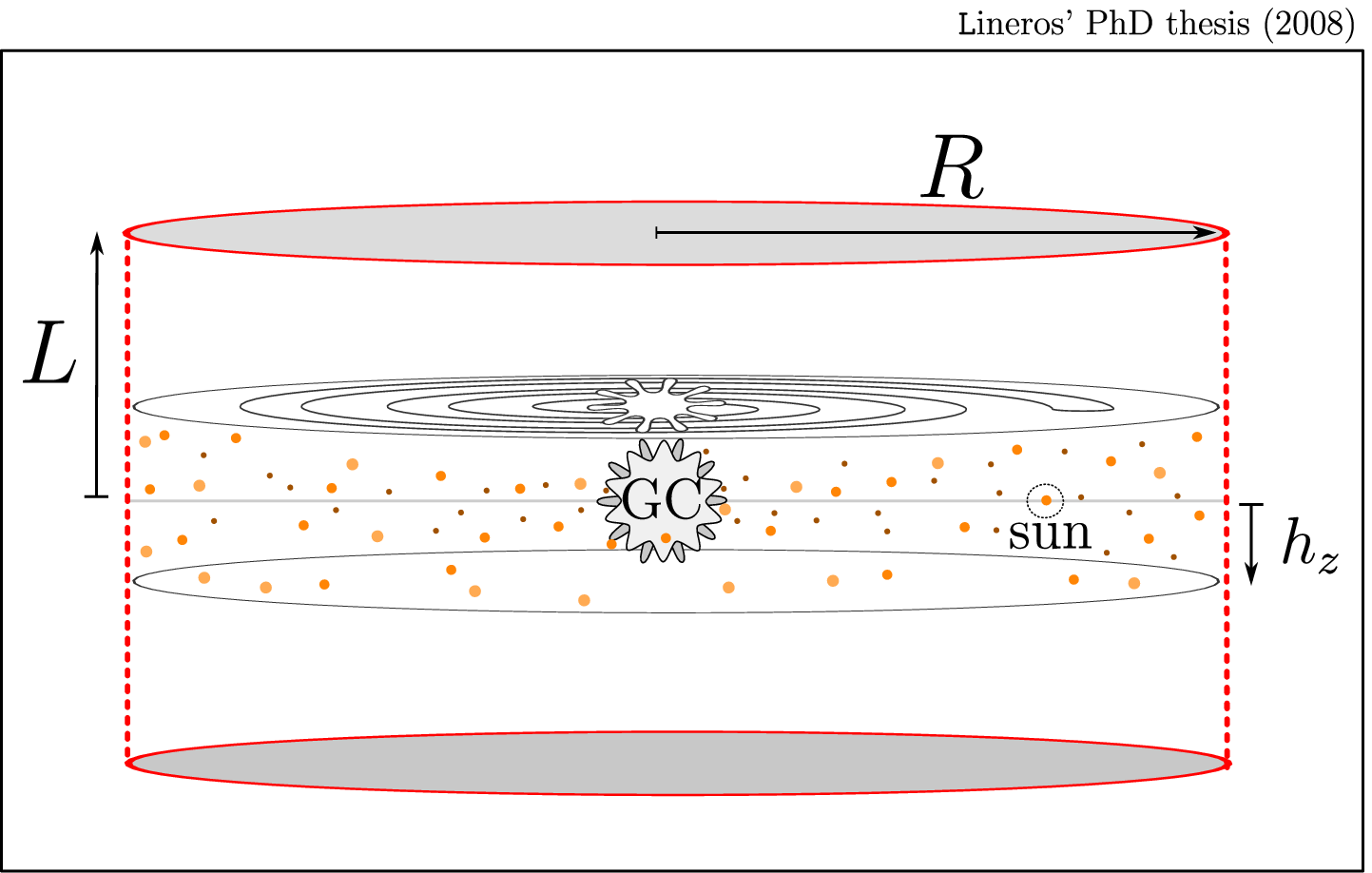}
 \caption{\label{f:pz-cyl} Propagation Zone geometry for the Milky Way. A cylinder of radius R ($= 20\;\tu{kpc}$) with thickness of $2\;L$ delimits the region where CR propagate. A small cylinder with same radius but with thickness $2\;h_z$ ($= 200\;\tu{pc}$) models the Galactic Plane. The Solar System is placed at the Galactic Plane with distance $r_{\astrosun} = 8.5 \; \tu{kpc}$ from the Galactic Center.}
\end{fig}
%
% 
% 

%%%%%%%%%%%%%%%%%%%%%%%%%%%%%%%%%%%%%%%%%%%%%%%%%%%%%%%%%%%%%%%%%%%%%%%%%%%%%%%%%%%%%%%%%%%%%%%%%%
\section{The Transport Equation}

The Transport Equation describes the physical effects involved in the CR propagation inside the PZ. In general terms, TE is based on a continuity equation for the CR density per unit of energy $(\psi)$ that written in terms of currents is simply:
\begin{eq}
 \partial \curr{}(t,\mb{x},\ener{}; \psi, \partial\psi) = s(t,\mb{x},\ener{}) \; ,
\end{eq}
where the currents $\curr{}(t,\mb{x},\ener{}; \psi, \partial\psi)$ denotes CR displacement in time, space and energy, 
\begin{eq}
 \partial \curr{\mb{x}} + \partial \curr{\ener{}} + \partial \curr{t} = s(t,\mb{x},\ener{}) \;,
\end{eq}
where $s(t,\mb{x},\ener{})$ are sources (or sink) which are independent of cosmic-ray evolution. \\

Diffusion due to random magnetic field and interaction with Galactic Wind (GW) are contained into the spatial current:
\begin{eq}
 \curr{\mb{x}} = - D_0 \partial_{\mb{x}} \psi + \mb{V} \psi \; ,
\end{eq}
where $D_0$ is the diffusion coefficient and $\mb{V}$ is the vector field of the GW. Usually diffusion is treated as a homogeneous function in space, dependent on the CR rigidity and the Lorentz $\beta$ factor \cite{Donato:2001ms,Maurin:2002ua} as,
\begin{eq}
 D_0(\ener{}) = K_0 \; \beta \; \times \; \left(\frac{\mathcal{R}}{\mathcal{R}_0}\right)^{\delta} \; ,
\end{eq}
where $\mathcal{R}_0$ is a rigidity scale, usually 1 \tu{GV}. In the case of ultrarelativistic particles it is just:
\begin{eq}
D_0(\ener{}) = K_0 \; \epsilon^{\delta} \; ,
\end{eq}
where $\epsilon = \ener{}/\ener{0}$ with $\ener{0} = 1 \; \tu{GeV}$.\\

There are some models to explain the form of the Galactic Wind field \cite{Ipavich:1972,Ipavich:1975,Webber:1992,Ptuskin:1997}, however it is still not well determined. The choice is to assume a GW field perpendicular to the Galactic Plane, 
\begin{eq}
\mb{V}(\mb{x}) = V_c(z) \hat{z} \; .
\end{eq}
Furthermore, it is assumed a constant behavior when we are out of the Galactic Plane,
\begin{eq}
	V_c(z) = V_c \; \tn{sign}(z) \; .
\end{eq}
Nevertheless, this choice may be not unique. In the work of Bloemen et al. \cite{Bloemen:1993A}, they have used a lineal profile, $V_c(z) = 3\; V_0\; z$, for modeling the GW. This profile has the advantage that in some cases, the TE can be solved analytically. \\ 

The energy--current ($\curr{\ener{}}$) is related to energy losses and gains,
\begin{eq}
 \curr{\ener{}} = \frac{d\ener{}}{dt}\;\psi = b_{\tn{loss}/\tn{gain}}(\ener{})\; \psi \; .
\end{eq}
In the galactic environment, CR are affected by processes like bremsstrahlung and ionization of IS gas, synchrotron radiation and inverse Compton scattering with photons and magnetic fields. As well, GW affects them producing adiabatic losses, \ie CR are cooled due to the expansion that GW drift produces in them, but just in zones where $\mb{V}$ variates spatially. On the other hand, CR are also allowed to gain energy, that is possible through reacceleration processes.  \\

The energy losses due to inverse Compton scattering and synchrotron radiation are calculated from the total radiation power \cite{Ginzburg:1965}: 
\begin{eq}
\frac{d\ener{}}{dt} = - \frac{16\pi}{3}\omega_{\tn{ph}} \times \frac{2\; e^4Z^4}{3 \mass{}^2 c^3} \; \gamma^2\beta^2 \; ,
\end{eq}
where $\omega_{\tn{ph}}$ is the photon density. A typical value for $\omega_{\tn{ph}}$ is $1 \; \tu{eV}/\tu{cm}^3$, which is a nominal value to include starlight, infrared and microwave radiation \cite{Berkey:1969}. In the case of ultrarelativistic positrons and electrons, the energy loss term is:
\begin{eq}
 b_{\tn{synch}}(\ener{}) = - \frac{\ener{}^2}{\tau_E \; \ener{0}} \; ,
\end{eq}
where $\ener{0} = 1 \; \tu{GeV}$ and $\tau_E = 10^{16} \; \tu{sec}$ is a time scale for the synchrotron energy loss.\\

Another process to consider is the adiabatic cooling due to the GW interaction. The energy loss term mainly depends of the divergence of the GW vector field,
\begin{eq}
 b_{\tn{adiab}}(\ener{},\mb{x}) = - \frac{\nabla \cdot \mb{V}(\mb{x})}{3} \; \frac{\mom{}^2}{\ener{}} \; ,
\end{eq}
that is reduced, in the TZPM context, to:
\begin{eq}
 b_{\tn{adiab}}(\ener{},\mb{x}) = \left\{ \begin{array}{cc} \displaystyle -\frac{1}{3} \; \partial_z V_c(z) \; \ener{} & \tn{inside GP} \\ \\0 & \tn{outside GP} \end{array} \right.\; ,
\end{eq}
for ultrarelativistic positrons and electron. Note that adiabatic cooling becomes important inside the GP: this puts in evidence how the inner structure of GW may take an important role in the propagation.\\

Other processes like bremsstrahlung, ionization and Coulomb losses, have contributions in the range above $\sim$1~\tu{GeV} smaller than the inverse Compton scattering and synchrotron radiation \cite{Strong:1998}. For example, the ionization energy losses for ultrarelativistic positrons (electrons) in neutral hydrogen and helium are:
\begin{eq}
 b_{\tn{ion}}(\ener{}) = - K_{\tn{ion}} \; \sum_{i=\tn{H,He}} \;Z_i n_i \left(3\log(\gamma)+ \zeta_i\right)
\end{eq}
where $Z_i$ is the atomic number, $n_i$ is the gas number density and the constant $K_{\tn{ion}} = 7.484 \times 10^{-18}\;\tu{GeV}\;\tu{cm}^3/\tu{s}$. The parameters $\zeta_i$ are related to the ionization potential, where the numerical values are: $\zeta_{\tn{H}} = 20.46$ and $\zeta_{\tn{He}} = 19.27$. Notice that for a electron energy of 10~\tu{GeV}. We got that $|b_{\tn{ion}}| \sim 10^{-16}\;\tu{GeV}/\tu{sec}$, which is two order of magnitud smaller than the same case but for energy losses due to synchroton radiation and inverse compton scattering $|b_{\tn{synch}}| \sim 10^{-14}\;\tu{GeV}/\tu{sec}$.
The same is illustrated in \citefig{f:eloss}, where time scale for those are shown. Smaller time scale interaction means bigger energy losses.\\

\begin{fig}
 \includegraphics[width=0.5\textwidth]{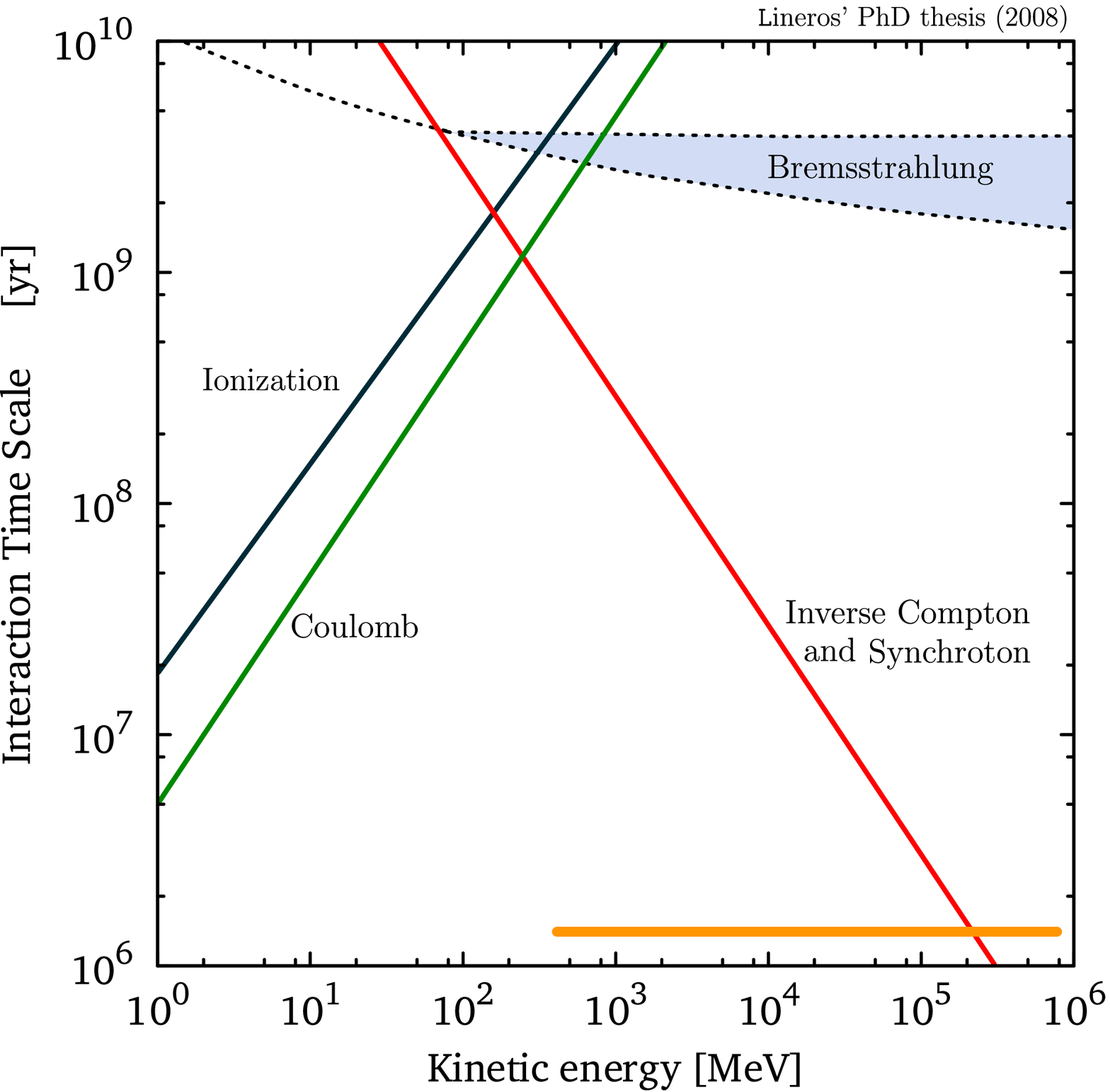}
 \caption{\label{f:eloss} Time scale related to energy loss processes for positrons and electrons as reported in \cite{Strong:1998}. In the energy range above 1 \tu{GeV} (orange line), energy losses are dominated by inverse Compton scattering and synchrotron radiation processes \cite{galprop:web}.}
\end{fig}

On the other hand, CR may gain energy through interactions with shockwaves. In the TE, this is included as an energy--gain term and an energy--diffusion term:
\begin{eq}
 \curr{\ener{},\tn{reac}} = \frac{1+\beta^2}{\ener{}} K_{ee} \; \psi - \beta^2 K_{ee} \; \partial_{\ener{}} \psi
\end{eq}
where $K_{ee}$ is the energy--diffusion coefficient, 
\begin{eq}
 K_{ee} = \frac{4}{3\delta(4-\delta^2)(4-\delta)} \; \frac{V_a^2 \mom{}^2}{D_0(\ener{})} \; ,
\end{eq}
which depends on the \emph{Alfv\'{e}n velocity} $(V_a)$ related to velocity of disturbances in the hydrodynamical plasma present in the ISM \cite{Seo:1994ApJ}. These expressions are simpler in the case of ultrarelativistic particles:
\begin{eq}
 \curr{\ener{},\tn{reac}} = \frac{2}{\ener{}} K_{ee} \; \psi - K_{ee} \; \partial_{\ener{}} \psi \quad \tn{and} \quad
 K_{ee} = \frac{4}{3\delta(4-\delta^2)(4-\delta)} \; \frac{V_a^2 \; \ener{0}^2}{K_0} \; \epsilon^{2-\delta} \; .
\end{eq}
Note that the inclusion of reacceleration processes in the TE produces a big change into the nature of the differential equation and also in the strategies to solve it \cite{Donato:2001ms}.\\ 

Time evolution of CR are important just in the case of a transient source. But for typical cases, like those discussed in the next chapter, time evolution can be safely neglected. In the energy range above 1~\tu{GeV} some of energy related processes can be safely neglected, as well.\\

In this context, the positron and electron transport equation (PETE) is reduced to:
\begin{eq}\label{e:pete1}
 -K_0 \epsilon^{\delta} \nabla^2 \psi - \frac{\partial}{\partial \epsilon} \left( \frac{\epsilon^2}{\tau_E} \psi \right) = s  \; .
\end{eq}
It includes the main processes (magnetic diffusion and inverse Compton scattering and synchrotron losses) that affect positrons and electron during their propagation inside the PZ.\\

% relation between the flux and the solution.
The cosmic--ray flux is the quantity that experiments are able to measure.
The solution of PETE gives just the information about the cosmic--ray density. 
The cosmic--ray flux, up to the Solar System borders, depends directly on the cosmic--ray density, 
\begin{eq}
 \frac{d\Phi_{\astrosun}}{d\ener{}}(\ener{}) = \frac{\beta c}{4 \pi} \; \psi(\mb{x}_{\astrosun}, \ener{}) \; .
\end{eq}
Let us warn that the evolution from the Solar System borders until the Earth can not be done through solving PETE, because the influence of Sun's activity makes this approach not reliable. Instead of this, the problem of propagation inside the Solar System can be treated in a simple way by modulating the flux, as will be discussed later in the chapter.\\

%
%
%%%%%%%%%%%%%%%%%%%%%%%%%%%%%%%%%%%%%%%%%%%%%%%%%%%%%%%%%%%%%%%%%%%%%%%%%%%%%%%%%%%%%%%%%%%%%%%%%
\section{PETE solutions}
A general solution of PETE is found using the Green function method. The Green function ($G$) is the solution for a point--like source in space and energy, 
\begin{eq}
 -\Lambda_0^2 \epsilon^{\delta} \nabla^{2} G - \frac{\partial}{\partial\epsilon} \left( \epsilon^{2} G \right) = \tau_E \; \delta^{3}(\mb{x}-\mb{x}_s)\delta(\epsilon - \epsilon_s) \; ,
\end{eq}
where $\Lambda_0^2 = K_0 \tau_e$ and formally $G=G(\mb{x},\mb{x}_s,\epsilon,\epsilon_s)$.\\

Furthermore, this type of differential equations can be solved by the method of separation of variables. The Green function and the Dirac delta on space can be expressed in terms of solutions of the Helmholtz equation ($\chi_{g}$),
\begin{eq}
 G(\mb{x},\mb{x}_s,\epsilon,\epsilon_s) &=& \sum_g \phi_g(\epsilon) \chi_g(\mb{x}) \; ,\\ 
\delta^3(\mb{x}-\mb{x}_s) &=& \sum_g \chi_g^{\dagger} (\mb{x}_s) \chi_g(\mb{x}) \; ,
\end{eq}
where $\chi_g$ satisfies:
\begin{eq}
 \nabla^2 \chi_g(\mb{x}) + g^2 \chi_g(\mb{x}) = 0 \; .
\end{eq}

After the substitution into PETE, we found that the equation is just a set of ordinary differential equations labeled by the eigenvalue $g$:
\begin{eq}
 \Lambda_0^2 \epsilon^{\delta} g^2 \phi_g(\epsilon) - \frac{\partial}{\partial\epsilon} \left(\epsilon^2 \phi_g(\epsilon) \right) = \tau_E \; \chi_g^{\dagger} (\mb{x}_s) \delta(\epsilon - \epsilon_s) \; .
\end{eq}
When $\epsilon \neq \epsilon_s$, all the equations become homogeneous and with solution of the form:
\begin{eq}
 \phi_g(\epsilon) = \frac{a_g}{\epsilon^2} \exp\left(g^2 \Lambda_0^2 \frac{\epsilon^{\delta -1}}{\delta-1}\right) \; .
\end{eq}
A restriction arises when the non--homogeneous equations are integrated around $\epsilon_s$,
\begin{eq}
 \phi^{+}_g(\epsilon_s) - \phi^{-}_g(\epsilon_s) = - \frac{\tau_E}{\epsilon_s^2} \chi_g^{\dagger}(\mb{x}_s) \; ,
\end{eq}
which links the solutions for energies above and below of $\epsilon_s$. Furthermore, we expect that particles are not allowed to gain energy because our system describes propagation with energy losses only. This implies that:
% As our system describes propagation with energy losses, we expect that particles are not allow to gain energy. That implies:
\begin{eq}
	\phi^{+}_g(\epsilon_s) = 0 \quad \Longrightarrow \quad \phi^{-}_g(\epsilon_s) = \frac{\tau_E}{\epsilon_s^2} \; \chi_g^{\dagger}(\mb{x}_s) \; ,
\end{eq}
which fixes the initial conditions for all solutions. we obtain in that way that the solution of non--homogeneous equations is:
\begin{eq}
 \phi_g(\epsilon) = \frac{\tau_E}{\epsilon^2} \chi_g^{\dagger}(\mb{x}_s)  \exp\left(-\frac{1}{4} \;g^2 \lD^2\right) \; ,
\end{eq}
where 
\begin{eq}\label{e:def-ld}
 \lD = 2 \Lambda_0 \sqrt{\frac{\epsilon_s^{\delta-1} - \epsilon^{\delta-1}}{\delta-1}}
\end{eq}  
is called the \emph{diffusion length}. And finally, the Green function is:
\begin{eq}
 G(\mb{x},\mb{x}_s,\epsilon,\epsilon_s) = \frac{\tau_E}{\epsilon^2} \; \widetilde{G}(\mb{x},\mb{x}_s, \lD) \; ,
\end{eq}
which is proportional to the \emph{tilded} Green function:
\begin{eq} \label{e:green-helm}
 \widetilde{G}(\mb{x},\mb{x}_s, \lD) = \sum_g \chi_g^{\dagger}(\mb{x}_s) \chi_g(\mb{x}) \exp\left(-\frac{1}{4} \;g^2 \lD^2\right) \; .
\end{eq}
Notice it depends on energies through  $\lD$. This function can also be described as:
\begin{eq}\label{e:wei-lam}
 \widetilde{G}(\mb{x},\mb{x}_s, \lD) = \langle \mb{x}_s | w(\lD) | \mb{x} \rangle \; ,
\end{eq}
where $w(\lD)=\exp\left(-\frac{1}{4} \;g^2 \lD^2\right)$ is the weight for each state $\chi_g$. As well, this function satisfies:
\begin{eq}\label{e:lam-zero}
 \lim_{\lD \rightarrow 0} \widetilde{G}(\mb{x},\mb{x}_s,\lD) = \delta^{3}(\mb{x}-\mb{x}_s) \; ,
\end{eq}
which describes a zero--distance propagation limit. In other words, particles observed with $\ener{} = \ener{s}$ at $\mb{x}$ cannot come from anywhere, except from the source at same position.\\

The general solution of PETE is obtained by convolution between the source term and the Green function:
\begin{eq}\label{e:formal_pete_sol}
 \psi(\mb{x},\epsilon) = \int_{\epsilon}^{\infty} d\epsilon_s \int d^3 \mr{x}_s  \; q(\mb{x}_s,\epsilon_s) \;  G(\mb{x},\mb{x}_s,\epsilon,\epsilon_s) \; .
\end{eq}

%
%
%%%%%%%%%%%%%%%%%%%%%%%%%%%%%%%%%%%%%%%%%%%%%%%%%%%%%%%%%%%%%%%%%%%%%%%%%%%%%%%%%%%%%%%%%%%%%
\subsection{Solution in free space}

Solutions of Helmholtz equation in a three dimensional free space are:
\begin{eq}\label{e:fourier_c}
 \chi_k(\mb{x}) = \frac{1}{(2 \pi)^{3/2}} \exp\left(i\; \mb{k}\cdot \mb{x}\right) \; ,
\end{eq}
where the eigenvalue $k$ is the composition of the eigenvalues that correspond to each spatial dimension,
\begin{eq}
	k^2 = k_x^2 + k_y^2 + k_z^2 \; .
\end{eq}
In this case, the tilded Green function is obtained from:
\begin{eq}
 \widetilde{G}_{\tn{free}}(\mb{x}, \mb{x}_s, \lD) = \frac{1}{(2\pi)^3} \int d^3k \; \exp\left(i \mb{k} \cdot (\mb{x}-\mb{x}_s) - \frac{1}{4} k^2 \lD^2\right) ,
\end{eq}
note that this is separable into three identical gaussian integrals of the form:
\begin{eq}
 \int_{-\infty}^{\infty} dk \;\exp\left(-k^2 a^2 + i k b \right) = \frac{\sqrt{\pi}}{a} \exp\left(-\frac{b^2}{4 a^2} \right) \; ,
\end{eq}
one for each dimension. Each of them produces its own tilded Green functions, which are independent from the others,  
\begin{eq}
 \widetilde{G}^{\tn{1d}}_{\tu{free}}(x,x_s,\lD) = \frac{1}{\sqrt{\pi}\lD} \exp\left(-\frac{(x-x_s)^2}{\lD^2} \right) \; .
\end{eq}\newline

This independence helps to compute the tilded Green function in three dimensions:
\begin{eq}
 \widetilde{G}^{\tn{3d}}_{\tn{free}}(\mb{x},\mb{x}_s,\lD) &=& \widetilde{G}^{\tn{1d}}_{\tu{free}}(x,x_s,\lD) \; \widetilde{G}^{\tn{1d}}_{\tu{free}}(y,y_s,\lD) \;
\widetilde{G}^{\tn{1d}}_{\tu{free}}(z,z_s,\lD)\\ \nonumber \\
&=& \frac{1}{\pi^{3/2} \lD^3} \exp\left(- \frac{(\mb{x}-\mb{x}_s)^2}{ \lD^2} \right) \nonumber \; . 
\end{eq}
At this point, it becomes clearer the name given to $\lD$ because it takes the place of the diffusion length in the standard diffusion theory. Note that this procedure produces the same results as reported by Baltz and Edsj\"o \cite{Baltz:1998xv}.\\

Let us denote that $\lD$ is a key quantity in order to understand the propagation. In \citefig{f:ag-ms}, we see how the source distance modifies the value of $\widetilde{G}^{1d}_{free}$, but on the other hand the maximum value is reached when $\lD$ is of the order of the distance between the observer and the source. \\

\begin{fig}
	\includegraphics[angle=270, width=0.5\textwidth]{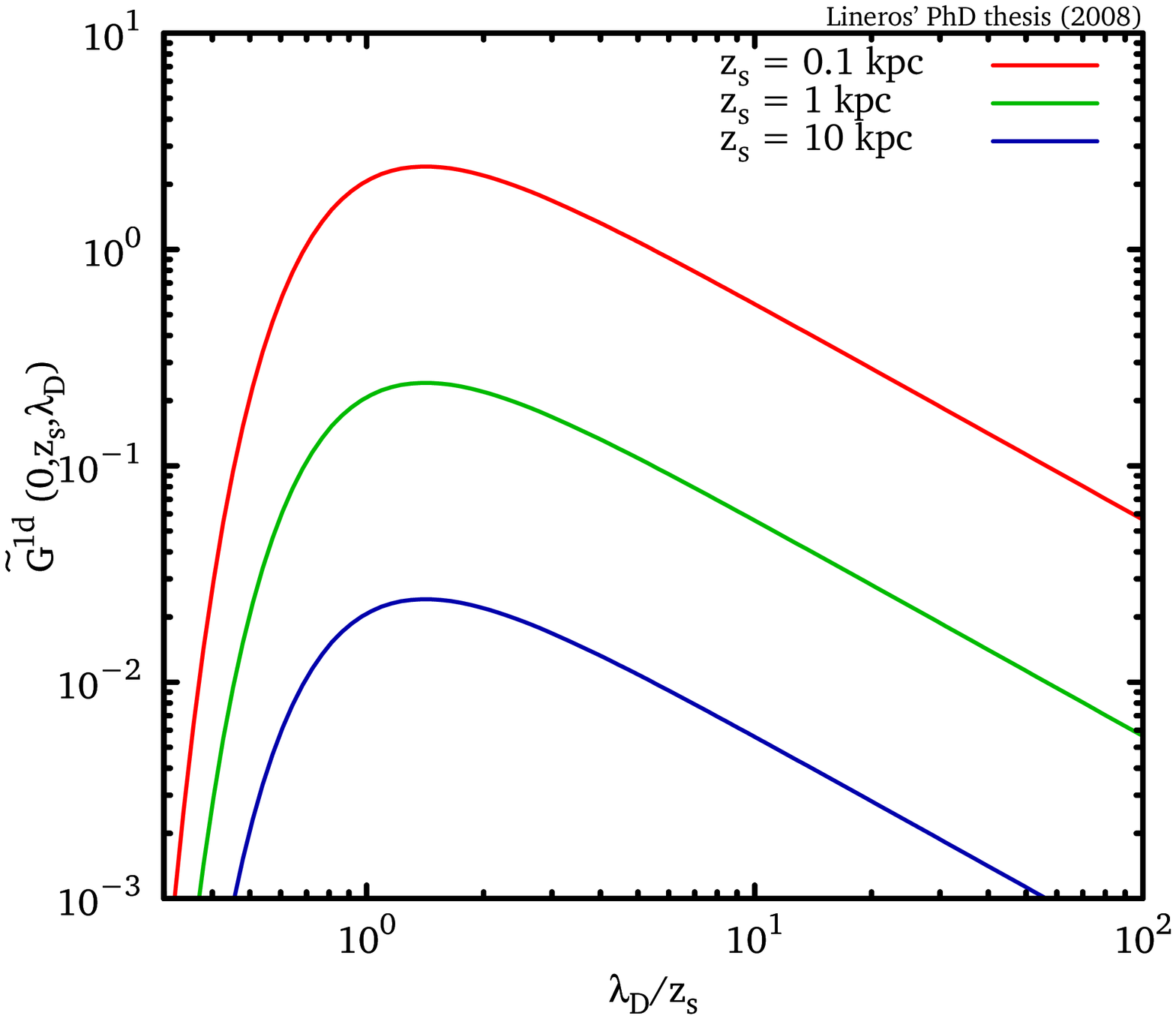}
	\caption{\label{f:ag-ms}$\widetilde{G}^{\tn{1d}}_{\tn{free}}$ versus $\lD/z_s$. The green function are smaller for farther sources. The maximum value is reached for $\lD$ of the order of the distance to the source.}
\end{fig}

As seen before, the solution of PETE depends on the convolution between the green function and the source term (\citeeq{e:formal_pete_sol}). By inspecting this, we note that the leading contributions to the solution come from a sphere of radius $\lD$ centered at $\mb{x}$.\\

%
%
%%%%%%%%%%%%%%%%%%%%%%%%%%%%%%%%%%%%%%%%%%%%%%%%%%%%%%%%%%%%%%%%%%
\subsection{Solution with boundary condition}
The TZPM considers propagation to occur in a finite volume, where \CR\ may escape if they arrive close to the boundaries.\\

A first case is to consider boundary conditions on the vertical axis, 
\begin{eq}
	\psi(x,y,z=\pm L) = 0 \; .
\end{eq}
As before, there is a orthonormal basis for $x$, $y$ dimensions, which is described by a continuous 2D Fourier space -- \citeeq{e:fourier_c}. However, to satisfy the boundary condition on $z$, a discrete version is required:
\begin{eq}
 \chi_{k_z}(z) = \frac{1}{\sqrt{L}}\left\{ 
\begin{array}{cc} 
\sin(k_{z,n} \; z) & n:\tn{even}\\
\cos(k_{z,n} \; z) & n:\tn{odd}
\end{array}\right. \; ,
\end{eq}
where the eigenvalues $k_z$ are:
\begin{eq}
 k_{z,n} = \frac{n \pi}{2 L} \quad n \in \mathbb{Z} \; .
\end{eq}

In the tilded Green function, two parts are identified. The first is similar to the free case and the second part satisfies the boundary conditions:
\begin{eq}
 \widetilde{G}(\mb{x},\mb{x}_s,\lD) = \left\{
 \widetilde{G}^{\tn{1d}}_{\tu{free}}(x,x_s,\lD) \;
 \widetilde{G}^{\tn{1d}}_{\tu{free}}(y,y_s,\lD) \right\} \; \times  \;
 \widetilde{\Sigma}(z,z_s,\lD) \; ,
\end{eq}
where $\widetilde{\Sigma}$ is calculated as a superposition of Fourier modes:
\begin{eq}
 \widetilde{\Sigma}_{\tu{F}}(z,z_s,\lD) &=& \frac{1}{L} \sum_{n:\tn{even}}^{\infty} \sin(k_{z,n} z_s) \sin(k_{z,n} z) \exp\left(-\frac{1}{4} k_{z,n}^2 \lD^2\right) \\ 
&&+ \frac{1}{L} \sum_{n:\tn{odd}}^{\infty} \cos(k_{z,n} z_s) \cos(k_{z,n} z) \exp\left(-\frac{1}{4} k_{z,n}^2 \lD^2\right) \nonumber \; .
\end{eq}

A different way to compute it is using the method of Image Charges (IC), where $\widetilde{\Sigma}$ is the superposition of many \mbox{free--space} Green functions \cite{Baltz:1998xv}:
\begin{eq}\label{e:sig-ic}
 \widetilde{\Sigma}_{\tn{IC}}(z,z_s,\lD) = \sum_{n=-\infty}^{n=\infty} (-1)^n \; \widetilde{G}^{\tn{1d}}_{\tu{free}}(z,z_{s,n},\lD)
\end{eq}
where $z_{s,n}$ denotes image charges positions, 
\begin{eq}
 z_{s,n} = 2 L \; n + (-1)^n z_s \; ,
\end{eq}
needed to satisfy boundary conditions.\\

\begin{fig}
	\includegraphics[angle=270, width=0.5\textwidth]{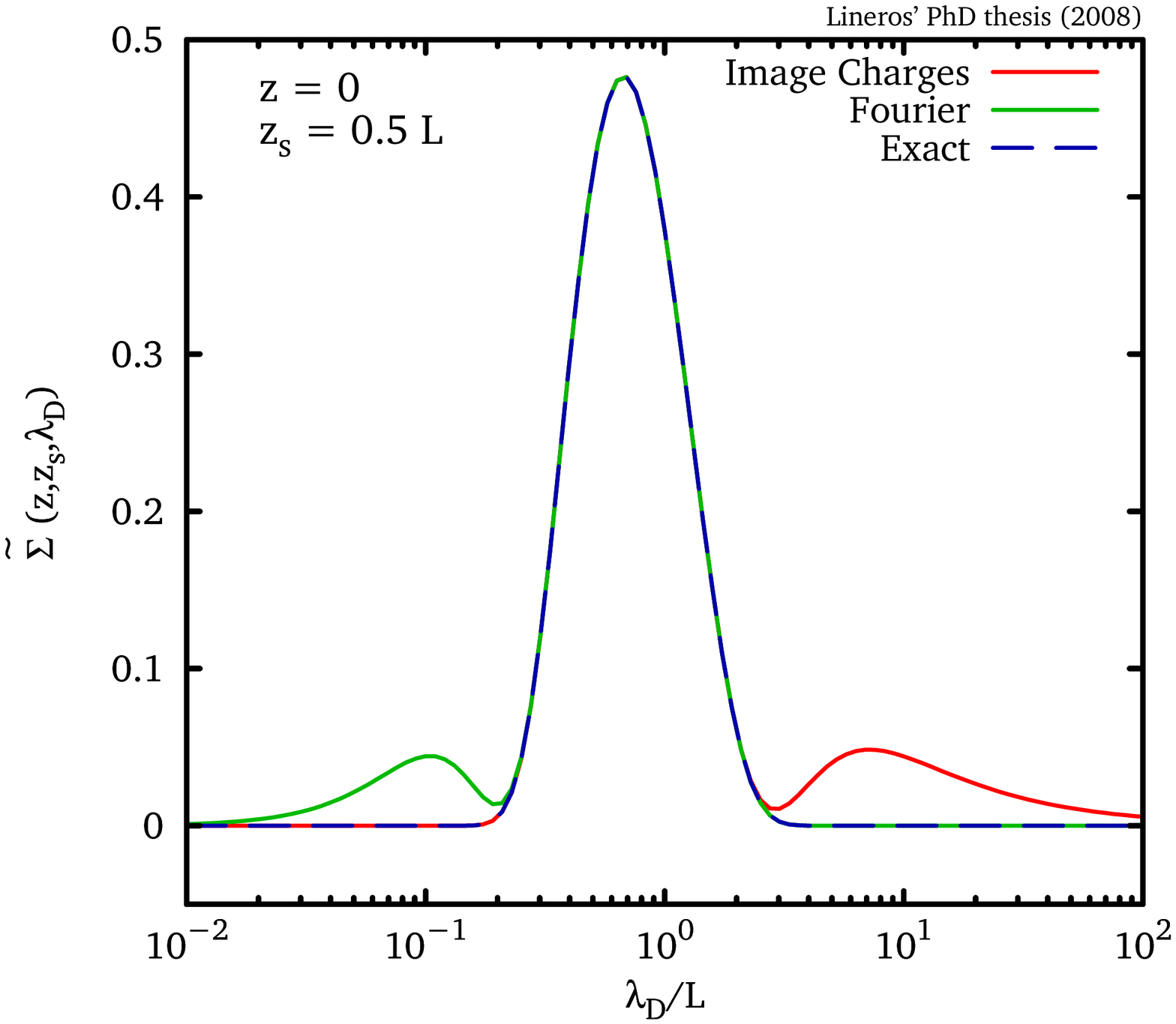}
	\caption{\label{f:ag-sig} $\widetilde{\Sigma}$ function versus $\lD/L$. The IC and Fourier method $\widetilde{\Sigma}$ are calculated with few terms to contrast those with an exact solution. It is shown that the IC method (red line) converges faster for $\lD$ values smaller than $L$ instead that the Fourier one (gree line) which converges for $\lD$ bigger that $L$.}
\end{fig}

Both methods are completely equivalent from the theoretical point of view. However, depending on the value of $\lD/L$, the rates of convergence are different (\citefig{f:ag-sig}). We found two regimes in which each method becomes the most efficient option: 
\begin{itemize}
 \item[i)] When $\lD < L$, the IC method is more efficient than the Fourier one. The sum converges after few terms because the diffusion length ($\lD$) is small enough that the propagation cannot reach the boundaries. On the contrary, the Fourier method becomes highly inefficient because the weight function (\citeeq{e:wei-lam}) becomes unity for many Fourier modes. That affects directly the rate of convergence because we need to sum too many terms.\\

 \item[ii)] When $\lD < L$, the Fourier method is the best option because Fourier modes with low eigenvalue mainly contribute to the sum and we need just to consider few of them.\\
\end{itemize}

%
% The regime $\lD < L$ produces $\widetilde{\Sigma}_{\tn{IC}}$ converges after summing few terms, instead of $\widetilde{\Sigma}_{\tn{F}}$. The Fourier method is inefficient at small $\lD$ because the weight function (\citeeq{e:wei-lam}) becomes closer than 1 for many Fourier modes; and to reach a convergent value, it is necessary to sum too many terms. The opposite behavior happens when $\lD>L$, the Fourier method becomes most efficient than IC method because low--eigenvalue states weight more in the sum.\\

In \citefig{f:ag-comp}, the effects of boundary conditions are shown. Boundary conditions induce the Green function $\widetilde{\Sigma}$ to decrease faster than the free-space one, when the diffusion length is bigger that $L$. This is related to the leaking of particles which arrive at the boundaries. For the same reason, both Green functions are similar when diffusion lengths are small, particles do not have time to propagates and reach the boundaries.\\

\begin{fig}
	\includegraphics[angle=270, width=0.5\textwidth]{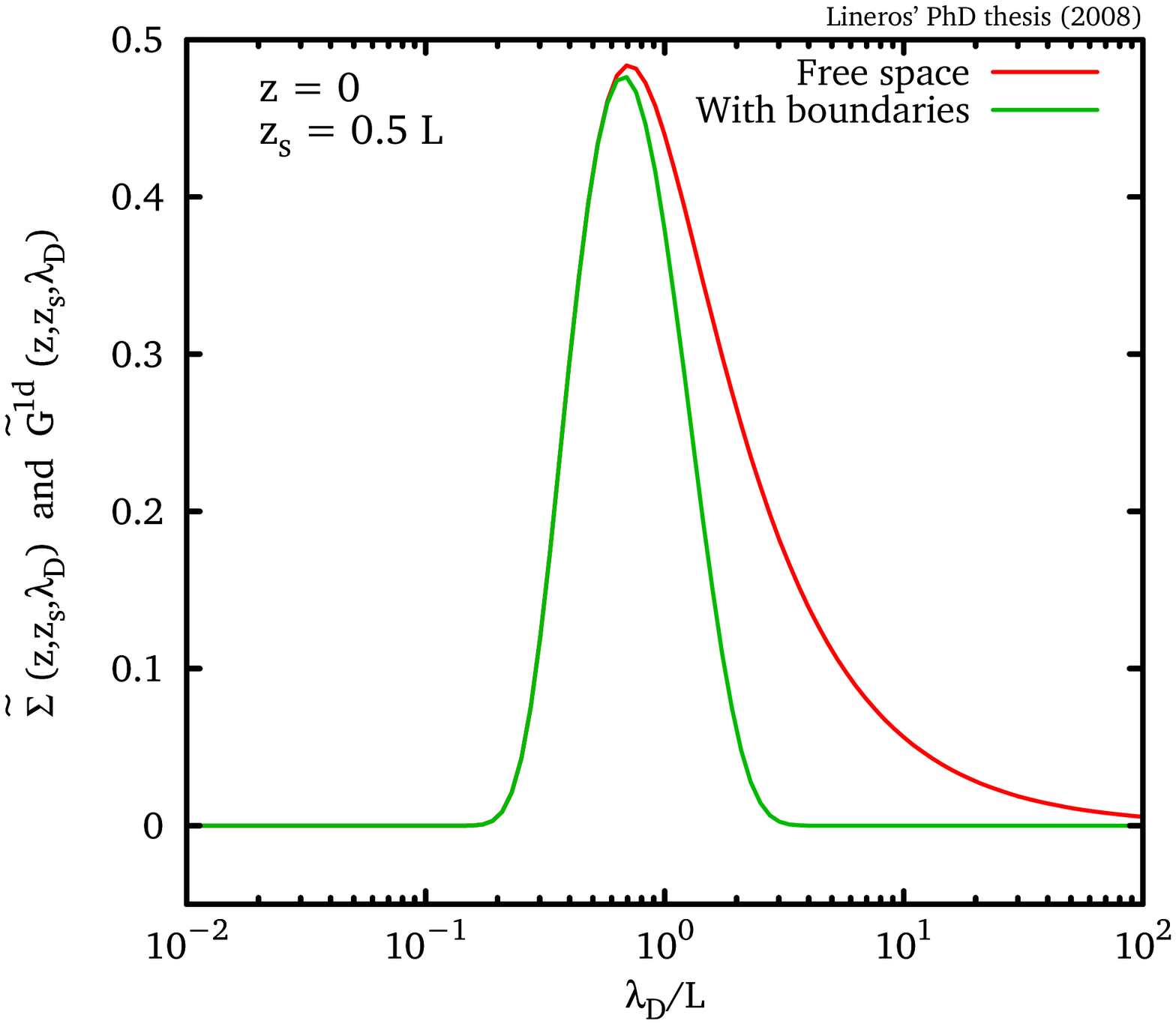}
	\caption{\label{f:ag-comp} $\widetilde{\Sigma}$ and $\widetilde{G}^{\tn{1d}}_{\tn{free}}$ functions versus $\lD/L$. The vertical boundary conditions produces a green function which decreases faster than the free--space case when $\lD$ grows. For diffusion distance smaller than $L$, both Green functions have same behavior.}
\end{fig}

%
%
%%%%%%%%%%%%%%%%%%%%%%%%%%%%%%%%%%%%%%%%%%%%%%%%%%%%%%%%%%%%%%%%%%%%%%%%%%%%%%%%%%%%
\subsection{Axial symmetric solution}

The PETE can be also solved for special geometries. Galaxies, as the Milky Way, tends to have axial symmetry with respect to the galatic center and oriented along the angular momentum direction. In such scenarios, it is expected that CR sources are located according to the symmetry.\\

In cylindrical coordinates an axial symmetric point-like source can be described as:
\begin{eq}
 q(r,z,\epsilon) = \frac{\delta(r-r_s)}{r} \; \delta(z-z_s) \; \delta(\epsilon-\epsilon_s) \; ,
\end{eq}
and in this case the source corresponds to a ring.\\

The function basis for the radial coordinate is based on Bessel functions,
\begin{eq}
\chi_{k_{r,i}}(r) = \frac{\sqrt{2}}{R \; J_1(\alpha_{0,i})} J_0(k_{r,i} r) \; ,
\end{eq}
where $\alpha_{0,i}$ is the $i$-th zero of Bessel function $J_0$. This basis also includes boundary conditions, where Green function vanishes at a characteristic radius $R$. This condition implies discrete eigenstates with eigenvalues:
\begin{eq}\label{e:rad-eig}
 k_{r,i} = \frac{\alpha_{0,i}}{R} \; .
\end{eq}\newline

Following the general procedure, the tilded Green function for the radial coordinate is:
\begin{eq}\label{e:ups_boun}
 \widetilde{\Upsilon} (r,r_s,\lD) = \sum_{i=1}^{\infty} 2\frac{J_0(k_{r,i} \; r_s) \; J_0(k_{r,i}\; r)}{\big(R \; J_1(\alpha_{0,i})\big)^2} \; \exp\left(-\frac{1}{4}k_{r,i}^2 \lD^2\right) \; .
\end{eq}
In the case with no radial boundary condition, this expression changes into a Bessel transform, which is not analytically solvable like in the cartesian case.\\

The Green function with boundary conditions on the radial and vertical axis is just the product of the previously seen Green functions:
\begin{eq}
 \widetilde{G}_{\tn{cyl}}(r,r_s,z,z_s,\lD) = \widetilde{\Upsilon} (r,r_s,\lD) \; \widetilde{\Sigma} (z,z_s,\lD) \; .
\end{eq}
Let us emphasize that $\widetilde{\Upsilon}$ behaves as $\widetilde{\Sigma}$ when $\lD$ is bigger than $R$. In a sense, particle leaking starts to be a dominant effect, producing a fast decrease in the Green function value. Generally $L \le R - r_{\astrosun}$, that means leaking in vertical axis will be even more dominant, because particles are closer to vertical boundaries than radial boundaries from Solar System position.\\

In this case, the solution of PETE is:
\begin{eq}
 \psi(r,z,\epsilon) = \frac{\tau_E}{\epsilon^2} \int_{\epsilon}^{\infty} d\epsilon_s \int_0^R r_s dr_s \int_{-L}^L dz_s\; q(r_s,z_s,\epsilon_s) \; \widetilde{G}_{\tn{cyl}}(r,r_s,z,z_s,\lD) \; ,
\end{eq}
where the region of integration is restricted to the zone inside the boundaries.\\

%
%
%%%%%%%%%%%%%%%%%%%%%%%%%%%%%%%%%%%%%%%%%%%%%%%%%%%%%%%%%%%%%%%%%%%%%%%%%%%%%%%%%%%%%%%%%%%%%%%%%%
\section{The Halo Function}
The Halo Function (HF) is an dimensionless function which encodes the information about spatial dependence of the source term. It can be calculated when energy and spatial dependence in the source terms can be separated as:
\begin{eq}
	q(\mb{x},\epsilon) = \kappa \times  f(\mb{x}) \times g(\epsilon) \; .
\end{eq}
Furthermore, $f(\mb{x})$ is an dimensionless function which is normalized to the Solar System sources density, 
\begin{eq}
	f(\mb{x}_{\astrosun}) = 1 \; ,
\end{eq}
where the Solar System position is located at 
\begin{eq}
 r_{\astrosun} = 8.5 \; \tu{kpc} \ \tn{and} \ z_{\astrosun} = 0 \; \tu{kpc} \; ,
\end{eq}
with respect to the Galactic Center. Let us remark that the normalization is just valid in cases of regular and non--vanishing distributions.\\

Similar to PETE's general solution, the HF is calculated as follows: 
\begin{eq}\label{e:hf-greenc}
 \widetilde{I}(\lD) = \int d^3 \mr{x}_s \; f(\mb{x}_s) \; \widetilde{G}(\mb{x}_{\astrosun},\mb{x}_s,\lD) \; ,
\end{eq}
which is a convolution between the spatial part of source and the Green function. Also, HF naturally satisfies:
\begin{eq}
 \lim_{\lD \rightarrow 0} \widetilde{I}(\lD) = 1 \; ,
\end{eq}
that is a direct consequence of the normalization of $f(\mb{x})$ and the \citeeq{e:lam-zero}.\\

The main aim of the halo function is to be used for calculating the PETE's solution:
% Nevertheless, its main aim is to be used as a step before to compute the PETE's solution:
\begin{eq}\label{e:flux-from-hf}
 \psi_{\astrosun}(\epsilon) = \kappa \; \frac{\tau_E}{\epsilon^2} \int_{\epsilon}^{\infty} d\epsilon_s \; g(\epsilon_s) \;  \widetilde{I}(\lD) \; ,
\end{eq}
where $\lD$ has an implicit dependence on $\epsilon$ and $\epsilon_s$ (\citeeq{e:def-ld}).\\

In any case, HF are functions of one variable. This is an advantage that improves the calculation speed for several different cases. For example, many sources may share same spatial distribution but different spectral shapes, in this case we need to calculate only once the HF and just perform many convolutions in energy without recomputing the HF.\\

A special case happens when source terms inject monoenergetic particles because the PETE's solutions become proportional to the HF,
\begin{eq}
 \psi_{\astrosun}(\epsilon) = \kappa \; \frac{\tau_E}{\epsilon^2} \; \widetilde{I}(\lD) \; .
\end{eq}
But in general rule, any PETE's solutions depend directly on the spatial distribution of sources.\\

An alternative approach to compute a HF instead of \citeeq{e:hf-greenc} is:
% The general approach to compute an HF is shown in \citeeq{e:hf-greenc}, but there is an equivalent expression,
\begin{eq} \label{e:hf-helm}
 \widetilde{I}(\lD) = \sum_g \; a_g^{\dagger} \; \chi_g(\mb{x}_{\astrosun})\; \exp\left(-\frac{1}{4}g^2\lD^2\right)\; ,
\end{eq}
which is obtained when the tilded Green function is decomposed in terms of solutions of the Helmholtz equation (\citeeq{e:green-helm}). The coefficients $a_g$,
\begin{eq}
 a_g = \int d^3\mr{x}_s \; f(\mb{x}_s) \; \chi_g(\mb{x}_s) \; ,
\end{eq}
are the projections of the spatial source distribution on the eigenstates of the Helmholtz equation.\\

In analogy to the tilded Green functions computed with Fourier and IC methods, the two previous versions of HF have different convergence rates. The HF in terms of eigenstates (\citeeq{e:hf-helm}) converges faster for big values of $\lD$. On the contrary, HF -- obtained by convolution -- converge faster for small values of $\lD$, but just in the case when the tilded Green function corresponds to the one obtained with the IC method.\\ 

%
%
%%%%%%%%%%%%%%%%%%%%%%%%%%%%%%%%%%%%%%%%%%%%%%%%%%%%%%%%%%%%%%%%%%%%%%%%%%%%%%%%%%%%%%%%%%%%%
\subsection{Halo function examples}

A very simple example occurs for homogeneous sources distributed in all the space. In this case, we get:
\begin{eq}
 \widetilde{I}_{\tn{free}}(\lD) = \int_{\tn{all space}} d^3 \mr{x}_s \; \widetilde{G}_{\tn{free}}^{\tn{3d}}(\mb{x}_{\astrosun},\mb{x}_s, \lD) = 1 \; ,
\end{eq}
when the source is confined into a box centered on $\mb{x}_{\astrosun}$ with edge lengths $2\;l_x$, $2\;l_y$ and $2\;l_z$ for each dimension, the correspondent HF is:
\begin{eq}
 \widetilde{I}_{\tn{box}}(\lD) = \erf\left(\frac{l_x}{\lD}\right) \times \erf\left(\frac{l_y}{\lD}\right) \times \erf\left(\frac{l_z}{\lD}\right) \; ,
\end{eq}
where $\erf(x)$ is the Gauss error function. In this case, the disappearance of the source, beyond the box limits, starts to manifest, decreasing HF value, from diffusion lengths bigger than half of the minimum side of the box. Also note that when the box sides become bigger than $\lD$, $\widetilde{I}_{\tn{box}}$ goes to $\widetilde{I}_{\tn{free}}$.\\

In the axial symmetric situation, we can calculate the HF for a cylinder with radius $R_{\tn{src}}$ ($<R$) and thickness of $2h_z$ ($<2L$). The homogeneous source distribution makes easy to split the HF into an exclusively-radial and -vertical HF. This is possible when the source term is composed by functions that depend only of one spatial dimension.\\

% For the axial symmetric case, we can compute the HF for a homogeneous disk with radius $R_{\tn{src}}$ and thickness of $2\;h_z$, vertically centered in $z=0$. All in a bounded region guided by $R$ and $2\;L_z$, directly related to the PZ. The homogeneous source makes easy to split the problem into a radial and a vertical independent problems, in fact that could be performed just in cases where spatial distribution is separable like independent functions for each dimension.\\

The radial part is described in the same way as a generic HF (\citeeq{e:hf-helm}), but in this case, it depends on Bessel functions:
\begin{eq}
 \widetilde{I}_{\tn{radial}}(\lD) = \frac{2\;R_{\tn{src}}}{R^2}\sum_{i=1}^{\infty}\; \frac{J_1(k_{r,i} \; R_{\tn{src}}) J_0(k_{r,i} \; r_{\astrosun})}{k_{r,i}\;J_1^2(k_{r,i} R)} \; \exp\left(-\frac{1}{4} k_{r,i}^2 \lD^2\right)
\end{eq}
where $k_{r,i}$ are the eigenvalues of the correspondent eigenstate (\citeeq{e:rad-eig}). In any case, the coefficients in the sum come from the radial integration over all the PZ, 
\begin{eq}
 \frac{\sqrt{2}}{R J_1(\alpha_{0,i})} \; \int_{0}^{R_{\tn{src}}} r_s dr_s \; J_0(k_{r,i} r_s) = \frac{\sqrt{2}}{k_{r,i}} \frac{R_{\tn{src}} \; J_1(k_{r,i}\; R_{\tn{src}})}{R \; J_1(k_{r,i}\; R)} \; .
\end{eq}

The vertical part is obtained for 2 cases. The first one is related to the Fourier method, which is similar to the previous examples:
\begin{eq}
 \widetilde{I}_{\tn{vertical,F}}(\lD) = \frac{1}{L} \; \sum_{n:\tn{odd}}^{\infty} \frac{2 \sin(k_{z,n} \; h_z)}{k_{z,n}} \; \exp\left(-\frac{1}{4} \; k_{z,n}^2 \; \lD^2\right) \; .
\end{eq}\newline
% this formula becomes simple because it is evaluated at $z_{\astrosun} = 0$.\\

A different approach is used to calculate the vertical part. In this case, the tilded Green function obtained with the IC method (\citeeq{e:sig-ic}) is used to perform a convolution with the spatial distribution, obtaining:
\begin{eq}
\widetilde{I}_{\tn{vertical,IC}}(\lD) = \frac{1}{2\lD} \; \sum_{n=-\infty}^{\infty} \erf\left(\frac{2 L n + (-1)^n h_z}{\lD}\right) - \erf\left(\frac{2 L n - (-1)^n h_z}{\lD}\right) \; .
\end{eq}
As seen previously in the comparison between IC and Fourier methods, these two vertical parts present the same feature respect to convergence speed. The one based on the Fourier method converges faster for bigger values of $\lD$ instead of $\widetilde{I}_{\tn{vertical,IC}}$, and viceversa. \\
Finally, the HF for a homogeneous source is:
\begin{eq}\label{e:hf-axial-homo}
 \widetilde{I}_{\tn{axial}}(\lD) = \widetilde{I}_{\tn{radial}}(\lD) \times \widetilde{I}_{\tn{vertical}}(\lD) \; ,
\end{eq}
which depends on the radial and vertical parts that were already calculated.\\

\begin{fig}
 \resizebox{\hsize}{!}{\includegraphics[angle=270,width=0.5\textwidth]{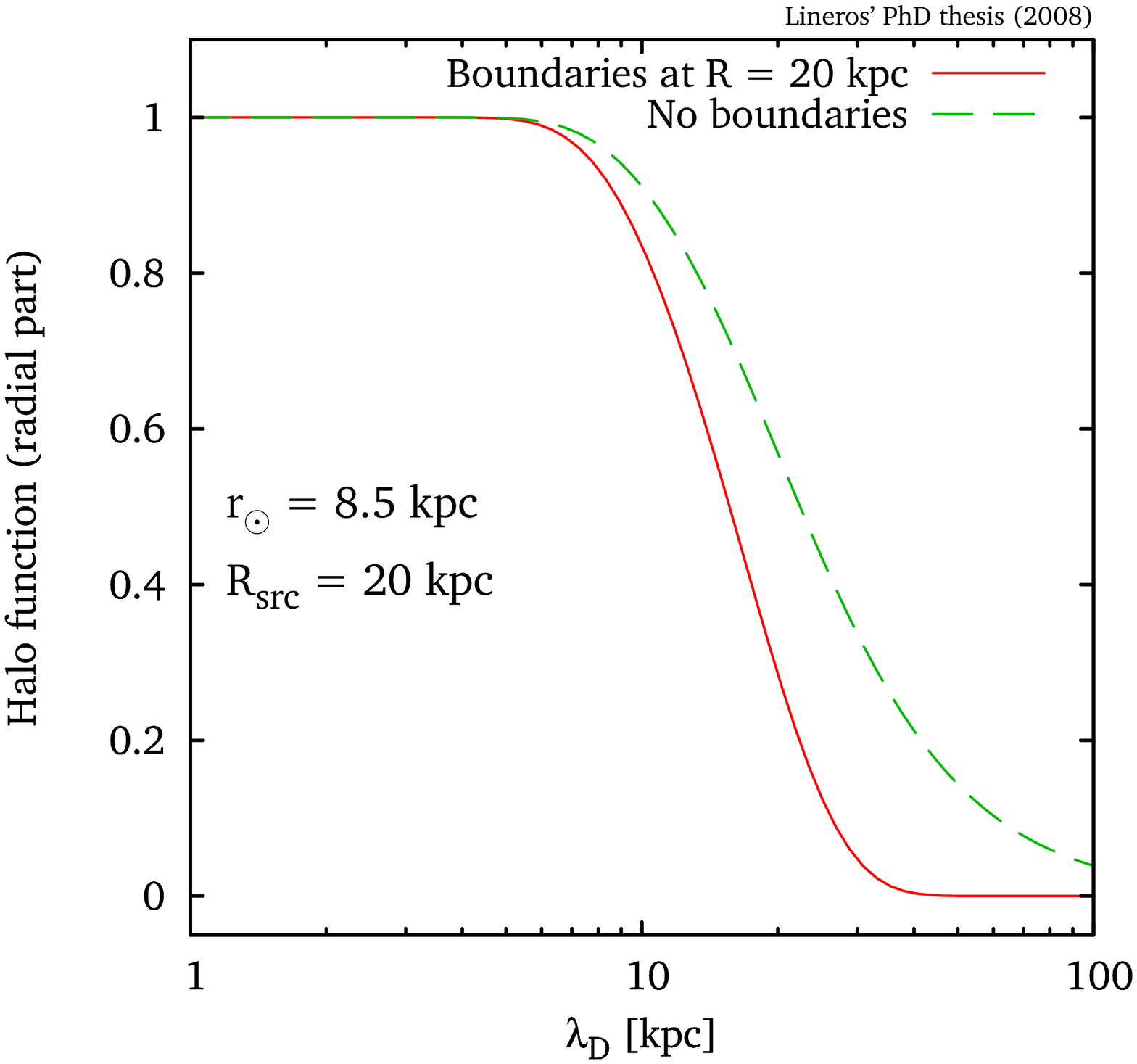}\includegraphics[angle=270,width=0.5\textwidth]{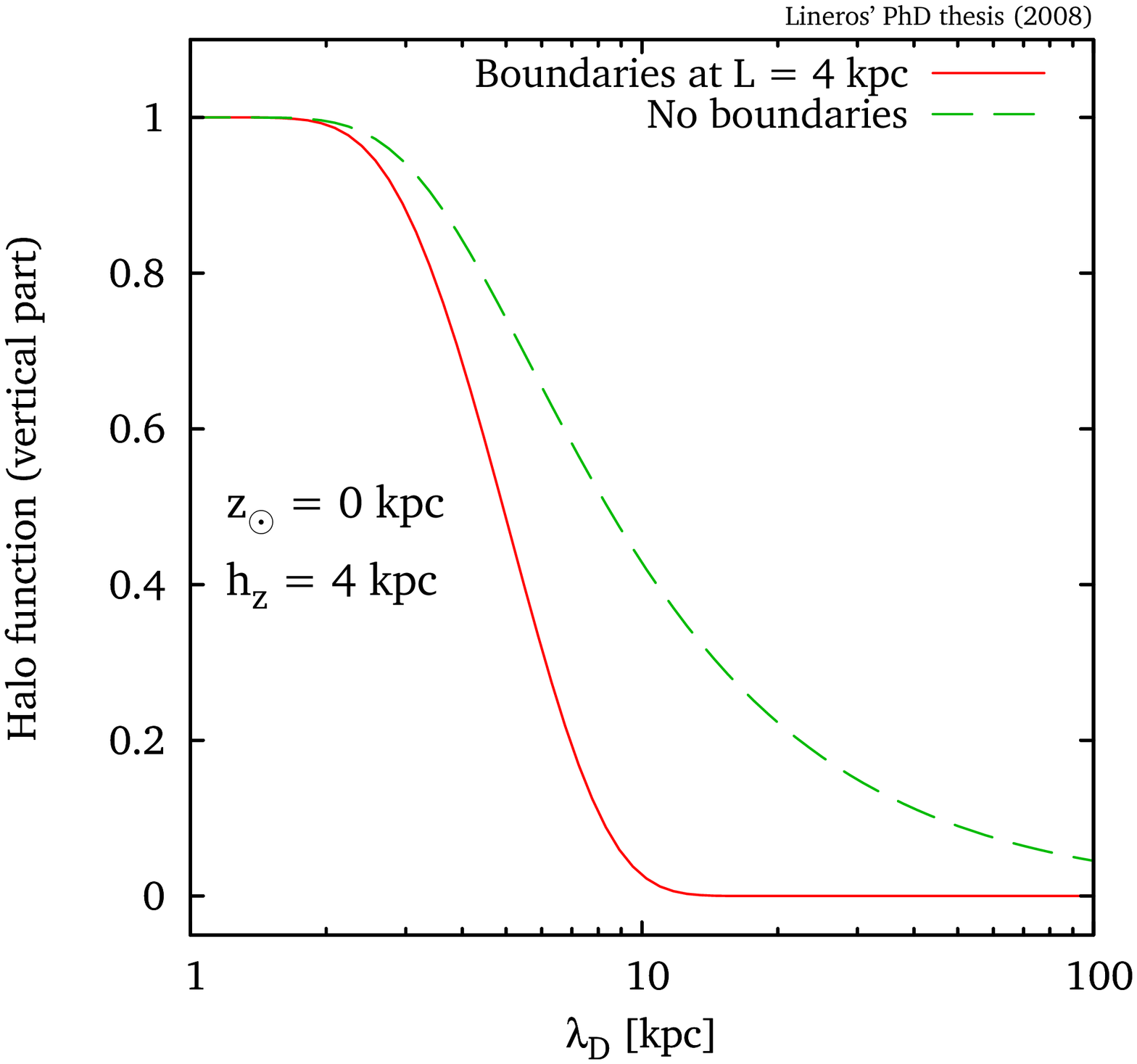}}
 \caption{\label{f:hf-bound-comp} Plot of the radial (right) and vertical (left) parts of axial symmetric HF versus $\lD$. The influence of boundary conditions produces that each part -- radial and vertical -- vanishes for values of $\lD$ larger than the closest boundary. For the radial part, the boundary effects start to dominate when $\lD \gtrsim 12 \; \tu{kpc}$. On the other hand, the same happends to the vertical part when $\lD \gtrsim 4\;\tu{kpc}$.}
\end{fig}

% % The effects of boundary conditions are also important when distributed sources are involved (we had previously studied the case of a punctual source). We saw that homogeneous sources let to separate the HF in two independent ones, which allows to visualize easily how big are these effects in comparison to the boundary--free case. \citefig{f:hf-bound-comp} shows the influence of radial and vertical boundaries. In the case of the radial part, it starts to decrease when $\lD$ is bigger than 10 \tu{kpc}, which is approximately the distance to the closest boundary (11.5 \tu{kpc}). Let us remind that Solar System is located at 8.5 \tu{kpc} from the GC in the Galactic plane. On the other hand, the vertical boundaries are closer (4 \tu{kpc}) explaining the decrease for $\lD$ bigger than 4 \tu{kpc}. \\  

When the radial and vertical HF are combined to form $\widetilde{I}_{\tn{axial}}$, the effect of vertical boundary conditions manifestates earlier if $L < R$ (\citefig{f:hf-homo}).
This produce that particles can reach first the vertical bondaries than the radial one.
Typically, $R$ is fixed to the Galactic radius (20~\tu{kpc}). This condition suggests that for $\lD$ of the order of $\sim 12\;\tu{kpc}$ the radial boundaries starts to affect the behaviour of HF.
Nevertheless, this limit depends on the source ditribution. If the sources are close to GC (the farthest point to any boundary), their contribution to the HF are less modified than HF for sources close to the boundaries.\\

\begin{fig}
 \includegraphics[width=0.5\textwidth]{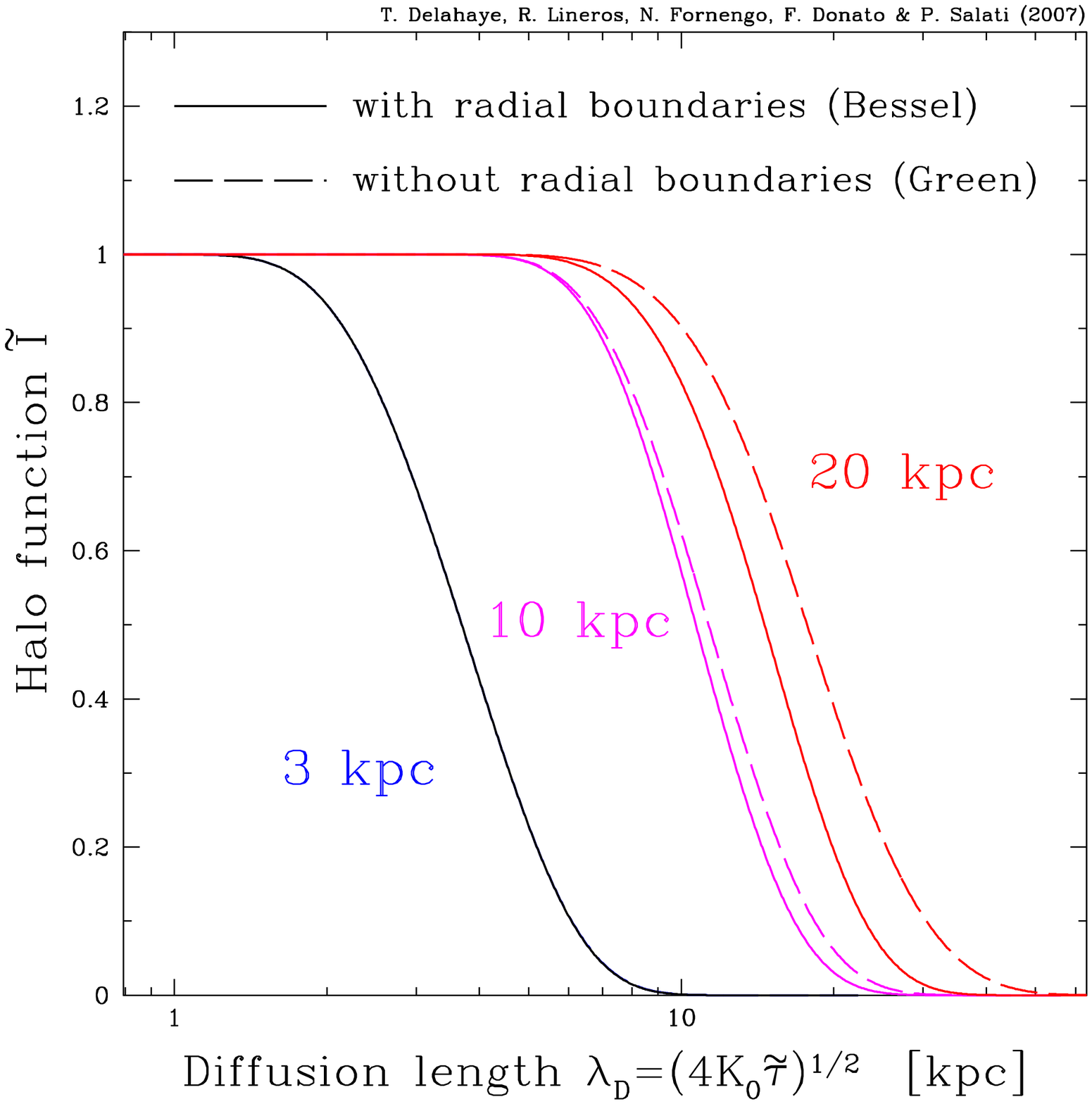}
 \caption{\label{f:hf-homo} Halo function versus diffusion length for a homogeneous sources for values of $L(=h_z)$ of 3, 10 and 20 \tu{kpc} ($R = 20 \; \tu{kpc}$). The effect of boundary condition in the PZ is reflected as a fast decreasing of HF at big values of $\lD$.}
\end{fig}

Some HF, from sources distributed in such a way that can not be separated as the previous example, are calculated following the general prescription (\citeeq{e:hf-helm}). In the case of axial symmetric sources the HF is:
\begin{eq}
 \widetilde{I}_{\tn{axial}}(\lD) = \sum_{i=1}^{\infty} \sum_{n:\tn{odd}}^{\infty} A_{i,j} \; J_0(k_{r,i} \; r_{\astrosun})  \; \exp\left(-\frac{1}{4} g^2_{i,n} \lD^2 \right) \; ,
\end{eq}
where the coefficients are:
\begin{eq}
 A_{i,j} = \frac{2}{L\; R^2 \; J_1^2(k_{r,i}\; R)} \; \int_{0}^{R} rdr\int_{-L}^{L} dz \; f(r,z)\; J_0(k_{r,i}\;r) \; \cos(k_{z,n}\;z) \; .
\end{eq}

The case of small $\lD$ is to calculate the HF neglecting radial boundary conditions:
\begin{eq}
\widetilde{I}_{\tn{axial}}(\lD) = \int_{0}^{R} r_s dr_s \int_{-L}^{L} dz_s \; f(r_s,z_s) \;  \widetilde{\Sigma}_{\tn{IC}}(z_{\astrosun},z_s,\lD) \; \widetilde{\Upsilon}_{\tn{free}}(r_{\astrosun},r_s,\lD)
\end{eq}
which depends on the function $\widetilde{\Sigma}_{\tn{IC}}$ (because it is fast to converge) and $\widetilde{\Upsilon}_{\tn{free}}$, which is a tilded Green function similar to $\widetilde{\Upsilon}$ (\citeeq{e:ups_boun}) but without radial boundary. This Green function has an analytical form, 
\begin{eq}
 \widetilde{\Upsilon}_{\tn{free}}(r,r_s,\lD) =  \frac{2}{\lD} \; \exp\left(-\frac{(r - r_s )^2}{\lD^2}\right) \; I_0^{\prime}\left(\frac{2\; r \; r_s}{\lD^2}\right) \;  
\end{eq}
where 
\begin{eq}
I_0^{\prime}(x) = e^{-x} \; I_0(x) \; ,
\end{eq}
which depends of the modified Bessel function $I_0(x)$ \cite{Arfken:2005}. In the regime of big $x$ values, that is equivalent to small $\lD$ values, this function is approximated by: 
\begin{eq}
 I_0^{\prime}(x) \longrightarrow \frac{1}{\sqrt{2\; \pi \;x}} \quad ( x \gg 1/4 ) \; .
\end{eq}
That establishes an upper limit on $\lD$,
\begin{eq}
 \sqrt{2 \; r_{\astrosun} \; r_s} \gg \frac{\lD}{2} \; ,
\end{eq}
where this approximation works.\\

%
%
%%%%%%%%%%%%%%%%%%%%%%%%%%%%%%%%%%%%%%%%%%%%%%%%%%%%%%%%%%%%%%%%%%%%%%%%%%%%%%%%%%%%%%%%%%%%%%%%%
\section{Parameters space.}

\begin{tab}
	\begin{tabular}{|c|ccccc|c|}
	\hline
	 label & $\delta$ & $K_0\;[\tu{kpc}^2/\tu{Myr}]$ & $L\;[\tu{kpc}]$ & $V_c\;[\tu{km}/\tu{sec}]$ & $V_a\;[\tu{km}/\tu{sec}]$ & $\chi^2_{\tn{B/C}}$ \\ 
	\hline \hline 
	 max & 0.46 & 0.0765 & 15 & 5 & 117.6 & 39.98 \\
   med & 0.70 & 0.0112 & 4 & 12 & 52.9 & 25.68 \\
   min & 0.85 & 0.0016 & 1 & 13.5 & 22.4 & 39.02 \\
  \hline
	\end{tabular}
	\caption{\label{t:prop-par} Astrophysical parameters giving the maximal, median and minimal antiproton flux compatible with B/C analysis \cite{Donato:2003xg}.}
\end{tab}

The model depends on 5 parameter. The medium height of PZ ($L$), the magnetic diffusion constant ($K_0$), the diffusion coefficient power index ($\delta$), the GW speed ($V_c$) and the Alfv\'en velocity ($V_a$) are determinated through the study of Nuclei CR measurements. \citefigg{f:un-plots}{f:un-p2} show the results presented by Maurin et al. \cite{Maurin:2001sj} based on a $\chi^2$ analysis to nuclei CR fluxes. To obtain that, they used the full propagation model, \ie with all processes included, and solving this by means of sofisticated numerical methods. \\

In the same spirit, Donato et al. \cite{Donato:2003xg} studied the problem of antiproton production obtaining constraints to the space of parameter (\citetab{t:prop-par}), which are totally compatible with B/C analysis.\\

The case of PETE is simpler, in the sense, it depends on just three parameters -- $L$, $K_0$ and $\delta$. Nevertheless, this reduced space of parameters is also described in the Maurin et al. framework because the leading processes, related to this three parameters, are taken into account for the positron and electron propagation.\\

\begin{fig}
 \includegraphics[width=0.7\textwidth]{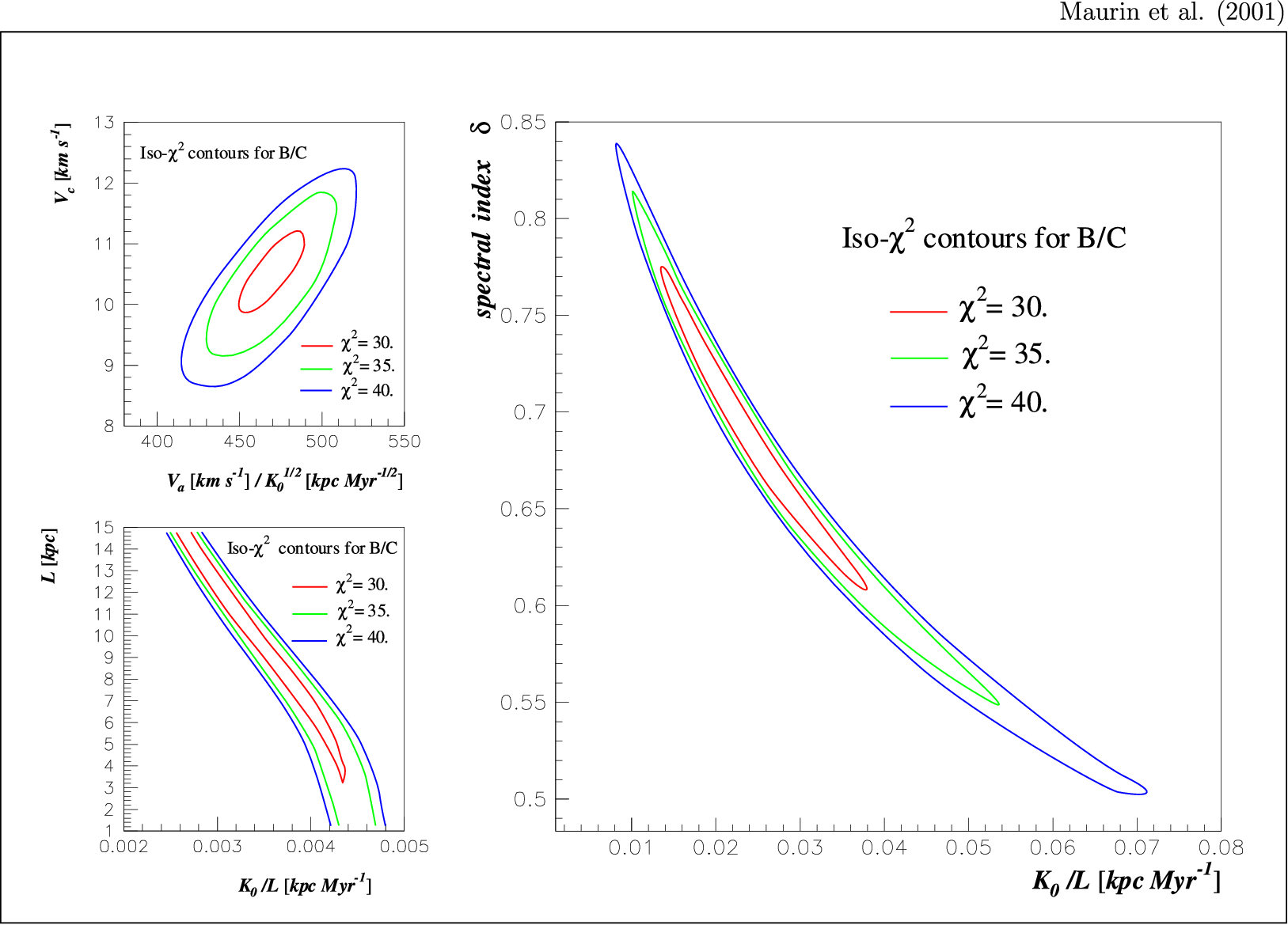}
\caption{\label{f:un-plots} $\chi^2$ contour iso-lines for some of the parameters of the TZPM~\cite{Maurin:2001sj}. The uncertainty bands for each parameters were obtained from the study of nuclei CR signals, specially in the B/C ratio case~\cite{Maurin:2001sj}. The two small plots show the $\chi^2$ contours when the power index $\delta$ is fixed to 0.6. The big plot shows the contours when $L = 3 \; \tu{kpc}$~\cite{Maurin:2001sj}.}
\end{fig}

\begin{fig}
 \includegraphics[width=0.7\textwidth]{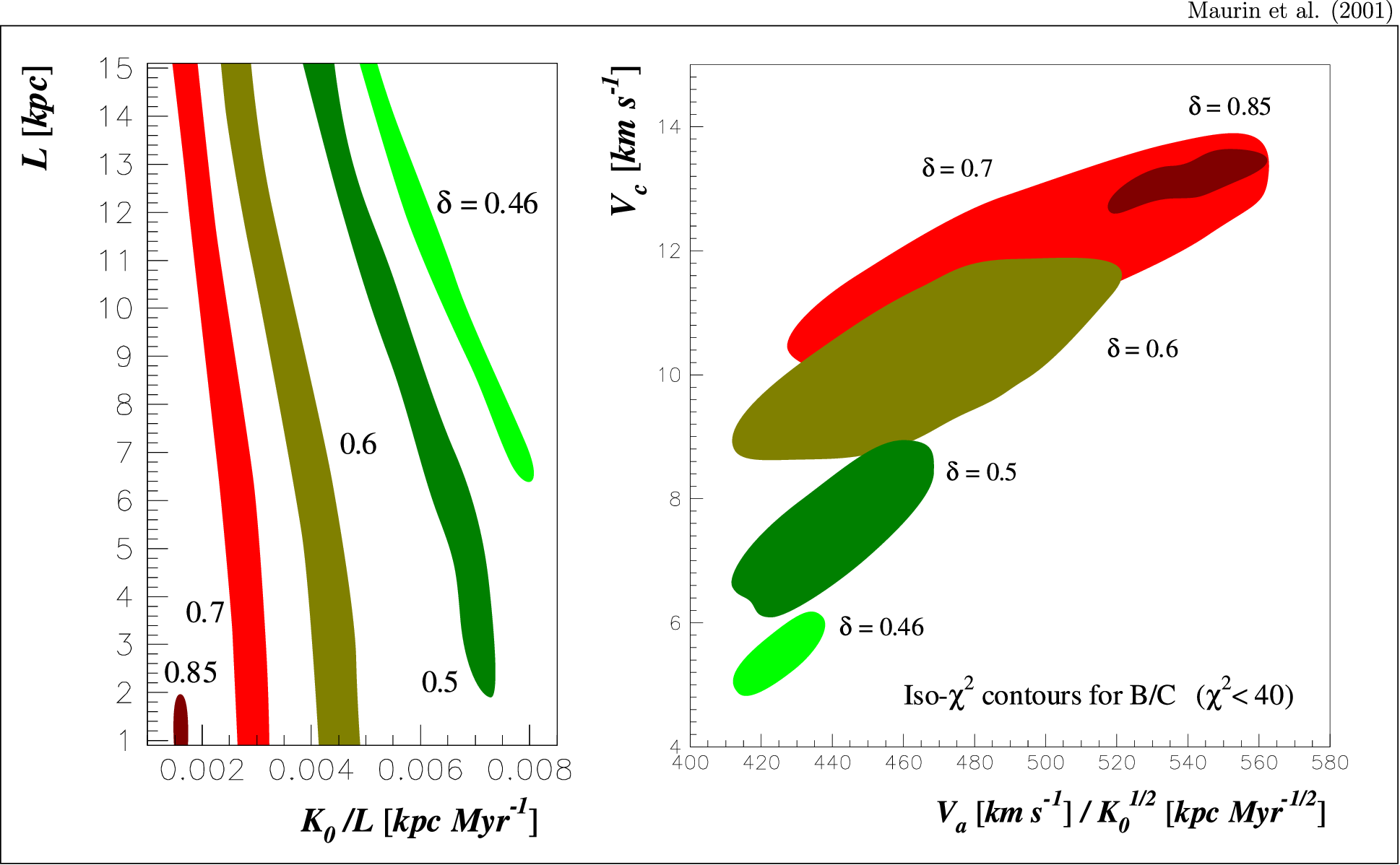}
\caption{\label{f:un-p2} Space of parameters of the TZPM compatible with B/C observations \cite{Maurin:2001sj}. All points plotted correspond to $\chi^2 < 40$ with a minimum of 24.}
\end{fig}

%
%
%%%%%%%%%%%%%%%%%%%%%%%%%%%%%%%%%%%%%%%%%%%%%%%%%%%%%%%%%%%%%%%%%%%%%%%%%%%%%%%%%%%%%%%%%%%%%%%%%
\section{Solar Modulation}
When \CR\ are closer to the Solar System, the Solar wind starts to pushing them away from the Solar neighborhood. The intensity in the cosmic--ray repulsion depends directly on Solar activity. The repulsion produces that \CR\ lose energy and makes harder that CR can reach the Earth~\cite{Fulks:1975}. \\

A simple approach is the Force--Field Modulation \cite{Perko:1983,McDonald:1981,Perko:1987A&A}, which models this phenomena as an electrical potential, which reduces CR energy and redistributes them. The \CR\ intensity observed at Earth's orbit ($J_{\earth}$) is related with the interstellar intensity ($J_{\infty}$) as follows:
\begin{eq}
	J_{\earth}(\ener{}) = \frac{\ener{}^2 - \mass{}^2}{\ener{\infty}^2 - \mass{}^2} \ J_{\infty}(\ener{\infty}),
\end{eq}
where $\ener{\infty}$ is the \CR\ energy at the boundary of the heliosphere.\\

Depending on the energy value, \CR\ are affected differently:
\begin{eq}
 \ener{\infty} = \left\{ \begin{array}{c@{\qquad}c} \displaystyle \mom{}\log\left(\frac{\mom{c} + \ener{c}}{\mom{} + \ener{}}\right) + \ener{} + \Phi & \ener{} < \ener{c} \\ \\
 \ener{} + \Phi & \ener{c} \le \ener{}\end{array} \right. ,
\end{eq}
where $\ener{c}$ represent the limit where those are affected by two different modulation regime. The low--energy regime, $\ener{} < \ener{c}$, is rarely used, however it is consistent with some studies about Solar Flares \cite{Perko:1983}. At a solar minimum, a typical value for $\mom{c}$ is $\sim 1 \ \tu{GeV}$ \cite{Perko:1987JGR}.\\

The commonly used regime ($\ener{c} \le \ener{}$) depends on an energy shift which is related to the modulation potential $\phi$,
\begin{eq}
 \Phi = |Z|e \phi ,
\end{eq}
where $|Z|e$ is the absolute value of \CR\ electrical charge. The modulation potential, in principle, should vary on time. It can be estimated from diffusion coefficient and solar wind velocity. Although it is typically considered as a free parameter that should be fixed in each experiment~\cite{Casadei2004ApJ}. \\

Temporal correlations between Neutron Monitors (NM) measurements, the Solar activity and intensity of cosmic--rays have been found~\cite{bartol:nm}. The information obtained in this type of analysis helps to improve the accuracy in the modulation potential $\phi$ determination.\\

% to demodulate the \CR\ signal and to estimate the interstellar \CR\ fluxes. In anycase, if this information is combined with \CR\ observation (\citetab{t:solmod_par}) it produces even more accurate estimation of the modulation potential.\\

\begin{tab}
	\begin{tabular}{|c|c|c|}
		\hline 
		Experiment & year & $\phi$ [\tu{MV}]  \\
		\hline \hline
		MASS 89 & 1989 & $1400\pm50$ \\
		MASS 91 & 1991 & $2000\pm200$ \\
		CAPRICE 94 & 1994 & $664\pm5$ \\
		HEAT 94 & 1994 & $650\pm50$ \\
		HEAT 95 & 1995 & $550\pm50$ \\
		BETS 97+98 &1997, 1998& $600\pm100$ \\
		AMS 98 & 1998 & $632\pm13$ \\
		\hline
	\end{tabular}
	\caption{\label{t:solmod_par} Estimated values of solar modulation parameter for a selected set of \CR\ experiments~\cite{Casadei2004ApJ}.}
\end{tab}

\cleardoublepage

\chapter{Positrons from DM annihilation in the galactic halo}
\label{cha4}
\begin{prechap}
The production of positrons from DM annihilation is a very exciting possibility to look for galactic DM. In order to study the positron signal it is necessary to study the propagation of positrons and the astrophysical uncertainties related to the propagation modeling.\\

This chapter is based on our work~\cite{Delahaye:2007fr}. \\
\end{prechap}

\section{Overview}

Secondary positrons and electrons are produced in the Galaxy from the interaction of nuclei CR on the ISM \cite{Moskalenko:1997gh} and are an important tool for the comprehension of cosmic-ray propagation. Data on the cosmic positron flux (often reported in terms of the positron fraction) have
been collected by several experiments \cite{Barwick:1997ig,Ahlen:1994ct,Alcaraz:2000PhLB,Aguilar:2007,Boezio:2000,Grimani:2002yz}. \\

% In particular, the HEAT data \cite{Barwick:1997ig} mildly indicate a possible excess of the positron fraction for energies above 10~\tu{GeV} and with respect to the available calculations for the secondary component \cite{Moskalenko:1997gh}.\\

The HEAT data \cite{Barwick:1997ig} have mildly indicated a possible excess of the positron fraction for energies above 10~\tu{GeV} with respect to the available calculations for the secondary component \cite{Moskalenko:1997gh}. But recently, in October 2008, the lastest results of PAMELA experiment~\cite{Boezio:2004jx} have confirmed this feature~\cite{Adriani:2008zr} in the fraction. \\

Different astrophysical contributions to the positron fraction in the 10~\tu{GeV} region have been explored \cite{Barwick:1997ig}, but only accurate and energy extended data could confirm the presence of a bump in the positron fraction and its physical interpretation. Alternatively, it has been conjectured that the possible excess of positrons found in the HEAT data could be due to the presence of DM annihilation in the galactic halo \cite{Baltz:1998xv,Hooper:2004bq}. Although, this interpretation is limited by the uncertainties in the halo structure and in the cosmic ray propagation modeling.\\

% Recently, it has been shown that the boost factor due to substructures in the DM halo depends on the positron energy and on the statistical properties of the DM distribution \cite{Lavalle:2006vb}. In addition, it has been pointed out that its numerical values is quite modest \cite{Lavalle:1900wn}.\\

In this chapter, the propagation of the positrons related DM annihilations are inspected through the solutions of the TZPM in connection with the study of the uncertainties due to propagation models compatibles with B/C measurements \cite{Maurin:2001sj}. To do this, the halo functions (\citecha{cha3}) for some of the standard DM distributions have been calculated, as well, typical spectra of DM annihilation in beyond the Standard Model theories have been used to study the effects of the propagation uncertainties on the positron signal.\\

% II PROPAGATION AND DM SOURCE
%\newpage
%\cleardoublepage
\section{Halo function and DM source}

%the particle physics part
%
We are here interested in primary positrons, namely the ones that are produced by the pair annihilations of DM particles.
According to the various supersymmetric theories, the annihilation of a DM pair leads either to the direct creation of an electron-positron pair or to the production of many species subsequently decaying into photons, neutrinos, hadrons and positrons. We have considered four possible annihilation channels which appear in any model of weakly interacting massive particles (WIMP).\\
The first one is the direct production of a $e^+ e^-$ pair and is actually generic for theories with extra-dimensions like the UED models \cite{Cheng:2002iz,Servant:2002aq,Appelquist:2000nn}. The energy of the positron line corresponds to the mass of the DM species.
We have alternatively considered annihilations into $W^{+} W^{-}$, $\tau^+ \tau^-$ and $b\bar{b}$ pairs. These unstable particles decay and produce showers which may contain positrons with a continuous energy spectrum.\\
%

%
%%%%%%%%%%%%%%%%%%%%%%%%%%%%%%%%%%%%%%%%%%%%%%%%%%%%%%%%%%%%%%%%%%%%%%%%%%%
%%%%%%%%%%%%%%%%%%%%%%%%%%%%%%%%%%%%%%%%%%%%%%%%%%%%%%%%%%%%%%%%%%%%%%%%%%%
\begin{tab}
{\begin{tabular}{|l|c|c|c|c|}
\hline
Halo model & $\alpha$ & $\beta$ & $\gamma$ & $r_s$ \tu{kpc} \\
\hline \hline
Cored isothermal~\cite{Bahcall:1980fb}
&  2  &  2  &  0  &  5  \\
Navarro, Frenk \& White~\cite{Navarro:1996gj}
&        1        &        3        &        1        &        20       \\
Moore~\cite{Diemand:2004wh}
&        1.5      &        3        &        1.3      &        30       \\
\hline
\end{tabular}}
\caption{\label{tab:indices}Dark matter distribution profiles in the Milky Way.}
\end{tab}
%%%%%%%%%%%%%%%%%%%%%%%%%%%%%%%%%%%%%%%%%%%%%%%%%%%%%%%%%%%%%%%%%%%%%%%%%%%
%%%%%%%%%%%%%%%%%%%%%%%%%%%%%%%%%%%%%%%%%%%%%%%%%%%%%%%%%%%%%%%%%%%%%%%%%%%
%

Whichever the annihilation channel, the source term is generically written as
\begin{eq}\label{source}
s\left({\mathbf x},\ener{e^+}\right) = \eta \; \langle\sigma v\rangle \; \left( {\displaystyle \frac{\rho(\mb{x})}{\mass{\chi}}} \right)^{2} \; \dnde{e^+} \; ,
\end{eq}
where the coefficient $\eta$ is a quantum term which depends on the particle being or not self--conjugate~: for instance, for a fermion it equals $1/2$ or $1/4$ depending on whether the WIMP is a self--conjugate particle or not. In what follows, we have considered a self--conjugate DM candidate (like the neutralino is supersymmetric theories) and taken $\eta = 1/2$.
The annihilation cross section is averaged over the momenta of the incoming DM particles to yield $\langle\sigma v\rangle$ , the value of which depends on the specific SUSY model and is constrained by cosmology \cite{Arina:2007}. We have actually taken here a benchmark value of $2.1 \times 10^{-26}$~\tu{cm}$^{3}$~\tu{sec}$^{-1}$ which leads to a relic abundance of $\Omega_{\chi} h^{2} \sim 0.14$ (in agreement with the WMAP observations~\cite{Spergel:2006hy,Hinshaw:2008kr}) under the hypothesis of dominant s--wave annihilation and by means of the relation:
\begin{eq}
\Omega h^{2} = 8.5 \cdot 10^{-11} \,\frac{g^{1/2}_{\star}(x_{f})}{g_{\star S}(x_{f})} \;\frac{\tu{GeV}^{-2}}{x_{f^{-1}}\langle \sigma v\rangle} = \frac{3 \cdot 10^{-27} \; \tu{cm}^{3} \; \tu{s}^{-1}}{\langle\sigma v\rangle}
\end{eq}
where $x_{f}=\mass{\chi}/T_{f}\simeq (20\div 25)$ with $T_{f}$ the freeze--out temperature and where $g_{\star}(x_{f})$ and $g_{\star S}(x_{f})$ denote the effective number of degrees of freedom of the energy and entropy density at freeze--out, respectively.\\ 

The DM mass $m_{\chi}$ is unknown. In the case of neutralinos, theoretical arguments as well as the LEP and WMAP results constrain this mass to range from a few \tu{GeV} \cite{Bottino:2002ry,Bottino:2003iu,Belanger:2002nr,Hooper:2002nq} up to a few \tu{TeV}. Keeping in mind the positron HEAT excess, we have chosen a neutralino mass of 100 \tu{GeV}. We have also analyzed the positron signal yielded by a significantly heavier DM particle of 500 \tu{GeV}.\\

Finally, all of the details in the energy distribution of the positrons $\left(\displaystyle\dnde{e^+}\right)$ produced in a single WIMP annihilation  is described in the second part of \citecha{cha2}.\\

%\vskip 0.1cm
%\noindent
%the astrophysical part
%
The only astronomical ingredient in the source term~(\citeeq{source}) is the DM distribution $\rho(\mb{x})$ inside the Milky Way halo. We have considered the generic profile 
\begin{eq} \label{eq:profile} 
\rho(r) = \rho_{\astrosun} \; \left( {\displaystyle \frac{r_{\astrosun}}{r}} \right)^{\gamma} \; \left( {\displaystyle \frac{1 \, + \, (r_{\astrosun}/r_{s})^{\alpha}} {1 \, + \, (r/r_{s})^{\alpha}}} \right)^{(\beta - \gamma) / \alpha} \;\; , 
\end{eq} 
where $r_{\astrosun} = 8.5$~\tu{kpc} is the galactocentric distance of the solar system. Notice that $r$ denotes here the radius in spherical coordinates. The solar neighborhood DM density has been set equal to $\rho_{\astrosun} = 0.3$~\tu{GeV}~\tu{cm}$^{-3}$ \cite{Berezinsky:1992}.
We have discussed three profiles: an isothermal cored distribution~\cite{Bahcall:1980fb} for which $r_{s}$ is the radius of the central core, the Navarro, Frenk and White profile~\cite{Navarro:1996gj} (hereafter NFW) and Moore's model~\cite{Diemand:2004wh}. The NFW and Moore profiles have been numerically established thanks to N-body simulations. In the case of the Moore profile, the index
$\gamma$ lies between 1 and 1.5 and we have chosen a value of 1.3~(\citetab{tab:indices}).\\

The possible presence of DM substructures inside those smooth distributions enhances the annihilation signals by the so-called boost factor, whose value is still open to debate~\cite{Baltz:2001ir,Kane:2001fz,Jeannerot:1999yn}. Nevertheless, it has recently been shown that the boost factor due to substructures in the DM halo depends on the positron energy and on the statistical properties of the DM distribution \cite{Lavalle:2006vb}. In addition, it has been pointed out that its numerical values is quite modest \cite{Lavalle:1900wn}, being of the order of 10-30.\\

%\vskip 0.1cm
%\noindent
%the astrophysical part
%
The positron flux at the Solar System is expressed as 
\begin{eq} \label{eq:fluxe}
\fluxe(\epsilon) = \frac{\beta\;c}{4 \pi} \; \left(\kappa \; \frac{\tau_{E}}{\epsilon^{2}} \,
\int_{\epsilon}^{\infty} d\epsilon_{s} \; \dnde{e^+}(\epsilon_{s}) \; \widetilde{I}(\lD)  \right)\; ,
\end{eq} 
where the information related to particle physics is contained in 
\begin{eq} 
\kappa = \eta \; \langle \sigma v \rangle \; \left( {\displaystyle
\frac{\rho_{\astrosun}}{\mass{\chi}}} \right)^{2} \; ,
\end{eq} 
and the halo functions ($\widetilde{I}$) for the different DM distribution are calculated using the methods described in \citecha{cha2}. \\

\begin{fig}
\resizebox{\textwidth}{!}{\includegraphics[width=0.5\textwidth]{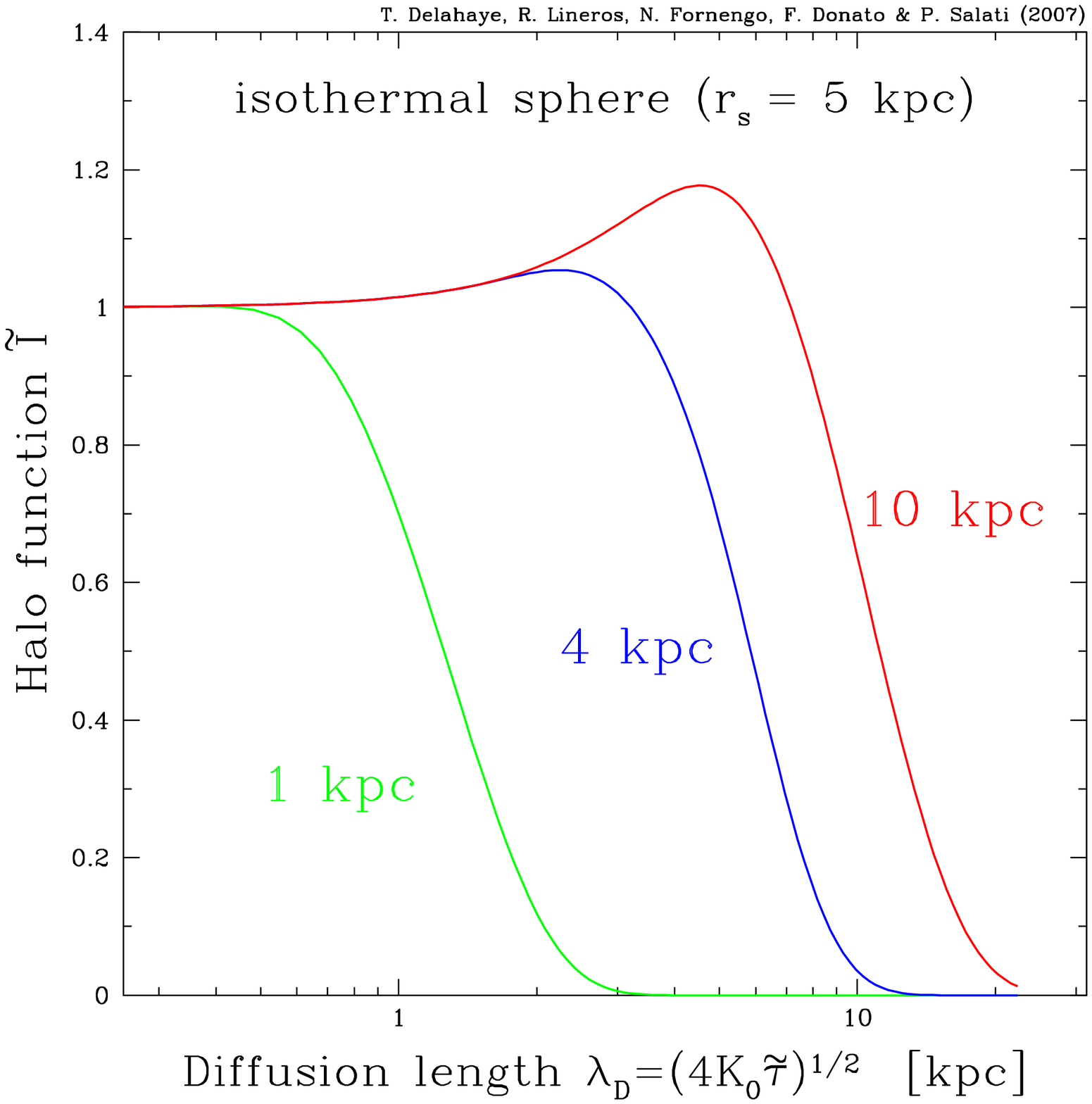}
\includegraphics[width=0.5\textwidth]{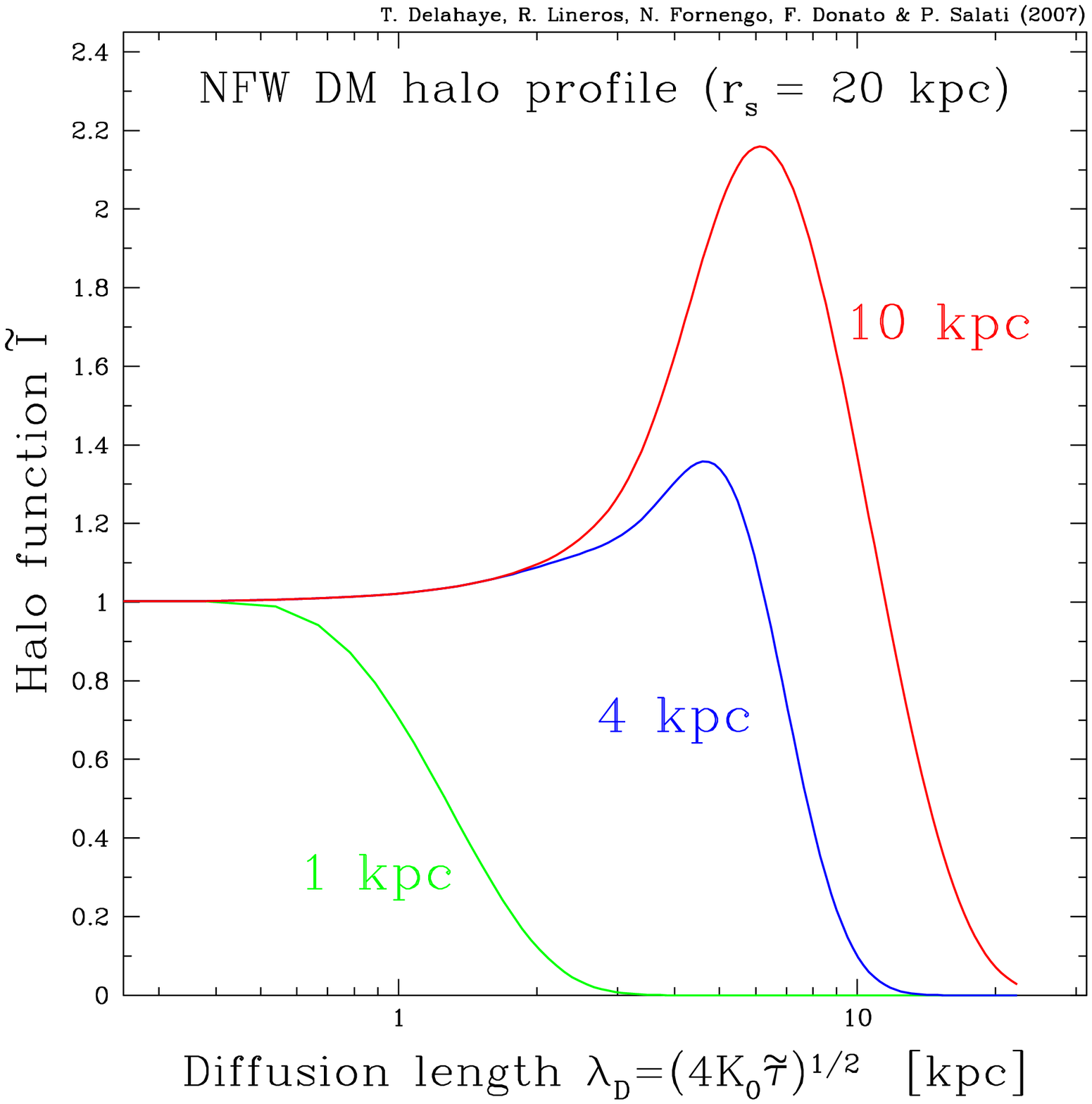}}
\caption{\label{f:hf-dmprof} The halo convolution $\tilde{I}$ is plotted as a function of the diffusion length $\lD$ for various values of the PZ half-thickness $L$. The left panel features the case of an isothermal DM distribution whereas a NFW profile has been assumed in the right panel (\citetab{tab:indices}). When $L$ is large enough for the positron horizon to reach the galactic center and its denser DM distribution, a maximum appears in the curves for $\lD \sim r_{\astrosun}$.}
\end{fig}

% description of f:hf-dmprof

\citefig{f:hf-dmprof} shows the effect of the overdensity at GC on the halo function. The two DM profiles are specially different in the central region, that becomes evident in the shape of each HF. When $\lD$ is around 8.5~\tu{kpc} -- \ie the distance to the galactic center --  both HF present an increment that depends on how big is their respective DM profile in the GC surroundings. Furthermore, the value of $L$ takes a very important role in the HF behaviors. When $L$ is small, both HF are similar because particles from the central region may escape earlier out of the PZ. On the other hand, a bigger $L$  increase the chances of particles to arrive until the Solar System.\\

However, we insist again on the fact that the true argument of the HF, whatever the approach followed to derive it, is the diffusion length $\lD$. The HF encodes the information relevant to cosmic ray propagation through the height $L$ of the diffusive zone, the normalization $K_{0}$ of the diffusion coefficient and its spectral index $\delta$. It is also the only relevant quantity concerning the DM distribution. The analysis of the various astrophysical uncertainties that may affect the positron signal of annihilating WIMPs will therefore be achieved by studying the behavior of $\widetilde{I}$.\\

% 
%
%
%%%%%%%%%%%%%%%%%%%%%%%%%%%%%%%%%%%%%%%%%%%%%%%%%%%%%%%%%%%%%%%%%%%%%%%%%%%%%%%%%%%%%%%%%%%%%%%%%%%%%%%%%%%%%%%
\subsubsection{The central divergence}

Numerically derived DM profiles -- NFW and Moore -- exhibit a divergence at the center of the Milky Way. The density increases like $r^{- \gamma}$ for small radii but cannot exceed the critical value for which the WIMP annihilation timescale is comparable to the age of the galactic bulge. The saturation of the density typically occurs within a sphere of $\sim 10^{-7}$~\tu{pc}, a much shorter distance than the size of the space increment in the numerical simulations. \\

Fortunately, this numerical difficulty can be eluded by noticing that the positrons Green propagator does not vary much over the central \dm\ distribution. This led us to replace inside a sphere of radius $r_{0}$ the $r^{- \gamma}$ cusp with the smoother profile 
\begin{eq}
\rho^{*}(r) = \rho_{0} \left( 1 + a_{1} j_0\big(\pi\frac{r}{r_0}\big) + a_{2}j_0\big(2 \pi \frac{r}{r_0}\big) \right)^{1/2} ,
\label{e:profile_smooth} 
\end{eq}
where $j_0$ denotes spherical Bessel function of first kind \cite{Arfken:2005}. The coefficients $\rho_{0}$, $a_{1}$ and $a_{2}$ are obtained by requiring that both the smooth density $\rho^{*}$ and its first derivative ${d\rho^{*}}/{dr}$ are continuous at $r_{0}$. The other crucial condition is the conservation of the total number of annihilations within $r_{0}$ as the diverging cusp $\rho \propto r^{- \gamma}$ is replaced by the distribution $\rho^{*}$. These conditions imply that $\rho_{0} \equiv \rho(r_{0})$ whereas
\begin{eq}
a_1 &=& a_2 + 2\gamma, \\
a_2 &=& 8 \gamma \left( \frac{\pi^{2} - 9 + 6 \gamma}{9(3 - 2\gamma)} \right).
\end{eq}

In \citefig{f:plat-smooth}, the halo integral $\widetilde{I}$ is plotted as a function of the diffusion length $\lD$ in the case of the Moore profile and assuming a PZ half-thickness $L = 10$ \tu{kpc}. Within the radius $r_{0}$, the central \dm\ divergence has been replaced either by a plateau with constant density $\rho_{0}$ or by the renormalized profile (\citeeq{e:profile_smooth}).
In the case of the plateau, the maximum which $\widetilde{I}$ reaches for a diffusion length $\lD \sim 7$ \tu{kpc} is underestimated even if values as small as 100 \tu{pc} are assumed for $r_{0}$. The larger that radius, the fewer the annihilations taking place within $r_{0}$ as compared to the Moore cusp and the worse the miscalculation of the halo integral. Getting the correct result featured by the solid red line would require a plateau radius so small that it would need to much numerical precision.\\
On the contrary, we observe that the halo integral $\widetilde{I}$ is stable with respect to a change of $r_{0}$ when the renormalized density $\rho^{*}$ is used.\\

\begin{fig}
\includegraphics[width=0.5\textwidth]{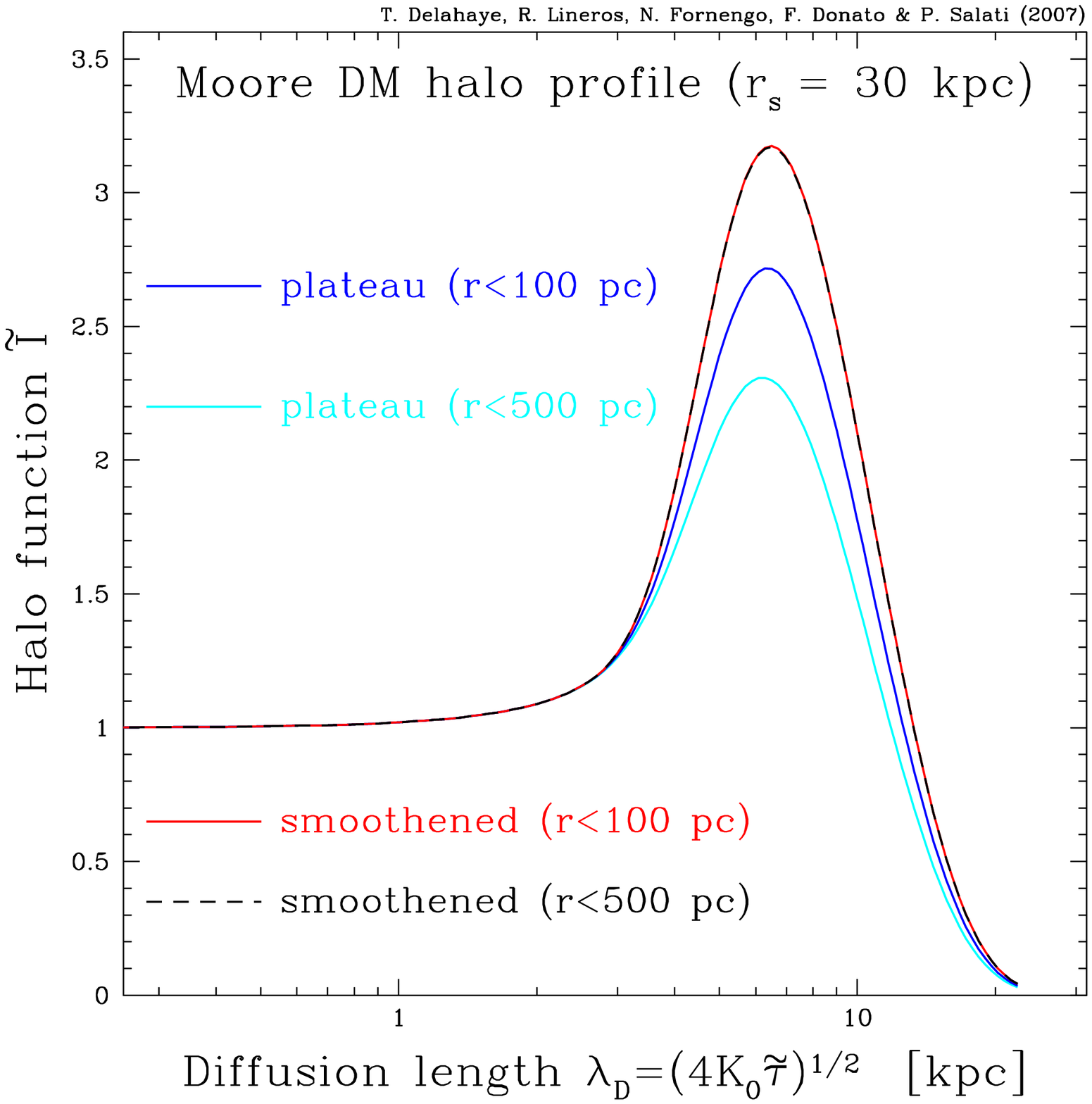}
\caption{\label{f:plat-smooth}
Same plot as \citefig{f:hf-dmprof}, the central DM profile within a radius $r_{0}$ is either a plateau at constant density $\rho_{0}$ or the smooth distribution $\rho^{*}$ of \citeeq{e:profile_smooth}.
In the former case, the bump which $\widetilde{I}$ exhibits is significantly underestimated even for values of $r_{0}$ as small as 100~\tu{pc} -- solid dark blue -- and drops as larger values are considered -- solid light blue. On the contrary, if the DM cusp is replaced by the smooth profile $\rho^{*}$, the halo integral no longer depends on the renormalization radius $r_{0}$ and the solid red and long-dashed black curves are superimposed on each other.}
\end{fig}
%

% III PROPAGATION UNCERTAINTIES ON THE HALO INTEGRAL

%\newpage
%\cleardoublepage
%
%  CHAPTER 4 - SECTION POSITRON FLUXES
%

\section{Propagation uncertainties on the halo integral}

% Following the prescription which has been given in the previous section, we can calculate accurately and quickly the halo integral $\widetilde{I}$ using either the Green propagator method or the Bessel expansion technique according to the typical diffusion length $\lD$.

To compute the uncertainties in the positron signal, the best strategy is to compute first the HF for each DM profile, and in that way, we are equipped with a rapid enough method for scanning the $\sim$ 1,600 different cosmic ray propagation models that have been found compatible \cite{Maurin:2001sj} with the B/C measurements. Each model is characterized by the half-thickness $L$ of the diffusion zone and by the normalization $K_{0}$ and spectral index $\delta$ of the space diffusion coefficient.  A large variation in these parameters is found in \cite{Maurin:2001sj} and yet they all lead to the same B/C ratio. The height $L$ of the PZ lies in the range from 1 to 15 \tu{kpc}. Values of the spectral index $\delta$ extend from 0.46 to 0.85 whereas the ratio ${K_{0}}/{L}$ varies from $10^{-3}$ to $8 \times 10^{-3}$ \tu{kpc} \tu{Myr}$^{-1}$ (See \citecha{cha3}).\\

%\vskip 0.1cm
%\noindent
% In this section, we analyze the sensitivity of the positron halo integral $\widetilde{I}$ with respect to galactic propagation. We would like eventually to gauge the astrophysical uncertainties which may affect the predictions on the positron DM signal.\\
%
% A similar investigation -- with only the propagation configurations that survive the B/C test -- has already been carried out for secondary \cite{Donato:2001ms} and primary \cite{Donato:2003xg} antiprotons. In the later case, three specific sets of parameters have been derived corresponding to minimal, medium and maximal antiproton fluxes -- see \citetab{t:prop-par}.\\

In the work about primary antiprotons performed by Donato et al. \cite{Donato:2003xg}, they found that the space of parameter obtained from the antiproton flux analysis is still compatible with the B/C analysis. Nevertheless, it is almost impossible to know if the same space produces equivalent results in the positron signal. In principle, nuclei CR and positrons (and electrons) are very different particles from the point of view of the physics of propagation. A first work on this topic was done by Hooper et al. \cite{Hooper:2004bq}, but they scanned the TZPM parameters independently without any connection with other CR species analysis.\\

% Do these configurations play the same role for positrons? Can we single out a few propagation models which could be used  later on to derive the minimal or the maximal positron flux without performing an entire scan over the parameter space?\\
%
% These questions have not been addressed in the pioneering investigation of \cite{Hooper:2004bq} where the cosmic ray propagation parameters have indeed been varied but independently of each other and without any connection to the B/C ratio.\\

%
% FIGURE
\begin{fig}
\resizebox{\textwidth}{!}{\includegraphics[width=\textwidth]{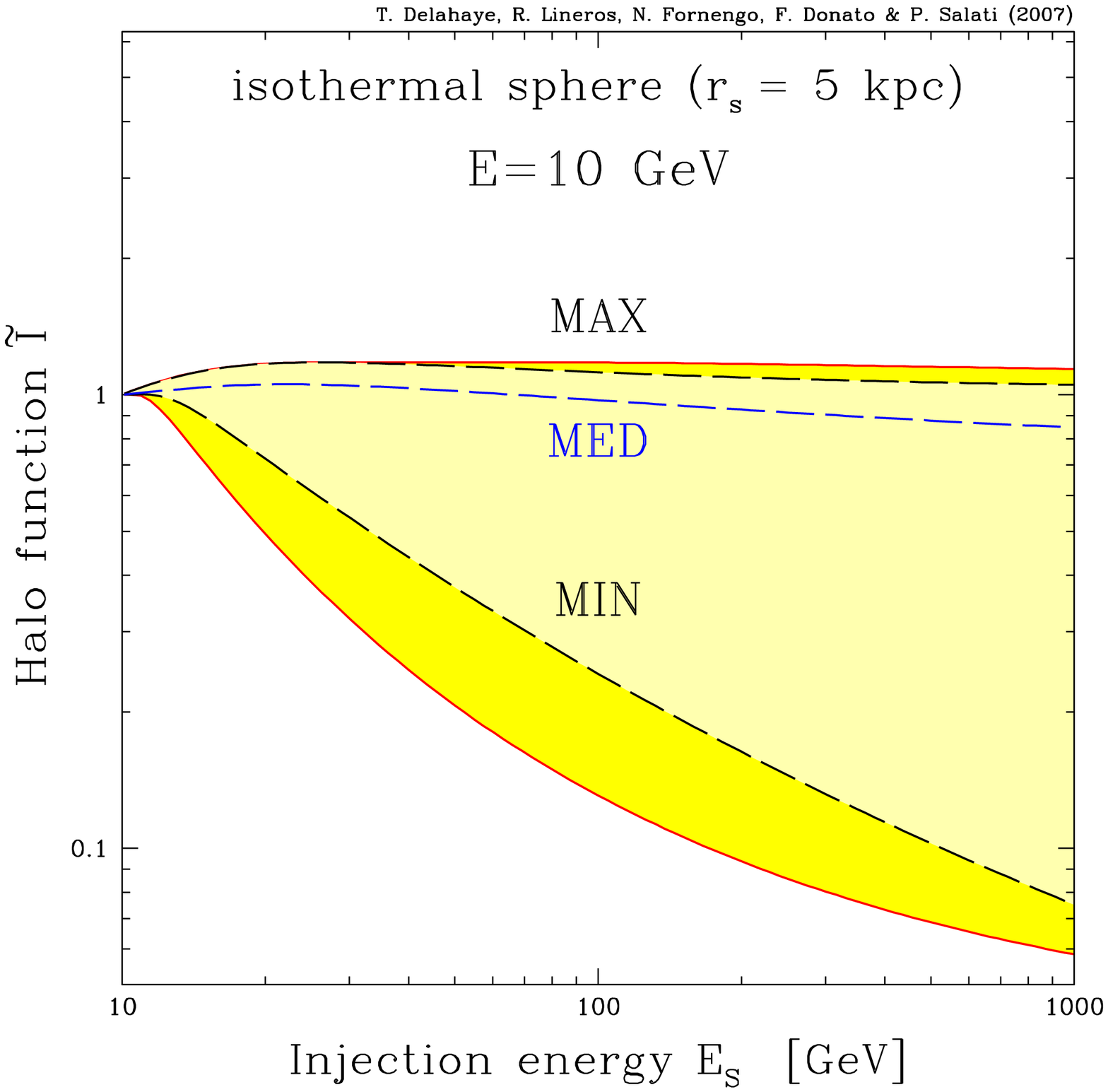}
\includegraphics[width=\textwidth]{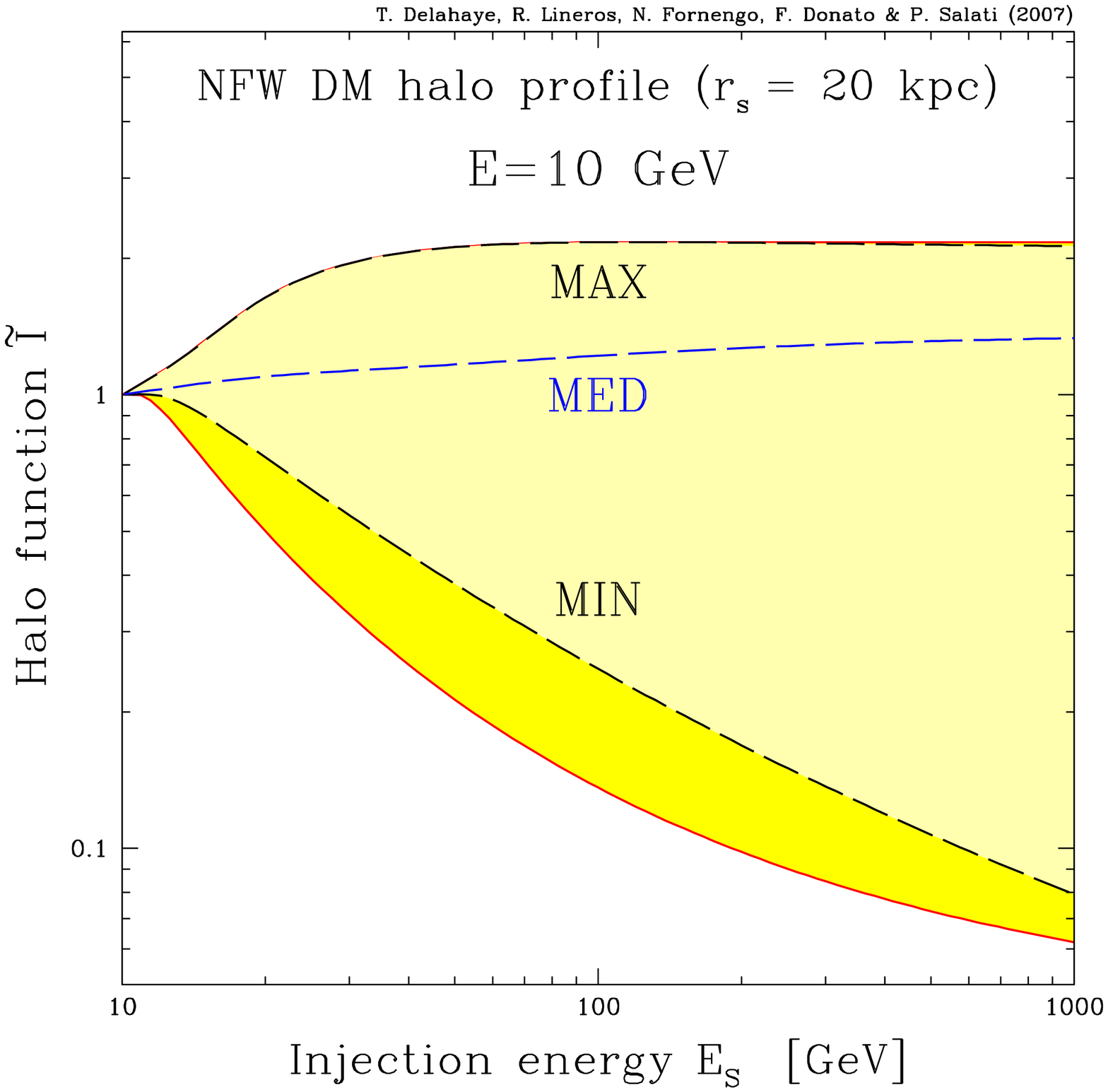}
\includegraphics[width=\textwidth]{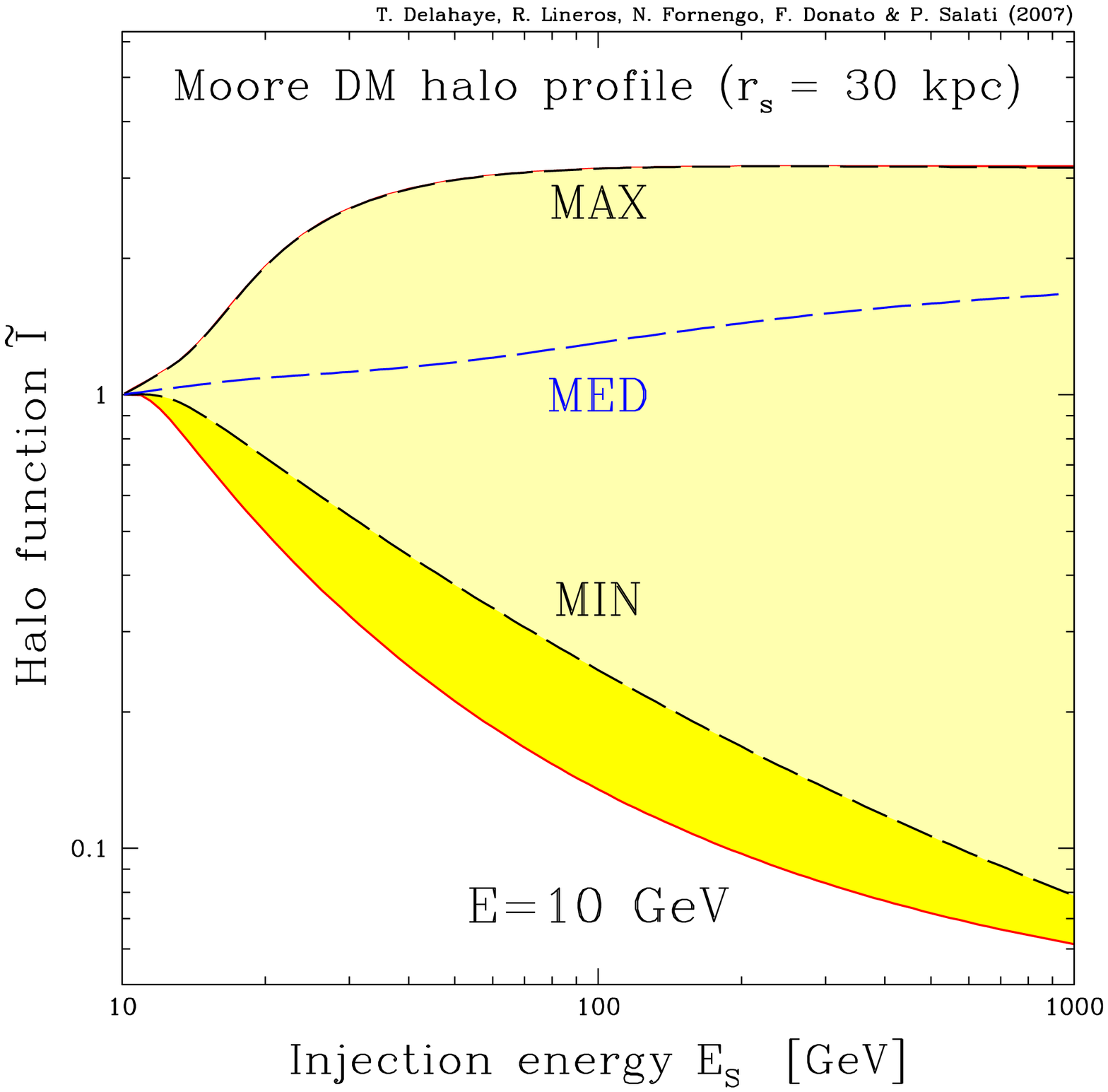}}
\caption{\label{fig5} In each panel, the halo integral $\widetilde{I}$ is plotted as a function of the positron injection energy $\ener{s}$ whereas the energy $\ener{}$ at the Earth is fixed at 10~\tu{GeV}. The galactic DM halo profiles of \citetab{tab:indices} are featured.
The curves labeled as MED correspond to the choice of cosmic ray propagation parameters which best-fit the B/C ratio \cite{Maurin:2001sj}. The MAX and MIN configurations correspond to the cases which were identified to produce the maximal and minimal DM antiproton fluxes \cite{Donato:2003xg}, while the entire colored band corresponds to the complete set of propagation models compatible with the B/C analysis \cite{Maurin:2001sj}. The MAX, MED and MIN configuration are given in \citetab{t:prop-par}.} 
\end{fig}
% FIGURE
%

%
%

%
%
%
In \citefig{fig5}, we have set the positron detection energy $\ener{}$ at 10~\tu{GeV} and varied the injection energy $\ener{s}$ from 10~\tu{GeV} up to 1~\tu{TeV}. The three panels correspond to the DM halo profiles of \citetab{tab:indices}. For each value of the injection energy $\ener{s}$, we have performed a complete scan over the 1,600 different configurations mentioned above and have found the maximal and minimal values of the HF $\widetilde{I}$ with the corresponding sets of propagation parameters. In each panel, the resulting uncertainty band corresponds to the yellow region extending between the two solid red lines. The lighter yellow domain is demarcated by the long-dashed black curves labeled MIN and MAX and has a smaller spread. The MED configuration is featured by the long-dashed blue line.\\
%

% FIGURE
\begin{fig}
\includegraphics[width=0.5\textwidth]{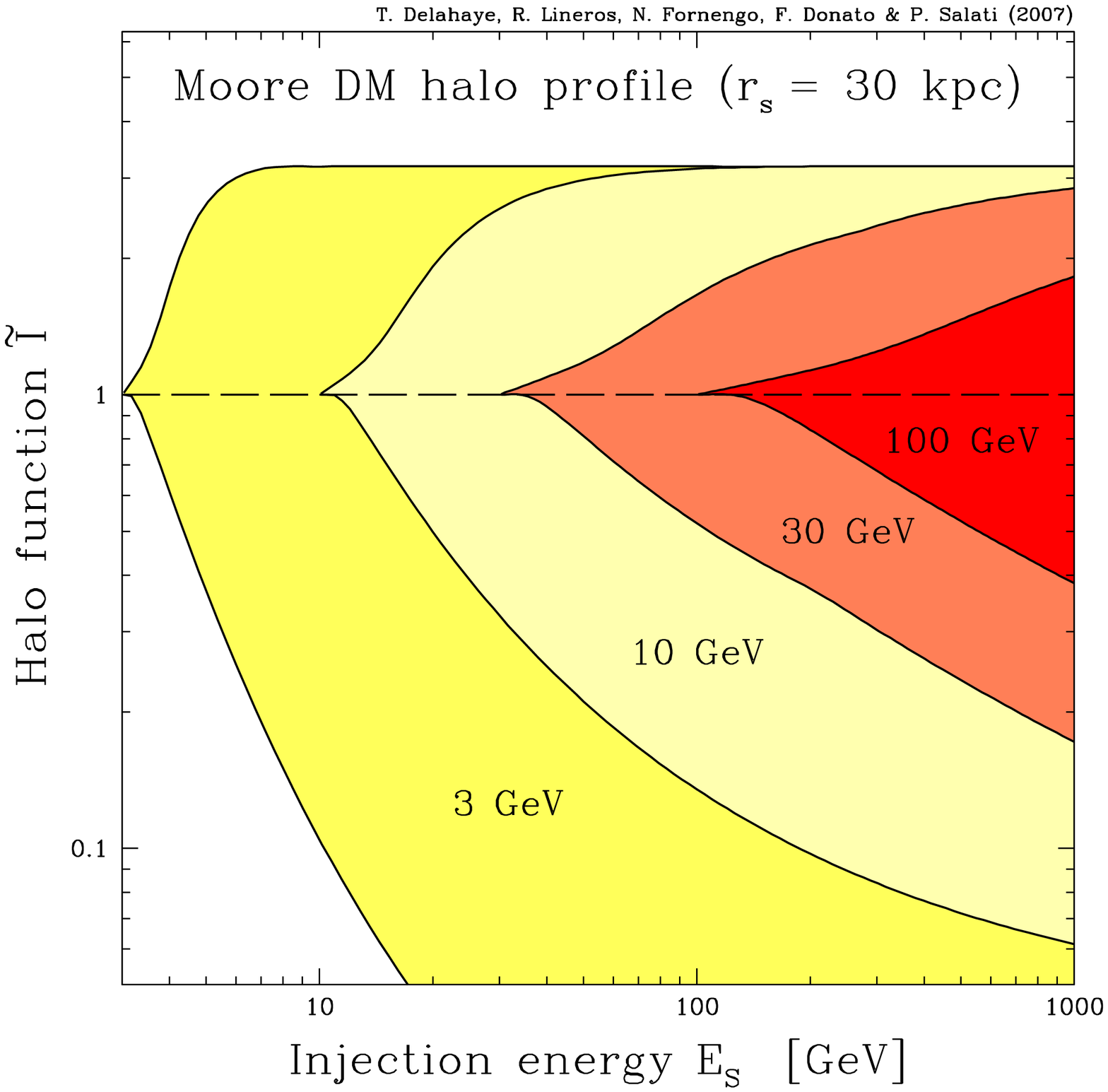}
\caption{\label{fig6} Same plot as \citefig{fig5} the Moore DM profile has been selected. Four values of the positron detection energy $\ener{}$ have been assumed. The flag-like structure of this figure results from the widening of the uncertainty band as the detection energy $\ener{}$ is decreased.}
\end{fig}
% FIGURE

In \citefig{fig6}, the Moore profile has been chosen  with four different values of the detection energy $\ener{}$. The corresponding uncertainty bands are coded with different colors and encompass each other as $\ener{}$ increases.\\
As $\ener{s}$ gets close to $\ener{}$, we observe that each uncertainty domain shrinks. In that regime, the diffusion length $\lD$ is very small and the positron horizon probes only the solar neighborhood where the DM density is given by $\rho_{\odot}$. Hence the flag--like structure of \citefig{fig6} and a halo integral $\widetilde{I}$ of order unity whatever the propagation model.\\

%\vskip 0.1cm
%\noindent
As is clear from \citefig{f:hf-dmprof}, a small half-thickness $L$ of the PZ combined with a large diffusion length $\lD$ implies a small positron HF $\widetilde{I}$. The lower boundaries of the various uncertainty bands in \citefig{fig5} and \fref{fig6} correspond therefore to parameter sets with $L = 1$ \tu{kpc}. Large values of $\lD$ are obtained when both the normalization $K_{0}$ and the spectral index $\delta$ are large (\citeeq{e:def-ld}). However both conditions cannot be satisfied together once the B/C constraints are applied. For a large normalization $K_{0}$, only small values of $\delta$ are allowed and vice versa. For small values of the detection energy $\ener{}$, the spectral index $\delta$ has little influence on $\lD$ and the configuration which minimizes the halo integral $\widetilde{I}$ corresponds to the large normalization $K_{0} = 5.95 \times 10^{-3}$ \tu{kpc}$^{2}$ \tu{Myr}$^{-1}$ and the rather small $\delta = 0.55$. For large values of $\ener{}$, the spectral index $\delta$ becomes more important than $K_{0}$ in the control of $\lD$. That is why in \citefig{fig6}, the lower bound of the red uncertainty domain corresponds now to the small normalization $K_{0} = 1.65 \times 10^{-3}$ \tu{kpc}$^{2}$ \tu{Myr}$^{-1}$ and the large spectral index $\delta = 0.85$. Notice that this set of parameters is very close to the MIN configuration of \citetab{t:prop-par}. For intermediate values of $\ener{}$, the situation becomes more complex. We find in particular that for $\ener{} = 30$~\tu{GeV}, the halo integral $\widetilde{I}$ is minimal for the former set of parameters as long as $\ener{s} \leq 200$~\tu{GeV} and for the later set as soon as $\ener{s} \geq 230$~\tu{GeV}. In between, a third propagation model comes into play with the intermediate values $K_{0} = 2.55 \times 10^{-3}$ \tu{kpc}$^{2}$ \tu{Myr}$^{-1}$ and $\delta = 0.75$. It is not possible therefore to single out one particular combination of $K_{0}$ and $\delta$ which would lead to the minimal value of the HF and of the positron DM signal. The MIN configuration which appeared in the antiproton analysis has no equivalent for positrons.\\

\begin{fig}
 \includegraphics[angle=270,width=0.5\textwidth]{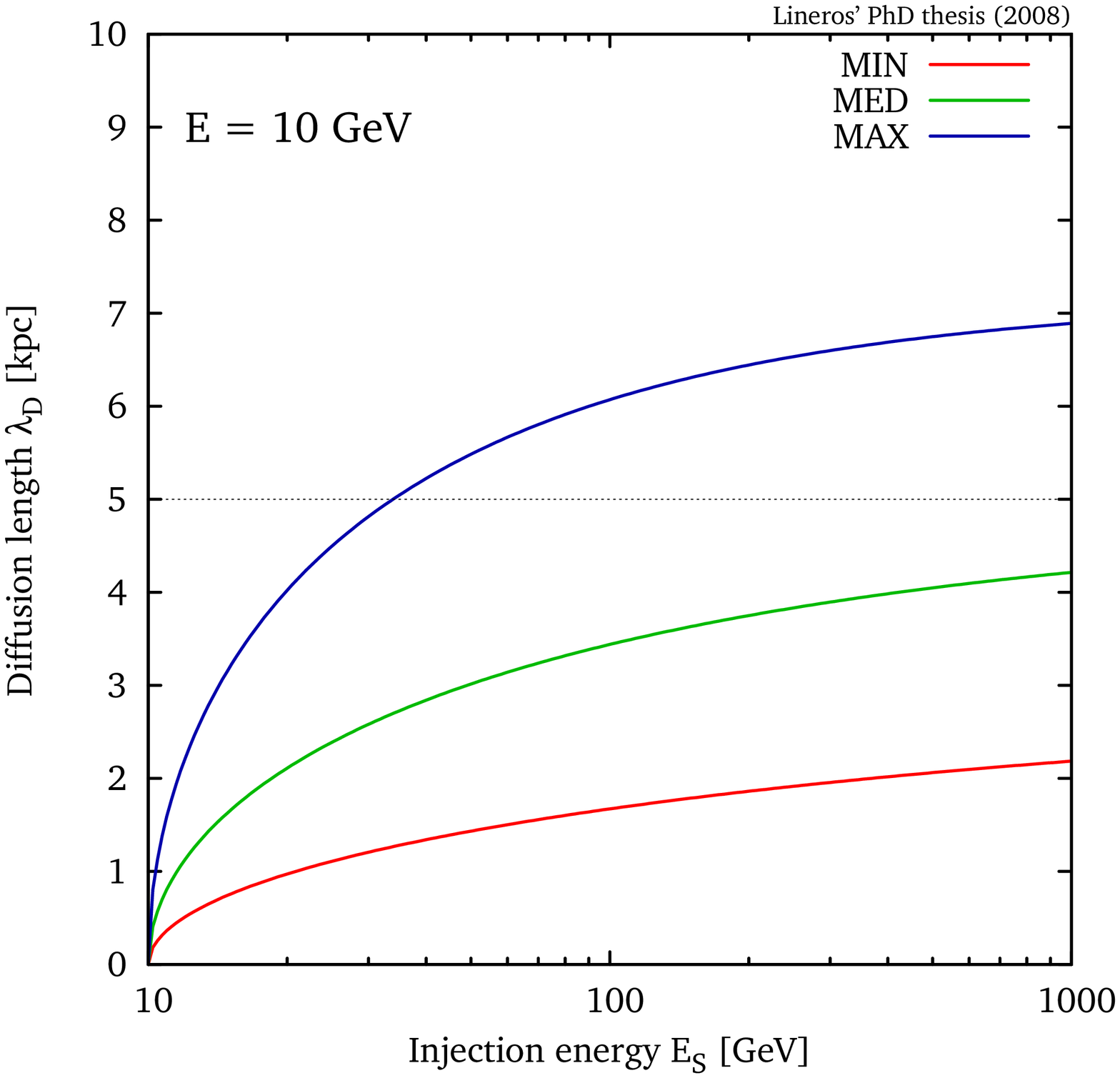}
 \caption{\label{f:lambdaBC} Diffusion length $\lD$ versus injection energy. The line at 5~\tu{kpc} has been traced to denote that $\lD$ bigger than this limit represent the range where the central DM overdensities start to appear. The two extreme B/C analysis configurations and the best--fit MED (\citetab{t:prop-par}) have been used to calculated their associated $\lD$ , showing how the MAX configuration covers longer distances instead of MED and MIN.}
\end{fig}

The same conclusion holds, even more strongly, in the case of the upper boundaries of the uncertainty bands.
Whatever the DM halo profile, the panels of \citefig{f:hf-dmprof} feature a peak in the halo function $\widetilde{I}$ for large values of $L$ and for a specific diffusion length $\lDM \sim 7$ \tu{kpc}. At fixed $\ener{}$ and $\ener{s}$, we anticipate that the maximal value for $\widetilde{I}$ will be reached for $L = 15$ \tu{kpc} and for a diffusion length $\lD$ as close as possible to the peak value $\lDM$. Two regimes can be considered at this stage:\\

\begin{itemize}
\item[(i)] When $\lD < \lDM$. This condition is satisfied when $\ener{}$ and $\ener{s}$ are close enough. Remember that $\ener{} = \ener{s}$ implies $\lD = 0$ for any set of $K_0$ and $\delta$ (\citeeq{e:def-ld}). Just to clarify the following discussion, \citefig{f:lambdaBC} shows the evolution of $\lD$ for the parameter sets MAX, MED and MIN of the B/C analysis. The objective is to look for the largest possible values of $\lD$ that maximize $\widetilde{I}$. This is fulfilled at least by two propagation models.
For small $\ener{}$, the large normalization $K_{0} = 7.65 \times 10^{-2}$ \tu{kpc}$^{2}$ \tu{Myr}$^{-1}$ is preferred with $\delta = 0.46$. We recognize the MAX configuration of \citetab{t:prop-par} and understand why the long-dashed black curves labeled MAX in the panels of \citefig{fig5} are superimposed on the solid red upper boundaries.
For large $\ener{}$, the spectral index $\delta$ dominates the diffusion length $\lD$ and takes over the normalization $K_{0}$ of the diffusion coefficient. The best model which maximizes $\widetilde{I}$ becomes then $\delta = 0.75$ and $K_{0} = 2.175 \times 10^{-2}$ \tu{kpc}$^{2}$ \tu{Myr}$^{-1}$.\\

\item[(ii)] When $\ener{}$ and $\ener{s}$ are far enough, the diffusion length $\lD$ may reach the critical value $\lDM$,  from values higher than $\lDM$, for at least one propagation model which therefore maximizes the halo integral.
As $\ener{}$ and $\ener{s}$ are varied, the peak value of $\widetilde{I}$ is always reached when a scan through the space of parameters is performed. This peak value corresponds to the maximum of the halo integral, hence a horizontal upper boundary for each of the uncertainty bands of \citefig{fig5} and \fref{fig6}.
The set that leads to $\lD = \lDM$ is different for each combination of $\ener{}$ and $\ener{s}$ and is not unique. In the case of the NFW DM profile of \citefig{fig5}, the halo integral $\widetilde{I}$ is maximized by more than 30 models above $\ener{s} = 120$~\tu{GeV}.\\
\end{itemize}

The complexity of this analysis confirms that the propagation configurations selected by B/C do not play the same role for primary antiprotons and positrons. The two species experience the propagation phenomena, and in particular energy losses, with different intensities. As pointed out in Maurin et al.~\cite{Maurin:2002uc}, the average distance traveled by a positron is sensibly lower than the one experienced by an antiproton produced in the halo.\\

% IV POSITRON FLUXES
%\newpage
%\cleardoublepage
%
%  CHAPTER 4 - SECTION POSITRON FLUXES
%

\section{Positron fluxes}

% TABLE
% 
% 
\begin{tab}
	\begin{tabular}{|c||c|c|c|}
		\hline
		Model  & $\delta$ & $K_0 \; [\tu{kpc}^2/\tu{Myr}]$ & $L\;[\tu{kpc}]$ \\
		\hline \hline
		MED & 0.70 &  0.0112 & 4  \\
		M1  & 0.46 &  0.0765 & 15  \\
		M2  & 0.55 &  0.00595 & 1 \\
		\hline
	\end{tabular}
\caption{\label{tab:model} Typical combinations of cosmic ray propagation parameters that are compatible with the B/C analysis \cite{Maurin:2001sj}. The model MED has been borrowed from \citetab{t:prop-par}. Models M1 and M2 respectively maximize and minimize the positron flux for energies above 10 GeV -- the precise extent of which depends on the mass of the DM particle, on the annihilation channel and also on the DM profile.
Note that M1 is the same as MAX in \citetab{t:prop-par} but this is coincidental.}
\end{tab}

In the previous section we discussed the solutions of PETE and the astrophysical uncertainties on the HF. At this point, the next step is to apply this analysis on the positron signal, flux and fraction. 
% Now that we have discussed in detail the solution of the propagation equation, and have identified and quantified the astrophysical uncertainties on the halo integral $\widetilde{I}$, we are ready to apply our analysis to the theoretical predictions for the positron signal at the Earth position. The positron flux is obtained through \citeeq{eq:fluxe}. 
As previously said, we do not adopt specific DM candidates, but do instead discuss the signals arising from a DM particle which annihilates into a pure final state. To do that, we consider four different specific DM annihilation channels: direct $e^+ e^-$ production as well as $W^{+} W^{-}$, $b \bar{b}$ and $\tau^+ \tau^-$. The DM annihilation cross section is fixed at the value $2.1~\times~10^{-26}\;\tu{cm}^{3}\;\tu{sec}^{-1}$ and we consider the cases of a DM species with mass of 100~\tu{GeV} and of 500~\tu{GeV}. \\

Generic DM candidates, for instance a neutralino or a sneutrino in supersymmetric models, or the lightest Kaluza--Klein particle in models with extra--dimensions, will entail annihilation processes with specific branching ratios into one or more of these benchmark cases. The positron flux in these more general situations would simply be a superposition of the results for each specific annihilation channel, weighted by the relevant branching ratios and normalized by the actual annihilation cross section.\\

%
%%%%
\begin{fig}
\includegraphics[angle=270, width=\textwidth]{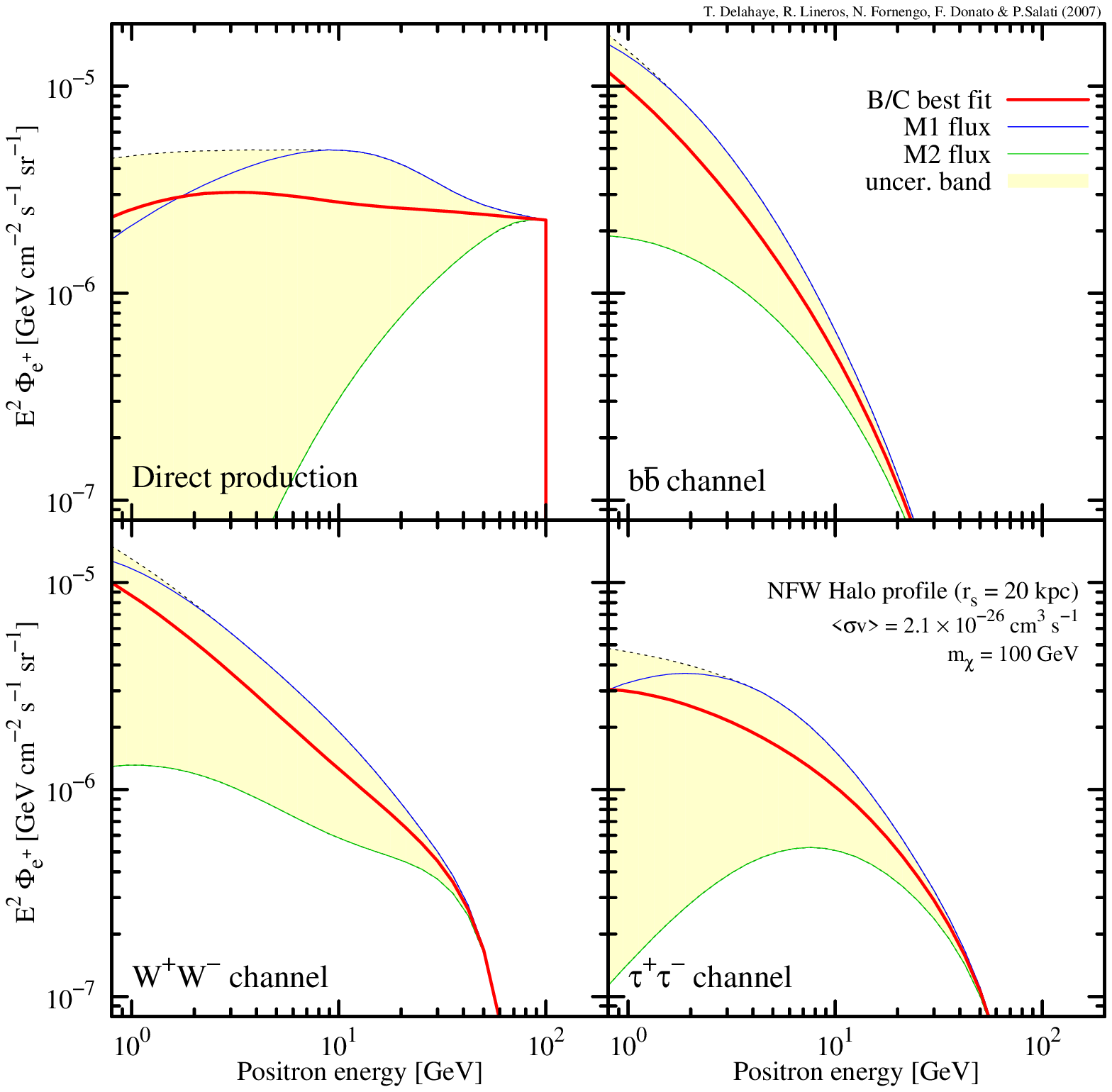}
\caption{ \label{fig:f1-100gev} Positron flux $E^2 \Phi_{e^+}$ versus the positron energy $\ener{}$, for a DM particle with a mass of 100 \tu{GeV} and for a NFW profile (\citetab{tab:indices}).
The four panels refer to different annihilation final states~: direct $e^+ e^-$ production (top left), $b\bar{b}$ (top right), $W^+ W^-$ (bottom left) and $\tau^+ \tau^-$ (bottom right). In each panel, the thick solid [red] curve refers to the best--fit choice (MED) of the astrophysical parameters. The upper [blue] and lower [green] thin solid lines correspond respectively to the astrophysical configurations which provide here the maximal (M1) and minimal (M2) flux -- though only for energies above a few GeV in the case of (M1).
The colored [yellow] area features the total uncertainty band arising from positron propagation.}
\end{fig}
%%%
%

%
%%%
\begin{fig}
\includegraphics[angle=270, width=\textwidth]{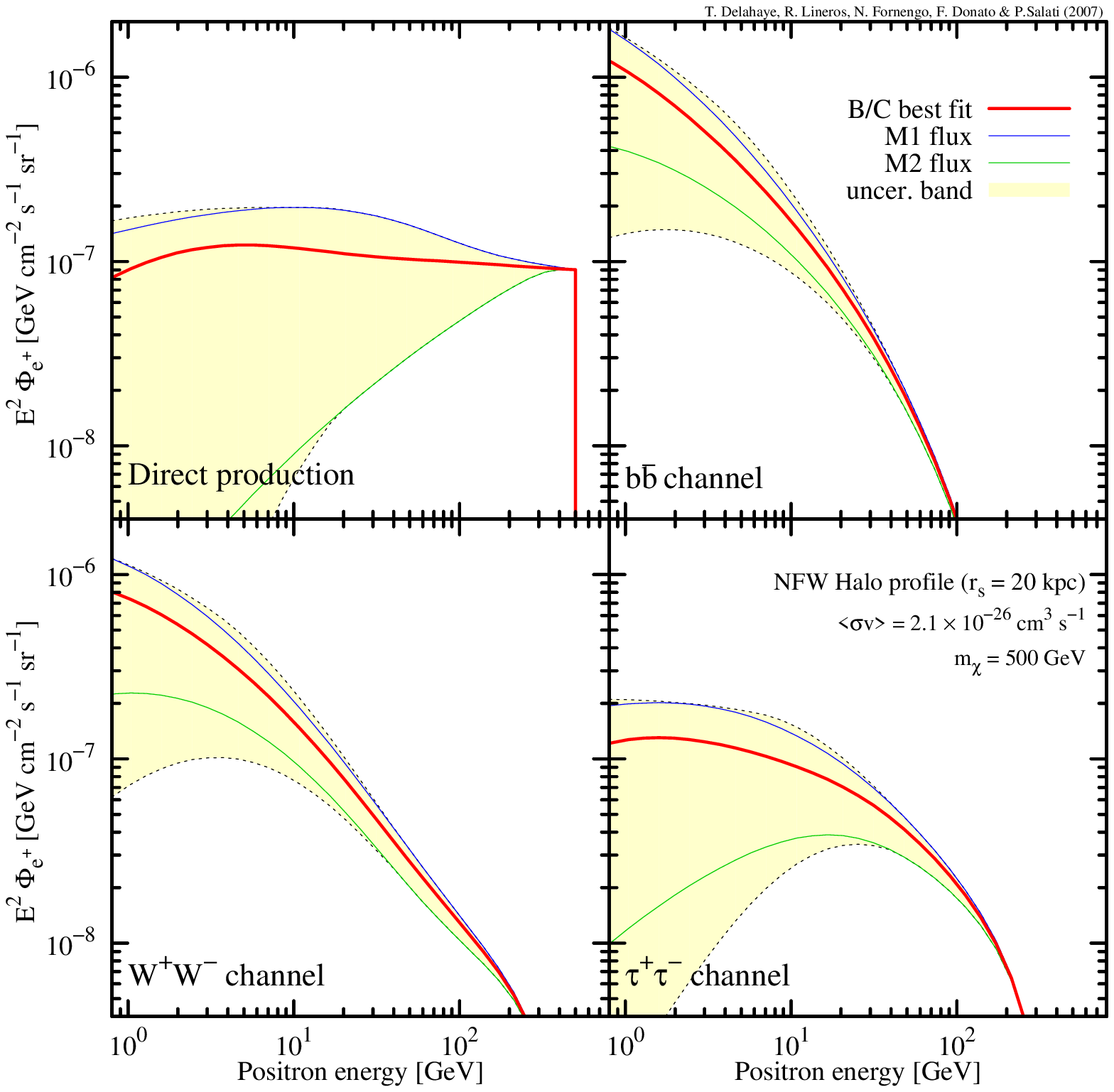}
\caption{\label{fig:f1-500gev} Same plot as in \citefig{fig:f1-100gev} but with a DM particle mass of $500$ \tu{GeV}.}
\end{fig}
%%%
%

%
% FIGURE 7 DISCUSSION
%
In \citefig{fig:f1-100gev}, the propagated positron flux~$\fluxe$ (times the positron energy squared) is featured as a function of $\ener{}$ for a 100 \tu{GeV} DM particle and a NFW density profile.
%
%The four panels refer to the four annihilation final states listed
%above. In each panel, the thick solid median [red] curve refers to
%the best--fit choice (MED) of the astrophysical parameters see
%Tab.~\fref{tab:pbar}. The upper [blue] and lower [green] thin solid
%lines refer to the astrophysical configurations which provide the
%maximal (M1) and minimal (M2) flux for energies above 10~\tu{GeV}, see
%Tab.~\fref{tab:model}.
%
The colored [yellow] area corresponds to the total uncertainty band arising from positron propagation.
In all panels, the uncertainties increases at low positron energy. This may be understood as a consequence of the behavior of the HF~$\widetilde{I}$ which was analyzed previously. Since positrons produced at energy $\ener{s}$ and detected at energy $\ener{}$ come from a sphere of radius $\lD$. That positron sphere grows as $\ener{}$ decreases and so does the uncertainty band. 
On the other hand, the details of galactic propagation become less important in the flux determination when positrons are originated closer to the Solar System. This is depicted in the HF because it goes to unity independently from the propagation parameters.\\

% As positrons originate further from the Earth, the details of galactic propagation become more important in the determination of the positron flux. On the contrary, high--energy positrons are produced locally and the halo integral~$\widetilde{I}$ becomes unity whatever the astrophysical parameters.
%
Notice also that the uncertainty band can be sizeable and depends significantly on the positron spectrum at production $\left(\displaystyle\dnde{e^+}\right)$. In the case of the $e^{+} e^{-}$ line of the upper left panel, the positron flux~$\fluxe$ exhibits a strongly increasing uncertainty as $\ener{}$ is decreased from $\mass{\chi}$ down to 1~\tu{GeV}. That uncertainty is one order of magnitude at $\ener{} = 10$~\tu{GeV}, and becomes larger than 2 orders of magnitude below 1~\tu{GeV}. 
Once again, the positron sphere argument may help to understand the situation. At fixed detected energy $\ener{}$, the radius $\lD$ increases with the injected energy $\ener{s}$. We therefore anticipate a wider uncertainty band as the source spectrum gets harder. This trend is clearly present in the panels of \citefig{fig:f1-100gev}. Actually direct production is affected by the largest uncertainty, followed by the $\tau^+ \tau^-$ and $W^+ W^-$ channels where a positron is produced either directly from the $W^+$ or from the leptonic decays.
In the $b\bar{b}$ case, which is here representative of all quark channels, a softer spectrum is produced since positrons arise mostly from the decays of charged pions originating from the quark hadronization. Most of the positrons have already a low energy $\ener{s}$ at injection and since they are detected at an energy $\ener{} \sim \ener{s}$, they tend to have been produced not too far from the Earth, obtaining smaller dependence to the propagation uncertainties. \\

The astrophysical configuration M2 (\citetab{tab:model}) provides the minimal positron flux. It corresponds to the lower boundaries of the yellow uncertainty bands of \citefig{fig:f1-100gev}. The M1 configuration maximizes the flux at high energies. For direct production and to a lesser extent for the $\tau^+ \tau^-$ channel, that configuration does not reproduce the upper envelope of the uncertainty band in the low energy tail of the flux. As discussed in the previous section, the response of $\fluxe$ to the propagation parameters depends on the detected energy $\ener{}$ in such a way that the maximal value cannot be reached for a single astrophysical configuration.
Finally, taking as a reference the median flux, the uncertainty bands extend more towards small values of the flux. In all channels, the maximal flux is typically a factor of $\sim$~1.5--2 times larger than the median prediction. The minimal flux features larger deviations with a factor of 5 for the $b\bar{b}$~channel at $E = 1$~\tu{GeV}, of 10 for $W^+ W^-$ and of 30 for $\tau^+ \tau^-$.\\

%
% FIGURE 8 DISCUSSION
%
\citefig{fig:f1-500gev} is similar to \citefig{fig:f1-100gev} but with a heavier DM species of 500~\tu{GeV} instead of 100~\tu{GeV}.
Since the mass $\mass{\chi}$ is larger, on average the injected energy $\ener{s}$ is higher, too. Notice that at fixed positron energy $\ener{}$, the radius $\lD$ of the positron sphere increases with $\ener{s}$. And then, we anticipate that the propagated fluxes are affected by larger uncertainties for heavy DM particles. Again, the maximal flux does not exceed twice the median flux, while the minimal configurations are significantly depressed. At the reference energy $\ener{} = 1$~\tu{GeV}, reductions by a factor of 10 between the median and minimal predictions are obtained for the $b\bar{b}$ channel and amount to a factor of 20 in the $W^+ W^-$ case. They reach up to 2 orders of magnitude for the direct positron production. \\

In this large DM mass regime, the astrophysical configuration M2 does not reproduce by far the lower bound of the uncertainty band as it did for the 100~\tu{GeV} case. That draws two possibilities.
%The message is therefore twofold.
%which affects the flux at the Earth requires
\begin{itemize}
\item[(i)] Once the full positron spectrum (not each single channel) at the source is chosen, the correct determination of the uncertainty in the flux requires a full scan of the propagation parameter space for each energy $\ener{}$. The use of representative astrophysical configurations such as M1 and M2 would not provide the correct uncertainty over the entire range of positron energy $\ener{}$.\\

\item[(ii)] Specific predictions have to be performed for a given model of DM particle and a fixed set of astrophysical parameters. This is why fits to the experimental data should be performed for each propagation configuration over the entire range of the measured positron energies $\ener{}$. The best fit should correspond to a unique set of astrophysical parameters. This procedure is the only way to reproduce properly the correct and specific spectral shape of the flux.\\
\end{itemize}

%
%%%
\begin{fig}
\includegraphics[angle=270, width=\textwidth]{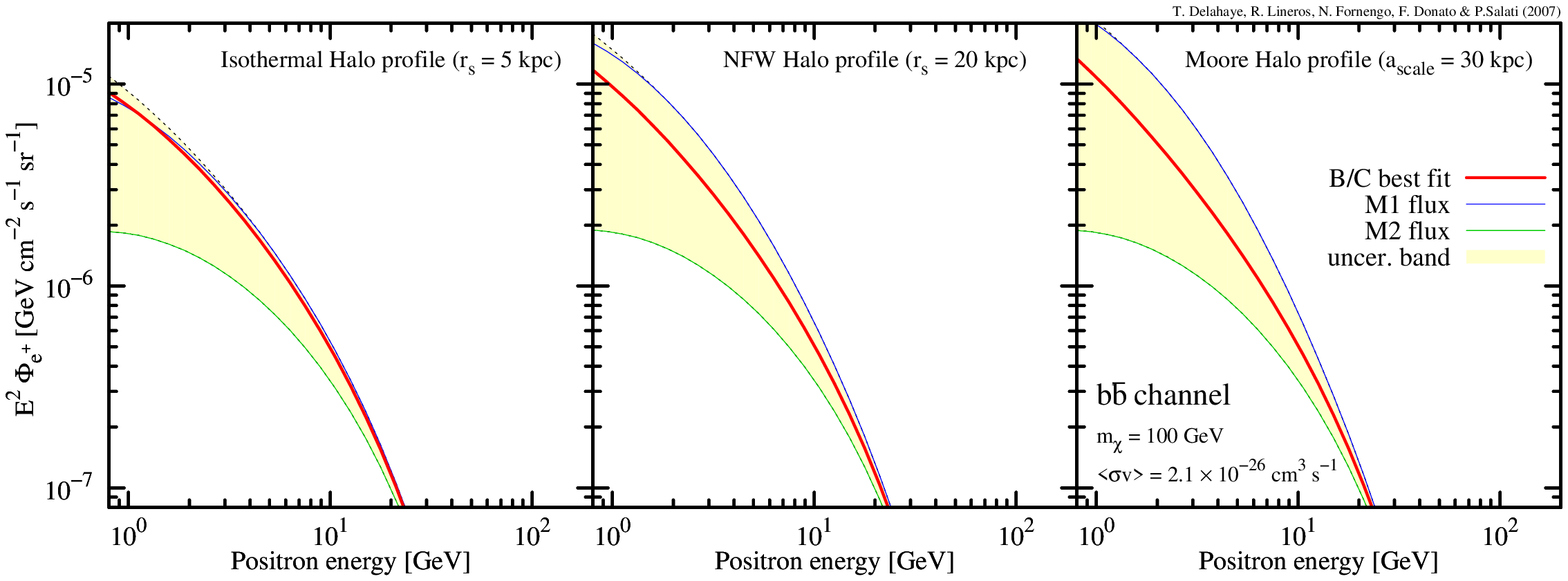}
\includegraphics[angle=270, width=\textwidth]{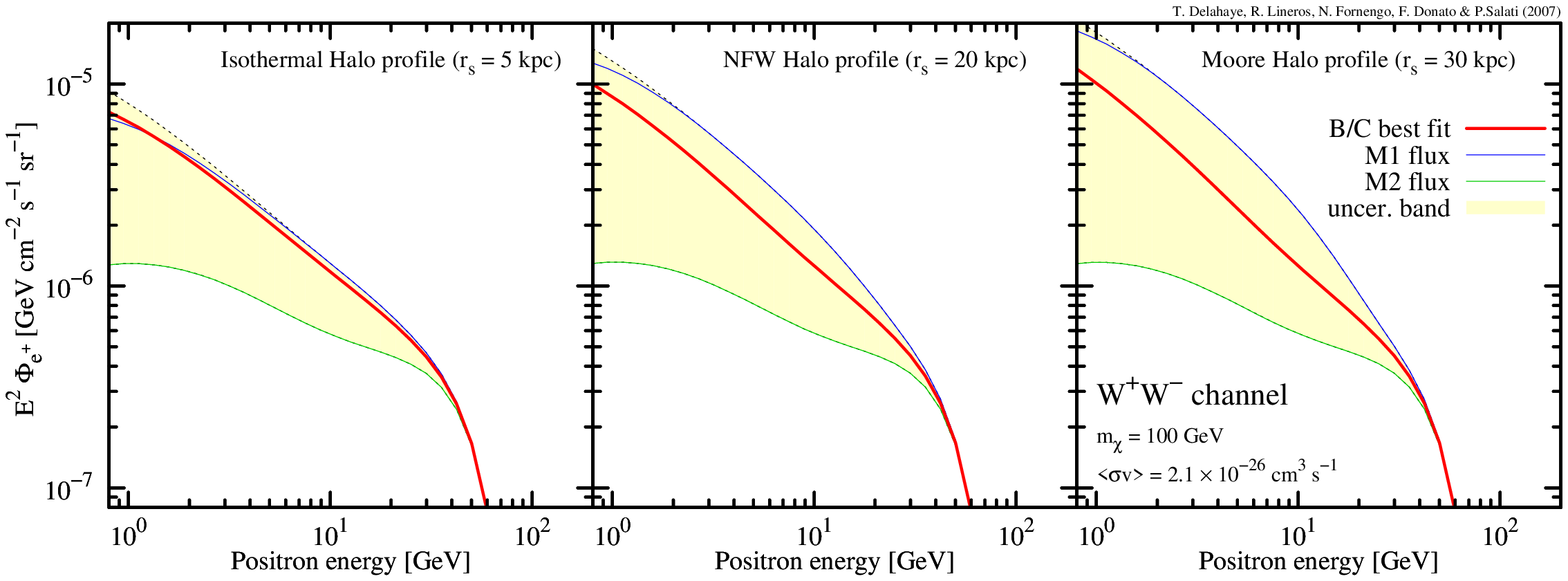}
\caption{\label{fig:f2-bb} Positron flux $\ener{}^2 \Phi_{e^+}$ versus the positron energy $\ener{}$, for a DM particle mass of 100~\tu{GeV} and for different halo density profiles~: cored isothermal sphere~\cite{Bahcall:1980fb} (left panels), NFW~\cite{Navarro:1996gj} (central panels) and Moore~\cite{Diemand:2004wh} (right panels) (\citetab{tab:indices}).
The upper and lower rows correspond respectively to a $b \bar{b}$ and $W^{+} W^{-}$ annihilation channel. In each panel, the thick solid [red] curve refers to the best--fit choice (MED) of the astrophysical parameters. The upper [blue] and lower [green] thin solid lines stand for the astrophysical configurations M1 and M2 of \citetab{tab:model}. 
The colored [yellow] area indicates the total uncertainty band arising from positron propagation.}
\end{fig}
%%%
%

%
%%%
\begin{fig}
\includegraphics[angle=270, width=\textwidth]{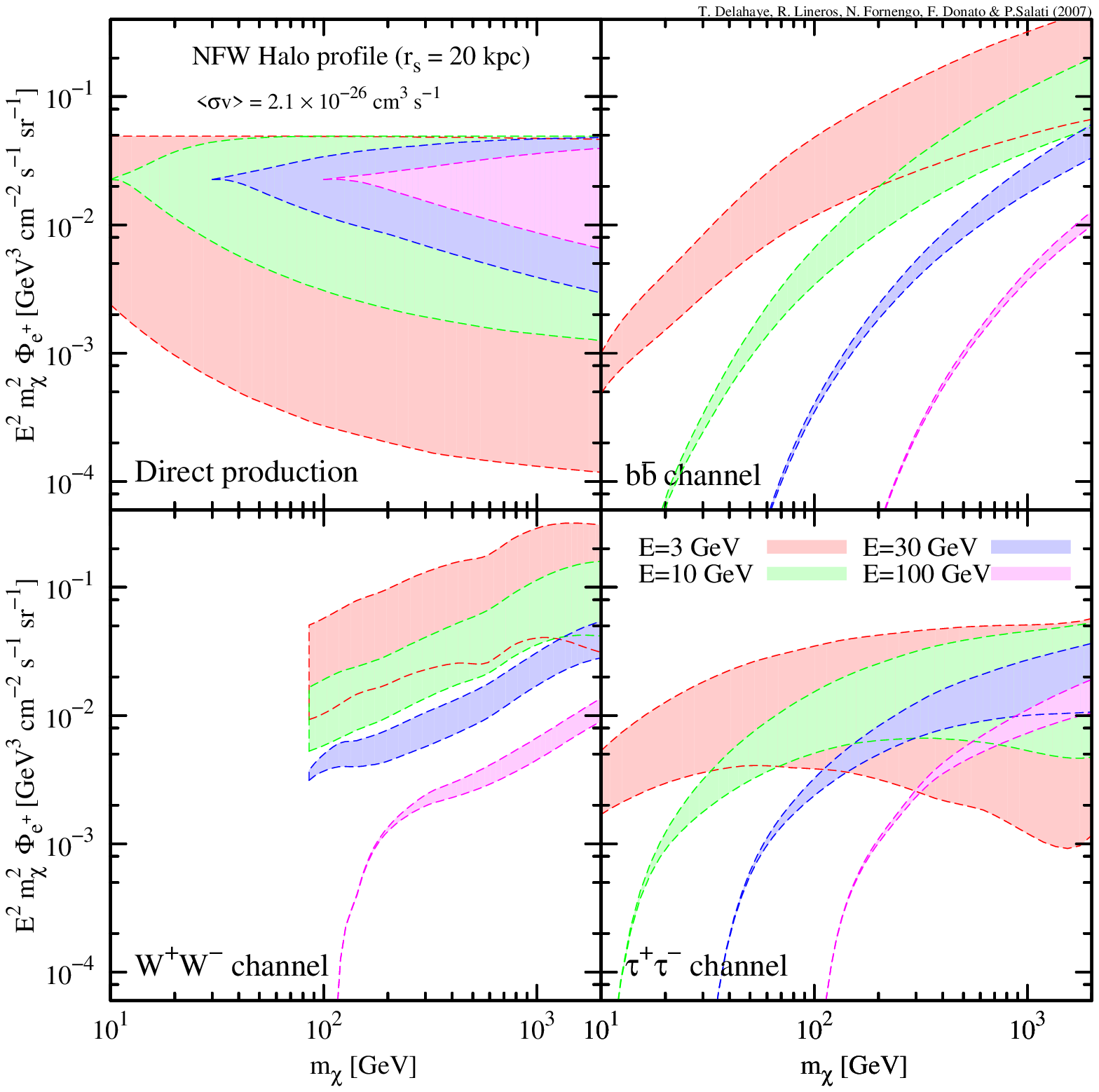}
\caption{\label{fig:f5-fluxmass} For fixed values of the detected energy $\ener{}$, the uncertainty bands on the positron flux $\ener{}^2 \mass{\chi}^2 \Phi_{e^+}$ are shown as a function of the mass $\mass{\chi}$ of the DM particle. The energies considered in the figure are $\ener{} = 3$, 10, 30 and 100~\tu{GeV}. Each band refers to one of those values and starts at $\mass{\chi} = \ener{}$.}
\end{fig}
%%%
%

%
% FIGURE 9 DISCUSSION
%
The effect induced by different DM profiles is presented in \citefig{fig:f2-bb}, where the positron fluxes for the $b\bar{b}$ and $W^+ W^-$ channels are reproduced for the DM distributions of \citetab{tab:indices}. The mass of the DM  particle is fixed at $\mass{\chi} = 100\;\tu{GeV}$.
Notice how steeper profiles entail larger uncertainties, especially for the upper bound. This is mostly due to the fact that for large values of $L$ the positron flux is more sensitive to the central region of the Galaxy, where singular profiles like the NFW and Moore distributions have larger densities and therefore induce larger annihilation rates.
On the contrary, the lower envelope of the uncertainty band is not affected by the variation of the halo profile. In this case, with typically small heights~$L$, positrons reach the solar system from closer regions, where the three halo distributions are very similar and do not allow to probe the central part of the Milky Way.\\

%
% FIGURE 10 DISCUSSION
%
\citefig{fig:f5-fluxmass} depicts the information on the positron flux uncertainty from a different perspective. The flux~$\fluxe$ and its uncertainty band are now featured for fixed values of the detected energy~$\ener{}$ whereas the DM particle mass is now varied. The flux~$\fluxe$ is actually rescaled by the product $\ener{}^2~\mass{\chi}^2~\fluxe$ for visual convenience. Each band corresponds to a specific detected energy $\ener{}$ and consequently starts at $\mass{\chi} = \ener{}$.\\

In the case of the $W^+ W^-$ channel, the bands start at $\mass{\chi} = m_W$ because this channel is closed for DM masses below that threshold. 
The behavior of these bands can be understood from \citefig{fig6}, where the halo function~$\widetilde{I}$ is plotted for the same detected energies, as a function of the injection energy~$\ener{S}$. In the case of direct positron production, there is a simple link between the two figures, because the source spectrum in this case is just a line at $\ener{s}~=~\mass{\chi}$. For the other channels the situation is more involved since we have a continuous injection spectrum with specific features as discussed above. 
The main information which can be obtained from \citefig{fig:f5-fluxmass} is that at fixed detection energy, the larger the DM mass, the larger the uncertainty. Let us take for instance a detection energy of $E = 3$~\tu{GeV}. For direct production, where $\ener{s} = \mass{\chi}$, increasing the DM mass translates into a larger radius $\lD$ of the positron sphere. As a consequence, the uncertainty band enlarges for increasing masses. This occurs for all the annihilation channels, but is less pronounced for soft spectra as in the $b\bar{b}$ case. Similar conclusions hold for all the other values of $\ener{}$.\\

%
%%%
\begin{fig}
\includegraphics[angle=270, width=\textwidth]{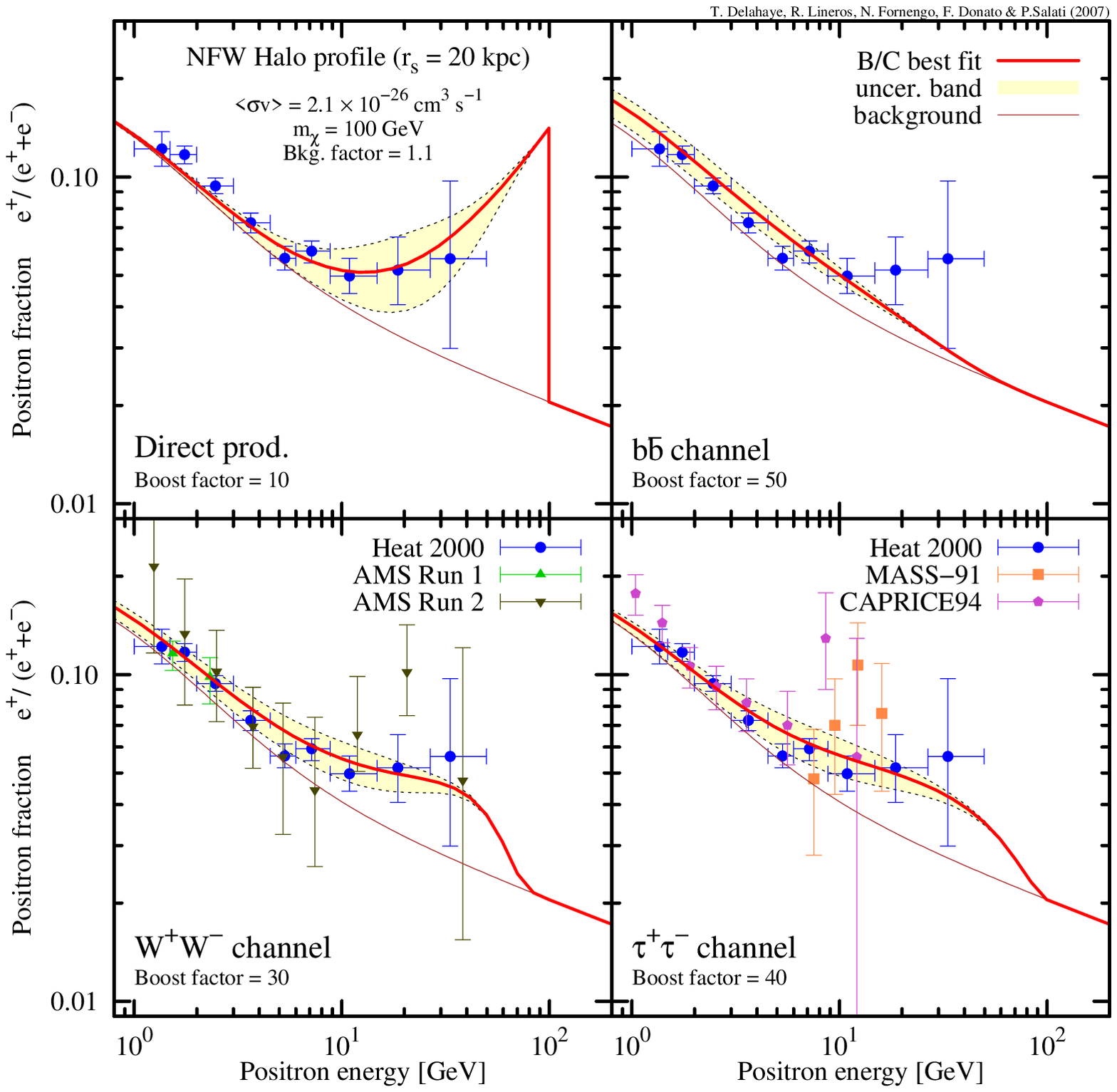}
\caption{\label{fig:f3-heat-pf-100gev} Positron fraction $e^+/(e^- + e^+)$ versus the positron detection energy $\ener{}$. Notations are as in \citefig{fig:f1-100gev}. In each panel, the thin [brown] solid line stands for the background \cite{Baltz:1998xv, Moskalenko:1997gh} whereas the thick solid [red] curve refers to the total positron flux where the signal is calculated with the best--fit choice (MED) of the astrophysical parameters.
Experimental data from HEAT~\cite{Barwick:1997ig}, AMS~\cite{Alcaraz:2000PhLB,Aguilar:2007}, CAPRICE~\cite{Boezio:2000} and MASS~\cite{Grimani:2002yz} are also plotted.}
\end{fig}
%%%
%

%
%%%
\begin{fig}
\includegraphics[angle=270, width=\textwidth]{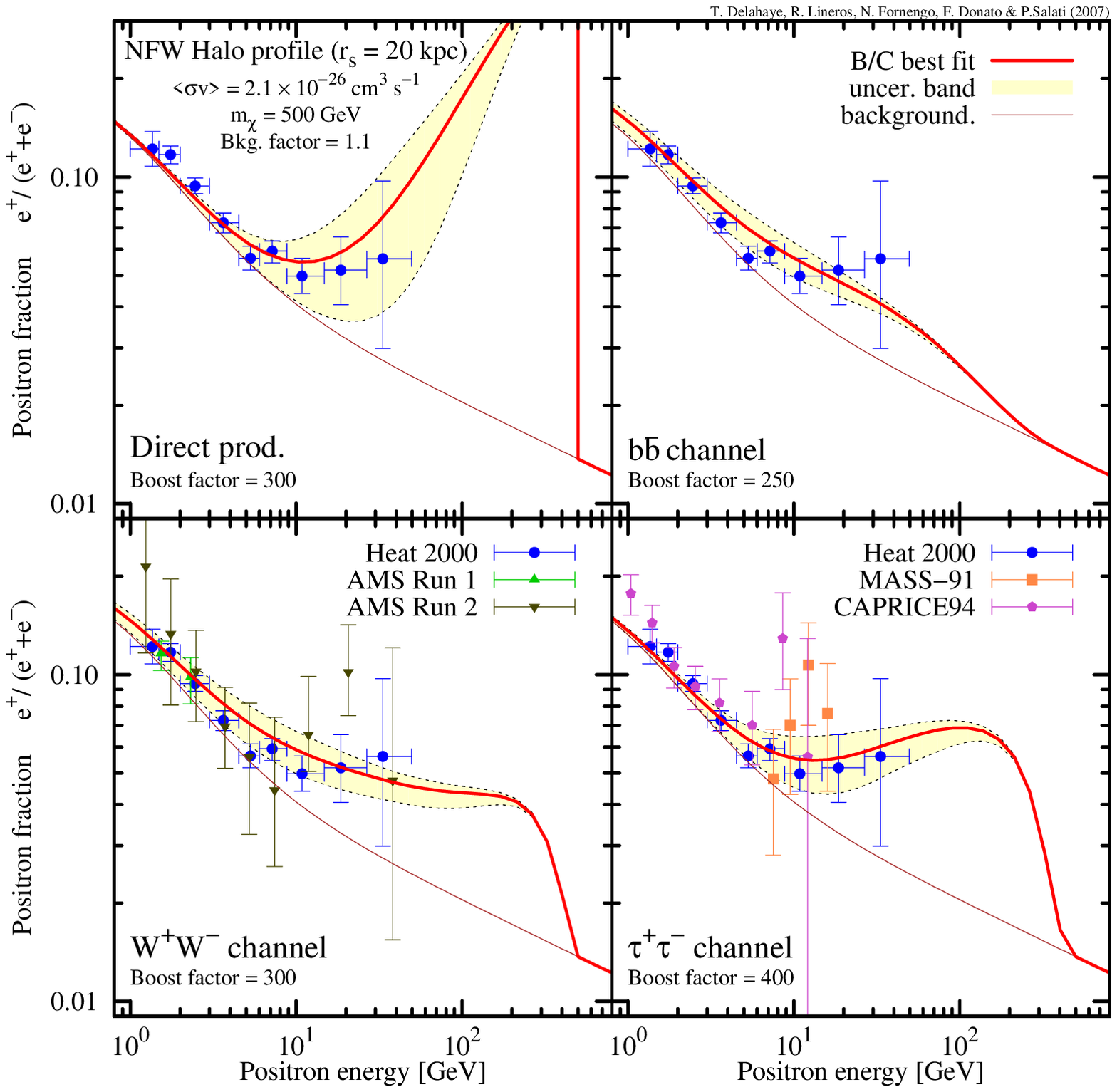}
\caption{\label{fig:f3-heat-pf-500gev} Same plot as in \citefig{fig:f3-heat-pf-100gev} but with a mass of the DM particle of $500$~\tu{GeV}.}
\end{fig}
%%%
%

%
% INTRODUCTION OF BALTZ PARAMETERIZATIONS
%

In addition, to compute a full positron signal, we need to deal with the secondary positrons and the full electron signal. To simplify the situation we used the results of Moskalenko et at.~\cite{Moskalenko:1997gh}, which were parameterized by Baltz et al.~\cite{Baltz:1998xv} obtaining function for these fluxes:
\begin{eq}\label{e:fit-bkg}
 \left(\frac{d\Phi}{d\ener{}}\right)^{e^-}_{\tu{prim.}} &=& \frac{0.16\;\epsilon^{-1.1}}{1\;+\;11\;\epsilon^{0.9}\;+\;3.2\;\epsilon^{2.15}} \;, \\ \nonumber \\
 \left(\frac{d\Phi}{d\ener{}}\right)^{e^-}_{\tu{sec.}} &=& \frac{0.7\;\epsilon^{0.7}}{1+110\;\epsilon^{1.5}+600\;\epsilon^{2.9}+580\;\epsilon^{4.2}} \;, \\ \nonumber \\
 \left(\frac{d\Phi}{d\ener{}}\right)^{e^+}_{\tu{sec.}} &=& \frac{4.5\;\epsilon^{0.7}}{1+650\;\epsilon^{2.3}+1500\;\epsilon^{4.2}} \;,
\end{eq}
where $\epsilon = \ener{}/1\;\tu{GeV}$ and all those are expressed in units of $\tu{GeV}^{-1}\;\tu{cm}^{-2}\;\tu{sec}^{-1}\;\tu{sr}^{-1}$. Notice that the full electron signal is composed by the addition of secondaries and primaries,
\begin{eq}
 \Phi_{e^-}^{\tn{TOT}} = \left(\frac{d\Phi}{d\ener{}}\right)^{e^-}_{\tu{prim.}} + \left(\frac{d\Phi}{d\ener{}}\right)^{e^-}_{\tu{sec.}} \; .
\end{eq}
\\

%
% FIGURE 11, 12 and 13 DISCUSSION
%
Comparison with available data is presented in \citefig{fig:f3-heat-pf-100gev}, \fref{fig:f3-heat-pf-500gev} and \fref{fig:f3-heat-flux-500gev}.
%
% FIGURE 11 DISCUSSION
%
In \citefig{fig:f3-heat-pf-100gev}, the positron fraction 
\begin{eq} \label{fraction} 
 \frac{e^+}{e^+ \; + \; e^-} \equiv \frac{\fluxe^{\tn{TOT}}}{\Phi_{e^-}^{\tn{TOT}} \, + \, \fluxe^{\tn{TOT}}}
\end{eq}
is plotted as a function of the positron energy $\ener{}$. The total positron flux $\fluxe^{\tn{TOT}}$ at the Earth encompasses the annihilation signal and a background component -- see the thin solid [brown] lines.
The mass of the DM particle is 100~\tu{GeV} and a NFW profile has been assumed. The data from HEAT~\cite{Barwick:1997ig}, AMS~\cite{Alcaraz:2000PhLB,Aguilar:2007}, CAPRICE~\cite{Boezio:2000} and MASS~\cite{Grimani:2002yz} are indications of a possible excess of the positron fraction for energies above 10~\tu{GeV}. Those measurements may be compared to the thick solid [red] line that corresponds to the MED configuration. In order to get a reasonable agreement between our results and the observations, the annihilation signal has been boosted by an energy--independent factor ranging from 10 to 50 as indicated in each panel.
%
%A more detailed statistical analysis will be presented
%elsewhere.
%
At the same time, the positron background -- for which we do not have an error estimate yet -- has been shifted upwards from its reference value of Baltz et al. (\citeeq{e:fit-bkg}) by a small amount of 10\%.\\

%
% Discussion of the GOOD agreement between the data and the predictions.
%
As is clear in the upper left panel, the case of direct production offers a very good agreement with the potential HEAT excess. Notice how well all the data points lie within the uncertainty band. A boost factor of 10 is enough to obtain an excellent agreement between the measurements and the median flux. A smaller value would be required for a flux at the upper envelope of the uncertainty band. The $W^+ W^-$ and $\tau^+ \tau^-$ channels may also reproduce reasonably well the observations, especially once the uncertainty is taken into account, but they need larger boost factors of the order of 30 to 40. On the contrary, softer production channels, like the $b\bar{b}$ case, are unable to match the features of the putative HEAT excess for this value of the DM particle mass. 
%
% Discussion of the spread of the uncertainty band.
%
For all annihilation channels, the uncertainty bands get thinner at high energies for reasons explained above. They surprisingly tend to shrink also at low energies, a regime where the positron horizon is the furthest and where the details of galactic propagation are expected to be the most important. Actually, the annihilation signal turns out to be completely swamped in the positron background. In particular, the signal from direct production stands up over the background only for energies larger than 5~\tu{GeV}. The corresponding uncertainty on the positron fraction is at most of the order of 50\% for energies between 10 and 20~\tu{GeV}. In the other cases, the uncertainty bands are even thinner. \\

%
% Necessity of deriving the uncertainty on the background itself.
%
% Beware finally of the positron background which should also be affected by uncertainties due to secondary production processes and propagation. These uncertainties are not currently available and there is clearly a need to estimate them in order to properly shape theoretical predictions and to perform better study of the current and forthcoming data. Such an investigation would involve a comprehensive analysis and is out of the scope of the present article.\\

Note that the positron background also should be affected by uncertainties due to secondary production processes and propagation. This chapter is just devoted to study the uncertainties at the level of primary positron production. However, the study of uncertainties of the secondary production is in \citecha{cha5}.\\

%
%%%
\begin{fig}
\includegraphics[angle=270, width=\textwidth]{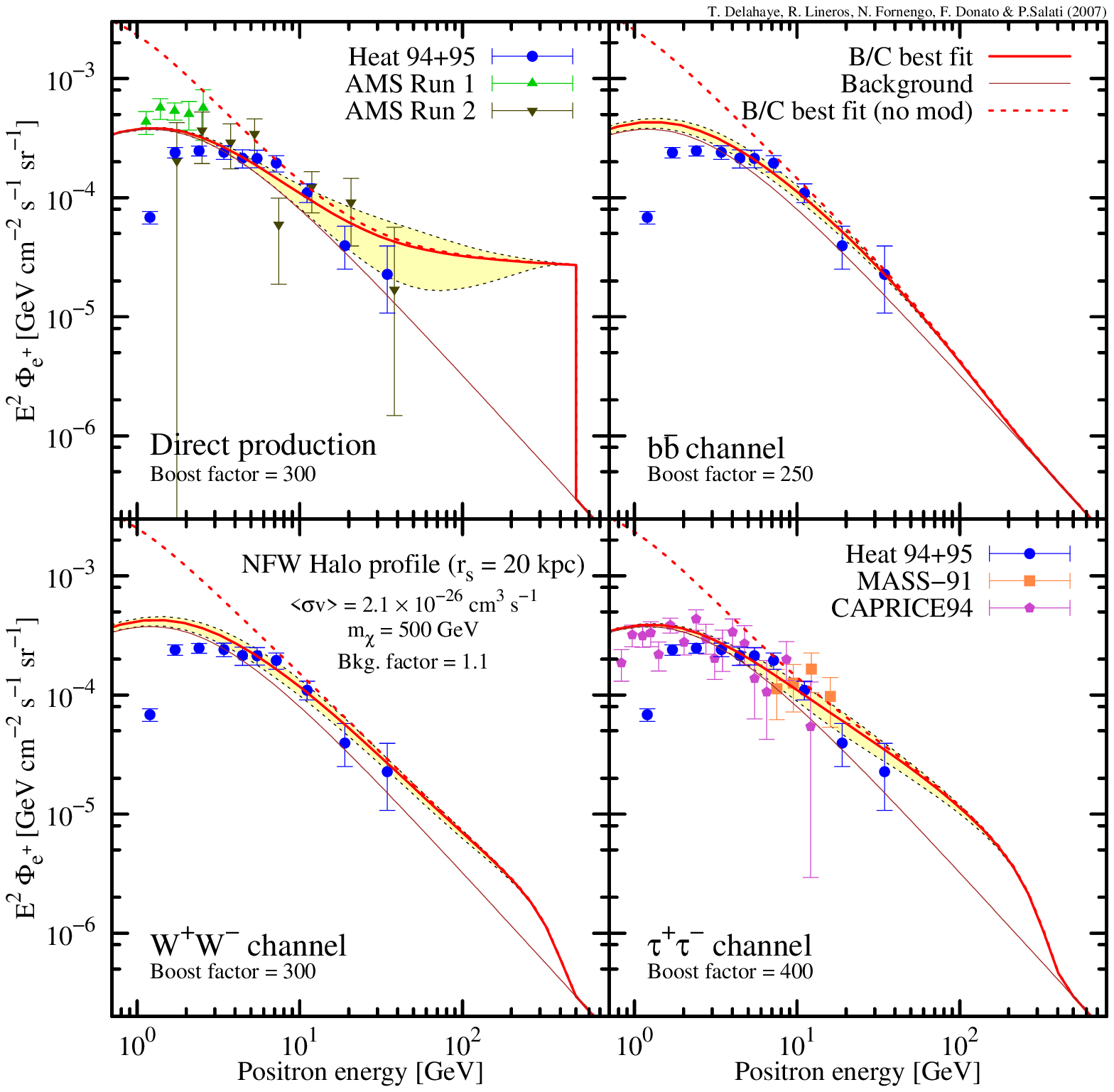}
\caption{\label{fig:f3-heat-flux-500gev} Positron flux $E^2 \Phi_{e^+}$ (not fraction) versus the positron energy $\ener{}$, for a 500~\tu{GeV} DM particle. Notations are the same as in \citefig{fig:f3-heat-pf-100gev}. 
Experimental data from HEAT~\cite{Barwick:1997ig}, AMS~\cite{Alcaraz:2000PhLB,Aguilar:2007}, CAPRICE~\cite{Boezio:2000} and MASS~\cite{Grimani:2002yz} are plotted.}
\end{fig}
%%%
%

%
% FIGURE 12 DISCUSSION
%
Somehow different is the situation for larger masses of the DM candidate. \citefig{fig:f3-heat-pf-500gev} features the same
information as \citefig{fig:f3-heat-pf-100gev}, but now for $\mass{\chi} = 500$~\tu{GeV}. In this case, all the annihilation channels manage to reproduce the experimental data, even the softest one $b\bar{b}$. For direct production, the positron fraction is very large at energies above 40~\tu{GeV}, where no data are currently available. This feature would be a very clear signature of DM annihilating directly into $e^+ e^-$ pairs, with strong implications also on the nature of the DM candidate. For instance, bosonic dark matter would be strongly preferred, since Majorana fermionic DM, like the neutralino, possesses a very depressed cross section into light fermions because of helicity suppression in the non--relativistic regime. 
Astrophysical uncertainties on the signal in this case show up more clearly than for the case of a lighter DM species, but still they are not very large. The drawback of having a heavier relic is that now the boost factors required to match the data are quite large. In \citefig{fig:f3-heat-pf-500gev} they range from 250 for the soft channel to 400 for the $\tau^+ \tau^-$ case. Such large boost factors appear to be disfavored, on the basis of the recent analysis of Lavalle et al.~\cite{Lavalle:2006vb,Lavalle:1900wn}.\\

%
% FIGURE 13 DISCUSSION
%
In \citefig{fig:f3-heat-flux-500gev}, the positron flux (not the fraction) is compared to the available experimental data for a 500~\tu{GeV} DM particle and a NFW profile.
The solid thin [brown] line features the positron background which we shifted upwards by 10\% with respect to the reference value of Baltz et al.~\cite{Baltz:1998xv}.
The thick solid [red] line encompasses both that background and the annihilation signal which we calculated with the best--fit choice (MED) of the astrophysical parameters. Both curves have been derived assuming solar modulation implemented through the force field approximation with a Fisk potential $\phi_{\tn{F}}$ of 500~\tu{MV} (\citecha{cha3}).
The dashed [red] line instead corresponds to the total positron interstellar flux without solar modulation. Notice that this curve is superimposed on the thick [red] line above $\sim$ 10~\tu{GeV}, a regime where cosmic ray propagation is no longer affected by the solar wind. 
A reasonably good agreement between the theoretical predictions and the data is obtained, especially once the theoretical uncertainties on the annihilation signal are taken into account. Notice that the spread of each uncertainty band is fairly limited as we already pointed out for the positron fraction. The reasons are the same. \\

Prospects for the future missions are shown in \citefig{fig:f4-pam-100gev} and \fref{fig:f7-ams-flux-500gev}.
%
% FIGURE 14 DISCUSSION
%
In \citefig{fig:f4-pam-100gev}, a 100~\tu{GeV} DM particle and a NFW halo profile have been assumed. 
The median [red] curve corresponds to the prediction for the best--fit MED choice of astrophysical parameters whereas the upper [blue] and lower [green] lines correspond respectively to the M1 and M2 propagation models (\citetab{tab:model}).
Since we are dealing with predictions which will eventually be compared to the measurements performed over an entire range of positron energies, we have to choose specific sets of propagation parameters. 
The upper and lower curves therefore do not represent the maximal uncertainty at each energy -- though they may do so in some limited energy range -- but instead they are ``true'' predictions for a specific set of propagation parameters.
\citefig{fig:f4-pam-100gev} summarizes our estimate of the capabilities of the PAMELA detector~\cite{Boezio:2004jx} after 3 years of running. We only plotted statistical errors. We reached the remarkable conclusion that not only will PAMELA have the capability to disentangle the signal from the background, but also to distinguish among different astrophysical models, especially for hard spectra.
Our conclusion still holds for the $b\bar{b}$ soft spectrum for which the M1, MED and M2 curves of the upper right panel differ one from each other by more than a few standard deviations. PAMELA could be able to select among them, even when systematical errors are included. \\

%
% FIGURE 15 DISCUSSION
%
In \citefig{fig:f7-ams-flux-500gev}, the case of a 500~\tu{GeV} DM particle is confronted with the sensitivity of AMS-02 for a 3--year flight.
The possibility to disentangle the signal from the background is also clearly manifest here, even once the astrophysical uncertainties are included -- provided though that boost factors of the order of 200 to 400 are possible.
But, unless direct production is the dominant channel, a clear distinction among the various astrophysical models will be very difficult because the M1 and M2 configurations are closer to the MED curve now than in the previous case of a lighter DM species. Comparison between \citefig{fig:f4-pam-100gev} and \citefig{fig:f7-ams-flux-500gev} clearly exhibits that at least below the \tu{TeV} scale, the effect of the mass $\mass{\chi}$ should not limit the capability of disentangling the annihilation signal from the background. More problematic is our potential to distinguish among different astrophysical models when the DM mass sizeably exceeds the 100~\tu{GeV} scale.\\

%
%%%
\begin{fig}
\includegraphics[angle=270, width=\textwidth]{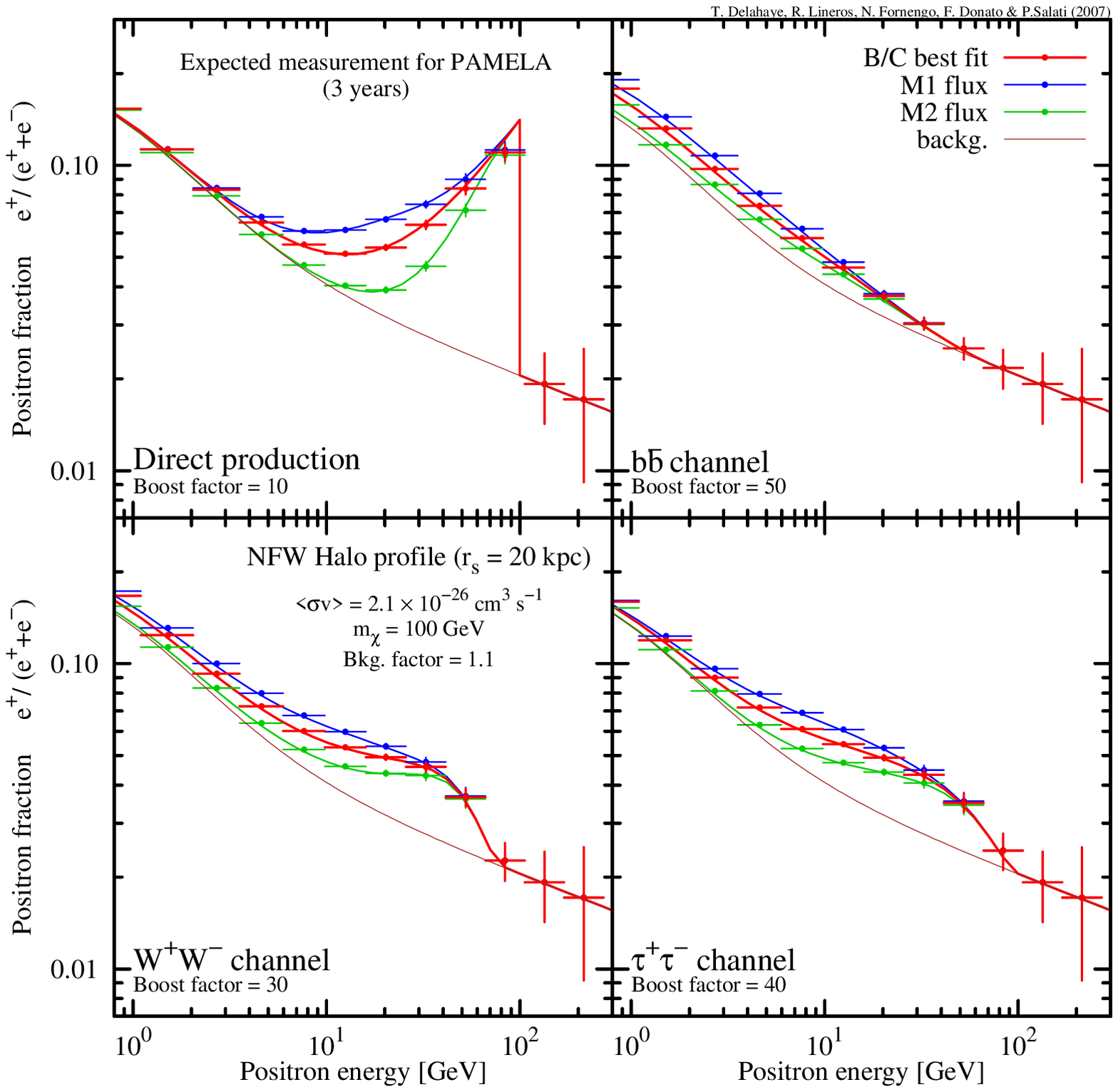}
\caption{\label{fig:f4-pam-100gev} Predictions for PAMELA for a 3--year mission.
The positron fraction $e^+/(e^- + e^+)$ and its statistical uncertainty are plotted against the positron energy $\ener{}$ for a 100~\tu{GeV} DM particle and a NFW profile. 
Notations are the same as in \citefig{fig:f3-heat-pf-100gev}.
The thick solid curves refer respectively to the total positron flux calculated with the M1 (upper [blue]), MED (median [red]) and M2 (lower [green]) sets of propagation parameters.}
\end{fig}
%%%
%

%
%%%
\begin{fig}
\includegraphics[angle=270, width=\textwidth]{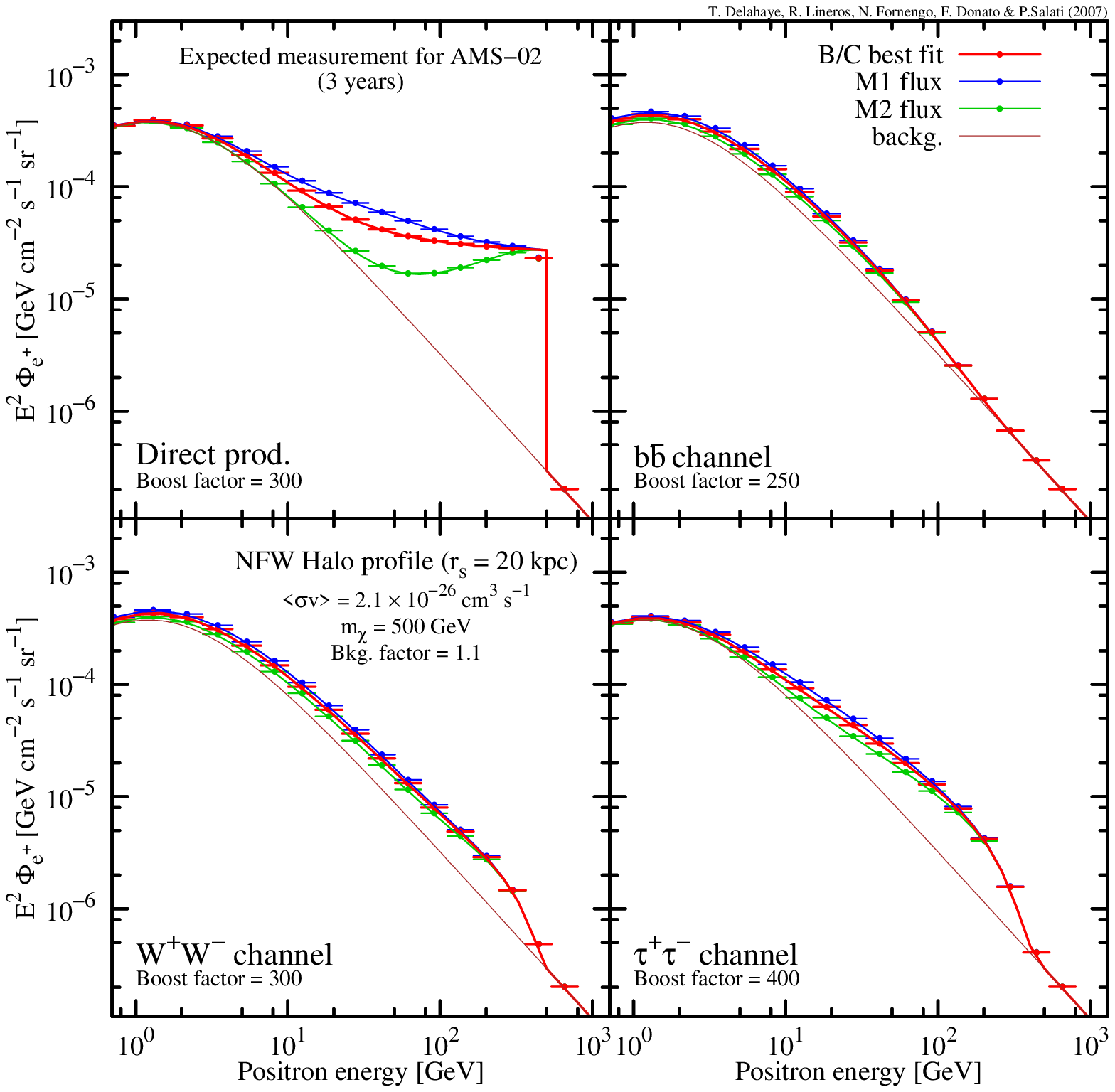}
\caption{\label{fig:f7-ams-flux-500gev} Predictions for AMS-02 for a 3--year mission.
The positron flux $\ener{}^2 \Phi_{e^+}$ and its statistical uncertainty are featured as
a function of the positron energy $\ener{}$ for a 500~\tu{GeV} DM species and a NFW profile.
Notations are the same as in \citefig{fig:f3-heat-flux-500gev}.
The thick solid curves refer respectively to the total positron flux calculated with the M1 (upper [blue]), MED (median [red]) and M2 (lower [green]) sets of propagation parameters.}
\end{fig}
%%%
%

% NOTE ABOUT PAMELA
\paragraph{Note added:}
As the time of our publication~\cite{Delahaye:2007fr}, on which this chapter is based, the PAMELA data of the positron fraction were not yet released. Nevertheless, most of the above analysis is still valid with the current state of the positron excess feature, which will be discussed in the next chapter in conection with the calculation of the positron background.\\

\cleardoublepage
\chapter{Secondary positron flux at Earth}
\label{cha5}

\begin{prechap}
Positrons in the galactic environment can be produced in many ways. The main component in the positron signal comes from nuclei cosmic--rays interactions with the interstellar medium. \\
%
% The propagation in the interstellar space is common to all cosmic--ray species, which helps to estimate in a better way the positron signal calculated from general considerations.\\

This chapter is partially based on our work \cite{Delahaye:2008}. \\
\end{prechap}

\section{Overview}

Cosmic positrons are created by interaction of cosmic-ray nuclei on interstellar matter 
and propagate in a diffusive mode, because of their interaction with the turbulent component of the galactic magnetic field. The expected flux of positrons can be calculated from the observed cosmic ray--nuclei fluxes, using the relevant nuclear physics and solving the transport equation.\\

The HEAT experiment \cite{Barwick:1997ig} showed that the positron fraction possibly exhibits an unexpected bump in the 10~\tu{GeV} region of the spectrum. Although it is not excluded that this bump could be due to some unknown systematic effect,  the HEAT result has triggered a lot of explanations. For instance, Moskalenko et al.~\cite{Moskalenko:1997gh} suggested that an interstellar  nucleon spectrum harder than the expected one could explain the excess. A lot  of works also focused on the dark matter hypothesis, the bump being due to a primary contribution from the annihilation of dark matter particles. The positron excess expected in this framework is very uncertain, because the nature of dark matter is not known, and because the propagation of positrons involves physical quantities that are also not currently precisely known.\\

The related astrophysical uncertainties were calculated and quantified in~\cite{Delahaye:2007fr}, where it has been shown that they may be sizeable, especially in the low energy part of the spectrum, a
property which is common also to the antiproton \cite{Donato:2003xg} and the antideuteron \cite{Donato:2000,Donato:2008} signals.\\

For positrons, sizeable fluxes from dark matter annihilation are typically possible if dark matter  overdensities are locally present, a fact that is usually coded into the  so--called ``boost factor''. A detailed analysis on the admissible boost factors for positrons and  antiprotons has been performed by Lavalle et al.~\cite{Lavalle:2008}, who have shown  that boost factors are typically confined to be less than about a factor of  10--20. Computing the antimatter fluxes directly in the frame of a cosmological  N-body simulation leads to the same conclusions~\cite{Lavale:2008b}. \\

The PAMELA experiment~\cite{Boezio:2004jx}  has just released its first results on the positron fraction for energies ranging from 1.5 GeV to 100 GeV and with a large statistics~\cite{Adriani:2008zr}. The positron fraction is observed to steadily rise for energies above 10 GeV, reinforcing the possibility that an excess is actually present.
% 
% It is therefore timely and  crucial to have a novel analysis of the positron flux, including a good estimation  of the accuracy of the theoretical determination. 
%
The calculation of the uncertainties affecting the standard  spallation--induced positron population is in fact especially important when unexpected distortions are observed in experimental data, in order to properly address the issue. This, together with the recent analysis on the positron signal from dark matter annihilation and its astrophysical uncertainties \cite{Delahaye:2007fr}, will set the proper basis to discuss in details the experimental results.\\

The uncertainties on the positron flux have several origins:
\begin{itemize}
 \item[i)] The cosmic--ray nuclei  measurements come with their experimental uncertainties, which then affect the  predictions of induced secondary fluxes, like positrons.
 \item[ii)] Various  modelings of the nuclear cross sections involved in the positron  production mechanism are available, and they do not exactly match each other,  implying a range of theoretical variation.
 \item[iii)] The uncertainties in the  propagation parameters involved in the transport equation have been thoroughly  studied \cite{Maurin:2001sj}.
\end{itemize}

% The uncertainties on the positron flux have several origins. First, the cosmic--ray nuclei  measurements come with their experimental uncertainties, which then affect the  predictions of induced secondary fluxes, like positrons. Second, various  modelings of the nuclear cross sections involved in the positron  production mechanism are available, and they do not exactly match each other,  implying a range of theoretical variation. Third, the uncertainties in the  propagation parameters involved in the transport equation have been thoroughly  studied \cite{Maurin:2001sj}: a detailed analysis on their impact on  the secondary positron flux is therefore needed.\\

In this chapter, we study the production of secondary positrons which come from the interaction of proton and alpha particles cosmic-rays with the interstellar medium formed by Hydrogen and Helium. As well, the propagation of positrons is modeled according to the TZPM. Finally, we calculate and analyze the positron flux and fraction comparing them with available experimental data.
In each part the uncertainties related to propagation, nuclear cross section and cosmic--rays measurement were studied.\\

%
% II PRODUCTION&PROPAGATION OF POSITRONS BY SPALATIONS
%
%\newpage
%\cleardoublepage
\section{Production of secondary positrons.}

Secondary positrons are produced by interactions of Nuclei CR, principally protons and alpha particles, with the ISM composed mainly by Hydrogen and Helium.\\

The source term of PETE (\citeeq{e:pete1}) is generically described by:
\begin{eq}
 s_{\tu{cab}}(\mb{x},\ener{\tu{c}}) = 4\pi \; n_{\tu{a}} \Phi_{\tu{b}} \frac{d\sigma}{d\ener{\tu{c}}}(\tu{a}+\tu{b} \rightarrow \tu{c}+X) \; ,
\end{eq}
where $n_{\tu{a}}$ is the number density of species \tu{a}, $\Phi_{\tu{b}}$ is the total flux of species \tu{b} per solid angle and ${\displaystyle \frac{d\sigma}{d\ener{}}}$ represents the differential cross section for the specific process.\\

In the case of secondary positrons, the source term related to the scattering between protons and Hydrogen is more complex than the former one, because we are in presence of a proton flux which continuously bombards the ISM. In this case, the source term is: 
\begin{eq}
	q_{p\tn{H}}(\mb{x},\ener{e^+}) = 4\pi \; n_{\tn{H}}(\mb{x}) \int d\ener{p}\; \Phi_{p}(\mb{x},\ener{p}) \;  \frac{d\ics_{p\tn{H}}}{d\ener{e^+}} (\ener{p},\ener{e^+})\;,
\end{eq}
where $n_H$ is the Hydrogen number density, $\ics$ represents the inclusive cross section of the process $p + p \rightarrow e^{+} + X$ discussed and calculated in \citecha{cha2}. Let us specify that the inclusive cross section, in our case, comes from three invariant cross section parameterizations Badhwar et al.~\cite{Badhwar:1977zf}, Tan and Ng~\cite{Tan:1984ha} and Kamae et al.~\cite{Kamae:2004xx}.\\

The proton flux (per solid angle) $\Phi_{p}$ -- in principle -- depends on the position, although the effects of magnetic diffusion help to homogenize the CR density, in addition to that, positrons (electrons) just can propagate not so far from the production point due to the huge energy losses. For this reason, we can assume that the proton (alpha particles) CR density is homogeneous in all the PZ~\cite{Delahaye:2008}.
To avoid the problem of modeling the propagation of Nuclei CR in the Galaxy, we used flux parameterizations of proton and alpha particles based on experimental data. The Shikaze et al. parameterization~\cite{Shikaze:2006je},
\begin{eq}
 \Phi(\ener{}) = A \; \beta^{P_1} \; \left(\frac{\mathcal{R}}{1\;\tu{GV}}\right) ^{-P_2} \qquad \big(\tu{m}^{-2}\;\tu{sec}^{-1}\;\tu{sr}^{-1}\; (\tu{GeV}/\tu{n})^{-1}\big)  \; ,
\end{eq}
is a function of the rigidity and is based on BESS experiment results. On the other hand, the Donato et al. parameterization~\cite{Donato:2001ms},
\begin{eq}
 \Phi(\ener{}) = N\;\left(\frac{\ener{\tu{kinetic}}}{1\;\tu{GeV}/\tu{n}}\right)^{-\gamma} \; ,
\end{eq}
is a function of the kinetic energy per nucleon and is based on data from BESS and AMS experiments. Notice that the functional forms are quite different, but both describe well enough the available data. The values of parameter are given in \citetab{t:nuclei-cr-param}.\\

\begin{tab}
 \begin{tabular}{|c|c|c|c|}
  \hline
  \multicolumn{3}{|c|}{Donato et al. \cite{Donato:2001ms} parameterization} \\
  \hline
  & proton & alpha \\
	\hline 
  $N$ & $1.3249 \times 10^{4}$ & $7.21 \times 10^{3}$ \\
  $\gamma$ & 2.72 & 2.74 \\
  \hline \hline
  \multicolumn{3}{|c|}{Shikaze et al. \cite{Shikaze:2006je} parameterization} \\
	\hline 
   & proton & alpha \\
	\hline 
  $A$ & $(1.94 \pm 0.13)\times 10^{4}$ & $(7.10 \pm 0.56)\times 10^{3}$ \\
  $P_1$ & $0.70 \pm 0.52$ & $0.50 \pm 0.31$ \\
	$P_2$ & $2.76 \pm 0.03$ & $2.78 \pm 0.03$ \\
  \hline
 \end{tabular}
\caption{\label{t:nuclei-cr-param} Values of parameters for Donato et al. and Shikaze et al. parameterizations of the proton and alpha particles fluxes. The units of fluxes for each parameterization are $\tu{m}^{-2}\;\tu{sec}^{-1}\;\tu{sr}^{-1}\; (\tu{GeV}/\tu{n})^{-1}$.}
\end{tab}
%
% 
% 

%
% discussion about f:src-secondaries
%

\citefig{f:src-secondaries} shows the effects produced by different proton flux parameterizations. 
The curves calculated with the Shikaze's parameterization are qualitative equivalent to the ones calculated with the Donato's parameterization.
On the other hand, most of the differences come from the nuclear cross section parameterizations. We see that Kamae's parameterization produces lower values of the source term instead of Tan and Ng and Badhwar ones, which have a lower relative error among them. 
At this level, all these differences can be considered like uncertainties which come from the nuclear physics context, because all of those were produced from equivalent experimental data.\\

\begin{fig}
 \includegraphics[width=0.5\textwidth]{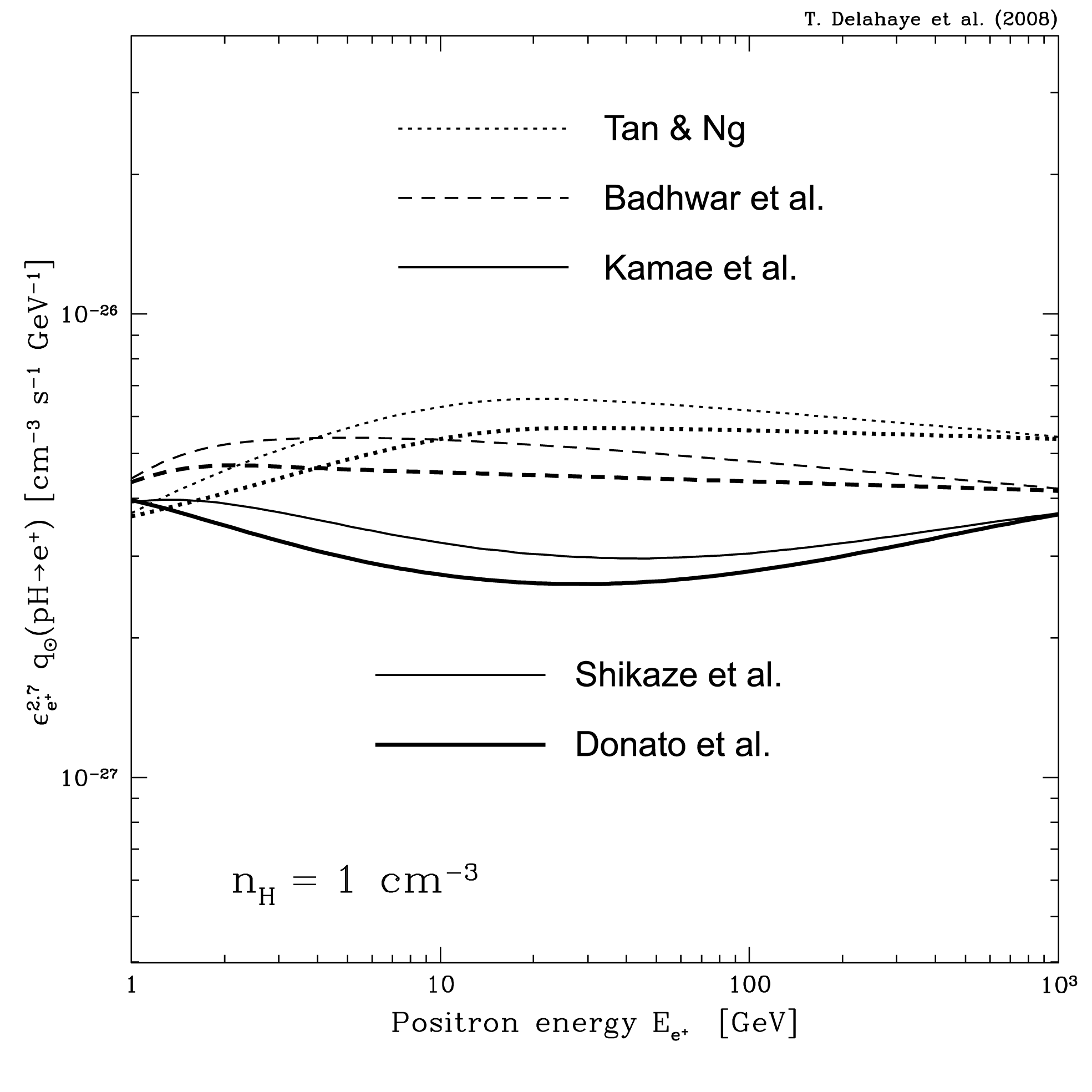}
\caption{\label{f:src-secondaries} Positron source term $\ener{}^{2.7} q_{\astrosun}$ versus positron energy.}
\end{fig}
%
%
%

% the bigger source term
The complete source term, which considers the interaction of protons, alpha particles with Hydrogen and Helium, has the same shape that the proton-Hydrogen case:
\begin{eq}\label{e:sec-srcterm}
	q_{\tn{full}}(\mb{x},\ener{e^+}) = 4\pi \sum_{i=\tn{H},\tn{He}} \sum_{j=p,\alpha}  \; n_{i}(\mb{x}) \int d\ener{j}\; \Phi_{j}(\ener{j}) \;  \frac{d\ics_{ji}}{d\ener{e^+}} (\ener{j},\ener{e^+})\;,
\end{eq}
where the inclusive cross sections for processes $p$+He, $\alpha$+H and $\alpha$+He have been described according to scaling factors $s_f$,
\begin{eq}
 \frac{d\ics_{ij}}{d\ener{e^+}} (\ener{j},\ener{e^+}) =  s_f \; \frac{d\ics_{p\tn{H}}}{d\ener{e^+}} (\frac{\ener{j}}{A_j},\ener{e^+}) \; ,
\end{eq}
where $A_j$ is the mass number of the incident particle. 
%
% taking about the scaling factors
The scaling factors could be quite controversial in the nuclear physics field, however, those reproduce good enough the available experimental data. Basically, we used two of the most known scaling factors. The Orth et al. scaling factor \cite{Orth:1976}, 
\begin{eq}
 s_f = \left(A_i^{3/8} + A_j^{3/8} - 1\right)^2\;,
\end{eq}
and the one proposed by Norbury et al. \cite{Norbury:2006hp},
\begin{eq}
 s_f = \big(A_i A_j\big)^{\frac{2.2}{3}} \; .
\end{eq}
Numerically speaking, both scaling factors produce slightly different values. For example the Orth's scaling factor is 2.54 instead of 2.76 from Norbury's factor for the case of p+He. That difference is screened in the final result due to the low intensity of alpha particles -- in comparison to proton flux --  and the low density of Helium in the ISM.\\

%taking about source distribution.
%
Another point that remains unseen is the spatial distribution of Hydrogen and Helium in the ISM. The interstellar gas principally is concentrated in the Galactic Plane. In first approximation, the gas is homogeneously distributed with densities $n_{\tn{H}} = 0.9\; \tu{cm}^{-3}$ and  $n_{\tn{He}} = 0.1\; \tu{cm}^{-3}$ \cite{Ferriere:2001rg}. Many structures are also present in the GP like the spiral arms, the galactic bulge, among others. However, positrons and electrons come from local distances, where the gas density is more or less constant, for these reason the gas in the galactic plane is considered homogeneous in the \emph{thin disk} of the TZPM (\citecha{cha3}), \ie in a cylinder of height $2 h_z$. In other words, the source term for secondary positron is the \citeeq{e:sec-srcterm} but with a gas density distribution given by:
\begin{eq}
 n_{\tn{H}/\tn{He}}(\mb{x}) = \left\{\begin{array}{cc} n_{\tn{H}/\tn{He}} & |z|\leq h_z \\ 0 & \tn{other cases} \end{array}\right.\; ,
\end{eq}
where $n_{\tn{H}/\tn{He}}$ correspond to the values previously given.\\

\section{Propagation of secondary positrons.}

As it was already discussed in the previous chapters, specially in \citecha{cha3}, the propagation of positrons is modeled according to the TZPM. The HF associated to the secondary positron source term (\citeeq{e:sec-srcterm}) is based on the HF for homogeneous sources and is easier to calculate (\citeeq{e:hf-axial-homo}). In our case, $R$ and $R_{\tn{src}}$ are set to 20~\tn{kpc} and $h_z$ to 100~\tn{pc}.\\

\begin{fig}
 \resizebox{\textwidth}{!}{\includegraphics[angle=270,width=0.5\textwidth]{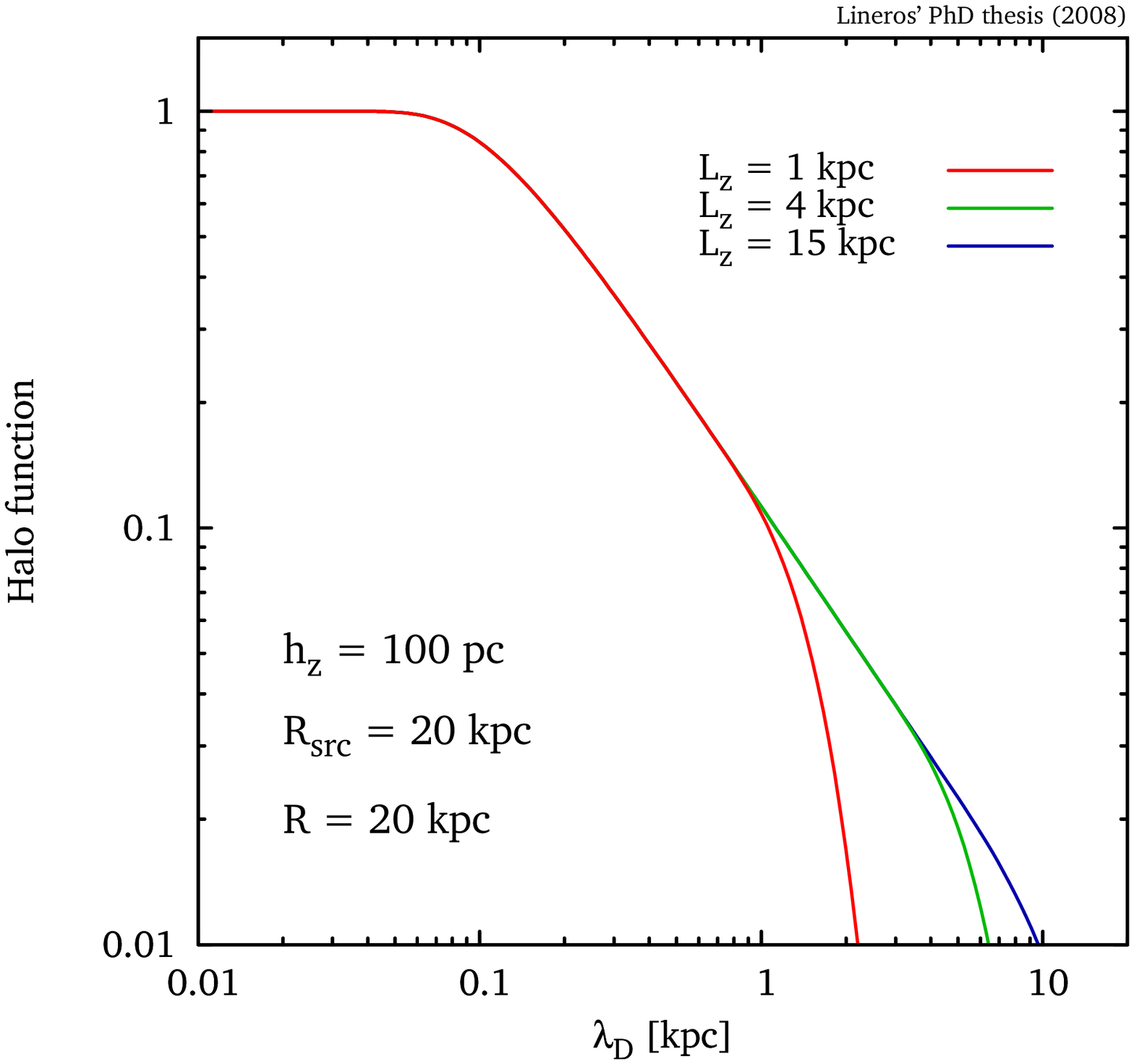}\includegraphics[angle=270,width=0.5\textwidth]{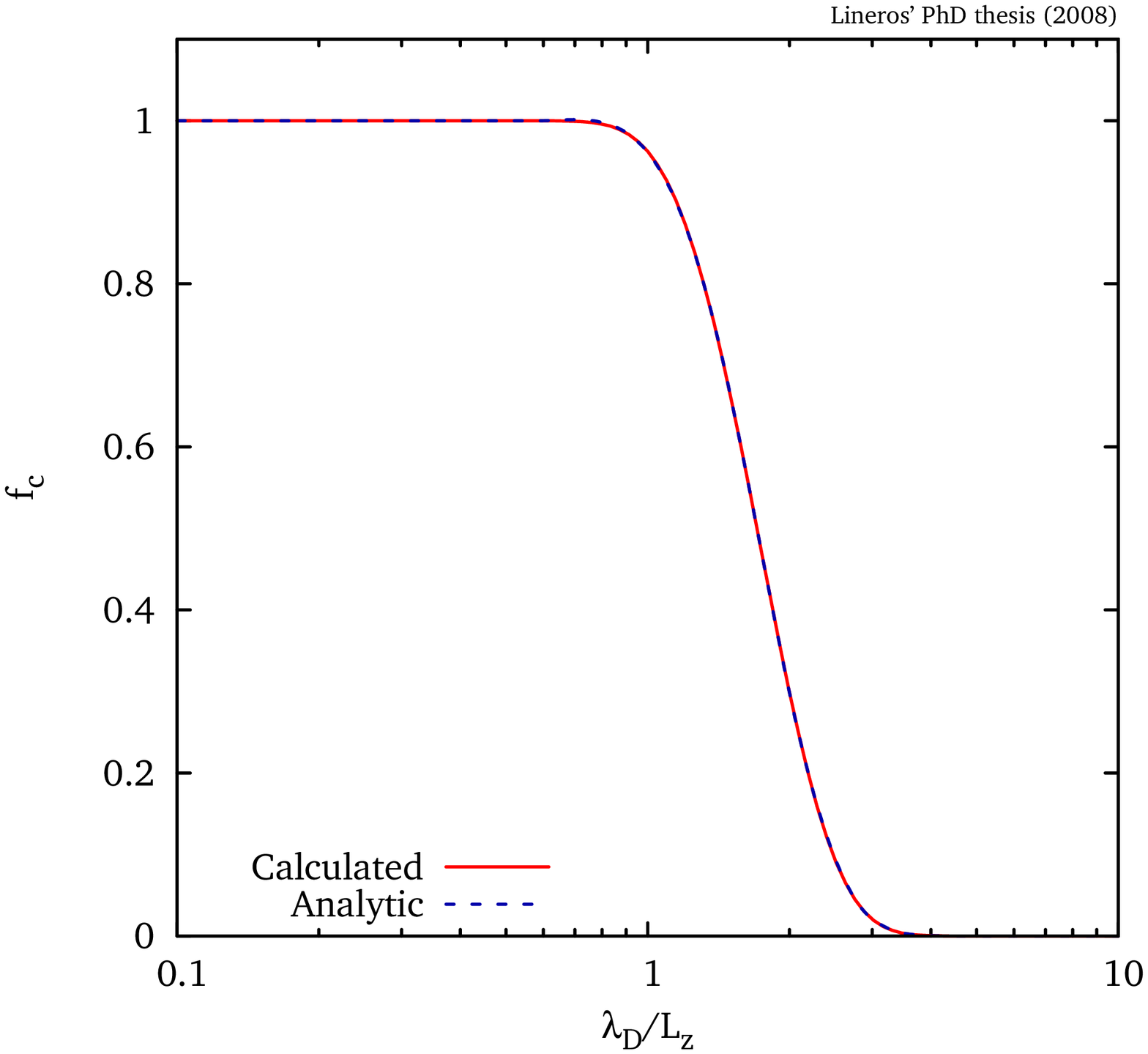}}
 \caption{\label{f:hf-second} Halo function for the secondary positron source term versus the diffusion length $\lD$ (left). Corrective function $f_c$ versus $\lD/L_z$. Both the numerically calculated function and the analytical fit are plotted (right).}
\end{fig}
%
%
%

% explanation and discussion plot halofunction.
\citefig{f:hf-second} show the halo function for the secondary positron source term. This HF presents a fast decreasing behavior for values of $\lD$ bigger than $h_z$, which is related to the propagation volume and the amount of sources inside it. In other words, when $\lD < h_z$ positrons come from a propagation sphere with radius $\lD$ which is full of sources. When $\lD$ goes beyond this limit, the sphere starts to be partially full.\\

Unlike the dark matter HF case, the effects of vertical boundaries play a less important role. Those still modify the HF  for big values of $\lD$, enhancing its decreasing behavior. Nevertheless, this effect appears in the range where HF is small, expecting a smaller contribution in the flux (\citefig{f:hf-second}). The three curves show how vertical boundaries speeds up the decreasing behaviors when $\lD$ is bigger than $L_z$. As well, all the curves have a common behavior until boundary effects appear.
This feature, that they have in common, can be used to produce an unique representation of them.\\
%But that feature can be used and we can take advantage of this symmetry.\\

% describing corrective function
The corrective function $f_c$ is a transfer function from the boundary--free HF respect to the one with vertical boundaries (right-panel in \citefig{f:hf-second}),
\begin{eq}
 \widetilde{I}_{\tn{sec.}}(\lD,L_z) = f_c(\lD/L_z) \times \widetilde{I}_{\tn{sec.}}(\lD,L_z = \infty) \;.
\end{eq}
It appears to be common to any HF, in the specific case of secondary positrons. Also, this nice property helps to improve the speed in flux calculations, because the only function that should be computed is the boundary--free HF. The corrective function can be parameterized as follows (\citefig{f:hf-second}),
\begin{eq}
 f_c(x) = \left\{\begin{array}{cc} 1 & x \leq x_r \\ \exp\left(a_1 t + a_2 t^2 + a_3 t^3 + a_4 t^4 + a_5 t^5 \right) & x_r < x \leq 3 \\ \exp\left(b_0 + b_1 t + b_2 t^2  \right) & 3<x\end{array} \right. \; ,
\end{eq}
where $t = x-x_r$ and the values of the parameters are given in \citetab{t:par-fc}.\\

\begin{tab}
 \begin{tabular}{|c|ccccc|ccc|}
 \hline
  \multicolumn{9}{|c|}{Parameters $f_c$}\\
 \hline \hline
 $x_r$ & $a_1$ & $a_2$ & $a_3$ & $a_4$ & $a_5$ & $b_0$ & $b_1$ & $b_2$ \\
 \hline
 0.6 & 0.0212 & 0.0267 & -1 & 0.5 & -0.0868 & 0.9713 & -0.5046 & -0.6229 \\
 \hline
 \end{tabular}
 \caption{\label{t:par-fc} Parameter values for the fit of the $f_c$ function.}
\end{tab}
%
%
%

% doing the interface from hf to flux.

Up to this point, the secondary positron propagation has been discussed through the analisis of the halo function. 
The following step is to calculate the flux at the Solar System position. As it was seen in \citeeq{e:flux-from-hf}, the flux is obtained by the convolution in energy between the source term and the halo function. 
However this procedure should be performed with proper care due to the fast changes that the halo function has. According to \citefig{f:src-secondaries}, the source term can be roughly assumed as a power-law with spectral index -2.7. On the other hand, \citefig{f:hf-second} shows that the HF is equal to 1 when $\lD \lesssim 0.1\;\tu{kpc}$ and starts to decrease as $\lD$ grows. This feature, when translated into energies, produces an extremely fast change in the function to be integrated, as \citefig{f:lambdaBC} shows.\\

% inverse transformation 
From the value of $\lD$ and the observed energy $\epsilon$, it is possible to infer the value of the injected positron energy $\epsilon_s$: 
\begin{eq}
 \epsilon_s = \left(\epsilon^{\delta-1} + (\delta-1)\frac{\lD^2}{\Lambda_0^2} \right)^{1/(\delta-1)}\; ,
\end{eq}
where $\Lambda_0^2 = K_0\tau_E$. For example, using the MED set (\citetab{tab:model}), a diffusion length of 0.1~\tu{kpc} and observed energies of 1 and 10~\tu{GeV}, we obtain injection energies of 1.003 and 10.056~\tu{GeV} respectively. That means the main contribution of the source term come from the range between the observed energy and the injected energy at 0.1~\tu{kpc}. After that limit, the halo function and the source term decreases fast, contributing less to the final results.\\

%
% III THE POSITRON FLUX AND UNCERTAINTIES
%
%\newpage
%\cleardoublepage
\section{Positron flux and uncertainties.}

Current experiments, as HEAT~\cite{Barwick:1997ig}, AMS~\cite{Aguilar:2007} among others, are able to measure the positron and electron fluxes that arrive to the Earth. 
As well, we are able to compute the positron and electron flux using the propagation model and sources that we discussed in last sections.\\

% FIGURE
%
%
\begin{fig}
 \includegraphics[width=\textwidth]{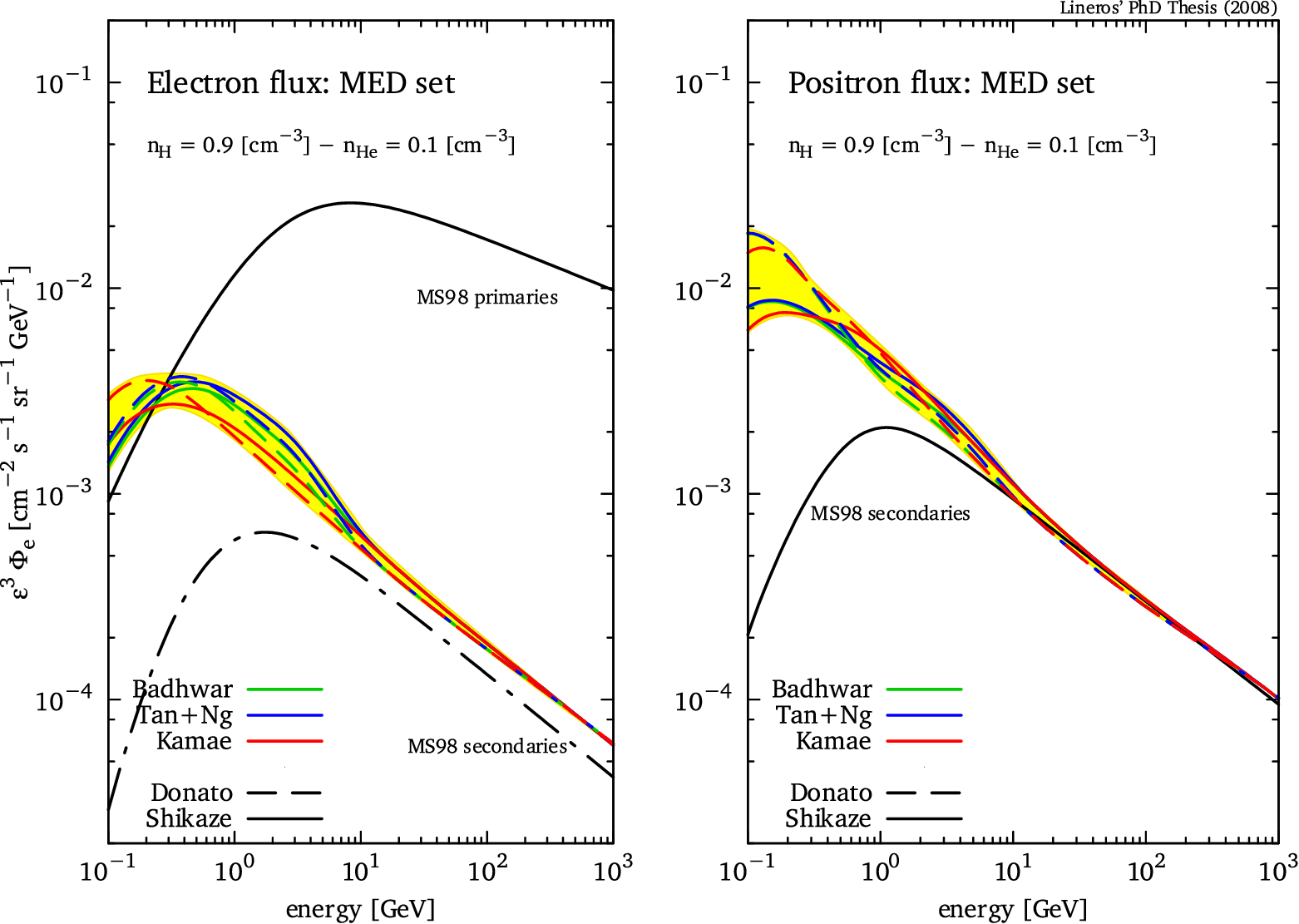}
 \caption{\label{f:ele-pos-flux} Interstellar electron and positron fluxes $\ener{}^{3}\Phi_e$ for the MED set (\citetab{t:prop-par}) versus energy. The curves correspond to fluxes calculated from Nuclei CR interactions with the ISM. Each curve represents a different nuclear cross section and Nuclei CR flux parameterization. Also the Strong et al. (\citeeq{e:fit-bkg}) flux parameterizations are shown. The uncertainty band related to those parameterization is plotted (yellow band) as well.}
\end{fig}
%
%
%

% DISCUSSION FIG f:ele-pos-flux
\citefig{f:ele-pos-flux} shows the positron and electron fluxes calculated using the two nuclei CR parameterization (Donato et al.~\cite{Donato:2001ms} and Shikaze et al.~\cite{Shikaze:2006je}) and the three nuclear cross section parameterizations (Tan and Ng~\cite{Tan:1984ha}, Kamae et al.~\cite{Kamae:2004xx} and Badhwar et al.~\cite{Badhwar:1977zf}). 
The yellow band represent the uncertainties from nuclear cross sections and nuclei CR observations.\\
%In principle, all of those were obtained from experimental data, which means each one is equivalent to estimate the positron flux. \\

%
For the different calculated fluxes, we observed that at high energy ($>$ 100~\tu{GeV}) all fluxes converges into one common curve. This is produced by a common asymptotic behavior of all parameterizations. 
On the other hand, differences arise at low energies because smaller differences from high energies start to accumulate when the flux is calculated. Let us to remark that the secondary positron and electron fluxes depend on an integration on energy from infinity down to the observed energy.\\

% COMPARISON WITH MS98
Moreover, in \citefig{f:ele-pos-flux} the reference curves of Baltz et al. parameterization (\citeeq{e:fit-bkg}), which come from the results of Strong and Moskalenko~\cite{Moskalenko:1997gh}, are plotted. We note how positron calculated with our model exhibits same power--law behavior with power index around -3.5 like Strong et al.
For energies below 10~\tu{GeV}, bigger differences appear. However, that is the range of energies dominated by the Solar modulation, which indirectly reduces the differences among the modulated fluxes.
The secondary electron flux presents bigger difference with respect to secondaries from Strong et al. However, the secondaries contribution to electron flux is highly diluted into the sea of primary electron flux. 
This establishes a first motivation for further improving calculations in other to estimate the primary electron flux, as well.\\

% FIGURE 
%
%
\begin{fig}
 \resizebox{\textwidth}{!}{\includegraphics[angle=270]{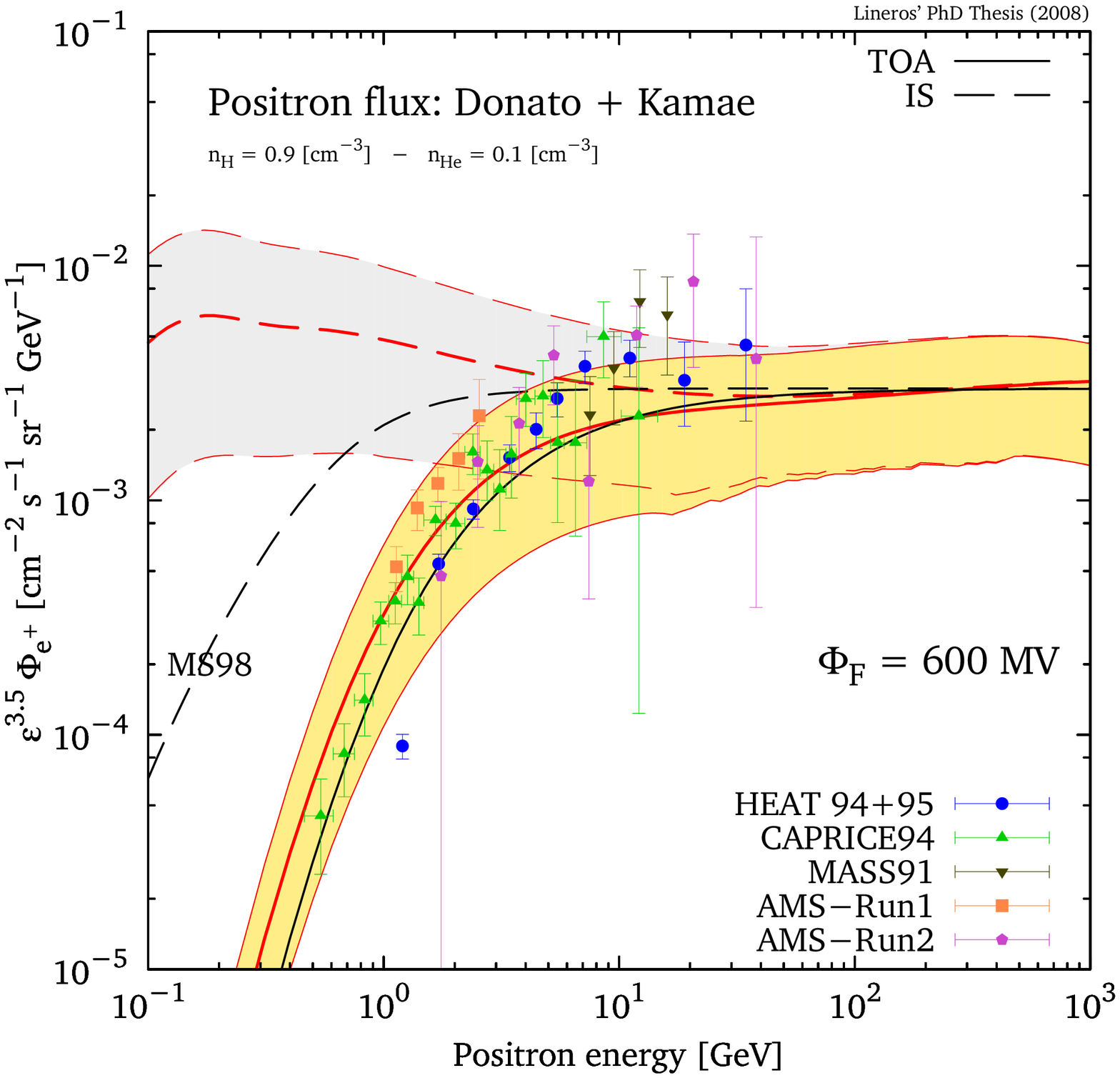}\includegraphics[angle=270]{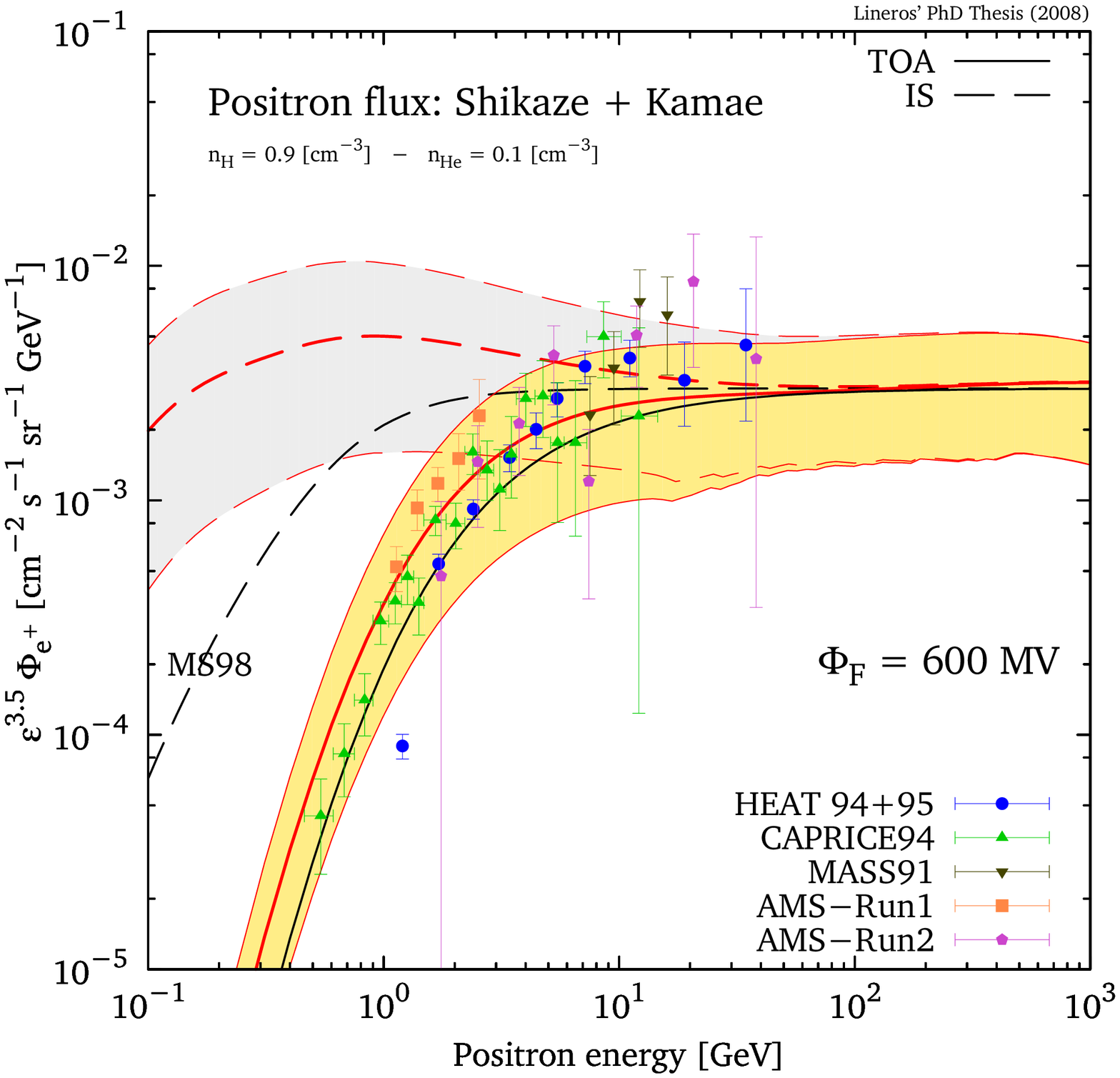}}
 \caption{\label{f:sec-flux-kamae} Secondary positron flux $\ener{}^{3.5} \fluxe$ versus positron energy. Positron fluxes were calculated using proton and alpha CR fluxes from Donato et al.~\cite{Donato:2001ms} and Shikaze et al.~\cite{Shikaze:2006je} with the Kamae et al. cross section parameterization~\cite{Kamae:2004xx}. }
\end{fig}
%
%
%

% FIGURE
%
%
\begin{fig}
 \resizebox{\textwidth}{!}{\includegraphics[angle=270]{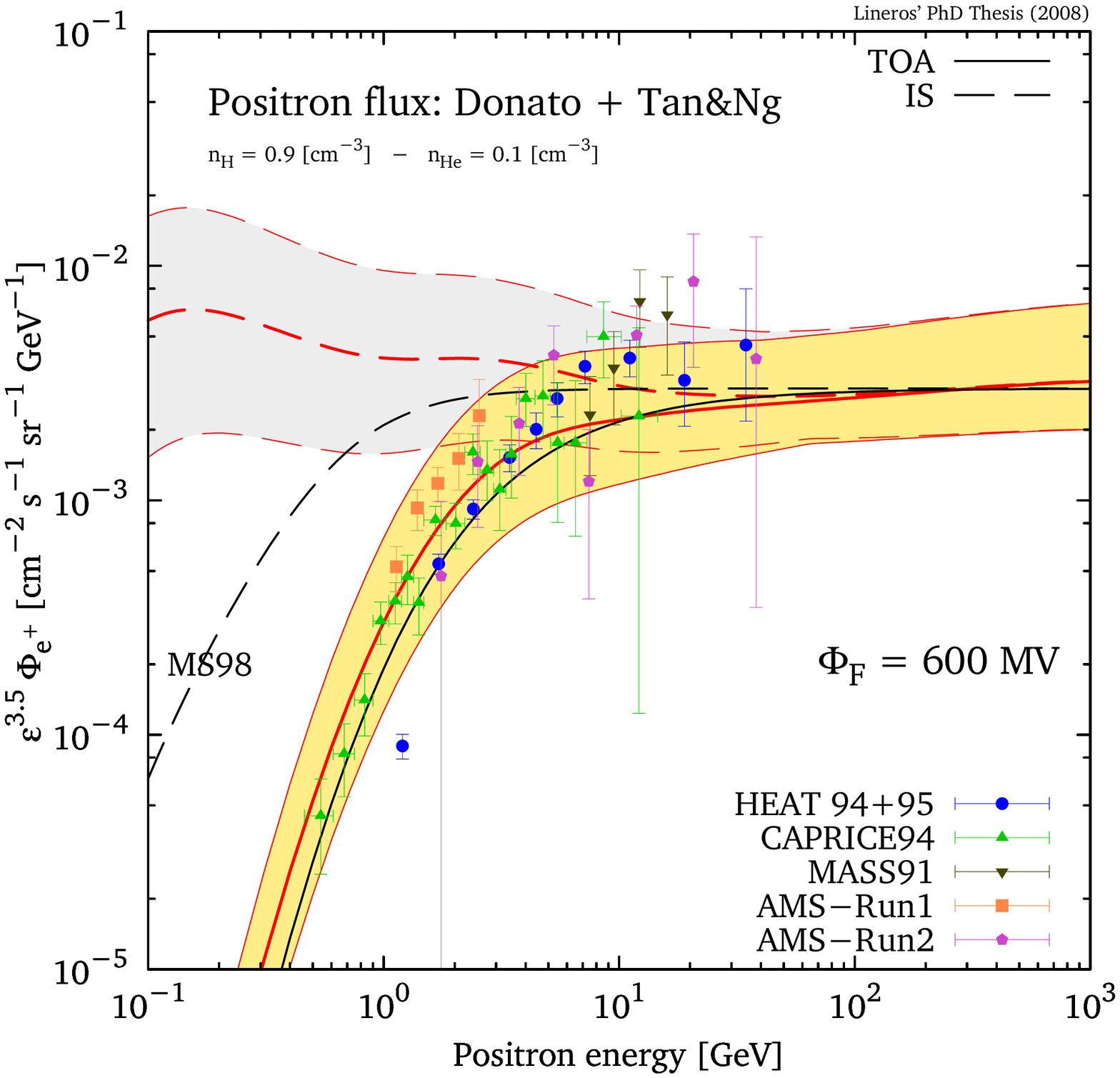}\includegraphics[angle=270]{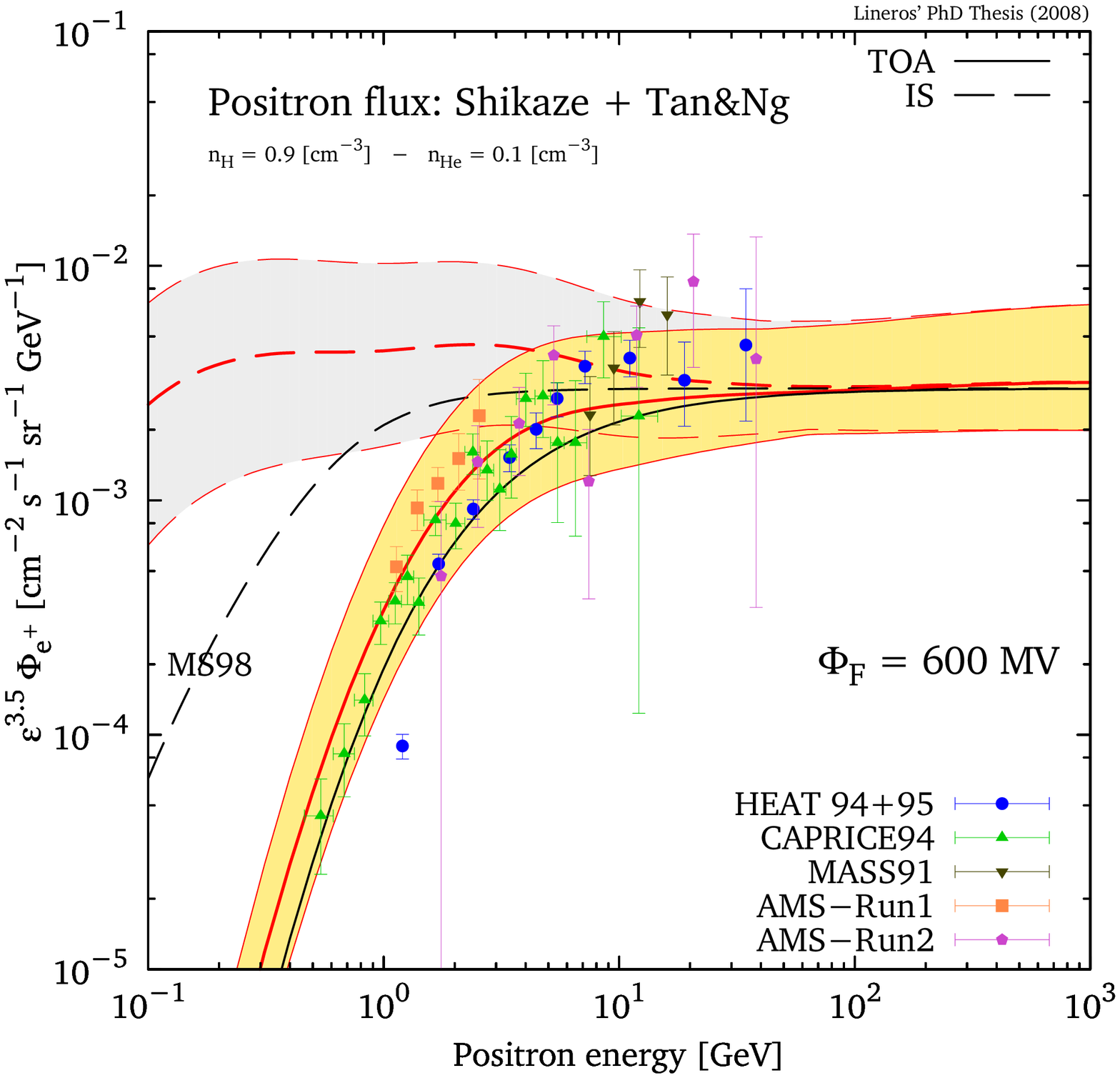}}
 \caption{\label{f:sec-flux-tanng} Secondary positron flux $\ener{}^{3.5} \fluxe$ versus positron energy. Positron fluxes were calculated using proton and alpha CR fluxes from Donato et al.~\cite{Donato:2001ms} and Shikaze et al.~\cite{Shikaze:2006je} with the Tan et al. cross section parameterization~\cite{Tan:1984ha}.}
\end{fig}
%
%
%

% INCLUDING B/C UNCERTAINTIES
Extra information is obtained from the study and analysis of the uncertainty in the propagation model. In a similar way to what was done in the case of dark matter annihilation, we calculate the positron flux considering the 1600 parameter configurations compatibles with the B/C measurement \cite{Maurin:2001sj}.\\

% DISCUSSION BOTH FIGURES f:sec-flux-kamae and f:sec-flux-tanng

\citefig{f:sec-flux-kamae} and \citefig{f:sec-flux-tanng} show the positron flux $\ener{}^{3.5}\Phi_{e^{+}}$ calculated for the MED set and in the solar-modulated  and -unmodulated regime (solid and dashed red lines, respectively), the reference curves for Strong et al. secondary positrons are plotted for the same regimes.\\

% intro uncertainties
In both figures, uncertainty bands associated to different propagation model are included. The upper and lower bounds in each band correspond to the maximum and the minimum fluxes obtained among all the parameter space. In a similar way that in the case of DM annihilation, there are not unique parameter sets which maximize or minimize the flux in the whole range of energies. \\

% experimental data
% Moreover, the latest experimental data of HEAT~\cite{Barwick:1997ig}, AMS~\cite{Aguilar:2007}, CAPRICE~\cite{Boezio:2000} and MASS~\cite{Grimani:2002yz} have been included in the figures with the objective to used them for comparison with the calculated fluxes.
%
The first interesting issue that we notice is how the experimental data is nicely contained inside the uncertainty band, this fact corroborates the consistency of the propagation model among different species of cosmic rays. Furthermore, we note that the fluxes at low-energy ($<$ 10~\tu{GeV}) present similar values due to the solar modulation effects.\\

%Furthermore, these figures shows the effect of solar modulation in the reduction of differences among fluxes in the low--energy range ($<$ 10~\tu{GeV}).\\

% comparison between all parameterization
From the figures, we see how different nuclei CR flux parameterizations do not modify too much the positron flux. For example, in \citefig{f:sec-flux-kamae}, we compare the positron fluxes calculated with the same cross section parameterization, this case Kamae's one, but with each of the CR flux parameterizations, Donato's and Shikaze's one. The same situation occurs in \citefig{f:sec-flux-tanng}.\\

The major differences arise in the low-energy range but just for interstellar fluxes. These differences cannot be easily seen in the modulated signal because the Solar modulation generate an energy shift equivalent to 0.6~\tu{GeV} respect to the energy of the interstellar flux. This energy shift is enough to allow us to see at Earth just the range of energies ($>$ 0.8~\tu{GeV}) where fluxes are very similar one to each other.\\ 

% do a cross comparison
A cross comparison between \citefig{f:sec-flux-kamae} and \citefig{f:sec-flux-tanng} gives information on how cross section parameterizations (Kamae's and Tan's ones) changes the estimation of positron flux. As we observe, the differences among those are almost negligible if we compare them with the effect of propagation uncertainties.\\

% conclusion of the section.
Uncertainties from nuclear cross section and CR flux parameterizations show to be not so important as the uncertainties in the propagation. In any case, the former one would be important in an analysis which involves multimessengers correlations. \\

%
% IV POSITRON FRACTION REVISITED
%
%\newpage
%\cleardoublepage
\section{Positron fraction analysis.}

% INTRO
The observable, that actual experiments are able to release first, is the positron fraction (\citeeq{fraction}) in which the systematic errors are easier to manage and require less data processing than the flux determination case.
% which requires less data processing in order to reduce the systematical errors instead of fluxes determination. 
%
The most controversial issue is the presence of an excess in the positron fraction, that was noted by Moskalenko et al.~\cite{Moskalenko:1997gh}, when they derived predictions for the electron and positron fluxes. This excess has tried  to be explained by many ways, for example, invoking contribution from DM annihilation (e.g. \cite{Baltz:1998xv}) or contribution from nearby pulsars (e.g. \cite{Grimani:2007, Yuksel:2008rf}).\\

% INTRO  PAMELA
In the later October 2008, the PAMELA experiment released the data for the positron fraction~\cite{Adriani:2008zr}, which would confirm a similar feature that HEAT experiment~\cite{Barwick:1997ig} had slightly observed. 
Nevertheless, it is still unclear if this feature comes from a undiscovered source of positrons or maybe lay in other type of explanation.
Another perspective is to think in the possibility that this feature is connected somehow to the electron component instead of the positron one.\\

% into our calculations and discus the figure and how it was obtained.
In the previous sections, we discussed and explained how to calculate the secondary positron flux. As well, we showed that current flux measurements are reproduced for same setting of the TZPM that reproduce B/C~\cite{Maurin:2001sj}, proton and antiproton CR measurements~\cite{Donato:2003xg}.\\

% description CASADEI & BINDI

In the work of Casadei et al.~\cite{Casadei2004ApJ}, they estimated the interstellar electron flux by combining the available data and producing a global fit. They performed on every dataset a demodulation, by inverting the modulation process described by Perko et al.~\cite{Perko:1987A&A}. As well, to be able to demodulate the signal, they studied correlations between solar activity, other species of CR and information from neutron monitors~\cite{bartol:nm}, in order to estimate the modulation parameter.
% value of fits.
The interstellar electron flux was modeled by a power--law function valid from 3~\tu{GeV} up to 2~\tu{TeV}:
\begin{eq}
 \Phi^{e^-}_{\tn{Casadei}}(\ener{e^{-}}) = N \; \left(\frac{\ener{e^{-}}}{1\;\tu{GeV}}\right)^{\gamma}\; ,
\end{eq}
where $\gamma = -3.44 \pm 0.05$ and $N = (412.3 \pm 22.8)\ \tu{m}^{-2}\;\tu{sr}^{-1}\;\tu{sec}^{-1}\;\tu{GeV}^{-1}$.\\

% FIGURE
%
%
\begin{fig}
\resizebox{\textwidth}{!}{\includegraphics[width=\textwidth]{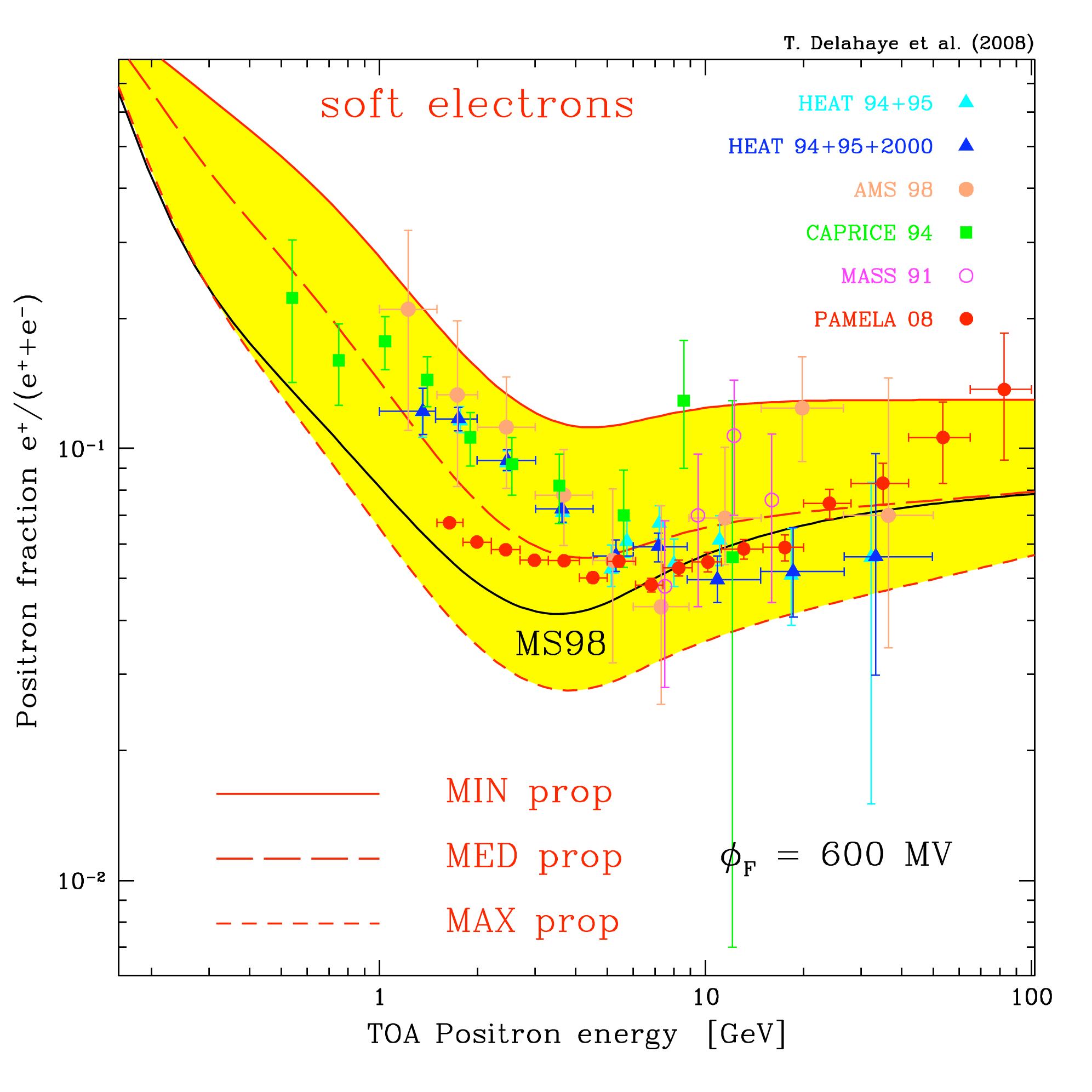}\includegraphics[width=\textwidth]{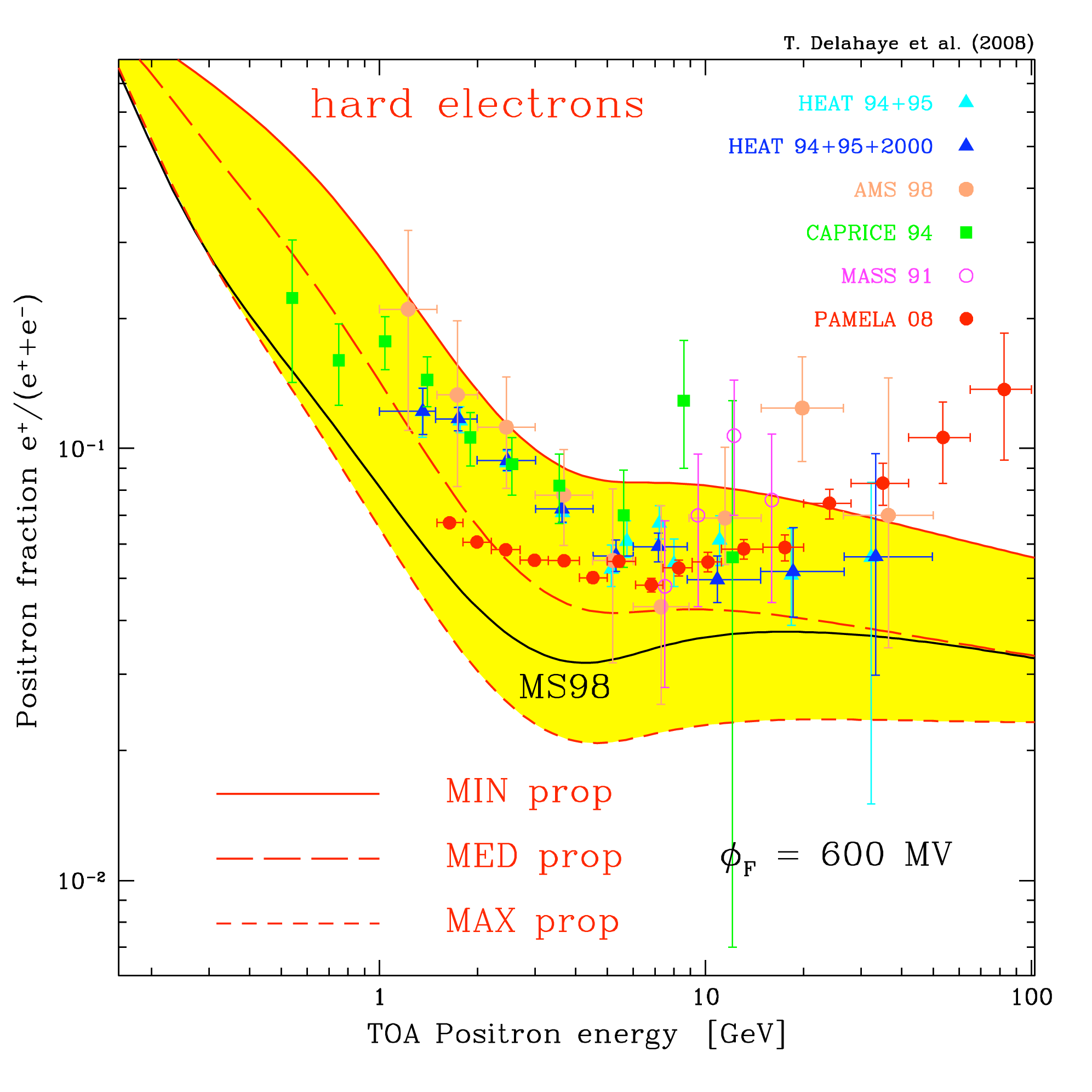}}
	\caption{\label{f:fit-exp-heat} Positron fraction as a function of the positron energy, for a soft and hard electron spectrum. Data are taken from CAPRICE~\cite{Boezio:2000}, HEAT~\cite{Barwick:1997ig}, AMS~\cite{Aguilar:2007}, MASS~\cite{Grimani:2002yz} and PAMELA~\cite{Adriani:2008zr}.}
\end{fig}
%
%
%

% intro for figures
\citefig{f:fit-exp-heat} shows the positron fraction obtained with the two extreme cases for the  electron flux, soft and hard. The soft electron flux has a power index of -3.54 (left) and the hard electron flux has one of -3.34 (right).
These two situations correspond to a variation of $3\sigma$ with respect to Casadei et al. reported values.\\
% These extremes correspond to 3$\sigma$ from Casadei et al. index. In that sense, we cover the extreme situations where electron flux could behave.\\

% description soft
In the case of a soft electron flux, a steeper behavior on the positron fraction is obtained. At high energies, the positron fraction tends to increase its value, and this is produced by the faster reduction in the electron flux. When we include the uncertainty band related to the propagation model, most of the experimental data goes inside this.
% description hard
On the other hand, in the case of a hard electron flux, the positron fraction in the high energy range gets reduced. 
This feature associated to a possible positron excess is supported better by the observations in the case of hard electron fluxes than in the case of soft electron fluxes.\\
%The unknown feature related to the positron excess became stronger in this situation. \\

% description general.
These extreme behaviors remark the important role of electrons in the positron fraction. As well, with the recent release of PAMELA data~\cite{Adriani:2008zr} the feature -- above 10~\tu{GeV} -- has been observed, but the results at lower energies need to be analyzed carefully.\\

\subsubsection{An alternative analysis.}
% TABLE
%
%
\begin{tab}
	\begin{tabular}{|c|ccc|c|}
	\hline
	Experiment & $n_0\;[\tu{m}^{-2}\;\tu{sr}^{-1}\;\tu{s}^{-1}\;\tu{GeV}^{-1}]$ & $\gamma$ & $\phi\;[\tu{GeV}]$ & $\chi^2/\tn{d.o.f.}$ \\
	\hline \hline
	AMS & 0.238 & 3.86 & 0.574 & 0.152 \\
	HEAT & 0.354 & 4.03 & 1.019 & 0.237 \\
  BETS & 0.2029 & 3.201 & 0.1122 & 0.7985 \\
  \hline
	\end{tabular}
	\caption{\label{t:fit-par} Parameters values for best fits for electron flux of AMS~\cite{Aguilar:2007}, HEAT~\cite{Barwick:1997ig} and BETS~\cite{Torii:2001ApJ} experiments.}
\end{tab}

A different approach is to analyze separately the electron fluxes of some experiments. We focused on the AMS~\cite{Aguilar:2007}, HEAT~\cite{Barwick:1997ig} and BETS~\cite{Torii:2001ApJ} data, because those show different characteristics. For example, AMS and HEAT are more precise in the range of low--energy ($<$ 30~\tu{GeV}) instead of BETS which has measured the high--energy range giving valuable information about the IS flux.\\

% how we did the fit.
For these three experiments, we performed fits based on the Top--of--Atmosphere (TOA) electron data, using the functional form:
\begin{eq}
 \Phi^{\tn{TOA}}(\ener{}) = n_0 \left( \frac{\ener{}+\phi}{10\;\tu{GeV}} \right)^{-\gamma} \; \left(\frac{\ener{}}{\ener{}+\phi}\right)^{2} \times \frac{\log(\ener{}/\phi)}{\log(10\;\tu{GeV}/\phi)} \; ,
\end{eq}
which depends on three parameters and it is clearly inspired by a modulated power--law function.
% explaining the shape
The values of the parameters are given in \citetab{t:fit-par}. Notice that the value of $\chi^2/\tn{d.o.f.}$ for each fit is lower that one, which indicates that the data can be nicely fitted with this functional form.\\

Probably the logarithmic term cannot be explained by some physical phenomenon. However, functions with that form behave like power--laws with variable power index. To clarify this point, the power index can be extracted from a function $f(x) = x^{\gamma}$ doing:
\begin{eq}
 \tn{power index}(x) = \frac{x f^{'}(x)}{f(x)} \quad \rightarrow \quad \tn{power index}(x) = \gamma \; .
\end{eq}
Applying the same method to $x^{\gamma}\log(x/\phi)$ , we obtain:
\begin{eq}
 \tn{power index}(x) = \gamma + \frac{1}{\log(x/\phi)}\; ,
\end{eq}
which shows that the power index changes slowly when $x \gg \phi$ and it asymptotically converges to $\gamma$.\\

% FIGURE f:fits-comp
%
%
\begin{fig}
 \includegraphics[angle=270, width=0.5\textwidth]{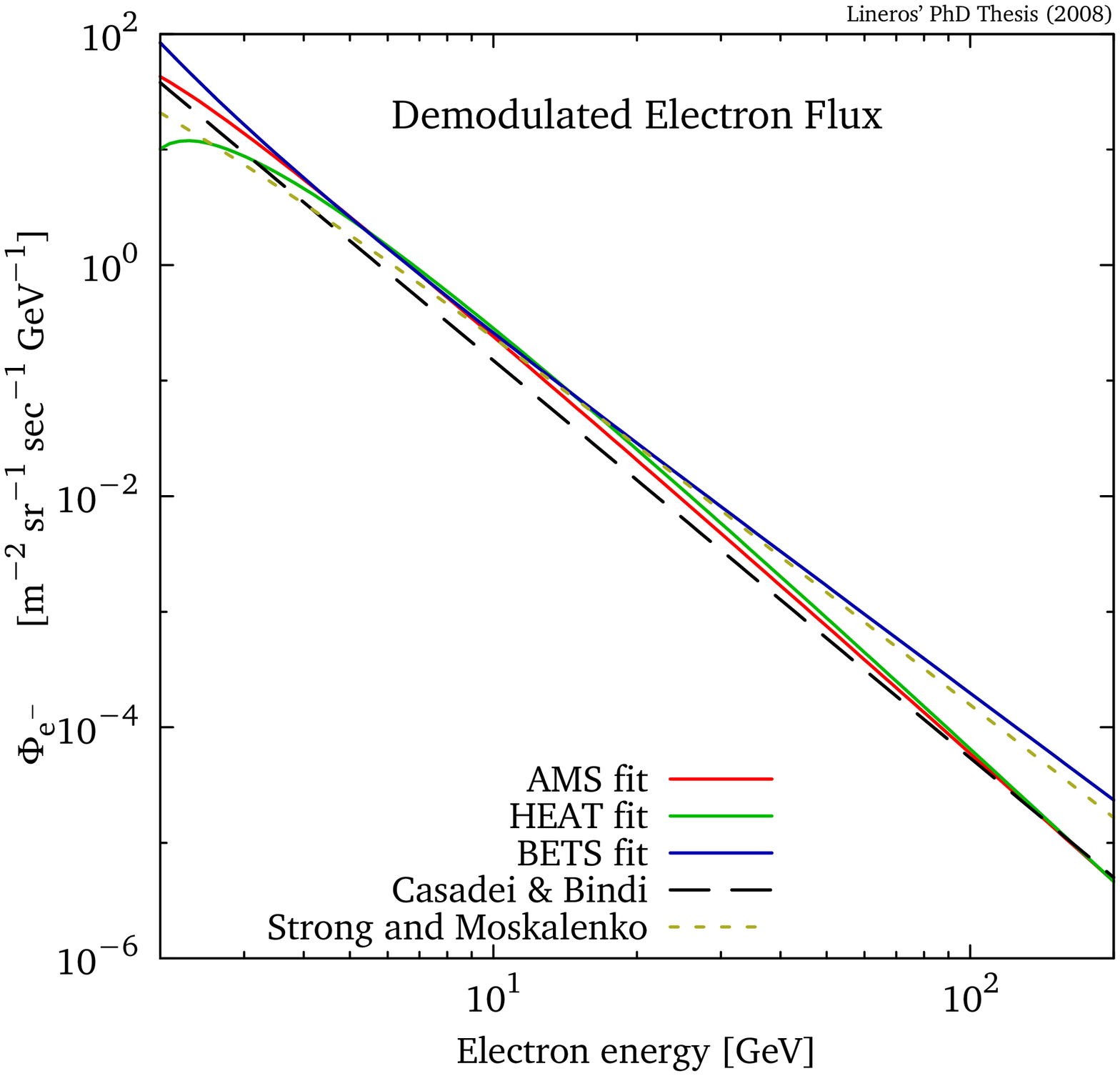}
 \caption{\label{f:fits-comp} Demodulated (interstellar) electron flux versus electron energy. The fits of data from AMS~\cite{Aguilar:2007}, HEAT~\cite{Barwick:1997ig} and BETS~\cite{Torii:2001ApJ}, have been demodulated according to modulation procedure used by Casadei et al.~\cite{Casadei2004ApJ}. 
The global fit obtained by Casadei et al. and the Strong et al.~\cite{Moskalenko:1997gh} flux estimation are shown as references.
The AMS and HEAT fits are the ones with a behavior similar to Casadei et. al. On the other hand, BETS fit produce larger differences.}
\end{fig}
%
%
%

% DISCUSSION OF FIGURE f:fits-comp
\citefig{f:fits-comp} shows the fits to electron data of AMS, HEAT, BETS and the Casadei et al. fit. In order to compare them, each flux was demodulated according to modulation parameters suggested by Casadei et al. (\citetab{t:solmod_par}). We notice that all curves exhibit similar behaviors except the one associated to BETS experiment, which present a harder spectra compared to the other experiments. Let us note that these small differences may produce big effects into the positron fraction.\\

% explaining how calculate the positron fraction in this case.

We compute the positron fraction using the fitted fluxes. Nevertheless, we are not interested in processing the fluxes as Casadei et al. did. Instead of that, just the positron flux is modulated, because the electron fluxes represent the TOA data which already contain modulation.\\

% Discussion of figure f:fit-exp-comp.
\citefig{f:fit-exp-comp} shows the fits for each experiment and the correspondent positron fraction. In the plots for electron flux -- left column -- the $1\sigma$ uncertainty bands (lighterblue) in the parameter determination are included. We can see that the fit for HEAT presents a narrow uncertainty band because its data are more accurate. The fit for BETS presents wider uncertainties at low-energy because there are not enough data in that range. The same situation occurs for the fit of AMS data, but in the high energy case.\\

% DISCUSSION RIGHT COLUMN
The right column of \citefig{f:fit-exp-comp} shows the positron fraction calculated with the positron flux obtained with the MED set (red line) and with the uncertainties in the propagation (darker blue band). Also, the fits for the data of AMS, HEAT and BETS, are used to calculate the fraction. The uncertainty band related to the parameter determination (lighter blue band) is included respect to the MED set situation.\\

% ANALYSIS OF THE FRACTIONS
Depending on the experiment used to determinate the electron flux, we notice that the high energy range tail of the positron fraction may sizeably change. In the case of HEAT and using a modulated positron flux at 800~\tu{MV}, the calculated positron fraction above 10~\tu{GeV} seems to be not as high as the experimental data, which means that a positron excess would be present.\\ 

The same situation occurs when the AMS electron flux fit and the positron flux modulated at 600~\tu{MV} are used to calculate the positron fraction. In this case, uncertainties in the parameter determination of the electron flux makes hard to determine if the excess is present or not. 
On the other hand, the excess is favored when the BETS electron fit is considered.\\

% FIGURE f:fit-exp-comp
%
%
\begin{fig}
 \resizebox{0.88\textwidth}{!}{\includegraphics[angle=270, width=\textwidth]{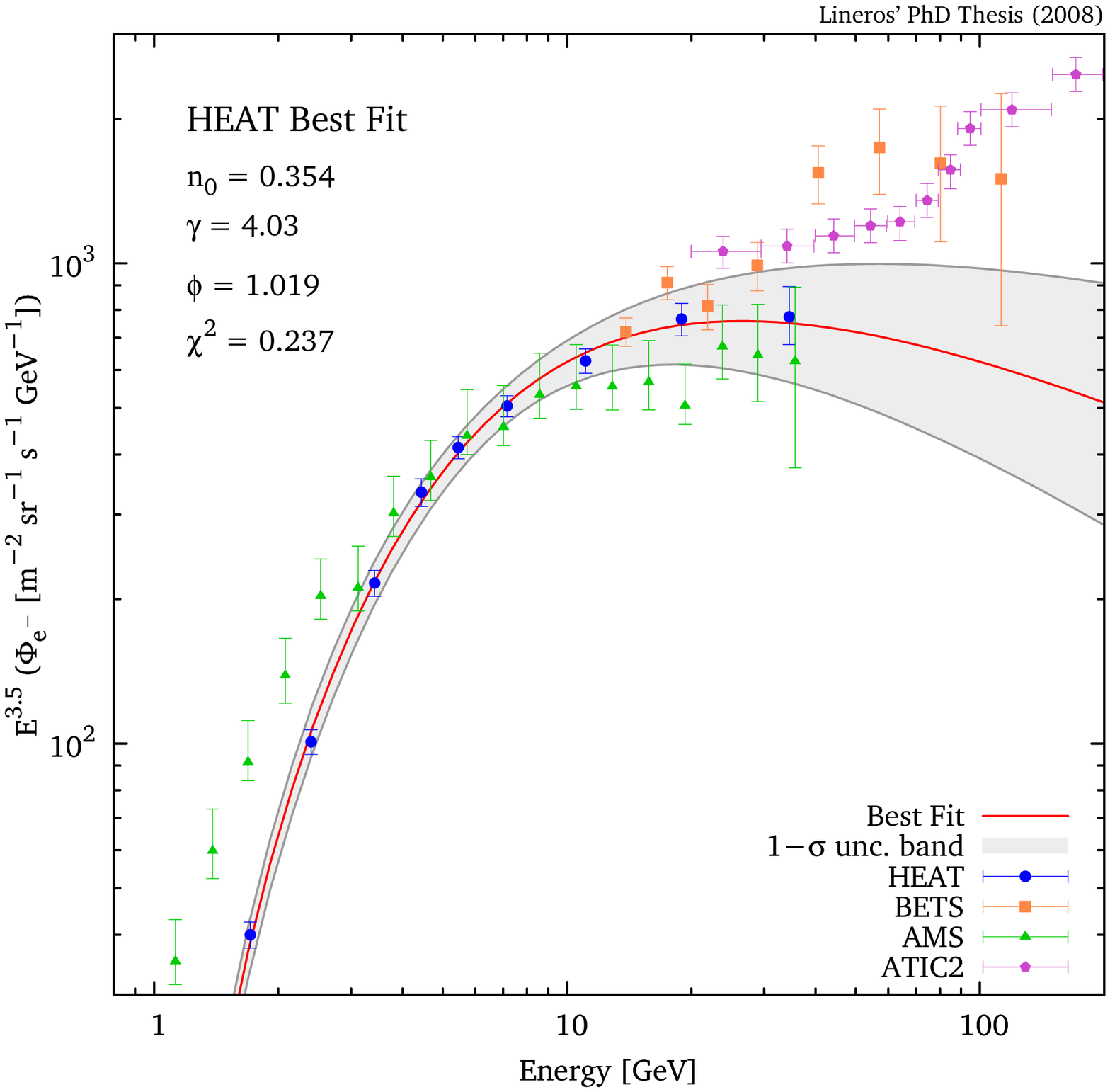}\includegraphics[angle=270, width=\textwidth]{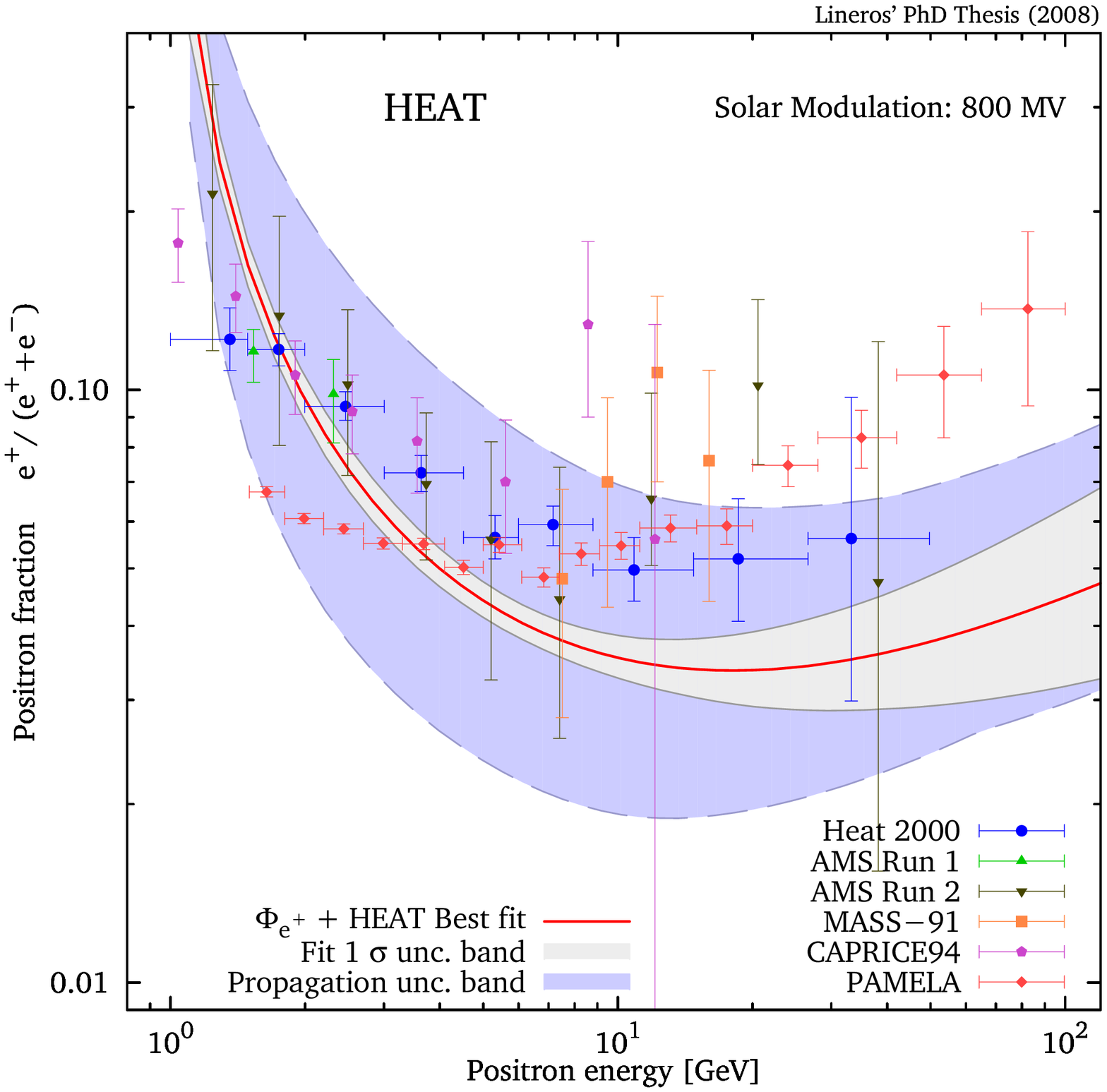}}\\
 \resizebox{0.88\textwidth}{!}{ \includegraphics[angle=270, width=\textwidth]{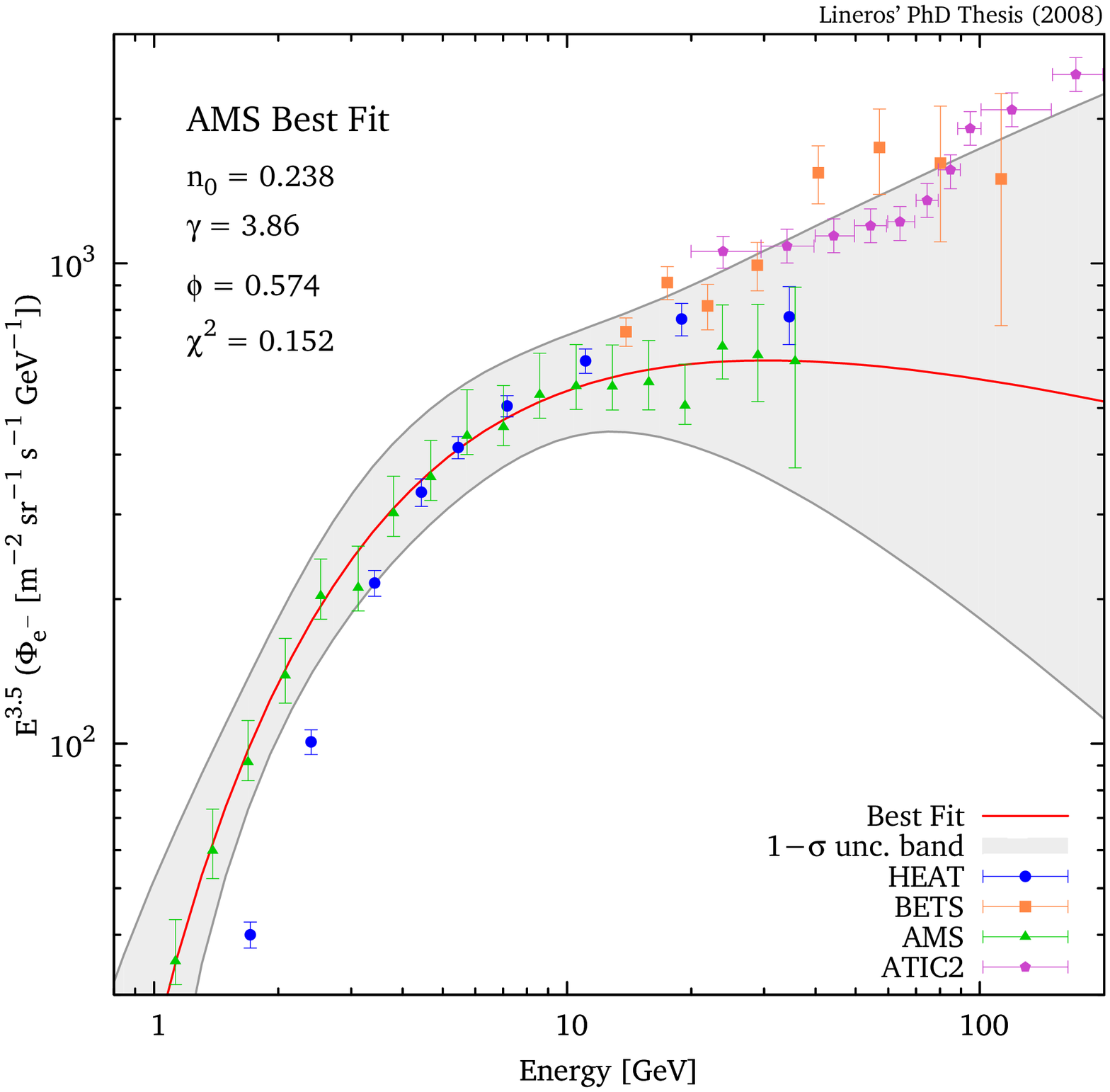}\includegraphics[angle=270, width=\textwidth]{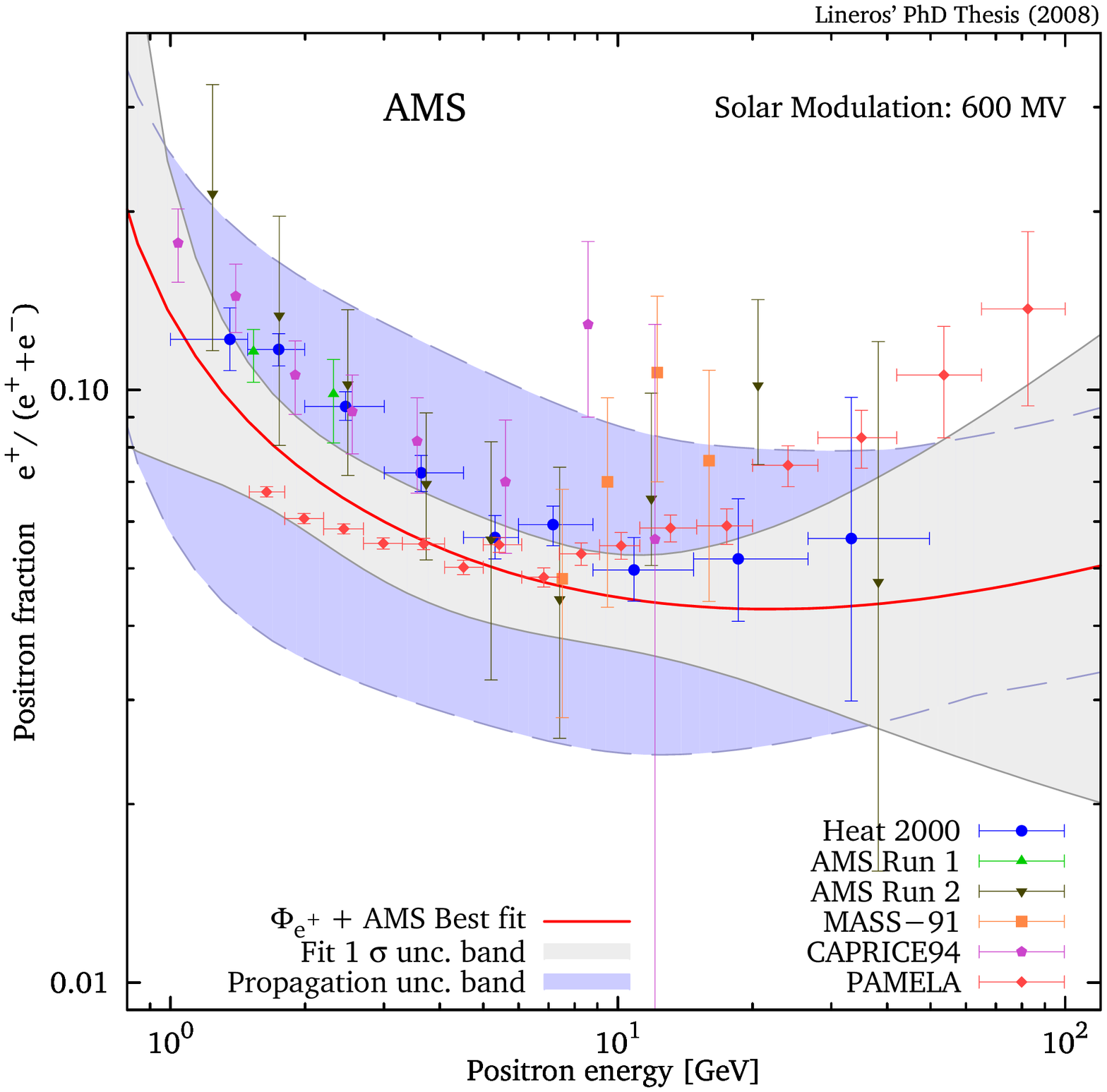}}\\
 \resizebox{0.88\textwidth}{!}{\includegraphics[angle=270, width=\textwidth]{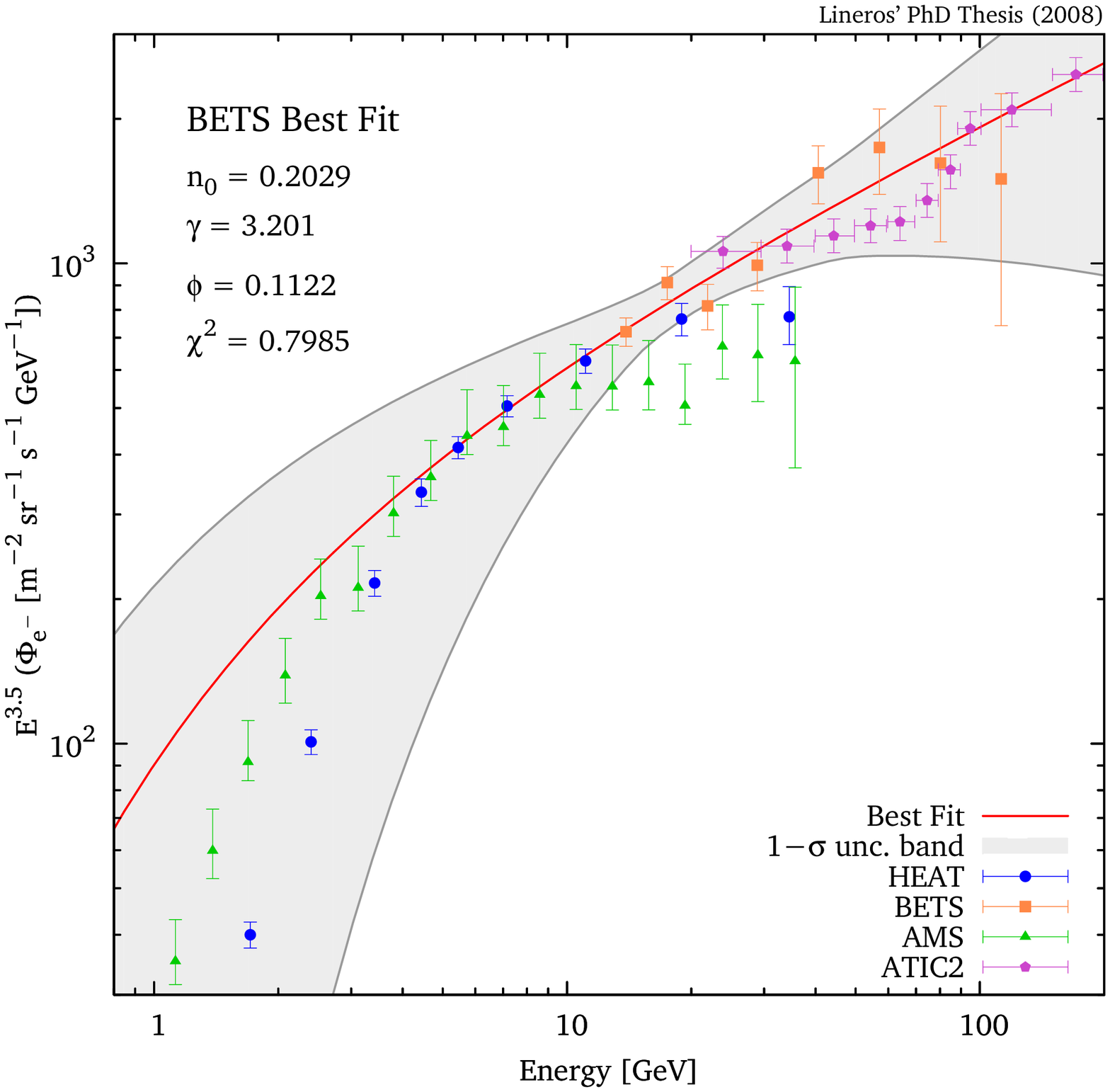}\includegraphics[angle=270, width=\textwidth]{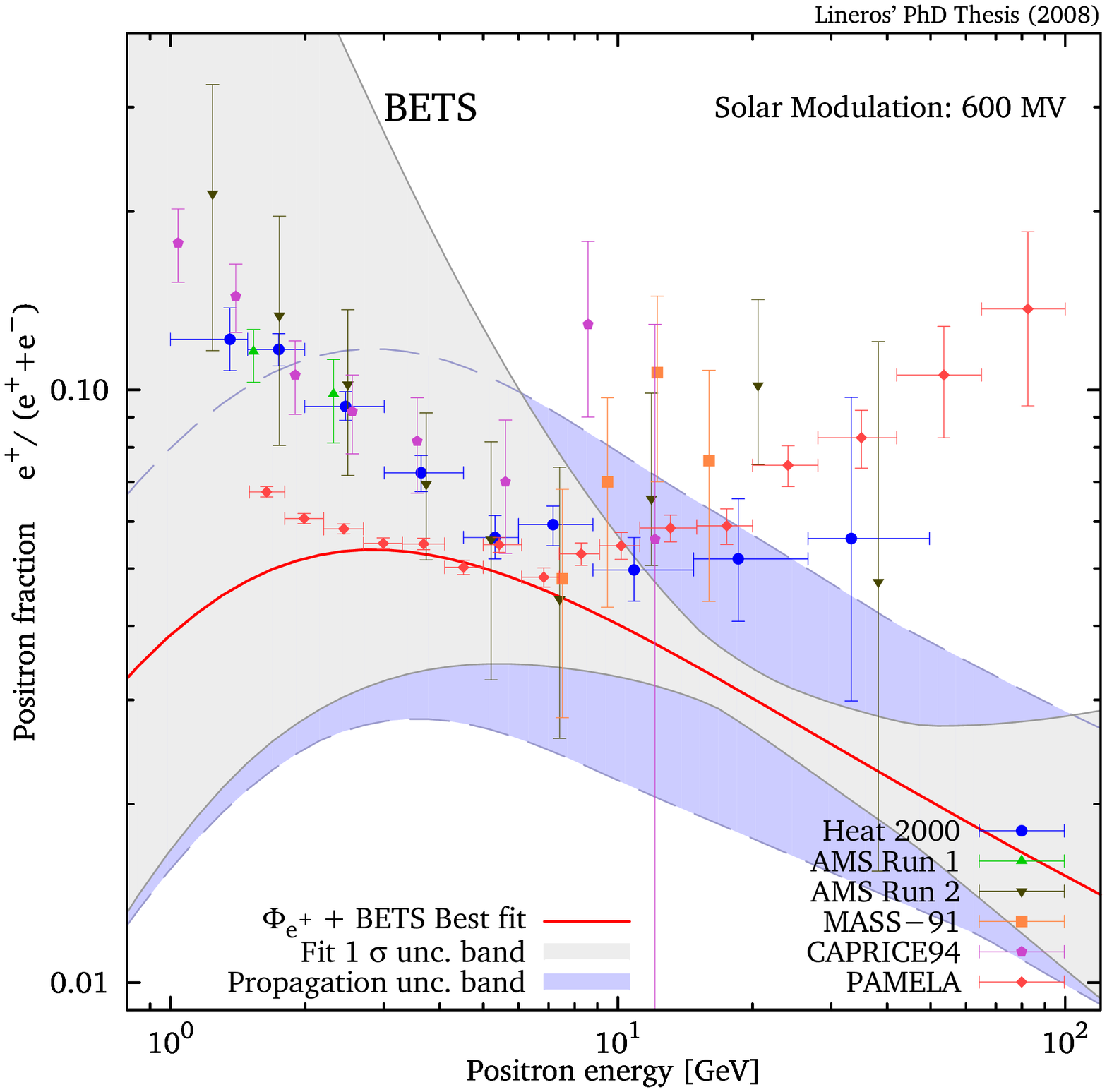}}
 \caption{\label{f:fit-exp-comp}Electron flux $\ener{}^3 \fluxe$ (left) and positron fraction (right) versus energy. Each of the positron fraction has been calculated by using the specific data fit of HEAT~\cite{Barwick:1997ig}, AMS~\cite{Aguilar:2007} and BETS~\cite{Torii:2001ApJ}, which are shown in the left column. The positron fractions were calculated using the positron flux for the MED set -- red line -- (See \citetab{tab:model}). The uncertainties  related to the propagation (darker blue) and the 1$\sigma$ uncertainty (lighter blue) band for each fit are included. Experimental data were taken from CAPRICE~\cite{Boezio:2000}, HEAT~\cite{Barwick:1997ig}, AMS~\cite{Aguilar:2007}, MASS~\cite{Grimani:2002yz}, PAMELA~\cite{Adriani:2008zr} and ATIC-2~\cite{Chang:2008zz}}
\end{fig}
%
% 
% 

% WRITE A KIND OF SUMMARY AND CONCLUSION.
The positron excess problem, that was noted as a deviation from Strong and Moskalenko predictions~\cite{Moskalenko:1997gh}, has attracted the attention of the scientific community. Most of the possible solutions invoke the existence of sources like dark matter annihilation, primordial black holes and nearby pulsars.
% The possible sources that may explain the positron excess 
Nevertheless, the positron excess might not be exclusively related to an unknown contribution to the positron signal. This can be related to the electron signal as well. We showed how different data sets from some single experiments may sizeably modify the predictions for the positron fraction. This does not discard the contribution of new sources: on the contrary, this motivates to study deeper and consistent the full picture, including the theoretical estimation of the electron flux.\\

\cleardoublepage
\chapter{Conclusions}
\label{conc}

% general aims.
In this thesis, we studied different aspects of positron cosmic--ray physics. 
The main aims were the study of signals related to galactic dark matter annihilation and secondary positrons.
The propagation of positrons were studied, as well as the effect of uncertainties related to the cosmic--ray propagation. 
We described and estimated the impact of uncertainties on the positron flux.\\

% chapter 2 (small)
In the galactic enviroment, the positron production takes place in many different ways. We described and studied two of the most important processes (\citecha{cha2}). The first one is the production of positrons and electrons in nuclear spallations in which positrons and electrons are mainly produced through the decay chains of charged pions and kaons.
The second mechanism of production is related to dark matter annihiliation. We studied and characterized positron and electron energy spectra obtained from neutral initial states. The method consists in the production of particle--antiparticle pairs in which the energy spectra are generated from the simulation of decay chains and hadronization process based on Lund's PYTHIA Montecarlo~\cite{Sjostrand:2000wi}. For the case of more complex intital states, we calculated them in terms of former spectra using the rules from the Standard Model and the Two Higgs Doublet Model.\\

% chapter 3 (small)
The propagation of positrons, and in general of any type of cosmic--rays, was treated according to the Two--Zone Propagation Model~\cite{Maurin:2001sj}. We solved the positron--electron transport equation using the Green function method and we defined the concept of Halo--Function, which is the response of the propagation domain to the distribution of sources. We calculated the halo functions for three dark matter distributions - cored isothermal~\cite{Bahcall:1980fb}, NFW~\cite{Navarro:1996gj} and Moore's~\cite{Diemand:2004wh} profiles - and for the source distribution of secondary positron production.
% go deeper into the halo--function
The halo--function concept naturally includes a definition of the diffusion lenght $\lD$ (\citeeq{e:def-ld}), which relates the propagation distances with energy evolution of cosmic--rays.
When energy losses related to inverse Compton scattering and synchroton radiation dominate the transport equation, the halo--function depends exclusively of $\lD$ and $L$. This fact improves the speed in the calculations of the positron signal, as well, it makes accesible to realize scans over the propagation parameters.\\

% geometry and boundaries
The geometry and boundaries of the propagation zone are important and affect the solutions of the transport equation. The Two--Zone Propagation Model considers a propagation zone described by a cylinder with radius 20~\tu{kpc} and thickness $2L$, which describes the extension of galactic magnetic fields reponsables of cosmic--ray diffusion.
We calculated Green functions for cartesian and cylindrical coordinates with and without boundary conditions. We found that the presence of boundaries speeds up the leaking of particles from the propagation zone, that is reflected in the halo function vanishing when $\lD$ is larger than the distance to the closest boundary.\\
%

% uncertainties B/C
The propagation uncertainties on the halo function have been calculated for the $\sim$ 1,600 different cosmic ray propagation models that have been found compatible \cite{Maurin:2001sj} with the B/C measurements. These uncertainties are strongly dependent on the source and detection energies, $\ener{s}$ and $\ener{}$.  As $\ener{s}$ gets close to $\ener{}$, we observe that each uncertainty domain shrinks. In that regime, the diffusion length $\lD$ is very small and the positron horizon probes only the solar neighborhood. In the opposite case, the uncertainty can be as large as one order of magnitude or even more.  As positrons originate further from the Earth, the details of galactic propagation become more important in the determination of the positron flux. On the contrary, high--energy positrons are produced locally and the halo function~$\tilde I$ becomes unity whatever the astrophysical parameters.\\

% chapter 4 (long)

In the case of dark matter annihilation, we inspected specifically the positron fluxes for a 100~\tu{GeV} DM particle annihilating into four typical channels: direct production of positrons, $\tau^{+}\tau^{-}$, $W^{+}W^{-}$ and $b\bar{b}$ pair. Uncertainties due to propagation on the positron flux are one order of magnitude at 1~\tu{GeV} and a factor of two at 10~\tu{GeV} and above. We find an increasing uncertainty for harder source spectra, heavier DM and steeper profiles.\\

The comparison with current data shows that the possible HEAT excess is reproduced for DM annihilating mostly into gauge bosons or directly into a positron--electron pair, and the agreement is not limited by the astrophysical uncertainties. 
A boost factor of 10 is enough to obtain an excellent agreement between the measurements and the median flux, for a 100~\tu{GeV} DM particle. A smaller value would be required for a flux at the upper envelope of the uncertainty band.\\

% prediction for PAMELA
We have finally drawn prospects for two interesting 3--year flight space missions, like PAMELA, already in operation and which has recently released its first data~\cite{Adriani:2008zr}, and the future AMS-02. We reach the remarkable conclusion that not only will PAMELA have the capability to disentangle the signal from the background, but it will also distinguish among different astrophysical models, especially for hard spectra. For AMS-02 the possibility to disentangle the signal from the background is also clearly manifest. We also wish to remind that improved experimental results on cosmic ray nuclei, expecially on the B/C ratio, will be instrumental to improve the determination  of the parameters of the propagation models, and will therefore lead to sharper theoretical predictions. This in turn will lead to a more refined comparison with the experimental data on the positron flux. Moreover, a good determination of the unstable/stable nuclei abundances like the $^{10}$Be/$^{9}$Be ratio could shed some light on the local environment, which is certainly mostly relevant to the positrons.\\

% chapter 5 (long)
Furthermore, we studied the secondary positron signal and we have compared the various models available for the interstellar secondary positron production. It has been shown that more positrons are expected when the proton flux from Shikaze et al.~\cite{Shikaze:2006je} is used, as compared to the case proposed by Donato et al.~\cite{Donato:2001ms}. Moreover, for a given proton flux, the three positron production cross sections we have considered produce different result: below a few GeV, the parameterization of Badhwar et al.~\cite{Badhwar:1977zf} gives more positrons, whereas above a few GeV, the model of Tan and Ng~\cite{Tan:1984ha} predicts a larger positron production. At any energy, the parameterization of Kamae et al.~\cite{Kamae:2006bf} produces the smallest amount of positrons.\\

% Concerning the propagation of the positrons in the interstellar medium, we have used a Green function approach that led us to disregard convection and diffusive reacceleration: we have specifically included diffusion and energy losses due to inverse Compton scattering on cosmic microwave background photons and synchrotron radiation. Nevertheless, this analytical method allowed us to scan our $\sim 1600$ sets of propagation parameters compatible with boron to carbon ratio measurements, and therefore to attack the problem of determining astrophysical uncertainties on the positron flux predictions.

%
We showed that varying the diffusion parameters has not the same effect as for primary positrons \cite{Delahaye:2007fr}. For exotic positrons created in the Dark Matter halo, the thickness  of the slab $2L$ was the most relevant parameter because the increase of the diffusion zone implies the increase of the number of  sources, whereas for secondary positrons -- which are created in the galactic disk only -- the most relevant parameter is the diffusion constant $K_0$. Therefore we expect the sets of parameters that basically maximize the primary positron flux to minimize the flux of secondary positrons, and vice versa.\\

We also showed that, because of energy losses during propagation, most of the positrons detected at the Earth have been created in the nearby 2~\tu{kpc}: this is the reason why we could safely neglect the variation of proton flux in the galaxy.\\ 

Finally, and this is our most important result, our estimation of the positron flux is compatible with all available data. This does not mean that there is no exotic positron contribution, since we have not tried to fit the data with a single diffusion model. However this shows that one should be cautious before claiming that there is any excess in present data. As regards a possible excess in the positron fraction, we have also clearly shown that the electron flux actually plays a role which is as important as that of positrons. This might sound tautological because the positron fraction is no more than a ratio, but so many energy has been involved in this issue that we again strongly stress this simple fact.\\

The just released PAMELA data~\cite{Adriani:2008zr} show a clear rise of the positron fraction for energies above 10~\tu{GeV}. From our analysis, whose objective is an accurate determination of the positron flux, we derive the conclusion that the presence of an excess is clear in the case of a hard electron spectrum, while for a soft electron spectrum the rise of the positron fraction may be explained by the standard secondary production. Scanning over the various parameters at stake, we in fact find in general the PAMELA measurements in excess of what a pure secondary  component would yield. Nevertheless, if the electron spectrum is soft most of the PAMELA data points are aligned with our prediction~\cite{Delahaye:2008}. Notice that, also in that case, the two last energy bins feature an increase, but the experimental uncertainties are large there and a presence of an excess is, in this case, currently not statisticlly significant.\\

% final words
In this thesis, we have thus presented the methods and the practical tools to evaluate the primary positron fluxes in detailed propagation models. We have provided careful estimations of the underlying uncertainties and shown the extraordinary potentials of already running, or near to come, space detectors.\\

More insight in these issues will therefore require, from the theoretical side, a revised understanding also of the electron flux, including the determination of its uncertainties, and from the experimental side, the separate provision of the electron and positron fluxes, in order to better compare theoretical predictions with the data. In addition, the upcoming data on cosmic rays above 10~\tu{GeV} will allow us to reduce considerably the theoretical uncertainties on all cosmic ray fluxes and help us elucidate the experimental status of the so-called excess in the positron spectrum.\\

\cleardoublepage
\addcontentsline{toc}{chapter}{Financial Support}
\chapter*{Financial Support}
This thesis was possible with financial support from:\\
\begin{itemize}
	\item Comisi\'{o}n Nacional de Ciencia y Tecnolog\'{\i}a (CONICYT), Gobierno de Chile. Concurso BECAS-DOC-BIRF-2005-00. Programa de becas de postgrado - Becas de doctorado al extranjero con convenio BIRF.\\
	\item Universit{\`a} degli Studi di Torino.\\
	\item Istituto Nazionale di Fisica Nucleare (INFN), Sezione di Torino.\\
	\item International Doctorate on AstroParticle Physics (IDAPP) program.\\
\end{itemize}
\pagestyle{fancy}
\fancyhf{}
\fancyhead{}
\fancyfoot[RO,LE]{\thepage}
\fancyhead[RO,LE]{\bfseries FINANCIAL SUPPORT}

\cleardoublepage
\addcontentsline{toc}{chapter}{Acknowledgements}
\chapter*{Acknowledgements}
This work has been made possible by the support of many people. It is my pleasure to acknowledge their contributions at this point.\\

First of all, I would like to acknowledge my advisor Professor Nicolao Fornengo, Fiorenza Donato and the members of the \emph{Astroparticle and Neutrino Physics Group} for all the support that they have gave me during these  three years. 
As well, I would like to acknowledge Professor Pierre Salati and the group in Annecy for all the fruitful work and ideas, in spite of the short period that I could stay there.\\

I very much appreciated the support of my family. They have been fundamental for me, they have trusted in me, gave me energy, inspiration and great moments.\\

Of course, I want to thank to all my comrades and friends: Mario, Dario, Hamed, Georgia, Victoria, Karla, Fabio, Alessandro, Carla, Riccardo, Jan, Maria Pilar,  Diogo, Mariana, Antonio, Michael, Lucia, Juan Carlos, David, St\`ephanie, among many others. 
They have always shared with me stories, experiences and fun. They taught me a lot. \\[4ex]

Thank you very much!\\

\newpage
\subsection*{The Anti--Acknowledgements}
Maybe, this type of section is uncommonly included in a thesis.
However, I would like to remark the displeasure situation that foreigners from outside of European Community, students and researchers, have to cope in Italy. \\

This situation regards to extreme--long delays in the release of \textbf{permits of stay}. 
These permits allow to foreigners to stay and transit legally in Italy and in the SCHENGEN countries.\\

During my stay in Italy, the delays in the release of them have been more than 12 months per year,
which sound ridiculous if we consider that permits of stay are valid for periods no longer than one year.\\

This situation is the result of immigration laws and politics, which are currently in effect and make almost impossible to do research and collaborations with other European universities and research centers.\\

The law in discussion is referred as \textbf{REGOLAMENTO (CE) N. 1030/2002 DEL CONSIGLIO del 13 giugno 2002}, which is intended to homogenize procedures to obtain permits of stay.\\
\pagestyle{fancy}
\fancyhf{}
\fancyhead{}
\fancyfoot[RO,LE]{\thepage}
\fancyhead[RO,LE]{\bfseries ACKNOWLEDGEMENTS}

\cleardoublepage
\appendix

\pagestyle{fancy}
\fancyhf{}
\fancyhead{}
\fancyfoot[RO,LE]{\thepage}
\fancyhead[LE]{\bfseries\leftmark}
\fancyhead[RO]{\bfseries\rightmark}
 
\begin{appendices}
\chapter{The boosted spectra method}
\label{app1}
\begin{prechap}
Particle distribution functions  are easier to calculate in the center of mass frame. However, this situation is not the most common one. Doing some kinematical analysis, the distribution can be rewritten for different reference frames in terms of new kinematical variables.\\
\end{prechap}

The first aim of this method is to transform the energy (spatial) distribution function from a specific reference frame into another. In \citecha{cha2}, it is used for calculating the positron (electron) energy spectra in the case of dark matter annihilation processes.\\

Let us start taking an angular and energy dependent spectrum,
% The aim of this method is to convert energy spectra among different reference frames. For a case of angular and energy dependent spectrum, 
\begin{eq}
 f(\ener{},\cos{\theta}) = \frac{d^2 \rho}{d\ener{} d(\cos{\theta})} \; .
\end{eq}
Under change of frames, the spectrum shape would change although it keeps constant its integral.\\

The energy Lorentz transformation between two frames is:
\begin{eq}
 \ener{}' = \gamma (\ener{} - \beta \; \mom{} \cos{\theta}) \; ,
\end{eq}
which gives restrictions on how energies will redistribute. All physical configuration, in the prime frame, are considered when the spectrum is integrated angular and in energy,
% At this point, to get the energy spectrum in the prime frame, it is needed to consider all physical configuration. That is performed by integrating on energy and angle:
\begin{eq}
 f'(\ener{}') = \int d\ener{} \ d(\cos{\theta}) \ f(\ener{},\cos{\theta}) \ \delta(r(\ener{}';\ener{},\cos\theta)) \; ,
\end{eq}
where $r$ is a restriction obtained from Lorentz transformations. Angular integration gives:
% After angular integration, it is obtained that:
\begin{eq}
 f'(\ener{}') = \frac{1}{\gamma \beta} \int \frac{d\ener{}}{\mom{}} \ f(\ener{},\cos\theta^{*}) \; ,
\end{eq}
where,
\begin{equation}
 \cos\theta^{*} = \frac{\gamma \ener{} - \ener{}'}{\gamma \beta \; \mom{}}
\end{equation}
is the angle respect to the boost direction, which links energies between frames. A simple case is an isotropic spectrum, where the boosted spectrum is:
\begin{eq}
 f'_{\tn{iso}}(\ener{}') = \frac{1}{2 \gamma \beta} \int \frac{d\ener{}}{\mom{}} \ f_{\tn{iso}}(\ener{}) \; .
\end{eq}\newline

A very important point is about the integration limits on the original frame energy. Those limits come from inverse Lorentz transformations and from the maximum energy ($\ener{\max}$) allowed in the original frame (\citefig{app1:f1}). Then the range on energy in the original frame, which contributes to the spectrum at a fixed energy $\ener{}'$ is: 

\begin{eq}
 \epsilon_{\max} &=& \min\Big(\ener{\tn{max}}, \gamma(\ener{}' + \beta \; \mom{}')\Big) \\
 \epsilon_{\min} &=& \max\Big(m, \gamma(\ener{}' - \beta \; \mom{}')\Big) \nonumber
\end{eq}

\begin{fig}
\includegraphics[width=0.5\textwidth]{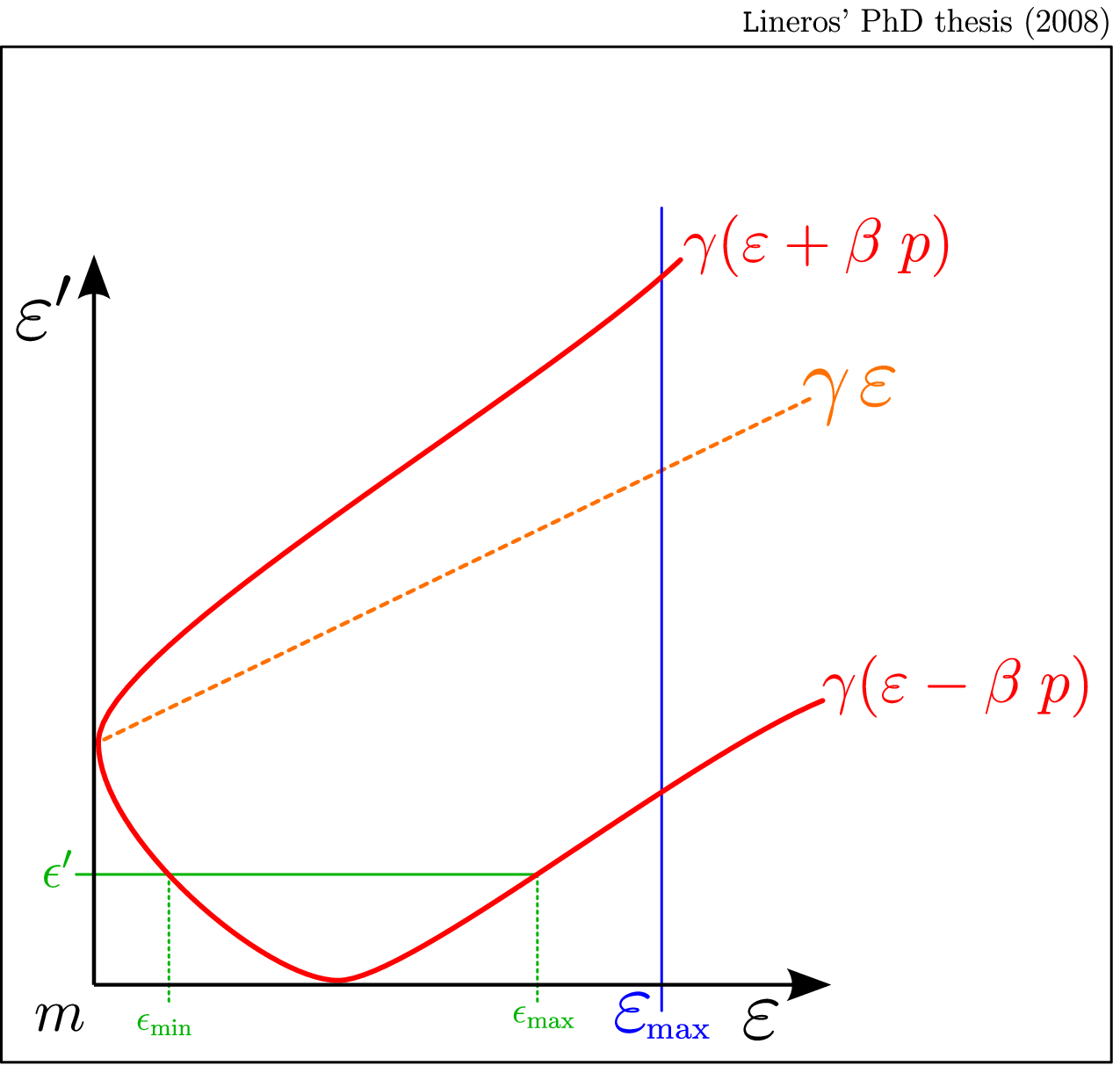}
\caption{\label{app1:f1}Schematic graphics $E^{\prime}$ vs $E$. It shows how to determine integration bounds $\epsilon_{\max,\min}$ for a fixed energy $\epsilon^{\prime}$ (in green). Maximum and minimum energies for the prime frame in terms of original frame energy (in red) are also shown.}
\end{fig}

The maximum energy in the prime frame, 
\begin{eq}
 \ener{\max}' = \gamma( \ener{\max} + \beta \; \mom{\max}) \ ,
\end{eq}
is obtained by transforming $\ener{\max}$. That provides a complete description for transforming any kind of spectrum.\\ 

\cleardoublepage

\chapter{Polarized Muon decay}
\label{app2}
\begin{prechap}
The Standard Model and Fermi theory give a precise description of muon decays. A common situation considers unpolarized muons decaying into electrons and neutrinos. However, there are processes where muons are produced fully polarized, as a consequence of left-right asymmetry present in electroweak interactions.\\
\end{prechap}

A production mechanism of electrons and positron is possible through the muon decay. Muons decay emitting two neutrinos and a positron or an electron (depending of muon's electrical charge). Usually muons are produced unpolarized, however when those come from a spin--0 state, they become fully polarized. That process modify the final electron or positron energy spectra. The polarized muon state induce an angular distribution respect to muon's propagation direction. \\

Positrons and electron produced in nuclear spallations are more affected by the muon's polarization effect, because the amount of charged pions and charged kaons produced enhance this effect and produce deviations from the unpolarized case.\\

In the muon rest frame, the electron energy and angular spectrum is \cite{PDBook}:
\begin{eq}
\frac{d^2 n}{dx dy} = f(x,y) = x^2 \big( 3-2x \pm y(2x -1) \big) = f_{I}(x,y) \pm f_{A}(x,y) \; ,
\end{eq}
where $x = 2\ener{}/\mass{\mu}$ and $y = \cos{\theta}$. Those are related with the electron energy and angle respect to the muon polarization vector.\\

It is important to remark that this spectrum is normalized to unity, 
\begin{eq}
 \int_{0}^{1} dx \int_{-1}^{1} dy \ f(x,y) = 1 \; ,
\end{eq}
in other words, it can be directly related to a probability density.\\

A property present in the muon rest frame is that the anisotropic term cancels when angular integration is performed,
\begin{eq}
 f(x) = \int_{-1}^{1} dy \ f(x,y) = 2x^2(3-2x) \; .
\end{eq}
% In this frame, the energy spectrum corresponds to unpolarized -- isotropic -- situation.\\
 
When muons are in motion, angular distribution becomes important and we need to consider a general approach. To solve that we used the boosted spectrum formulae (\citeapp{app1}). \\

% A general situation happends when muons are moving. For that, boosted spectrum formulae are needed (\citeapp{app1}). \\
In this context, the boosted energy spectrum is:
\begin{eq}
 f'(x') = \int_{x_{\min}}^{x_{\max}} dx \frac{f(x,y^{*})}{\gamma\beta x} \; ,
\end{eq}
where  
\begin{eq}
 y^{*} = \frac{\gamma x - x'}{\gamma \beta x} 
\end{eq}
is related with physical angle which gives the correct energy $x'$.\\

The integration limits are obtained from inverse Lorentz transformation:
\begin{eq}
 x_{\min} = \frac{x'}{x'_{\tn{sup}}} \quad ; \quad x_{\max} = \left\{ 
 \begin{array}{cr}
 x'/x'_{\tn{ch}} & 0 \le x' \le x'_{\tn{ch}} \\ \\
 1 &  x'_{\tn{ch}} \le x' \le x'_{\tn{sup}} \end{array}\right.
\end{eq}
where 
\begin{eq}
x'_{\tn{sup}} = \gamma(1+\beta) \quad\tn{and}\quad x'_{\tn{ch}} = \gamma(1-\beta) \; .
\end{eq}

For simplicity, electron boosted spectrum can be divided into an isotropic and an anisotropic term,  
\begin{eq}
 f'(x') = f_{I}'(x') \pm f'_{A}(x') \; .
\end{eq}\newline 

After the integration on $x$, the isotropic part becomes:
\begin{eq}
 f_{I}'(x') = \frac{2}{3} x'^2 \gamma [9 - 2 x' \gamma (3 + \beta^2)] \; ,
\end{eq}
when $ 0 \le x' \le x'_{\tn{ch}}$, and
\begin{eq}
f_{I}'(x') = \frac{5 + x'^2 x'^2_{\tn{ch}} [4x'x'_{\tn{ch}}-9]}{6\beta\gamma} \; ,
\end{eq}
when $x'_{\tn{ch}} \le x' \le x'_{\tn{sup}}$.\\

Also the anisotropic part is splitted into two expressions:
\begin{eq}
 f_{A}'(x') = \frac{2}{3} x'^2 \gamma\beta(8x'\gamma -3)
\end{eq}
and 
\begin{eq}
 f_{A}'(x') = \frac{1+x'^2 x'^2_{\tn{ch}}
[2 x' x'_{\tn{ch}}(1+3\beta) - 3(1 + 2\beta)]}{6 \beta^2 \gamma} \; ,
\end{eq}
for same the ranges than the isotropic case.\\

Let us clarify that we work in terms of dimensionless quantities. In order to get the energy spectrum in proper units, we need to do:
\begin{eq}
 \frac{d n}{d \ener{}} = \frac{2}{m_{\mu}}f\big(2\ener{}/m_{\mu}\big)
\end{eq}
which has units of [$\ener{}^{-1}$].\\

%%%%%%%%%%%%%%%%%%%%%%%%%%%%%%%%%%%%%%%%%%%%%%%%%%%%%%%%%%%%%%%%%%%%%%%%%%%%%%%%%%%%%%%%%%%%%%%%%%%%%%%%%
\section{Accumulated probability}
The accumulated probability is a useful tool to be used in Montecarlo programs because it can be used for event generation. By definition, 
\begin{eq}
 F(x') = \int_0^{x'} f(t') dt' \; ,
\end{eq}
gives a monotonic function with values which go from 0 to 1. 
Computing the accumulated probability for the isotropic and the anisotropic terms, we obtain that: 
\begin{eq}
 F_I(x') &=& \frac{1}{3}x'^3 \gamma[6 -x'(3+\beta^2)\gamma] \; ,\\ \nonumber \\
 F_A(x') &=& \frac{2}{3} x'^3 \gamma \beta[2x'\gamma -1] \; ,
\end{eq}
for $0 \le x' \le x'_{ch}$. And with values at $x'_{\tn{ch}}$:
\begin{eq}
 F_I(x'_{\tn{ch}}) = \frac{3 +\beta[3 + \beta(\beta-7)]}{3(1+\beta)^3} \quad \tn{and} \quad
F_A(x'_{\tn{ch}}) = \frac{2\beta(1-\beta)^2}{3(1+\beta)^3} \; .
\end{eq}\newline

Distribution function for $x'$ between $x'_{\tn{ch}}$ and $x'_{\tn{sup}}$ are:
\begin{eq}
 F_I(x') = F_I(x'_{\tn{ch}}) + \frac{5(x'-x'_{\tn{ch}}) + x'^2_{\tn{ch}}[
x'_{\tn{ch}}(x'^4-x'^4_{\tn{ch}}) - 3(x'^3-x'^3_{\tn{ch}})]}{6 \gamma \beta}
\end{eq}
\begin{eq}
 F_A(x') = F_A(x'_{\tn{ch}}) + \frac{2(x'-x'_{\tn{ch}}) + x'^2_{ch}[x'_{\tn{ch}}(1+3\beta)(x'^4-x'^4_{\tn{ch}}) - 2(1+2\beta)(x'^3-x'^3_{\tn{ch}})]}{12\gamma\beta^2}
\end{eq}
To produce electron (positron) energy spectrum in a Montecarlo, it is necessary to invert these function. This can be obtained by using numerical invertion methods.\\

\cleardoublepage

\chapter{Electrons and Positrons energy spectra from Pion decay}
\label{app3}
\begin{prechap}
The principal source of fully polarized muons are the charged pions. Those are spin--0 particles which are abundantly produced in nuclear scattering processes.\\
\end{prechap}

\begin{fig}
 \includegraphics[width=0.5\textwidth]{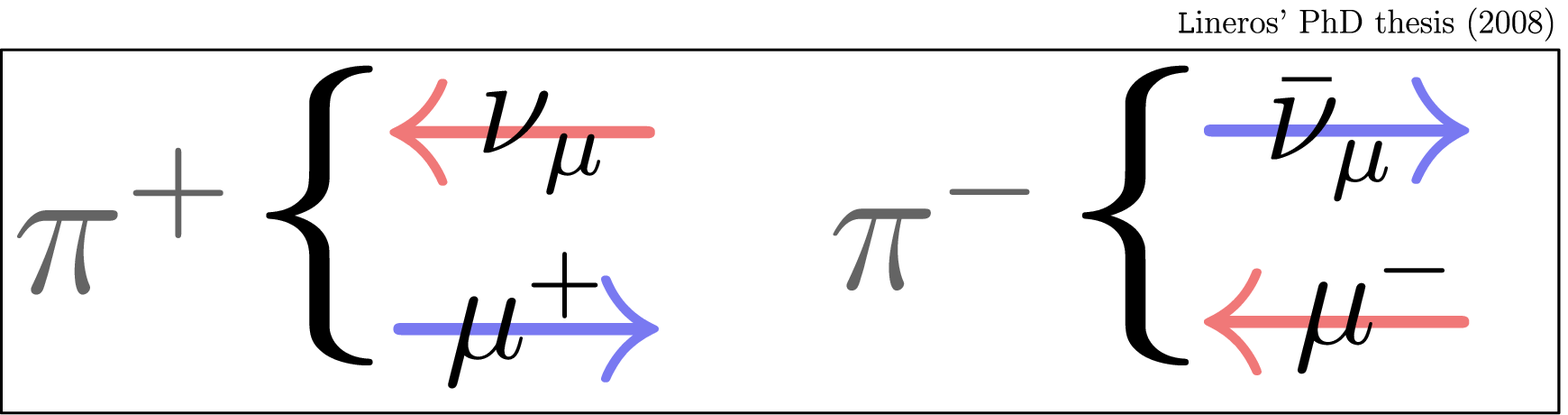}
 \caption{\label{app3:f1}Schematic plot to illustrate how the charged pions decay produces polarized muons. Blue arrow represent left-handed polarization and red arrows right-handed polarized particles}
\end{fig}

Charged pions decay mainly into muons with a probability of $99.98770 \pm 0.00004 \%$ \cite{PDBook}. Nevertheless, muons are produced fully polarized due to pions are spin--0 particles (See \citefig{app3:f1}). \\

Depending on the pion charge, electrons or positrons are more or less energetic with respect to unpolarized situation. This phenomenon should be taken into account in order to understand possible asymmetries present in this type of CR.\\

% As charged pion decays into two bodies modes, all produced muons are monoenergetic at pion rest frame. In this situation, muon energy spectrum will be:

In the pion's rest frame, all muons come from a two--body decay mode. In this frame, they are monoenergetic, and then, the energy spectrum associated is:
\begin{eq}
 \frac{d n}{d \ener{\mu}} = \delta\big( \ener{\mu} - \ener{}^{*}\big) = f_{\mu,\pi}\big(\ener{\mu}, \mass{\pi}\big) \ ,
\end{eq}
where $\ener{}^{*}$ is the muon energy in pion rest frame and its value is obtained from energy-momenta conservation, 
\begin{eq}\label{e:muon-ener}
 \mom{}^{*} = \frac{\mass{\pi}^2 - \mass{\mu}^2}{2 \mass{\pi}}  \quad ; \quad \ener{}^{*} = \sqrt{\mom{}^{*2} + m^2_{\mu}} \; .
\end{eq}\newline

The electron spectrum can be obtained as a convolution between pion decay -- into muons ($f_{\mu,\pi}$) -- and muon decay -- into electron ($f_{e,\mu}$),
\begin{eq}
 f_{e,\pi}\big(\ener{e}, \ener{\pi}\big) = \int d \ener{\mu} \ f_{\mu,\pi}\big(\ener{\mu}, \ener{\pi}\big) \  f_{e,\mu}\big(\ener{e}, \ener{\mu}\big) \ .
\end{eq} \newline

Muons are monoenergetically produced when pions are at rest (\citeeq{e:muon-ener}). In this limit the electron spectrum is simply:
\begin{eq}
f_{e,\pi}\big(\ener{e}, \mass{\pi}\big) =  f_{e,\mu}\big(\ener{e}, \ener{\mu}^{*}\big) \; .
\end{eq}
When pions are in motion, the boosted spectra method are applied to the muon spectrum . In this situation, the boost factor depends on pion energy $\gamma_{\pi} = \ener{\pi} / \mass{\pi}$, and then, the muon spectrum is:
\begin{eq}
 f_{\mu,\pi}\big(\ener{\mu}, \ener{\pi}\big) = \left\{ 
 \begin{array}{cc}
\displaystyle \frac{1}{2 \gamma_{\pi} \beta_{\pi} \ \mom{}^{*}} & \ener{\mu}^{\min} < \ener{\mu} < \ener{\mu}^{\max} \\ \\
0 & \tn{other cases}
 \end{array} \right. \; ,
\end{eq}
where energy bounds are got from boosting $\ener{}^{*}$,
\begin{eq}
\ener{\mu}^{\max} = \gamma_{\pi} ( \ener{}^{*} + \beta_{\pi} \ \mom{}^{*}) \quad \tn{and} \quad  \ener{\mu}^{\min} = \gamma_{\pi} ( \ener{}^{*} - \beta_{\pi} \ \mom{}^{*}) \; .
\end{eq}\newline

Finally, the electron spectrum is calculated through convolution of partial spectra:
\begin{eq}
 f_{e,\pi}\big(\ener{e}, \ener{\pi}\big) = \frac{1}{2 \gamma_{\pi} \beta_{\pi} \ \mom{}^{*}} \int_{\ener{\mu}^{\min}}^{\ener{\mu}^{\max}} d \ener{\mu} \  f_{e,\mu}\big(\ener{e}, \ener{\mu}\big) \; ,
\end{eq}
where the information on muon's polarization is inside $f_{e,\mu}$.\\ 

The maximum electron energy for a pion decay can be calculated from composition of Lorentz transformations from muon and pion rest frames, 
\begin{eq}
 \ener{e}^{\max} = \gamma_{\pi} (1+\beta_{\pi}) \; \gamma_{\mu}^{\max} (1+\beta_{\mu}^{\max}) \frac{\mass{\mu}}{2} \; .
\end{eq}

\section{Analytic formulae for polarized muons}
The electron spectrum from pion decay depends on the integral on muon energy. As it was seen in \citeapp{app2}, electron spectrum from polarized muons can be described by composition of isotropic and anisotropic terms,
\begin{eq}
 f_{e,\mu}\big(\ener{e},\ener{\mu}\big) = \frac{2}{\mass{\mu}} \left(f'_I(x') \pm f'_A(x')\right) \; ,
\end{eq}
where $x' = 2 \ener{e} / \mass{\mu}$ and prime denotes that those were calculated for a moving muon with boost factor of $\gamma = \ener{\mu}/\mass{\mu}$.\\ 

The integration on muon energy can be transformed in terms of the muon's $\gamma$ factor, 
\begin{eq}
 \int_{\ener{\mu}^{\min}}^{\ener{}^{\max}} d \ener{\mu} \  f_{e,\mu}\big(\ener{e}, \ener{\mu}\big) = 2 \int_{\gamma_{\min}}^{\gamma_{\max}} \ d\gamma \ \left(f'_I(x') \pm f'_A(x')\right) \; ,
\end{eq}
and then work in terms of dimensionless quantities.\\
 
An important point is about different behaviors present on $f'$. To work in a abstract way, it has been renamed the formulae for muon decay (showed in \citeapp{app2}) for each range of $x'$,
\begin{eq}
 0 \le x' < x'_{\tn{ch}} \Rightarrow  f'_{1}(x')  \quad ; \quad x'_{\tn{ch}} \le x' \le x'_{\tn{sup}} \Rightarrow f'_{2}(x') \ .
\end{eq}

When $\gamma$ varies, $x'_{\tn{ch}}$ and $x'_{\tn{sup}}$ vary too. Then
\begin{eq}
 \gamma^{*} = \frac{x'^2 +1}{2 x'}
\end{eq}
marks the critical point where $f'$ changes behavior. Depending on $x'$ value, the integral just considers one or both functions (\citefig{app3:f2}).\\

% FIGURE
% 
% 
\begin{fig}
 \includegraphics[width=0.5\textwidth]{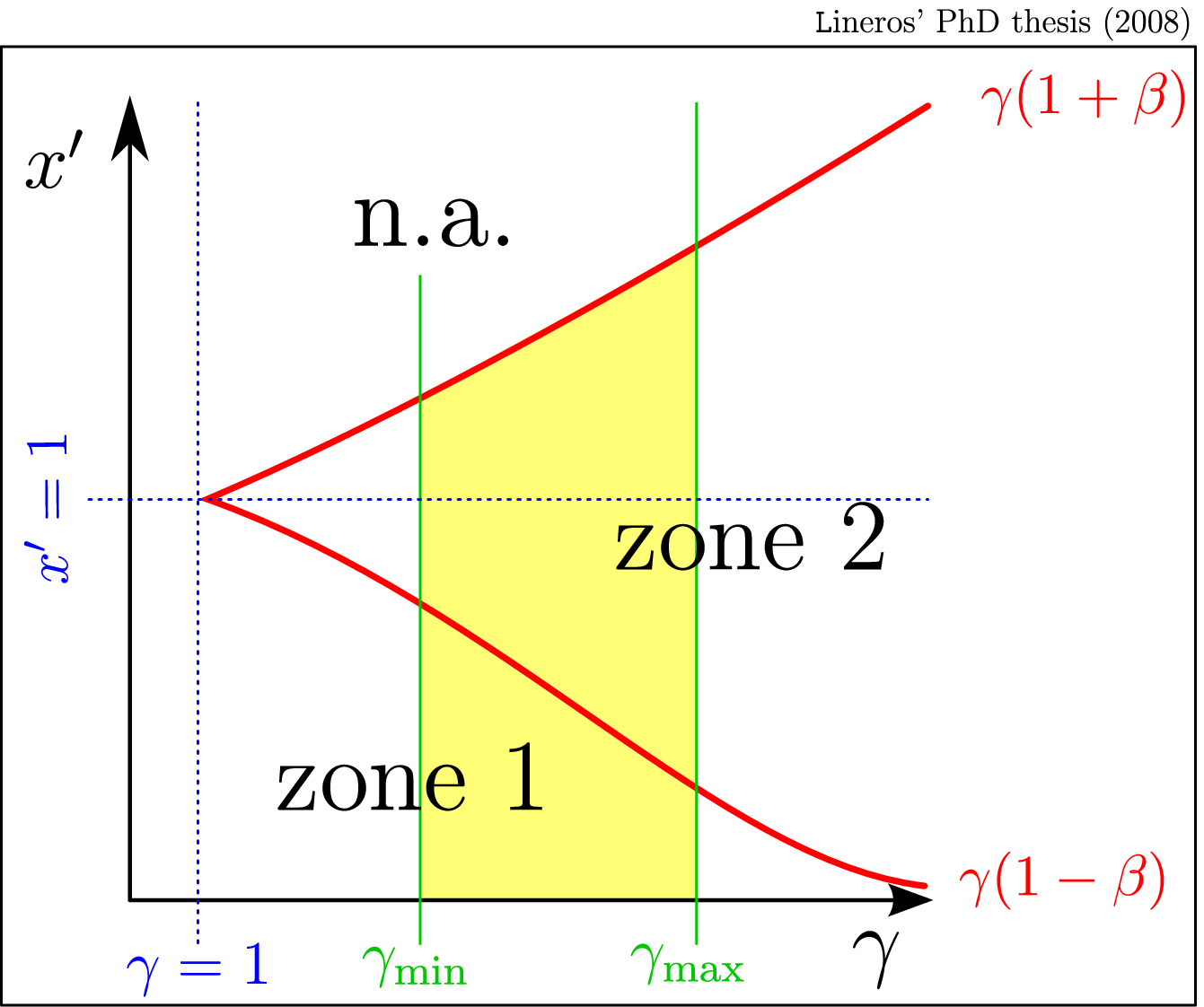}
 \caption{\label{app3:f2}Schematic plot $x'$ versus $\gamma$. Integral limits on muon's $\gamma$ and frontiers between zones (red line), where spectrum changes behavior, are shown.}
\end{fig}

For $x' \le 1$ and $\gamma_{\min} \le \gamma^{*} \le \gamma_{\max}$:
\begin{eq}
 \int d\gamma \ f' = \int_{\gamma_{\min}}^{\gamma^{*}} d\gamma \ f'_1 \ + \  \int^{\gamma_{\max}}_{\gamma^{*}} d\gamma \ f'_2
\end{eq}
Otherwise, it is the integral which corresponds to the proper sector.\\

For $x' > 1$:
\begin{eq}
 \int d\gamma \ f' = \int^{\gamma_{\max}}_{\max(\gamma^{*},\gamma_{\min})} d\gamma \ f'_2
\end{eq}
where there is contribution from $f'_2$ because region below $\gamma^{*}$ are not allowed.\\

Performing the integral on the muon's $\gamma$ factor over isotropic and anisotropic terms, a set of indefinite integrals are obtained. Where they should be evaluated in the proper integration limits.\\

The integration of isotropic term gives,
\begin{eq}
F_{I1}(x,\gamma) &=& \frac{x^2 \gamma}{9} \Big( 27 \gamma + 4x(3-4\gamma^2)\Big) \\
F_{I2}(x,\gamma) &=& \frac{1}{18} \Big(x^2 \{27 x'_{\tn{ch}} \gamma - 4 x \gamma (\beta -3 + 4 x'_{\tn{ch}} \gamma)\} + 15 \log(x'_{\tn{sup}}) \Big) \; ,
\end{eq}
and the anisotropic term is:
\begin{eq}
F_{A1}(x,\gamma) &=& \frac{x^2}{9} \Big( \gamma^2 \beta (16 x \gamma \beta^2 - 9) + 9 \log(x'_{\tn{sup}}) \Big) \\
F_{A2}(x,\gamma) &=& \frac{1}{36} \Big(2 x^2 \{9 x'_{\tn{ch}}\gamma - 8 x \gamma(2 x'_{\tn{ch}} + 2\beta-3) + 9 \log(x'_{\tn{sup}})\} \\
&&  + (3-9x^2) \log(\beta^2 \gamma^2) - 6 x^3 \log(\frac{\gamma+1}{\gamma-1})  \Big) \nonumber
\end{eq}

Finally, we have obtained all the ingredients to compute a realistic electron spectrum. Those are been used to replace muon decay subroutines in the Lund's PYTHIA montecarlo generator \cite{Sjostrand:2000wi}.\\

\cleardoublepage

\chapter{Electrons and Positrons energy spectra from Kaon decay}
\label{app4}
\begin{prechap}
The second in importance source of fully polarized muons are the charged kaons. Those are spin--0 particles which are abundantly produced in nuclear scattering processes, as charged pions.\\
\end{prechap}

The electron spectrum from charged Kaon decay can be more difficult to manage due to the various decay modes of Kaons (\citetab{app4:t1}). These modes are dominated by two body modes with a probability of $\sim 85 \%$. The remained modes are mainly composed by three body ones. As in pion decay, muons are produced fully polarized.\\

To calculate a complete electron spectrum, each decay mode can be treated separately and weighted according to its own probability, 
\begin{eq}
 f_{e,K} = \sum_{\tn{mode}} \tn{Br}(K\rightarrow \tn{mode}) \ f_{e,K}^{\tn{mode}} \; .
\end{eq}
In this way, each mode is independently studied.\\

% TABLE
% 
% 
\begin{tab}
 \begin{tabular}{|l|c|c|}
  \hline $K^{-}$ decay mode & Probability $\%$ & $\langle n_e \rangle$ produced \\
  \hline \hline $\mu \ \bar{\nu}_{\mu}$ & $63.44 \pm 0.14$ & 1 \\
  $\pi^{-} \pi^{0}$ & $20.92 \pm 0.12$ & 1 \\
  $\pi^{-} \pi^{-} \pi^{+}$ & $5.590 \pm 0.031$ & 2 \\
  $\pi^{0} \ e \ \bar{\nu}_{e}$ & $4.98 \pm 0.07$ & 1 \\
  $\pi^{0} \ \mu \ \bar{\nu}_{\mu}$ & $3.32 \pm 0.06$ & 1 \\
  $\pi^{-} \pi^{0} \pi^{0}$ & $1.757 \pm 0.024$ & 1 \\
  \hline Total & $100.0 \pm 0.445$& $1.06$\\ \hline
 \end{tabular}
 \caption{\label{app4:t1} Most important charge kaon modes, probability and number of electron associated to each of these \cite{PDBook}.}
\end{tab}

\section{Kaon two body modes}
Two body decay modes are easily described by analytical expressions. The electron (positron) spectrum, for isotropic two--body modes, is:
\begin{eq}
 f_{e,K}\big(\ener{e},\mass{K}\big) = f_{e,X}\big(\ener{e},\ener{X}^{*}\big) \; ,
\end{eq}
which is for Kaon at rest. On the other hand, for Kaons in motion with boost factor $\gamma = \ener{K}/\ener{K}$, the spectrum is:
 \begin{eq}
 f_{e,K}\big(\ener{e},\ener{K}\big) = \frac{1}{2 \gamma \beta \; \mom{X}^{*}} \int_{\ener{X}^{\min}}^{\ener{X}^{\max}} d\ener{X} \; f_{e,X}\big(\ener{e},\ener{X}\big) \; ,
\end{eq}
where particle $X$ refers muons or charged pions or any two--body mode. The integration limits are calculated from $\ener{X}^{*}$ boosted in the backward and forward direction respect to the boost direction, 
\begin{eq}
 \ener{X}^{\max} = \gamma \big(\ener{X}^{*} + \beta \; \mom{X}^{*} \big) \; , \; \ener{X}^{\min} = \gamma \big(\ener{X}^{*} - \beta \; \mom{X}^{*} \big) \; ,
\end{eq}
Where $\mom{}^{*}$ depends on which two--body modes is involved. For the mode \mbox{$K^{-} \rightarrow \mu \bar{\nu}_{\mu}$}, it is: 
\begin{eq}
\mom{\mu}^{*} = \frac{\mass{K}^2 - \mass{\mu}^2 }{2 \mass{K}}
\end{eq}
and for \mbox{$K^{-} \rightarrow \pi^0 \pi^{-}$} ,
\begin{eq}
\mom{\pi^{\pm}}^{*} = \frac{1}{2 \mass{K}}\sqrt{(\mass{K}^2 + \mass{\pi^{\pm}}^2 - \mass{\pi^0}^2)^2 - \mass{\pi^{\pm}}^2 \mass{K}^2} \ .
\end{eq}
Note that when Kaons decays into Pions, those modes are more complex than modes with muons. Due to that, it is needed to perform numerical integration.\\

\section{Kaon three body decay}
The next important contribution to electron spectrum are composed by three body modes. 

%------------------------------------------------------------------------------
\subsection{Mode $\pi^{0} \ e \ \bar{\nu}_{e}$}
This is a mode where energy spectra for positron and electron are equal. For an isotropic production, energy spectrum at Kaon rest frame is:
\begin{eq}
 f_{\pi e \nu}\big(\ener{e},\mass{K}\big) = a_0 \ener{e} \frac{E_{\max} - \ener{e}}{a_1 - \ener{e}} \; ,
\end{eq}
where $a_0 = 54.14 \ \tu{GeV}^{-2}$, $a_1 = 0.2468 \ \tu{GeV}$ and $E_{\max} = 0.2283 \ \tu{GeV}$.\\
 
The boosted version of this spectrum is:
\begin{eq}
 f_{\pi e \nu}\big(\ener{e},\ener{K}\big) = \frac{a_0}{2 \gamma \beta} \left(x_{\max} -x_{\min} + (a_1-E_{\max}) \log(\frac{x_{\max}-a_1}{x_{\min}-a_1}) \right) \ ,
\end{eq}
where $\gamma = \ener{K}/\mass{K}$ and
\begin{eq}
 x_{\min} &=& \frac{\ener{e}}{\gamma(1+\beta)} \\
 x_{\max} &=& \left\{ \begin{array}{ccl} \displaystyle \frac{\ener{e}}{\gamma(1-\beta)} & & 0 \le \ener{e} < \gamma(1-\beta) E_{\max} \\ \\
 E_{\max} & & \gamma(1-\beta) E_{\max} \le \ener{e} \le \gamma(1+\beta) E_{\max} \end{array} \right. \; .
\end{eq}
%---------------------------------------------------------------------------
\subsection{$\pi^{0} \ \mu \ \bar{\nu}_{\mu}$}
For this mode, it is needed to calculate the muon energy spectrum, where the maximum muon energy at kaon rest frame is:
\begin{eq}
 \ener{\mu,\max}^{*} = \frac{\ener{K}^2 + m_{\mu}^2 - m_{\pi^0}^2}{2 \ener{K}} = 0.2395 \ \tu{GeV} ,
\end{eq}
and the spectrum:
\begin{eq}
 f_{\pi \mu \nu, K}\big( \ener{\mu}, \ener{K} \big) = A \ p_{\mu} \frac{\left(\ener{\mu,\max}^{*} - \ener{\mu}\right)  }{a_0 - a_1 \ener{\mu} - \ener{\mu}^2} \; .
\end{eq}

\begin{eq}
 A &=& 1374.86 \ \tu{GeV}^{-1} \\
 a_0 &=& 0.5046 \ \tu{GeV}^{2} \\
 a_1 &=& 1 \ \tu{GeV}
\end{eq}

The boosted energy spectrum is:
\begin{eq}
 f_{\pi \mu \nu,K}\big( \varepsilon_{\mu}, \ener{K} \big) = \frac{A}{2 \gamma \beta} \ \big\{g_{\pi \mu \nu}(\epsilon_{\max}) - g_{\pi \mu \nu}(\epsilon_{\min}) \big\}
\end{eq}
where $g_{\pi \mu \nu}\big(\epsilon\big)$ is an auxiliary function defined:
\begin{eq}
 g_{\pi \mu \nu}\big(\epsilon\big) = \frac{1}{2} \log\left(a_0 - a_1 \epsilon - \epsilon^2\right) - \frac{a_1 + 2 \varepsilon_{\mu,\max}^{*}}{2\Delta_0} \log\left(\frac{\Delta_0 - a_1 - 2 \epsilon}{\Delta_0 + a_1 + 2 \epsilon}\right)
\end{eq}
where
\begin{eq}
 \Delta_0 = \sqrt{4 a_0 + a_1^2}
\end{eq}

In order to get the electron spectrum, it is needed to convolute this spectrum with the electron spectrum due to muon decay.

\begin{eq}
 f_{e,K}\big(\ener{e}, \ener{K}\big) =  \int_{\varepsilon_{\mu}^{\min}}^{\varepsilon_{\mu}^{\max}} d \varepsilon_{\mu} \ f_{\pi \mu \nu,K}\big(\varepsilon_{\mu},\ener{K}\big) \ f_{e,\mu}\big(\ener{e}, \varepsilon_{\mu}\big) \ ,
\end{eq}

In this case:
\begin{eq}
 \varepsilon_{\mu}^{\max} = \gamma \big( \varepsilon_{\mu,\max}^{*} + \beta \ p_{\mu,\max}^{*} \big) \quad ; \quad \varepsilon_{\mu}^{\min} = m_{\mu}
\end{eq}

%-----------------------------------------------------------------------------
\subsection{$\pi^{-} \pi^{-} \pi^{+}$}
This mode produces a couple of electrons and one positron and muon polarization effect are also present. As it is known, electron spectrum can be got from convolution between the pion spectrum from kaon decay and electron spectrum from pion decay (\citeapp{app3}),
\begin{eq}
 f_{e,K}\big(\ener{e}, \ener{K}\big) = 2 \int_{\ener{\pi}^{\min}}^{\ener{\pi}^{\max}} d \ener{\pi} \ f_{3 \pi,K}\big(\ener{\pi},\ener{K}\big) \ f_{e,\pi}\big(\ener{e}, \ener{\pi}\big) \ ,
\end{eq}
and it is convenient to work just on the 3 pions energy spectrum. Note that the factor 2 is related with the electron multiplicity. \\

At kaon rest frame, the maximum pion energy is:
\begin{eq}
 \varepsilon_{\max}^{*} = \frac{m^2_K - 3 m^2_{\pi}}{2 \ener{K}} = 0.18758 \ \tu{GeV}
\end{eq}
and the pion energy spectrum can be written as:
\begin{eq}
f_{3 \pi,K}\big(\ener{\pi},\ener{K}\big) = A \ p_{\pi} \left(\ener{\pi} - m_{\pi} \right)^{\delta}  \left(\varepsilon_{\max}^{*} - \ener{\pi} \right)^{\frac{1}{2} - \delta}
\end{eq}
where the parameters are:
\begin{eq}
 A = 2002.19 \ \tu{GeV}^{-2.5} \quad ; \quad \delta = 0.05
\end{eq}
for the actual measured masses.\\

The boosted version is obtained through:
\begin{eq}
 f_{3 \pi,K}\big(\ener{\pi},\ener{K}\big) = \frac{1}{2 \gamma \beta} \int_{\epsilon_{\min}}^{\epsilon_{\max}} \frac{d\epsilon}{p} \ f_{3\pi,K}\big(\epsilon, m_{K}\big) \ ,
\end{eq}
where 
\begin{eqnarray}
 \epsilon_{\min} &=& \gamma \left( \ener{\pi} - \beta \ p_{\pi}  \right) \\
 \epsilon_{\max} &=& \min\left( \ \varepsilon_{max}^{*}, \gamma ( \ener{\pi} + \beta \ p_{\pi}) \ \right)
\end{eqnarray}
and then, it can be written in a analytical form:
\begin{eq}
 f_{3 \pi,K}\big(\ener{\pi},\ener{K}\big) = \frac{A}{2 \gamma \beta} \  \left\{g_{3 \pi}\big(\epsilon_{\max}\big) - g_{3 \pi}\big(\epsilon_{\min}\big)\right\}
\end{eq}
by using the auxiliary function $g_{3\pi}$:
\begin{eq}
 g_{3 \pi}\big(\epsilon\big) = t(\epsilon) \sqrt{T_{max}} \left(\frac{t(\epsilon)}{T_{max}}\right)^{\delta} \ {}_2F_1 \left(\delta + 1; \delta - \frac{1}{2} ; \delta + 2; \frac{t(\epsilon)}{T_{max}} \right)
\end{eq}
where ${}_2F_1(a;b;c;z)$ is a hypergeometric function, $t(\epsilon)$ and $T_{\max}$ are the kinetic energy and the maximum kinetic energy for charged pions.\\

%--------------------------------------------------------------------------------
\subsection{$\pi^{-} \pi^{0} \pi^{0}$}
This mode is very similar to the previous seen. Kinematics restrictions are slightly different.\\
\begin{eq}
 \varepsilon^{*}_{\max} = \frac{m^2_K + m^2_{\pi^{\pm}}- 4 m^2_{\pi^{0}} }{2 \ener{K}} = 0.1926 \ \tu{GeV}
\end{eq}

The same parameterization is good enough to describe the energy spectrum at kaon rest frame:\\
\begin{eq}
f_{\pi^{\pm}2\pi^0,K}\big(\varepsilon_{\pi^{\pm}}, \ener{K} \big) = A \ p_{\pi^{\pm}} \ \left( \varepsilon_{\pi^{\pm}} - m_{\varepsilon_{\pi^{\pm}}} \right)^{\delta} \left(\varepsilon^{*}_{\max} - \varepsilon_{\pi^{\pm}}\right)^{\frac{1}{2} - \delta}
\end{eq}
where the parameters are:
\begin{eq}
 A = 1624.47 \ \tu{GeV}^{-2.5} \quad ; \quad \delta = 0.065
\end{eq}

And the boosted version is:
\begin{eq}
 f_{\pi^{\pm}2\pi^0,K}\big(\varepsilon_{\pi^{\pm}}, \ener{K} \big) = \frac{A}{2 \gamma\beta} \left\{ g_{3\pi}\big(\epsilon_{\max}\big) - g_{3\pi}\big(\epsilon_{\min}\big) \right\}
\end{eq}
where 
\begin{eqnarray}
 \epsilon_{\min} &=& \gamma \left( \ener{\pi} - \beta \ p_{\pi}  \right) \\
 \epsilon_{\max} &=& \min\left( \ \varepsilon_{max}^{*}, \gamma ( \ener{\pi} + \beta \ p_{\pi}) \ \right)
\end{eqnarray}
and $g_{3\pi}\big(\epsilon\big)$ is the same auxiliary function used for the mode with three charged pions.\\

\cleardoublepage

\chapter{Method of Averages: A method to produce accurate histograms}
\label{app5}
\begin{prechap}
The Method of Averages (\textbf{MoA}) is an algorithm which has been developed to estimate the proper binning rule to create a histogram. This method uses directly the data as an input and produces a binning rule where narrow bins are placed in high--density zones and wider bins in less dense zones.\\
\end{prechap}

%------------------------------------------------------------
\begin{tab}
	\begin{tabular}{|l|c|}
		\hline Model & number of bins \\
		\hline Sturges' formula \cite{Sturges:1926} & $log_2(n) + 1$ \\[2ex]
		Scott's choice \cite{Scott:1979} &  $\displaystyle \frac{x_{\max} - x_{\min}}{3.5 \ \sigma} \ n^{1/3}$\\[2ex]
		Freedman-Diaconis' choice \cite{Freedman:1981} & $\displaystyle \frac{x_{\max} - x_{\min}}{2 \ \tn{IQR}(x)} \ n^{1/3}$ \\[2ex]
		\hline
	\end{tabular}
	\caption{\label{app5:t1} Number of bins estimation formulas for Sturges, Scott and Freedman-Diaconis models \cite{Sturges:1926,Scott:1979,Freedman:1981}. For all models, $n$ is the number of data, $x_{\max/\min}$ are extreme values present, $\sigma$ is the standard deviation and $\tn{IQR}(x)$ is the interquartile range function.}
\end{tab}
%------------------------------------------------------------

There are many criteria to estimate the number of bins necessary to make a histogram (\citetab{app5:t1}). Nevertheless, in most of cases, there are big assumptions about how the data are distributed. As an example, Scott's choice \cite{Scott:1979} makes the assumption in which data follow a gaussian distribution, where bin widths are proportional to the standard deviation of the sample.\\

\section{Description}
The key in MoA is to analyze the sample in a simple way. For that, we classify extracted quantities in term of number of  passes, or times in which data were used. After reading the data once, the \emph{pass-1} quantities are the extreme values, $x_{\max/\min}$, and the average $\langle x \rangle_{p1}$. The \emph{pass-1} average gives information about the tendency point; for example, if the average is close to $x_{\min}$, that means data are roughly concentrated around the minimum. However, if the \emph{pass-1} average is close to the mid point between $x_{\max}$ and $x_{\min}$, we expect that data are spread in the whole range. Of course, the real information is inside the standard deviation, but that belongs to the \emph{pass-2} quantities group.\\

To get an accurate binning rule, we need to read data many times. To reduce the screening effect due to dense--data zones, before going to a next pass, we split the sample into subsets. Those are delimited by averages obtained in previous passes (\citefig{app5:f1}). For example, to compute \emph{pass-2} averages, we need to compute the averages of data in ranges $(x_{\min}, \langle x \rangle_{p1})$ and $(\langle x \rangle_{p1}, x_{\max})$.\\

\begin{fig}
	\includegraphics[angle=270,width=0.5\textwidth]{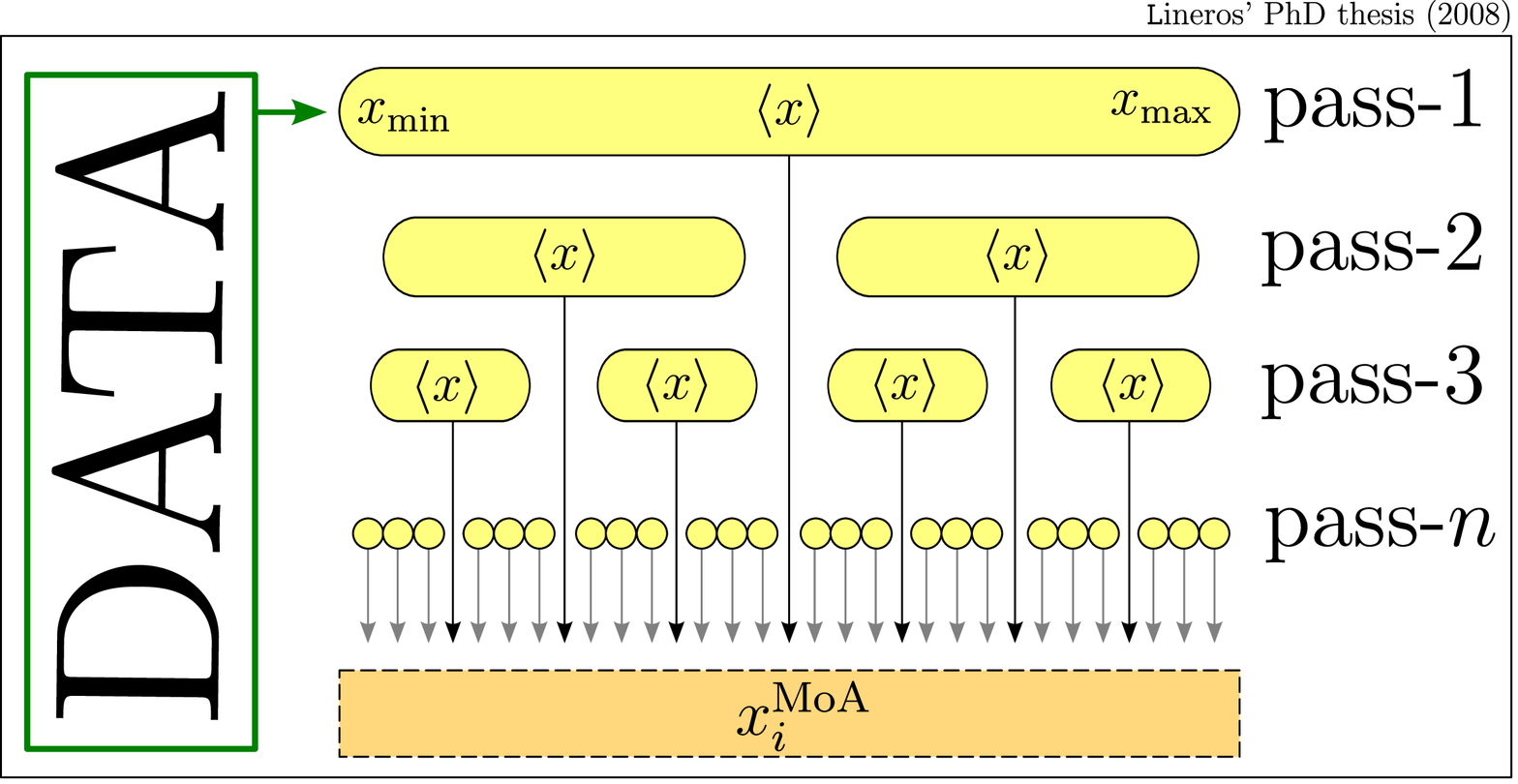}
	\caption{\label{app5:f1} Schematic view of creation of MoA--rule and how data splitting is performed.}
\end{fig}

At the end of $p$ passes, we obtain a rule composed by $n_{\tn{MoA}} = 2^p - 1$ bins. The choice of $p$ depends on the desiderate number of bins ($n_{\tn{bin}}$). It is recommendable to use a value of $p$ that satisfies:
\begin{equation}
n_{\tn{MoA}} < \frac{n_{\tn{bin}}}{2} \ ,
\end{equation}
to reduce the risk to skip one of the averages, which are very important point in our method.\\

Starting from the rule obtained with MoA, we generate the final rule using linear interpolation among MoA--rule points (\citefig{app5:f2}). The interpolation is done assuming that MoA--rule points are homogeneously and monotonically distributed in the range $(0,1)$ of a parameter $t$, where $t=0$ corresponds to $x_{\min}$ and $t=1$ to $x_{\max}$. For example, a MoA--rule composed by $n+1$ elements is mapped into $n+1$ values of $t$ according to:
\begin{equation}
	x^{\tn{MoA}}_i \rightarrow t^{\tn{MoA}}_i = \frac{i}{n} \quad \forall \ i \in \mathbb{Z} \ \slash \ 0 \leq i \leq n \ ,
\end{equation}
where the extremes are $x^{\tn{MoA}}_0 = x_{\min}$ and $x^{\tn{MoA}}_0 = x_{\max}$.\\

To get the final rule with $n_{\tn{bin}}$ bins, a homogeneous discretization on $t$ is performed and then the linear interpolation between MoA--rule points is easily managed.\\

\begin{fig}
	\includegraphics[width=0.5\textwidth]{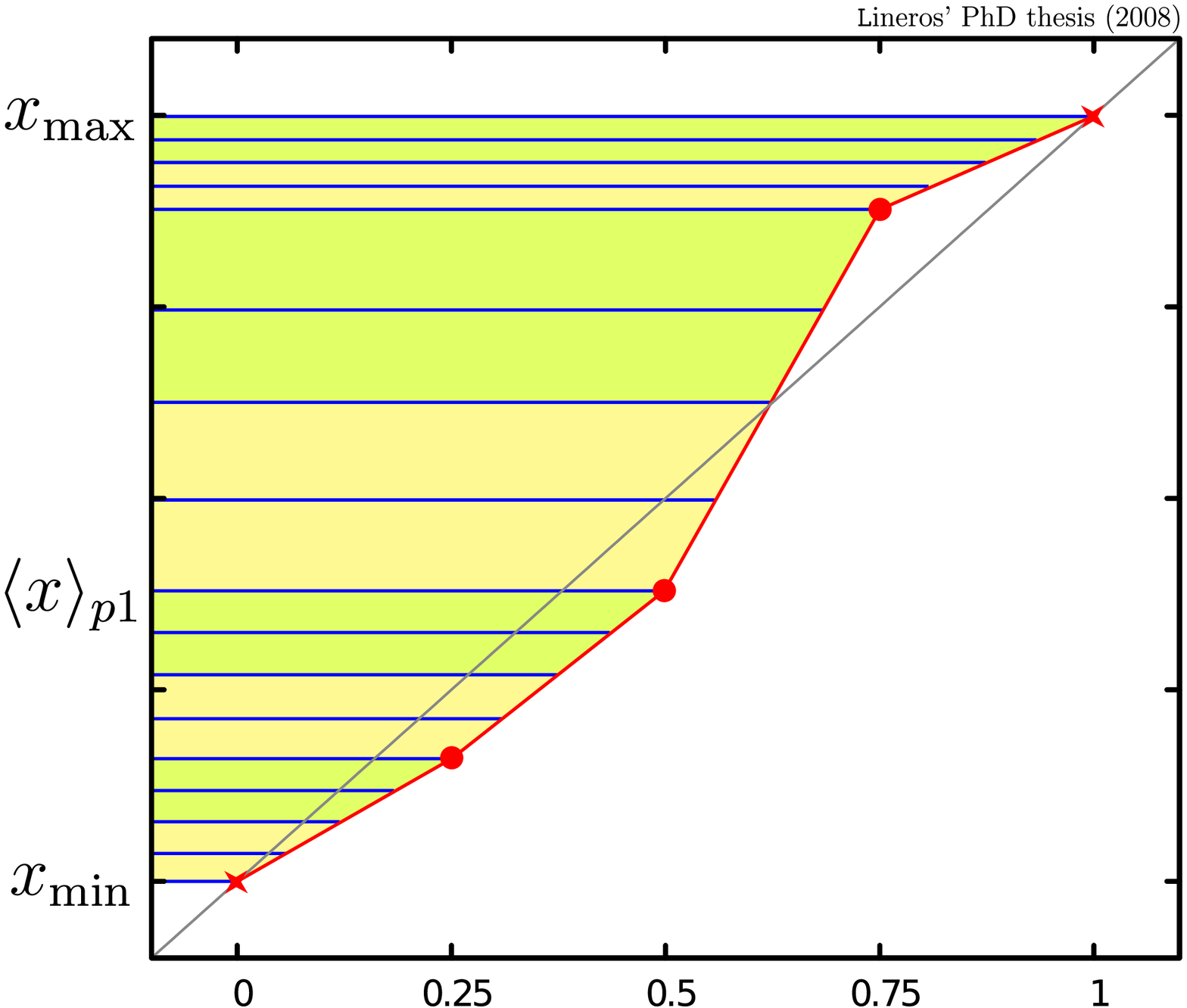}
	\caption{\label{app5:f2} Example of rule created with MoA and the composition of the final rule by using linear interpolation. In this figure: red dots are the averages and red crosses correspond to extreme values. Note that final rule presents a non homogeneous distribution.}
\end{fig}

The MoA can be easily tested and implemented. For example in \citefig{app5:f3}, the distribution for generating random events is:
\begin{eq} 
	\rho(x) = \alpha x^{-2} + \beta \exp{\left(- \frac{1}{2}(x-10)^2 \right)} + \beta \exp{ \left(-\frac{1}{2} (x-100)^2 \right)} \; ,
\end{eq}
which is a composition of a power--law distribution and two gaussian distributions, one centered in $x=10$ and the other in $x=100$ with the same standard deviation of 1, finally $\alpha$ and $\beta$ are just normalization constants. Usually, power--law distributions are better described when a uniform-log rule is used. On the other hand, gaussian distributions are usually well described with an uniform-linear rule. When both distribution are mixed, there is not an efficient binning rule based on generic rules. After both use the MoA to analyze the generated data, the binning rule produced describes well enough Gaussian and the power--law distributions. \\

\begin{fig}
	\includegraphics[angle=270,width=0.5\textwidth]{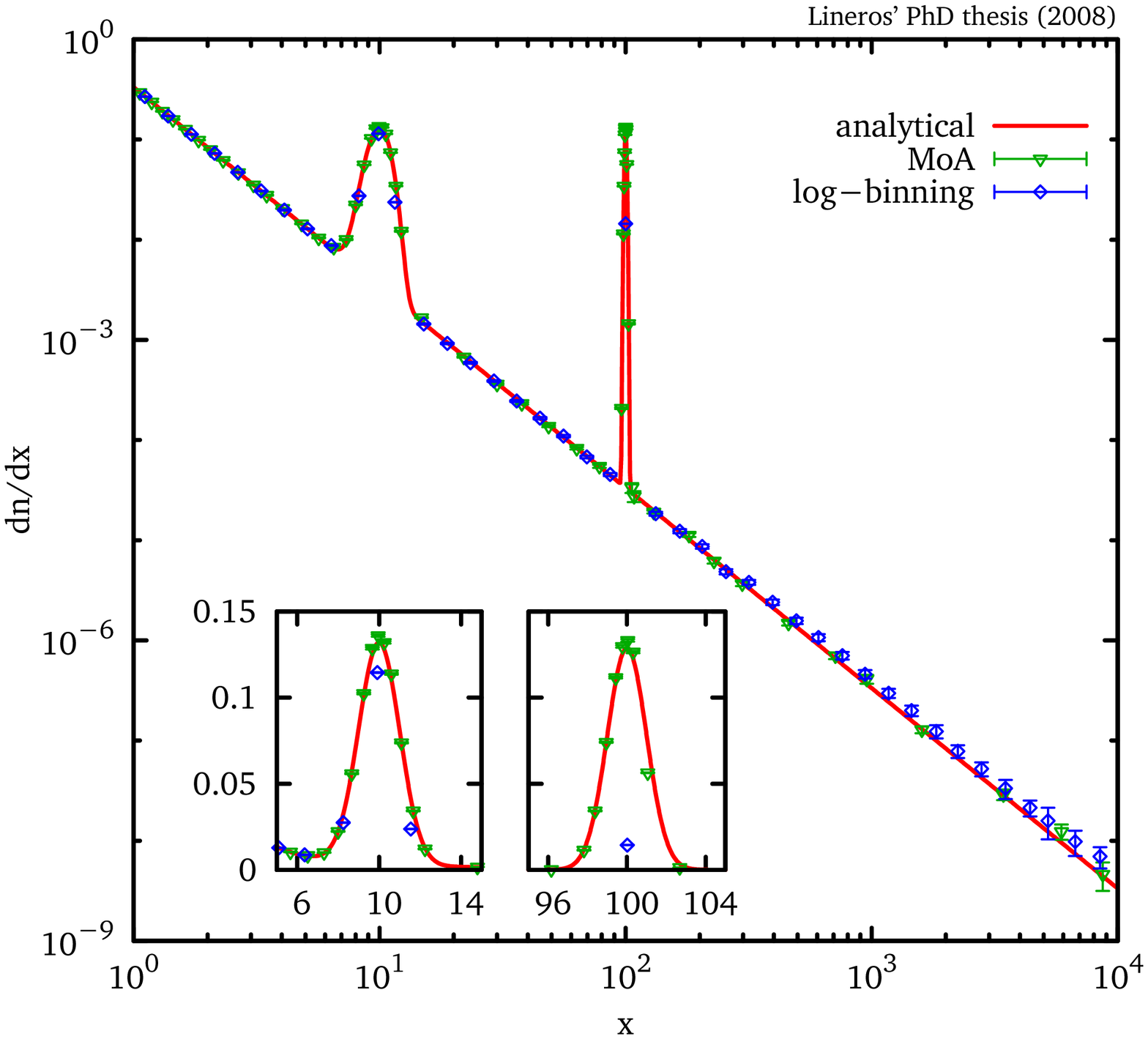}
	\caption{\label{app5:f3} Histograms made with a sample of $10^6$ events and 60 bins. An analytical distribution, MoA and uniform log-bin rule are  plotted. The analytical distribution is composed by three known distributions, a power-law ($\propto x^{-2}$) with two gaussian located in 10 and 100, all distribution have same weight. Binning rule obtained from MoA produce a more accurate solution than the uniform log rule, especially in the regions close to gaussians.}
\end{fig}

\section{Histogram creation}
Once the binning is produced, the next step is to scan the data and fill each bin of a histogram. To take advantage of the process of counting, we compute data--averages for each bin,
\begin{eq}
 x_{i+1/2} = \langle x \rangle_{x_{i+1}}^{x_i} \quad \tn{and} \quad \left(\frac{\Delta n}{\Delta x}\right)_{i+1/2} = \frac{n_{\tn{events}} \in (x_{i},x_{i+1})}{x_{i+1} - x_{i}}\; .
\end{eq}\newline

This process increase the accuracy in the histogram because the tendency points are computed for each bin instead of the mid point of the bins, which is the standard way.\\

This method gives good results in case of many combined distributions. Especially, this can be applied to produce particle energy spectra as in the case of particle decay chains, where due to particle boost energy spectra are deformed.\\

The MoA was used in the production of histograms related to DM annihilation (See \citecha{cha2}). We generate the data with Lund's PYTHIA~\cite{Sjostrand:2000wi} for several types of particles jets.\\

\cleardoublepage
\end{appendices}

\backmatter
\cleardoublepage
\addcontentsline{toc}{chapter}{List of Figures}
\listoffigures

\cleardoublepage
\addcontentsline{toc}{chapter}{List of Tables}
\listoftables

\cleardoublepage
\addcontentsline{toc}{chapter}{Bibliography}
\bibliographystyle{h-elsevier3}
\bibliography{bib/bibliography} % this is the reference to bibtex file!

\printindex

\end{document}